%% file: neutrinos.tex
\documentclass{article}
\usepackage{fancyhea}
\usepackage{amsmath,amssymb}
\usepackage{psfrag}
\usepackage{threeparttable}
\usepackage{epsfig}  
\usepackage{color}
\usepackage{graphicx}               % Standard graphics package  
\usepackage{url}
\usepackage[normalem]{ulem}
\usepackage{hyperref}
\usepackage[left=1.0in,top=1.0in,bottom=1.0in,right=1.0in]{geometry}
\input{rgb}
 \usepackage{epstopdf}
\usepackage{multirow} 

\newcommand{\boss}[2]{\ensuremath{\rlap{\kern-2.5pt\ensuremath{\overset{\scriptscriptstyle(-)}{\phantom{#1}}}}{\ensuremath{{#1}_{#2}}}}}

\begin{document}

\def\bibname{References}
\bibliographystyle{vitae}

\raggedbottom
\pagenumbering{roman}
 \parskip=8pt
% \setlength{\evensidemargin}{0pt}
% \setlength{\oddsidemargin}{0pt}
% \setlength{\marginparsep}{0.0in}
% \setlength{\marginparwidth}{0.0in}
% \marginparpush=0pt

% The content begins here 

\pagenumbering{arabic}
\renewcommand{\arraystretch}{1.25}
\addtolength{\arraycolsep}{-3pt}

%%    TEMPLATE for contributions to the Proceedings of the
%%       Workshop on Fundamental Physics at the Intensity Frontier
%%    
%%     
%%
%%
%%     Questions?  Send email to : hewett@slac.stanford.edu  %%   
%\def\Title#1{\begin{center} {\large {\bf #1}} \end{center} 
%%%%%%%%%%%%%%%%%%%%  Neutrino Chapter  %%%%%%%%%%%%%%%%%%%%%%%%%%%%%%%%%%%%

%-----------------------------------------------------------------------
%-----------------------------------------------------------------------

\begin{center}
\Huge
\textbf{Neutrinos}
\end{center}
%%%%%%%%%%%%%%%%%%%%%%%%%%%%%%%%%%%
\begin{center}\begin{boldmath}

\input{authors}

\end{boldmath}
\end{center}

\begin{abstract}
This document represents the response of the Intensity Frontier Neutrino Working Group to the Snowmass charge.  We summarize the current status of neutrino physics and identify many exciting future opportunities for studying the properties of neutrinos and for addressing important physics and astrophysics questions with neutrinos.
\end{abstract}

\tableofcontents

%-----------------------------------------------------------------------

%%%%%%%%%%%%%%%%%%%%%%%%%%%%%%%%%%%%%%%
\include{execsum}
\include{intro}
\include{nu1}

\include{nu2}
\include{nu3}

\include{nu4}

\include{nu5}

\include{nu6}

\include{nu7}

% Material previous 1.8 here?  Scientific judgement perhaps

%\include{conclusions}

\appendix
\include{glossary}

%-----------------------------------------------------------------------
%\begin{thebibliography}{99}

\bibliography{refs,onepagers}
%\end{thebibliography}

%-----------------------------------------------------------------------
\def\Discussion{\setlength{\parskip}{0.3cm}\setlength{\parindent}{0.0cm}
     \bigskip\bigskip      {\Large {\bf Discussion}} \bigskip}\def\speaker#1{{\bf #1:}\ }
\def\endDiscussion{} 
\end{document}

%% file: rgb.tex
\definecolor{AliceBlue}{rgb}{0.94,0.97,1.00}
\definecolor{AntiqueWhite1}{rgb}{1.00,0.94,0.86}
\definecolor{AntiqueWhite2}{rgb}{0.93,0.87,0.80}
\definecolor{AntiqueWhite3}{rgb}{0.80,0.75,0.69}
\definecolor{AntiqueWhite4}{rgb}{0.55,0.51,0.47}
\definecolor{AntiqueWhite}{rgb}{0.98,0.92,0.84}
\definecolor{BlanchedAlmond}{rgb}{1.00,0.92,0.80}
\definecolor{BlueViolet}{rgb}{0.54,0.17,0.89}
\definecolor{CadetBlue1}{rgb}{0.60,0.96,1.00}
\definecolor{CadetBlue2}{rgb}{0.56,0.90,0.93}
\definecolor{CadetBlue3}{rgb}{0.48,0.77,0.80}
\definecolor{CadetBlue4}{rgb}{0.33,0.53,0.55}
\definecolor{CadetBlue}{rgb}{0.37,0.62,0.63}
\definecolor{CornflowerBlue}{rgb}{0.39,0.58,0.93}
\definecolor{DarkBlue}{rgb}{0.00,0.00,0.55}
\definecolor{DarkCyan}{rgb}{0.00,0.55,0.55}
\definecolor{DarkGoldenrod1}{rgb}{1.00,0.73,0.06}
\definecolor{DarkGoldenrod2}{rgb}{0.93,0.68,0.05}
\definecolor{DarkGoldenrod3}{rgb}{0.80,0.58,0.05}
\definecolor{DarkGoldenrod4}{rgb}{0.55,0.40,0.03}
\definecolor{DarkGoldenrod}{rgb}{0.72,0.53,0.04}
\definecolor{DarkGray}{rgb}{0.66,0.66,0.66}
\definecolor{DarkGreen}{rgb}{0.00,0.39,0.00}
\definecolor{DarkGrey}{rgb}{0.66,0.66,0.66}
\definecolor{DarkKhaki}{rgb}{0.74,0.72,0.42}
\definecolor{DarkMagenta}{rgb}{0.55,0.00,0.55}
\definecolor{DarkOliveGreen1}{rgb}{0.79,1.00,0.44}
\definecolor{DarkOliveGreen2}{rgb}{0.74,0.93,0.41}
\definecolor{DarkOliveGreen3}{rgb}{0.64,0.80,0.35}
\definecolor{DarkOliveGreen4}{rgb}{0.43,0.55,0.24}
\definecolor{DarkOliveGreen}{rgb}{0.33,0.42,0.18}
\definecolor{DarkOrange1}{rgb}{1.00,0.50,0.00}
\definecolor{DarkOrange2}{rgb}{0.93,0.46,0.00}
\definecolor{DarkOrange3}{rgb}{0.80,0.40,0.00}
\definecolor{DarkOrange4}{rgb}{0.55,0.27,0.00}
\definecolor{DarkOrange}{rgb}{1.00,0.55,0.00}
\definecolor{DarkOrchid1}{rgb}{0.75,0.24,1.00}
\definecolor{DarkOrchid2}{rgb}{0.70,0.23,0.93}
\definecolor{DarkOrchid3}{rgb}{0.60,0.20,0.80}
\definecolor{DarkOrchid4}{rgb}{0.41,0.13,0.55}
\definecolor{DarkOrchid}{rgb}{0.60,0.20,0.80}
\definecolor{DarkRed}{rgb}{0.55,0.00,0.00}
\definecolor{DarkSalmon}{rgb}{0.91,0.59,0.48}
\definecolor{DarkSeaGreen1}{rgb}{0.76,1.00,0.76}
\definecolor{DarkSeaGreen2}{rgb}{0.71,0.93,0.71}
\definecolor{DarkSeaGreen3}{rgb}{0.61,0.80,0.61}
\definecolor{DarkSeaGreen4}{rgb}{0.41,0.55,0.41}
\definecolor{DarkSeaGreen}{rgb}{0.56,0.74,0.56}
\definecolor{DarkSlateBlue}{rgb}{0.28,0.24,0.55}
\definecolor{DarkSlateGray1}{rgb}{0.59,1.00,1.00}
\definecolor{DarkSlateGray2}{rgb}{0.55,0.93,0.93}
\definecolor{DarkSlateGray3}{rgb}{0.47,0.80,0.80}
\definecolor{DarkSlateGray4}{rgb}{0.32,0.55,0.55}
\definecolor{DarkSlateGray}{rgb}{0.18,0.31,0.31}
\definecolor{DarkSlateGrey}{rgb}{0.18,0.31,0.31}
\definecolor{DarkTurquoise}{rgb}{0.00,0.81,0.82}
\definecolor{DarkViolet}{rgb}{0.58,0.00,0.83}
\definecolor{DeepPink1}{rgb}{1.00,0.08,0.58}
\definecolor{DeepPink2}{rgb}{0.93,0.07,0.54}
\definecolor{DeepPink3}{rgb}{0.80,0.06,0.46}
\definecolor{DeepPink4}{rgb}{0.55,0.04,0.31}
\definecolor{DeepPink}{rgb}{1.00,0.08,0.58}
\definecolor{DeepSkyBlue1}{rgb}{0.00,0.75,1.00}
\definecolor{DeepSkyBlue2}{rgb}{0.00,0.70,0.93}
\definecolor{DeepSkyBlue3}{rgb}{0.00,0.60,0.80}
\definecolor{DeepSkyBlue4}{rgb}{0.00,0.41,0.55}
\definecolor{DeepSkyBlue}{rgb}{0.00,0.75,1.00}
\definecolor{DimGray}{rgb}{0.41,0.41,0.41}
\definecolor{DimGrey}{rgb}{0.41,0.41,0.41}
\definecolor{DodgerBlue1}{rgb}{0.12,0.56,1.00}
\definecolor{DodgerBlue2}{rgb}{0.11,0.53,0.93}
\definecolor{DodgerBlue3}{rgb}{0.09,0.45,0.80}
\definecolor{DodgerBlue4}{rgb}{0.06,0.31,0.55}
\definecolor{DodgerBlue}{rgb}{0.12,0.56,1.00}
\definecolor{FloralWhite}{rgb}{1.00,0.98,0.94}
\definecolor{ForestGreen}{rgb}{0.13,0.55,0.13}
\definecolor{GhostWhite}{rgb}{0.97,0.97,1.00}
\definecolor{GreenYellow}{rgb}{0.68,1.00,0.18}
\definecolor{HotPink1}{rgb}{1.00,0.43,0.71}
\definecolor{HotPink2}{rgb}{0.93,0.42,0.65}
\definecolor{HotPink3}{rgb}{0.80,0.38,0.56}
\definecolor{HotPink4}{rgb}{0.55,0.23,0.38}
\definecolor{HotPink}{rgb}{1.00,0.41,0.71}
\definecolor{IndianRed1}{rgb}{1.00,0.42,0.42}
\definecolor{IndianRed2}{rgb}{0.93,0.39,0.39}
\definecolor{IndianRed3}{rgb}{0.80,0.33,0.33}
\definecolor{IndianRed4}{rgb}{0.55,0.23,0.23}
\definecolor{IndianRed}{rgb}{0.80,0.36,0.36}
\definecolor{LavenderBlush1}{rgb}{1.00,0.94,0.96}
\definecolor{LavenderBlush2}{rgb}{0.93,0.88,0.90}
\definecolor{LavenderBlush3}{rgb}{0.80,0.76,0.77}
\definecolor{LavenderBlush4}{rgb}{0.55,0.51,0.53}
\definecolor{LavenderBlush}{rgb}{1.00,0.94,0.96}
\definecolor{LawnGreen}{rgb}{0.49,0.99,0.00}
\definecolor{LemonChiffon1}{rgb}{1.00,0.98,0.80}
\definecolor{LemonChiffon2}{rgb}{0.93,0.91,0.75}
\definecolor{LemonChiffon3}{rgb}{0.80,0.79,0.65}
\definecolor{LemonChiffon4}{rgb}{0.55,0.54,0.44}
\definecolor{LemonChiffon}{rgb}{1.00,0.98,0.80}
\definecolor{LightBlue1}{rgb}{0.75,0.94,1.00}
\definecolor{LightBlue2}{rgb}{0.70,0.87,0.93}
\definecolor{LightBlue3}{rgb}{0.60,0.75,0.80}
\definecolor{LightBlue4}{rgb}{0.41,0.51,0.55}
\definecolor{LightBlue}{rgb}{0.68,0.85,0.90}
\definecolor{LightCoral}{rgb}{0.94,0.50,0.50}
\definecolor{LightCyan1}{rgb}{0.88,1.00,1.00}
\definecolor{LightCyan2}{rgb}{0.82,0.93,0.93}
\definecolor{LightCyan3}{rgb}{0.71,0.80,0.80}
\definecolor{LightCyan4}{rgb}{0.48,0.55,0.55}
\definecolor{LightCyan}{rgb}{0.88,1.00,1.00}
\definecolor{LightGoldenrod1}{rgb}{1.00,0.93,0.55}
\definecolor{LightGoldenrod2}{rgb}{0.93,0.86,0.51}
\definecolor{LightGoldenrod3}{rgb}{0.80,0.75,0.44}
\definecolor{LightGoldenrod4}{rgb}{0.55,0.51,0.30}
\definecolor{LightGoldenrodYellow}{rgb}{0.98,0.98,0.82}
\definecolor{LightGoldenrod}{rgb}{0.93,0.87,0.51}
\definecolor{LightGray}{rgb}{0.83,0.83,0.83}
\definecolor{LightGreen}{rgb}{0.56,0.93,0.56}
\definecolor{LightGrey}{rgb}{0.83,0.83,0.83}
\definecolor{LightPink1}{rgb}{1.00,0.68,0.73}
\definecolor{LightPink2}{rgb}{0.93,0.64,0.68}
\definecolor{LightPink3}{rgb}{0.80,0.55,0.58}
\definecolor{LightPink4}{rgb}{0.55,0.37,0.40}
\definecolor{LightPink}{rgb}{1.00,0.71,0.76}
\definecolor{LightSalmon1}{rgb}{1.00,0.63,0.48}
\definecolor{LightSalmon2}{rgb}{0.93,0.58,0.45}
\definecolor{LightSalmon3}{rgb}{0.80,0.51,0.38}
\definecolor{LightSalmon4}{rgb}{0.55,0.34,0.26}
\definecolor{LightSalmon}{rgb}{1.00,0.63,0.48}
\definecolor{LightSeaGreen}{rgb}{0.13,0.70,0.67}
\definecolor{LightSkyBlue1}{rgb}{0.69,0.89,1.00}
\definecolor{LightSkyBlue2}{rgb}{0.64,0.83,0.93}
\definecolor{LightSkyBlue3}{rgb}{0.55,0.71,0.80}
\definecolor{LightSkyBlue4}{rgb}{0.38,0.48,0.55}
\definecolor{LightSkyBlue}{rgb}{0.53,0.81,0.98}
\definecolor{LightSlateBlue}{rgb}{0.52,0.44,1.00}
\definecolor{LightSlateGray}{rgb}{0.47,0.53,0.60}
\definecolor{LightSlateGrey}{rgb}{0.47,0.53,0.60}
\definecolor{LightSteelBlue1}{rgb}{0.79,0.88,1.00}
\definecolor{LightSteelBlue2}{rgb}{0.74,0.82,0.93}
\definecolor{LightSteelBlue3}{rgb}{0.64,0.71,0.80}
\definecolor{LightSteelBlue4}{rgb}{0.43,0.48,0.55}
\definecolor{LightSteelBlue}{rgb}{0.69,0.77,0.87}
\definecolor{LightYellow1}{rgb}{1.00,1.00,0.88}
\definecolor{LightYellow2}{rgb}{0.93,0.93,0.82}
\definecolor{LightYellow3}{rgb}{0.80,0.80,0.71}
\definecolor{LightYellow4}{rgb}{0.55,0.55,0.48}
\definecolor{LightYellow}{rgb}{1.00,1.00,0.88}
\definecolor{LimeGreen}{rgb}{0.20,0.80,0.20}
\definecolor{MediumAquamarine}{rgb}{0.40,0.80,0.67}
\definecolor{MediumBlue}{rgb}{0.00,0.00,0.80}
\definecolor{MediumOrchid1}{rgb}{0.88,0.40,1.00}
\definecolor{MediumOrchid2}{rgb}{0.82,0.37,0.93}
\definecolor{MediumOrchid3}{rgb}{0.71,0.32,0.80}
\definecolor{MediumOrchid4}{rgb}{0.48,0.22,0.55}
\definecolor{MediumOrchid}{rgb}{0.73,0.33,0.83}
\definecolor{MediumPurple1}{rgb}{0.67,0.51,1.00}
\definecolor{MediumPurple2}{rgb}{0.62,0.47,0.93}
\definecolor{MediumPurple3}{rgb}{0.54,0.41,0.80}
\definecolor{MediumPurple4}{rgb}{0.36,0.28,0.55}
\definecolor{MediumPurple}{rgb}{0.58,0.44,0.86}
\definecolor{MediumSeaGreen}{rgb}{0.24,0.70,0.44}
\definecolor{MediumSlateBlue}{rgb}{0.48,0.41,0.93}
\definecolor{MediumSpringGreen}{rgb}{0.00,0.98,0.60}
\definecolor{MediumTurquoise}{rgb}{0.28,0.82,0.80}
\definecolor{MediumVioletRed}{rgb}{0.78,0.08,0.52}
\definecolor{MidnightBlue}{rgb}{0.10,0.10,0.44}
\definecolor{MintCream}{rgb}{0.96,1.00,0.98}
\definecolor{MistyRose1}{rgb}{1.00,0.89,0.88}
\definecolor{MistyRose2}{rgb}{0.93,0.84,0.82}
\definecolor{MistyRose3}{rgb}{0.80,0.72,0.71}
\definecolor{MistyRose4}{rgb}{0.55,0.49,0.48}
\definecolor{MistyRose}{rgb}{1.00,0.89,0.88}
\definecolor{NavajoWhite1}{rgb}{1.00,0.87,0.68}
\definecolor{NavajoWhite2}{rgb}{0.93,0.81,0.63}
\definecolor{NavajoWhite3}{rgb}{0.80,0.70,0.55}
\definecolor{NavajoWhite4}{rgb}{0.55,0.47,0.37}
\definecolor{NavajoWhite}{rgb}{1.00,0.87,0.68}
\definecolor{NavyBlue}{rgb}{0.00,0.00,0.50}
\definecolor{OldLace}{rgb}{0.99,0.96,0.90}
\definecolor{OliveDrab1}{rgb}{0.75,1.00,0.24}
\definecolor{OliveDrab2}{rgb}{0.70,0.93,0.23}
\definecolor{OliveDrab3}{rgb}{0.60,0.80,0.20}
\definecolor{OliveDrab4}{rgb}{0.41,0.55,0.13}
\definecolor{OliveDrab}{rgb}{0.42,0.56,0.14}
\definecolor{OrangeRed1}{rgb}{1.00,0.27,0.00}
\definecolor{OrangeRed2}{rgb}{0.93,0.25,0.00}
\definecolor{OrangeRed3}{rgb}{0.80,0.22,0.00}
\definecolor{OrangeRed4}{rgb}{0.55,0.15,0.00}
\definecolor{OrangeRed}{rgb}{1.00,0.27,0.00}
\definecolor{PaleGoldenrod}{rgb}{0.93,0.91,0.67}
\definecolor{PaleGreen1}{rgb}{0.60,1.00,0.60}
\definecolor{PaleGreen2}{rgb}{0.56,0.93,0.56}
\definecolor{PaleGreen3}{rgb}{0.49,0.80,0.49}
\definecolor{PaleGreen4}{rgb}{0.33,0.55,0.33}
\definecolor{PaleGreen}{rgb}{0.60,0.98,0.60}
\definecolor{PaleTurquoise1}{rgb}{0.73,1.00,1.00}
\definecolor{PaleTurquoise2}{rgb}{0.68,0.93,0.93}
\definecolor{PaleTurquoise3}{rgb}{0.59,0.80,0.80}
\definecolor{PaleTurquoise4}{rgb}{0.40,0.55,0.55}
\definecolor{PaleTurquoise}{rgb}{0.69,0.93,0.93}
\definecolor{PaleVioletRed1}{rgb}{1.00,0.51,0.67}
\definecolor{PaleVioletRed2}{rgb}{0.93,0.47,0.62}
\definecolor{PaleVioletRed3}{rgb}{0.80,0.41,0.54}
\definecolor{PaleVioletRed4}{rgb}{0.55,0.28,0.36}
\definecolor{PaleVioletRed}{rgb}{0.86,0.44,0.58}
\definecolor{PapayaWhip}{rgb}{1.00,0.94,0.84}
\definecolor{PeachPuff1}{rgb}{1.00,0.85,0.73}
\definecolor{PeachPuff2}{rgb}{0.93,0.80,0.68}
\definecolor{PeachPuff3}{rgb}{0.80,0.69,0.58}
\definecolor{PeachPuff4}{rgb}{0.55,0.47,0.40}
\definecolor{PeachPuff}{rgb}{1.00,0.85,0.73}
\definecolor{PowderBlue}{rgb}{0.69,0.88,0.90}
\definecolor{RosyBrown1}{rgb}{1.00,0.76,0.76}
\definecolor{RosyBrown2}{rgb}{0.93,0.71,0.71}
\definecolor{RosyBrown3}{rgb}{0.80,0.61,0.61}
\definecolor{RosyBrown4}{rgb}{0.55,0.41,0.41}
\definecolor{RosyBrown}{rgb}{0.74,0.56,0.56}
\definecolor{RoyalBlue1}{rgb}{0.28,0.46,1.00}
\definecolor{RoyalBlue2}{rgb}{0.26,0.43,0.93}
\definecolor{RoyalBlue3}{rgb}{0.23,0.37,0.80}
\definecolor{RoyalBlue4}{rgb}{0.15,0.25,0.55}
\definecolor{RoyalBlue}{rgb}{0.25,0.41,0.88}
\definecolor{SaddleBrown}{rgb}{0.55,0.27,0.07}
\definecolor{SandyBrown}{rgb}{0.96,0.64,0.38}
\definecolor{SeaGreen1}{rgb}{0.33,1.00,0.62}
\definecolor{SeaGreen2}{rgb}{0.31,0.93,0.58}
\definecolor{SeaGreen3}{rgb}{0.26,0.80,0.50}
\definecolor{SeaGreen4}{rgb}{0.18,0.55,0.34}
\definecolor{SeaGreen}{rgb}{0.18,0.55,0.34}
\definecolor{SkyBlue1}{rgb}{0.53,0.81,1.00}
\definecolor{SkyBlue2}{rgb}{0.49,0.75,0.93}
\definecolor{SkyBlue3}{rgb}{0.42,0.65,0.80}
\definecolor{SkyBlue4}{rgb}{0.29,0.44,0.55}
\definecolor{SkyBlue}{rgb}{0.53,0.81,0.92}
\definecolor{SlateBlue1}{rgb}{0.51,0.44,1.00}
\definecolor{SlateBlue2}{rgb}{0.48,0.40,0.93}
\definecolor{SlateBlue3}{rgb}{0.41,0.35,0.80}
\definecolor{SlateBlue4}{rgb}{0.28,0.24,0.55}
\definecolor{SlateBlue}{rgb}{0.42,0.35,0.80}
\definecolor{SlateGray1}{rgb}{0.78,0.89,1.00}
\definecolor{SlateGray2}{rgb}{0.73,0.83,0.93}
\definecolor{SlateGray3}{rgb}{0.62,0.71,0.80}
\definecolor{SlateGray4}{rgb}{0.42,0.48,0.55}
\definecolor{SlateGray}{rgb}{0.44,0.50,0.56}
\definecolor{SlateGrey}{rgb}{0.44,0.50,0.56}
\definecolor{SpringGreen1}{rgb}{0.00,1.00,0.50}
\definecolor{SpringGreen2}{rgb}{0.00,0.93,0.46}
\definecolor{SpringGreen3}{rgb}{0.00,0.80,0.40}
\definecolor{SpringGreen4}{rgb}{0.00,0.55,0.27}
\definecolor{SpringGreen}{rgb}{0.00,1.00,0.50}
\definecolor{SteelBlue1}{rgb}{0.39,0.72,1.00}
\definecolor{SteelBlue2}{rgb}{0.36,0.67,0.93}
\definecolor{SteelBlue3}{rgb}{0.31,0.58,0.80}
\definecolor{SteelBlue4}{rgb}{0.21,0.39,0.55}
\definecolor{SteelBlue}{rgb}{0.27,0.51,0.71}
\definecolor{VioletRed1}{rgb}{1.00,0.24,0.59}
\definecolor{VioletRed2}{rgb}{0.93,0.23,0.55}
\definecolor{VioletRed3}{rgb}{0.80,0.20,0.47}
\definecolor{VioletRed4}{rgb}{0.55,0.13,0.32}
\definecolor{VioletRed}{rgb}{0.82,0.13,0.56}
\definecolor{WhiteSmoke}{rgb}{0.96,0.96,0.96}
\definecolor{YellowGreen}{rgb}{0.60,0.80,0.20}
\definecolor{aliceblue}{rgb}{0.94,0.97,1.00}
\definecolor{antiquewhite}{rgb}{0.98,0.92,0.84}
\definecolor{aquamarine1}{rgb}{0.50,1.00,0.83}
\definecolor{aquamarine2}{rgb}{0.46,0.93,0.78}
\definecolor{aquamarine3}{rgb}{0.40,0.80,0.67}
\definecolor{aquamarine4}{rgb}{0.27,0.55,0.45}
\definecolor{aquamarine}{rgb}{0.50,1.00,0.83}
\definecolor{azure1}{rgb}{0.94,1.00,1.00}
\definecolor{azure2}{rgb}{0.88,0.93,0.93}
\definecolor{azure3}{rgb}{0.76,0.80,0.80}
\definecolor{azure4}{rgb}{0.51,0.55,0.55}
\definecolor{azure}{rgb}{0.94,1.00,1.00}
\definecolor{beige}{rgb}{0.96,0.96,0.86}
\definecolor{bisque1}{rgb}{1.00,0.89,0.77}
\definecolor{bisque2}{rgb}{0.93,0.84,0.72}
\definecolor{bisque3}{rgb}{0.80,0.72,0.62}
\definecolor{bisque4}{rgb}{0.55,0.49,0.42}
\definecolor{bisque}{rgb}{1.00,0.89,0.77}
\definecolor{black}{rgb}{0.00,0.00,0.00}
\definecolor{blanchedalmond}{rgb}{1.00,0.92,0.80}
\definecolor{blue1}{rgb}{0.00,0.00,1.00}
\definecolor{blue2}{rgb}{0.00,0.00,0.93}
\definecolor{blue3}{rgb}{0.00,0.00,0.80}
\definecolor{blue4}{rgb}{0.00,0.00,0.55}
\definecolor{blueviolet}{rgb}{0.54,0.17,0.89}
\definecolor{blue}{rgb}{0.00,0.00,1.00}
\definecolor{brown1}{rgb}{1.00,0.25,0.25}
\definecolor{brown2}{rgb}{0.93,0.23,0.23}
\definecolor{brown3}{rgb}{0.80,0.20,0.20}
\definecolor{brown4}{rgb}{0.55,0.14,0.14}
\definecolor{brown}{rgb}{0.65,0.16,0.16}
\definecolor{burlywood1}{rgb}{1.00,0.83,0.61}
\definecolor{burlywood2}{rgb}{0.93,0.77,0.57}
\definecolor{burlywood3}{rgb}{0.80,0.67,0.49}
\definecolor{burlywood4}{rgb}{0.55,0.45,0.33}
\definecolor{burlywood}{rgb}{0.87,0.72,0.53}
\definecolor{cadetblue}{rgb}{0.37,0.62,0.63}
\definecolor{chartreuse1}{rgb}{0.50,1.00,0.00}
\definecolor{chartreuse2}{rgb}{0.46,0.93,0.00}
\definecolor{chartreuse3}{rgb}{0.40,0.80,0.00}
\definecolor{chartreuse4}{rgb}{0.27,0.55,0.00}
\definecolor{chartreuse}{rgb}{0.50,1.00,0.00}
\definecolor{chocolate1}{rgb}{1.00,0.50,0.14}
\definecolor{chocolate2}{rgb}{0.93,0.46,0.13}
\definecolor{chocolate3}{rgb}{0.80,0.40,0.11}
\definecolor{chocolate4}{rgb}{0.55,0.27,0.07}
\definecolor{chocolate}{rgb}{0.82,0.41,0.12}
\definecolor{coral1}{rgb}{1.00,0.45,0.34}
\definecolor{coral2}{rgb}{0.93,0.42,0.31}
\definecolor{coral3}{rgb}{0.80,0.36,0.27}
\definecolor{coral4}{rgb}{0.55,0.24,0.18}
\definecolor{coral}{rgb}{1.00,0.50,0.31}
\definecolor{cornflowerblue}{rgb}{0.39,0.58,0.93}
\definecolor{cornsilk1}{rgb}{1.00,0.97,0.86}
\definecolor{cornsilk2}{rgb}{0.93,0.91,0.80}
\definecolor{cornsilk3}{rgb}{0.80,0.78,0.69}
\definecolor{cornsilk4}{rgb}{0.55,0.53,0.47}
\definecolor{cornsilk}{rgb}{1.00,0.97,0.86}
\definecolor{cyan1}{rgb}{0.00,1.00,1.00}
\definecolor{cyan2}{rgb}{0.00,0.93,0.93}
\definecolor{cyan3}{rgb}{0.00,0.80,0.80}
\definecolor{cyan4}{rgb}{0.00,0.55,0.55}
\definecolor{cyan}{rgb}{0.00,1.00,1.00}
\definecolor{darkblue}{rgb}{0.00,0.00,0.55}
\definecolor{darkcyan}{rgb}{0.00,0.55,0.55}
\definecolor{darkgoldenrod}{rgb}{0.72,0.53,0.04}
\definecolor{darkgray}{rgb}{0.66,0.66,0.66}
\definecolor{darkgreen}{rgb}{0.00,0.39,0.00}
\definecolor{darkgrey}{rgb}{0.66,0.66,0.66}
\definecolor{darkkhaki}{rgb}{0.74,0.72,0.42}
\definecolor{darkmagenta}{rgb}{0.55,0.00,0.55}
\definecolor{darkolive}{rgb}{0.33,0.42,0.18}
\definecolor{darkorange}{rgb}{1.00,0.55,0.00}
\definecolor{darkorchid}{rgb}{0.60,0.20,0.80}
\definecolor{darkred}{rgb}{0.55,0.00,0.00}
\definecolor{darksalmon}{rgb}{0.91,0.59,0.48}
\definecolor{darksea}{rgb}{0.56,0.74,0.56}
\definecolor{darkslate}{rgb}{0.18,0.31,0.31}
\definecolor{darkslate}{rgb}{0.18,0.31,0.31}
\definecolor{darkslate}{rgb}{0.28,0.24,0.55}
\definecolor{darkturquoise}{rgb}{0.00,0.81,0.82}
\definecolor{darkviolet}{rgb}{0.58,0.00,0.83}
\definecolor{deeppink}{rgb}{1.00,0.08,0.58}
\definecolor{deepsky}{rgb}{0.00,0.75,1.00}
\definecolor{dimgray}{rgb}{0.41,0.41,0.41}
\definecolor{dimgrey}{rgb}{0.41,0.41,0.41}
\definecolor{dodgerblue}{rgb}{0.12,0.56,1.00}
\definecolor{firebrick1}{rgb}{1.00,0.19,0.19}
\definecolor{firebrick2}{rgb}{0.93,0.17,0.17}
\definecolor{firebrick3}{rgb}{0.80,0.15,0.15}
\definecolor{firebrick4}{rgb}{0.55,0.10,0.10}
\definecolor{firebrick}{rgb}{0.70,0.13,0.13}
\definecolor{floralwhite}{rgb}{1.00,0.98,0.94}
\definecolor{forestgreen}{rgb}{0.13,0.55,0.13}
\definecolor{gainsboro}{rgb}{0.86,0.86,0.86}
\definecolor{ghostwhite}{rgb}{0.97,0.97,1.00}
\definecolor{gold1}{rgb}{1.00,0.84,0.00}
\definecolor{gold2}{rgb}{0.93,0.79,0.00}
\definecolor{gold3}{rgb}{0.80,0.68,0.00}
\definecolor{gold4}{rgb}{0.55,0.46,0.00}
\definecolor{goldenrod1}{rgb}{1.00,0.76,0.15}
\definecolor{goldenrod2}{rgb}{0.93,0.71,0.13}
\definecolor{goldenrod3}{rgb}{0.80,0.61,0.11}
\definecolor{goldenrod4}{rgb}{0.55,0.41,0.08}
\definecolor{goldenrod}{rgb}{0.85,0.65,0.13}
\definecolor{gold}{rgb}{1.00,0.84,0.00}
\definecolor{gray0}{rgb}{0.00,0.00,0.00}
\definecolor{gray100}{rgb}{1.00,1.00,1.00}
\definecolor{gray10}{rgb}{0.10,0.10,0.10}
\definecolor{gray11}{rgb}{0.11,0.11,0.11}
\definecolor{gray12}{rgb}{0.12,0.12,0.12}
\definecolor{gray13}{rgb}{0.13,0.13,0.13}
\definecolor{gray14}{rgb}{0.14,0.14,0.14}
\definecolor{gray15}{rgb}{0.15,0.15,0.15}
\definecolor{gray16}{rgb}{0.16,0.16,0.16}
\definecolor{gray17}{rgb}{0.17,0.17,0.17}
\definecolor{gray18}{rgb}{0.18,0.18,0.18}
\definecolor{gray19}{rgb}{0.19,0.19,0.19}
\definecolor{gray1}{rgb}{0.01,0.01,0.01}
\definecolor{gray20}{rgb}{0.20,0.20,0.20}
\definecolor{gray21}{rgb}{0.21,0.21,0.21}
\definecolor{gray22}{rgb}{0.22,0.22,0.22}
\definecolor{gray23}{rgb}{0.23,0.23,0.23}
\definecolor{gray24}{rgb}{0.24,0.24,0.24}
\definecolor{gray25}{rgb}{0.25,0.25,0.25}
\definecolor{gray26}{rgb}{0.26,0.26,0.26}
\definecolor{gray27}{rgb}{0.27,0.27,0.27}
\definecolor{gray28}{rgb}{0.28,0.28,0.28}
\definecolor{gray29}{rgb}{0.29,0.29,0.29}
\definecolor{gray2}{rgb}{0.02,0.02,0.02}
\definecolor{gray30}{rgb}{0.30,0.30,0.30}
\definecolor{gray31}{rgb}{0.31,0.31,0.31}
\definecolor{gray32}{rgb}{0.32,0.32,0.32}
\definecolor{gray33}{rgb}{0.33,0.33,0.33}
\definecolor{gray34}{rgb}{0.34,0.34,0.34}
\definecolor{gray35}{rgb}{0.35,0.35,0.35}
\definecolor{gray36}{rgb}{0.36,0.36,0.36}
\definecolor{gray37}{rgb}{0.37,0.37,0.37}
\definecolor{gray38}{rgb}{0.38,0.38,0.38}
\definecolor{gray39}{rgb}{0.39,0.39,0.39}
\definecolor{gray3}{rgb}{0.03,0.03,0.03}
\definecolor{gray40}{rgb}{0.40,0.40,0.40}
\definecolor{gray41}{rgb}{0.41,0.41,0.41}
\definecolor{gray42}{rgb}{0.42,0.42,0.42}
\definecolor{gray43}{rgb}{0.43,0.43,0.43}
\definecolor{gray44}{rgb}{0.44,0.44,0.44}
\definecolor{gray45}{rgb}{0.45,0.45,0.45}
\definecolor{gray46}{rgb}{0.46,0.46,0.46}
\definecolor{gray47}{rgb}{0.47,0.47,0.47}
\definecolor{gray48}{rgb}{0.48,0.48,0.48}
\definecolor{gray49}{rgb}{0.49,0.49,0.49}
\definecolor{gray4}{rgb}{0.04,0.04,0.04}
\definecolor{gray50}{rgb}{0.50,0.50,0.50}
\definecolor{gray51}{rgb}{0.51,0.51,0.51}
\definecolor{gray52}{rgb}{0.52,0.52,0.52}
\definecolor{gray53}{rgb}{0.53,0.53,0.53}
\definecolor{gray54}{rgb}{0.54,0.54,0.54}
\definecolor{gray55}{rgb}{0.55,0.55,0.55}
\definecolor{gray56}{rgb}{0.56,0.56,0.56}
\definecolor{gray57}{rgb}{0.57,0.57,0.57}
\definecolor{gray58}{rgb}{0.58,0.58,0.58}
\definecolor{gray59}{rgb}{0.59,0.59,0.59}
\definecolor{gray5}{rgb}{0.05,0.05,0.05}
\definecolor{gray60}{rgb}{0.60,0.60,0.60}
\definecolor{gray61}{rgb}{0.61,0.61,0.61}
\definecolor{gray62}{rgb}{0.62,0.62,0.62}
\definecolor{gray63}{rgb}{0.63,0.63,0.63}
\definecolor{gray64}{rgb}{0.64,0.64,0.64}
\definecolor{gray65}{rgb}{0.65,0.65,0.65}
\definecolor{gray66}{rgb}{0.66,0.66,0.66}
\definecolor{gray67}{rgb}{0.67,0.67,0.67}
\definecolor{gray68}{rgb}{0.68,0.68,0.68}
\definecolor{gray69}{rgb}{0.69,0.69,0.69}
\definecolor{gray6}{rgb}{0.06,0.06,0.06}
\definecolor{gray70}{rgb}{0.70,0.70,0.70}
\definecolor{gray71}{rgb}{0.71,0.71,0.71}
\definecolor{gray72}{rgb}{0.72,0.72,0.72}
\definecolor{gray73}{rgb}{0.73,0.73,0.73}
\definecolor{gray74}{rgb}{0.74,0.74,0.74}
\definecolor{gray75}{rgb}{0.75,0.75,0.75}
\definecolor{gray76}{rgb}{0.76,0.76,0.76}
\definecolor{gray77}{rgb}{0.77,0.77,0.77}
\definecolor{gray78}{rgb}{0.78,0.78,0.78}
\definecolor{gray79}{rgb}{0.79,0.79,0.79}
\definecolor{gray7}{rgb}{0.07,0.07,0.07}
\definecolor{gray80}{rgb}{0.80,0.80,0.80}
\definecolor{gray81}{rgb}{0.81,0.81,0.81}
\definecolor{gray82}{rgb}{0.82,0.82,0.82}
\definecolor{gray83}{rgb}{0.83,0.83,0.83}
\definecolor{gray84}{rgb}{0.84,0.84,0.84}
\definecolor{gray85}{rgb}{0.85,0.85,0.85}
\definecolor{gray86}{rgb}{0.86,0.86,0.86}
\definecolor{gray87}{rgb}{0.87,0.87,0.87}
\definecolor{gray88}{rgb}{0.88,0.88,0.88}
\definecolor{gray89}{rgb}{0.89,0.89,0.89}
\definecolor{gray8}{rgb}{0.08,0.08,0.08}
\definecolor{gray90}{rgb}{0.90,0.90,0.90}
\definecolor{gray91}{rgb}{0.91,0.91,0.91}
\definecolor{gray92}{rgb}{0.92,0.92,0.92}
\definecolor{gray93}{rgb}{0.93,0.93,0.93}
\definecolor{gray94}{rgb}{0.94,0.94,0.94}
\definecolor{gray95}{rgb}{0.95,0.95,0.95}
\definecolor{gray96}{rgb}{0.96,0.96,0.96}
\definecolor{gray97}{rgb}{0.97,0.97,0.97}
\definecolor{gray98}{rgb}{0.98,0.98,0.98}
\definecolor{gray99}{rgb}{0.99,0.99,0.99}
\definecolor{gray9}{rgb}{0.09,0.09,0.09}
\definecolor{gray}{rgb}{0.75,0.75,0.75}
\definecolor{green1}{rgb}{0.00,1.00,0.00}
\definecolor{green2}{rgb}{0.00,0.93,0.00}
\definecolor{green3}{rgb}{0.00,0.80,0.00}
\definecolor{green4}{rgb}{0.00,0.55,0.00}
\definecolor{greenyellow}{rgb}{0.68,1.00,0.18}
\definecolor{green}{rgb}{0.00,1.00,0.00}
\definecolor{grey0}{rgb}{0.00,0.00,0.00}
\definecolor{grey100}{rgb}{1.00,1.00,1.00}
\definecolor{grey10}{rgb}{0.10,0.10,0.10}
\definecolor{grey11}{rgb}{0.11,0.11,0.11}
\definecolor{grey12}{rgb}{0.12,0.12,0.12}
\definecolor{grey13}{rgb}{0.13,0.13,0.13}
\definecolor{grey14}{rgb}{0.14,0.14,0.14}
\definecolor{grey15}{rgb}{0.15,0.15,0.15}
\definecolor{grey16}{rgb}{0.16,0.16,0.16}
\definecolor{grey17}{rgb}{0.17,0.17,0.17}
\definecolor{grey18}{rgb}{0.18,0.18,0.18}
\definecolor{grey19}{rgb}{0.19,0.19,0.19}
\definecolor{grey1}{rgb}{0.01,0.01,0.01}
\definecolor{grey20}{rgb}{0.20,0.20,0.20}
\definecolor{grey21}{rgb}{0.21,0.21,0.21}
\definecolor{grey22}{rgb}{0.22,0.22,0.22}
\definecolor{grey23}{rgb}{0.23,0.23,0.23}
\definecolor{grey24}{rgb}{0.24,0.24,0.24}
\definecolor{grey25}{rgb}{0.25,0.25,0.25}
\definecolor{grey26}{rgb}{0.26,0.26,0.26}
\definecolor{grey27}{rgb}{0.27,0.27,0.27}
\definecolor{grey28}{rgb}{0.28,0.28,0.28}
\definecolor{grey29}{rgb}{0.29,0.29,0.29}
\definecolor{grey2}{rgb}{0.02,0.02,0.02}
\definecolor{grey30}{rgb}{0.30,0.30,0.30}
\definecolor{grey31}{rgb}{0.31,0.31,0.31}
\definecolor{grey32}{rgb}{0.32,0.32,0.32}
\definecolor{grey33}{rgb}{0.33,0.33,0.33}
\definecolor{grey34}{rgb}{0.34,0.34,0.34}
\definecolor{grey35}{rgb}{0.35,0.35,0.35}
\definecolor{grey36}{rgb}{0.36,0.36,0.36}
\definecolor{grey37}{rgb}{0.37,0.37,0.37}
\definecolor{grey38}{rgb}{0.38,0.38,0.38}
\definecolor{grey39}{rgb}{0.39,0.39,0.39}
\definecolor{grey3}{rgb}{0.03,0.03,0.03}
\definecolor{grey40}{rgb}{0.40,0.40,0.40}
\definecolor{grey41}{rgb}{0.41,0.41,0.41}
\definecolor{grey42}{rgb}{0.42,0.42,0.42}
\definecolor{grey43}{rgb}{0.43,0.43,0.43}
\definecolor{grey44}{rgb}{0.44,0.44,0.44}
\definecolor{grey45}{rgb}{0.45,0.45,0.45}
\definecolor{grey46}{rgb}{0.46,0.46,0.46}
\definecolor{grey47}{rgb}{0.47,0.47,0.47}
\definecolor{grey48}{rgb}{0.48,0.48,0.48}
\definecolor{grey49}{rgb}{0.49,0.49,0.49}
\definecolor{grey4}{rgb}{0.04,0.04,0.04}
\definecolor{grey50}{rgb}{0.50,0.50,0.50}
\definecolor{grey51}{rgb}{0.51,0.51,0.51}
\definecolor{grey52}{rgb}{0.52,0.52,0.52}
\definecolor{grey53}{rgb}{0.53,0.53,0.53}
\definecolor{grey54}{rgb}{0.54,0.54,0.54}
\definecolor{grey55}{rgb}{0.55,0.55,0.55}
\definecolor{grey56}{rgb}{0.56,0.56,0.56}
\definecolor{grey57}{rgb}{0.57,0.57,0.57}
\definecolor{grey58}{rgb}{0.58,0.58,0.58}
\definecolor{grey59}{rgb}{0.59,0.59,0.59}
\definecolor{grey5}{rgb}{0.05,0.05,0.05}
\definecolor{grey60}{rgb}{0.60,0.60,0.60}
\definecolor{grey61}{rgb}{0.61,0.61,0.61}
\definecolor{grey62}{rgb}{0.62,0.62,0.62}
\definecolor{grey63}{rgb}{0.63,0.63,0.63}
\definecolor{grey64}{rgb}{0.64,0.64,0.64}
\definecolor{grey65}{rgb}{0.65,0.65,0.65}
\definecolor{grey66}{rgb}{0.66,0.66,0.66}
\definecolor{grey67}{rgb}{0.67,0.67,0.67}
\definecolor{grey68}{rgb}{0.68,0.68,0.68}
\definecolor{grey69}{rgb}{0.69,0.69,0.69}
\definecolor{grey6}{rgb}{0.06,0.06,0.06}
\definecolor{grey70}{rgb}{0.70,0.70,0.70}
\definecolor{grey71}{rgb}{0.71,0.71,0.71}
\definecolor{grey72}{rgb}{0.72,0.72,0.72}
\definecolor{grey73}{rgb}{0.73,0.73,0.73}
\definecolor{grey74}{rgb}{0.74,0.74,0.74}
\definecolor{grey75}{rgb}{0.75,0.75,0.75}
\definecolor{grey76}{rgb}{0.76,0.76,0.76}
\definecolor{grey77}{rgb}{0.77,0.77,0.77}
\definecolor{grey78}{rgb}{0.78,0.78,0.78}
\definecolor{grey79}{rgb}{0.79,0.79,0.79}
\definecolor{grey7}{rgb}{0.07,0.07,0.07}
\definecolor{grey80}{rgb}{0.80,0.80,0.80}
\definecolor{grey81}{rgb}{0.81,0.81,0.81}
\definecolor{grey82}{rgb}{0.82,0.82,0.82}
\definecolor{grey83}{rgb}{0.83,0.83,0.83}
\definecolor{grey84}{rgb}{0.84,0.84,0.84}
\definecolor{grey85}{rgb}{0.85,0.85,0.85}
\definecolor{grey86}{rgb}{0.86,0.86,0.86}
\definecolor{grey87}{rgb}{0.87,0.87,0.87}
\definecolor{grey88}{rgb}{0.88,0.88,0.88}
\definecolor{grey89}{rgb}{0.89,0.89,0.89}
\definecolor{grey8}{rgb}{0.08,0.08,0.08}
\definecolor{grey90}{rgb}{0.90,0.90,0.90}
\definecolor{grey91}{rgb}{0.91,0.91,0.91}
\definecolor{grey92}{rgb}{0.92,0.92,0.92}
\definecolor{grey93}{rgb}{0.93,0.93,0.93}
\definecolor{grey94}{rgb}{0.94,0.94,0.94}
\definecolor{grey95}{rgb}{0.95,0.95,0.95}
\definecolor{grey96}{rgb}{0.96,0.96,0.96}
\definecolor{grey97}{rgb}{0.97,0.97,0.97}
\definecolor{grey98}{rgb}{0.98,0.98,0.98}
\definecolor{grey99}{rgb}{0.99,0.99,0.99}
\definecolor{grey9}{rgb}{0.09,0.09,0.09}
\definecolor{grey}{rgb}{0.75,0.75,0.75}
\definecolor{honeydew1}{rgb}{0.94,1.00,0.94}
\definecolor{honeydew2}{rgb}{0.88,0.93,0.88}
\definecolor{honeydew3}{rgb}{0.76,0.80,0.76}
\definecolor{honeydew4}{rgb}{0.51,0.55,0.51}
\definecolor{honeydew}{rgb}{0.94,1.00,0.94}
\definecolor{hotpink}{rgb}{1.00,0.41,0.71}
\definecolor{indianred}{rgb}{0.80,0.36,0.36}
\definecolor{ivory1}{rgb}{1.00,1.00,0.94}
\definecolor{ivory2}{rgb}{0.93,0.93,0.88}
\definecolor{ivory3}{rgb}{0.80,0.80,0.76}
\definecolor{ivory4}{rgb}{0.55,0.55,0.51}
\definecolor{ivory}{rgb}{1.00,1.00,0.94}
\definecolor{khaki1}{rgb}{1.00,0.96,0.56}
\definecolor{khaki2}{rgb}{0.93,0.90,0.52}
\definecolor{khaki3}{rgb}{0.80,0.78,0.45}
\definecolor{khaki4}{rgb}{0.55,0.53,0.31}
\definecolor{khaki}{rgb}{0.94,0.90,0.55}
\definecolor{lavenderblush}{rgb}{1.00,0.94,0.96}
\definecolor{lavender}{rgb}{0.90,0.90,0.98}
\definecolor{lawngreen}{rgb}{0.49,0.99,0.00}
\definecolor{lemonchiffon}{rgb}{1.00,0.98,0.80}
\definecolor{lightblue}{rgb}{0.68,0.85,0.90}
\definecolor{lightcoral}{rgb}{0.94,0.50,0.50}
\definecolor{lightcyan}{rgb}{0.88,1.00,1.00}
\definecolor{lightgoldenrod}{rgb}{0.93,0.87,0.51}
\definecolor{lightgoldenrod}{rgb}{0.98,0.98,0.82}
\definecolor{lightgray}{rgb}{0.83,0.83,0.83}
\definecolor{lightgreen}{rgb}{0.56,0.93,0.56}
\definecolor{lightgrey}{rgb}{0.83,0.83,0.83}
\definecolor{lightpink}{rgb}{1.00,0.71,0.76}
\definecolor{lightsalmon}{rgb}{1.00,0.63,0.48}
\definecolor{lightsea}{rgb}{0.13,0.70,0.67}
\definecolor{lightsky}{rgb}{0.53,0.81,0.98}
\definecolor{lightslate}{rgb}{0.47,0.53,0.60}
\definecolor{lightslate}{rgb}{0.47,0.53,0.60}
\definecolor{lightslate}{rgb}{0.52,0.44,1.00}
\definecolor{lightsteel}{rgb}{0.69,0.77,0.87}
\definecolor{lightyellow}{rgb}{1.00,1.00,0.88}
\definecolor{limegreen}{rgb}{0.20,0.80,0.20}
\definecolor{linen}{rgb}{0.98,0.94,0.90}
\definecolor{magenta1}{rgb}{1.00,0.00,1.00}
\definecolor{magenta2}{rgb}{0.93,0.00,0.93}
\definecolor{magenta3}{rgb}{0.80,0.00,0.80}
\definecolor{magenta4}{rgb}{0.55,0.00,0.55}
\definecolor{magenta}{rgb}{1.00,0.00,1.00}
\definecolor{maroon1}{rgb}{1.00,0.20,0.70}
\definecolor{maroon2}{rgb}{0.93,0.19,0.65}
\definecolor{maroon3}{rgb}{0.80,0.16,0.56}
\definecolor{maroon4}{rgb}{0.55,0.11,0.38}
\definecolor{maroon}{rgb}{0.69,0.19,0.38}
\definecolor{mediumaquamarine}{rgb}{0.40,0.80,0.67}
\definecolor{mediumblue}{rgb}{0.00,0.00,0.80}
\definecolor{mediumorchid}{rgb}{0.73,0.33,0.83}
\definecolor{mediumpurple}{rgb}{0.58,0.44,0.86}
\definecolor{mediumsea}{rgb}{0.24,0.70,0.44}
\definecolor{mediumslate}{rgb}{0.48,0.41,0.93}
\definecolor{mediumspring}{rgb}{0.00,0.98,0.60}
\definecolor{mediumturquoise}{rgb}{0.28,0.82,0.80}
\definecolor{mediumviolet}{rgb}{0.78,0.08,0.52}
\definecolor{midnightblue}{rgb}{0.10,0.10,0.44}
\definecolor{mintcream}{rgb}{0.96,1.00,0.98}
\definecolor{mistyrose}{rgb}{1.00,0.89,0.88}
\definecolor{moccasin}{rgb}{1.00,0.89,0.71}
\definecolor{navajowhite}{rgb}{1.00,0.87,0.68}
\definecolor{navyblue}{rgb}{0.00,0.00,0.50}
\definecolor{navy}{rgb}{0.00,0.00,0.50}
\definecolor{oldlace}{rgb}{0.99,0.96,0.90}
\definecolor{olivedrab}{rgb}{0.42,0.56,0.14}
\definecolor{orange1}{rgb}{1.00,0.65,0.00}
\definecolor{orange2}{rgb}{0.93,0.60,0.00}
\definecolor{orange3}{rgb}{0.80,0.52,0.00}
\definecolor{orange4}{rgb}{0.55,0.35,0.00}
\definecolor{orangered}{rgb}{1.00,0.27,0.00}
\definecolor{orange}{rgb}{1.00,0.65,0.00}
\definecolor{orchid1}{rgb}{1.00,0.51,0.98}
\definecolor{orchid2}{rgb}{0.93,0.48,0.91}
\definecolor{orchid3}{rgb}{0.80,0.41,0.79}
\definecolor{orchid4}{rgb}{0.55,0.28,0.54}
\definecolor{orchid}{rgb}{0.85,0.44,0.84}
\definecolor{palegoldenrod}{rgb}{0.93,0.91,0.67}
\definecolor{palegreen}{rgb}{0.60,0.98,0.60}
\definecolor{paleturquoise}{rgb}{0.69,0.93,0.93}
\definecolor{paleviolet}{rgb}{0.86,0.44,0.58}
\definecolor{papayawhip}{rgb}{1.00,0.94,0.84}
\definecolor{peachpuff}{rgb}{1.00,0.85,0.73}
\definecolor{peru}{rgb}{0.80,0.52,0.25}
\definecolor{pink1}{rgb}{1.00,0.71,0.77}
\definecolor{pink2}{rgb}{0.93,0.66,0.72}
\definecolor{pink3}{rgb}{0.80,0.57,0.62}
\definecolor{pink4}{rgb}{0.55,0.39,0.42}
\definecolor{pink}{rgb}{1.00,0.75,0.80}
\definecolor{plum1}{rgb}{1.00,0.73,1.00}
\definecolor{plum2}{rgb}{0.93,0.68,0.93}
\definecolor{plum3}{rgb}{0.80,0.59,0.80}
\definecolor{plum4}{rgb}{0.55,0.40,0.55}
\definecolor{plum}{rgb}{0.87,0.63,0.87}
\definecolor{powderblue}{rgb}{0.69,0.88,0.90}
\definecolor{purple1}{rgb}{0.61,0.19,1.00}
\definecolor{purple2}{rgb}{0.57,0.17,0.93}
\definecolor{purple3}{rgb}{0.49,0.15,0.80}
\definecolor{purple4}{rgb}{0.33,0.10,0.55}
\definecolor{purple}{rgb}{0.63,0.13,0.94}
\definecolor{red1}{rgb}{1.00,0.00,0.00}
\definecolor{red2}{rgb}{0.93,0.00,0.00}
\definecolor{red3}{rgb}{0.80,0.00,0.00}
\definecolor{red4}{rgb}{0.55,0.00,0.00}
\definecolor{red}{rgb}{1.00,0.00,0.00}
\definecolor{rosybrown}{rgb}{0.74,0.56,0.56}
\definecolor{royalblue}{rgb}{0.25,0.41,0.88}
\definecolor{saddlebrown}{rgb}{0.55,0.27,0.07}
\definecolor{salmon1}{rgb}{1.00,0.55,0.41}
\definecolor{salmon2}{rgb}{0.93,0.51,0.38}
\definecolor{salmon3}{rgb}{0.80,0.44,0.33}
\definecolor{salmon4}{rgb}{0.55,0.30,0.22}
\definecolor{salmon}{rgb}{0.98,0.50,0.45}
\definecolor{sandybrown}{rgb}{0.96,0.64,0.38}
\definecolor{seagreen}{rgb}{0.18,0.55,0.34}
\definecolor{seashell1}{rgb}{1.00,0.96,0.93}
\definecolor{seashell2}{rgb}{0.93,0.90,0.87}
\definecolor{seashell3}{rgb}{0.80,0.77,0.75}
\definecolor{seashell4}{rgb}{0.55,0.53,0.51}
\definecolor{seashell}{rgb}{1.00,0.96,0.93}
\definecolor{sienna1}{rgb}{1.00,0.51,0.28}
\definecolor{sienna2}{rgb}{0.93,0.47,0.26}
\definecolor{sienna3}{rgb}{0.80,0.41,0.22}
\definecolor{sienna4}{rgb}{0.55,0.28,0.15}
\definecolor{sienna}{rgb}{0.63,0.32,0.18}
\definecolor{skyblue}{rgb}{0.53,0.81,0.92}
\definecolor{slateblue}{rgb}{0.42,0.35,0.80}
\definecolor{slategray}{rgb}{0.44,0.50,0.56}
\definecolor{slategrey}{rgb}{0.44,0.50,0.56}
\definecolor{snow1}{rgb}{1.00,0.98,0.98}
\definecolor{snow2}{rgb}{0.93,0.91,0.91}
\definecolor{snow3}{rgb}{0.80,0.79,0.79}
\definecolor{snow4}{rgb}{0.55,0.54,0.54}
\definecolor{snow}{rgb}{1.00,0.98,0.98}
\definecolor{springgreen}{rgb}{0.00,1.00,0.50}
\definecolor{steelblue}{rgb}{0.27,0.51,0.71}
\definecolor{tan1}{rgb}{1.00,0.65,0.31}
\definecolor{tan2}{rgb}{0.93,0.60,0.29}
\definecolor{tan3}{rgb}{0.80,0.52,0.25}
\definecolor{tan4}{rgb}{0.55,0.35,0.17}
\definecolor{tan}{rgb}{0.82,0.71,0.55}
\definecolor{thistle1}{rgb}{1.00,0.88,1.00}
\definecolor{thistle2}{rgb}{0.93,0.82,0.93}
\definecolor{thistle3}{rgb}{0.80,0.71,0.80}
\definecolor{thistle4}{rgb}{0.55,0.48,0.55}
\definecolor{thistle}{rgb}{0.85,0.75,0.85}
\definecolor{tomato1}{rgb}{1.00,0.39,0.28}
\definecolor{tomato2}{rgb}{0.93,0.36,0.26}
\definecolor{tomato3}{rgb}{0.80,0.31,0.22}
\definecolor{tomato4}{rgb}{0.55,0.21,0.15}
\definecolor{tomato}{rgb}{1.00,0.39,0.28}
\definecolor{turquoise1}{rgb}{0.00,0.96,1.00}
\definecolor{turquoise2}{rgb}{0.00,0.90,0.93}
\definecolor{turquoise3}{rgb}{0.00,0.77,0.80}
\definecolor{turquoise4}{rgb}{0.00,0.53,0.55}
\definecolor{turquoise}{rgb}{0.25,0.88,0.82}
\definecolor{violetred}{rgb}{0.82,0.13,0.56}
\definecolor{violet}{rgb}{0.93,0.51,0.93}
\definecolor{wheat1}{rgb}{1.00,0.91,0.73}
\definecolor{wheat2}{rgb}{0.93,0.85,0.68}
\definecolor{wheat3}{rgb}{0.80,0.73,0.59}
\definecolor{wheat4}{rgb}{0.55,0.49,0.40}
\definecolor{wheat}{rgb}{0.96,0.87,0.70}
\definecolor{whitesmoke}{rgb}{0.96,0.96,0.96}
\definecolor{white}{rgb}{1.00,1.00,1.00}
\definecolor{yellow1}{rgb}{1.00,1.00,0.00}
\definecolor{yellow2}{rgb}{0.93,0.93,0.00}
\definecolor{yellow3}{rgb}{0.80,0.80,0.00}
\definecolor{yellow4}{rgb}{0.55,0.55,0.00}
\definecolor{yellowgreen}{rgb}{0.60,0.80,0.20}
\definecolor{yellow}{rgb}{1.00,1.00,0.00}

%% file: authors.tex
\textbf{Conveners:} A.~de~Gouv\^ea$^{50}$, K.~Pitts$^{32}$, K.~Scholberg$^{26}$, G.P.~Zeller$^{27}$

\textbf{Subgroup Conveners:} J.~Alonso$^{46}$, A.~Bernstein$^{41}$, M.~Bishai$^{8}$, S.~Elliott$^{42}$, K.~Heeger$^{78}$, K.~Hoffman$^{43}$

% ...
$^{8}$Brookhaven National Laboratory, Upton, NY 11973-5000, USA\\ 
% ...
$^{26}$Duke University, Durham, NC 27708-0754, USA\\ 
$^{27}$Fermi National Accelerator Laboratory, Batavia, IL 60510, USA\\ 
% ...
$^{32}$University of Illinois, Urbana, IL 61801, USA\\ 
% ...
$^{41}$Lawrence Livermore National Laboratory, Livermore, CA, 94550, USA\\ 
$^{42}$Los Alamos National Laboratory, Los Alamos, NM 87545, USA\\ 
$^{43}$University of Maryland, College Park, MD 20742, USA\\ 
% ...
$^{46}$Massachusetts Institute of Technology, Cambridge, MA 02139, USA\\ 
% ...
$^{50}$Northwestern University, Evanston, IL  60208 USA\\ 
% ...
$^{78}$Yale University, New Haven, CT 06511-8962, USA\\

%% file: execsum.tex
\section{Executive summary}

Decades of experimental and observational scrutiny have revealed less than a handful of phenomena outside the standard model, among them evidence for dark energy and dark matter, and the existence of nonzero neutrino masses.
While many experiments continue to look for other new phenomena and deviations from standard model predictions, it is clear that continued detailed study of the neutrino sector is of the utmost importance.    
 
Compared to the other fermions, the elusive nature of the neutrinos has made them extremely difficult to study in detail.    In spite of the challenges, 
\textbf{neutrino physics has advanced quickly and dramatically since the end of the last century}.  Thanks to a remarkable suite of experiments and associated theoretical work, two previously unknown and closely related features of neutrinos now stand out clearly: neutrinos have mass and leptons mix with each other.
Starting from almost 
no knowledge of the neutrino masses or lepton mixing parameters twenty years ago,
we have built a robust, simple, three-flavor paradigm which successfully describes most of the data. 

Experiments with solar, atmospheric, reactor and accelerator neutrinos have established, beyond reasonable doubt, that a neutrino produced in a well-defined flavor state (say, a muon-type neutrino $\nu_{\mu}$) has a nonzero probability of being detected in a different flavor state  (say, an electron-type neutrino $\nu_e$). This flavor-changing probability depends on the neutrino energy and the distance traversed between the source and the detector. The only consistent explanation of nearly all neutrino data collected over the last two decades is a phenomenon referred to as ``neutrino-mass-induced flavor oscillation.''

\begin{figure}[ht]
\begin{center}
\includegraphics[width=1.0\textwidth]{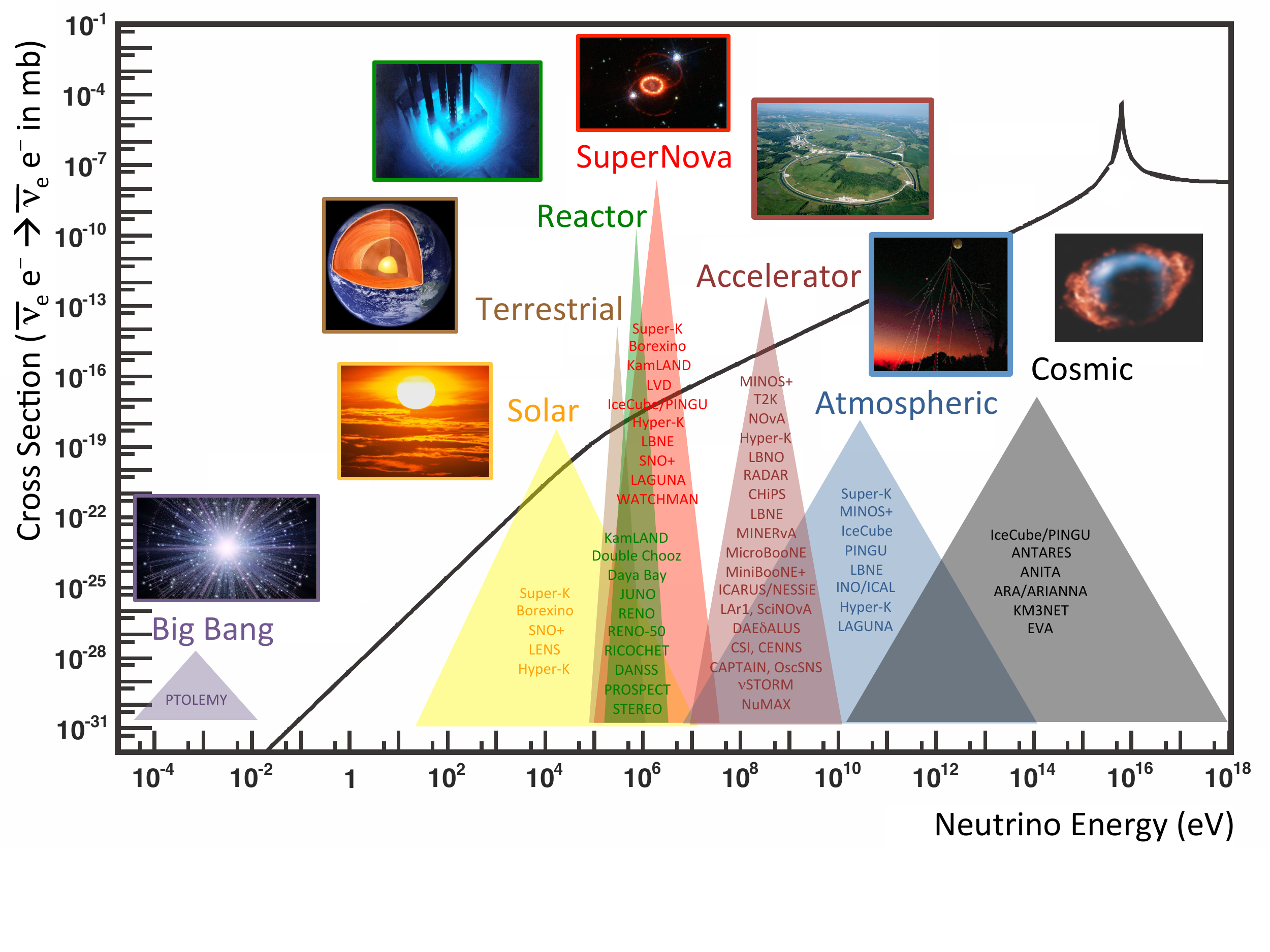}
\end{center}
\vskip-2.0cm
\caption{Neutrino interaction cross section as a function of energy, showing typical energy regimes accessible by different neutrino sources and experiments. The curve shows the scattering cross section for $\bar{\nu}_e \, e^- \rightarrow e^- \, \bar{\nu}_e$ on free electrons, for illustration. Plot modified from~\cite{Formaggio:2013kya}.}
\label{fig:sources}
\end{figure}

In two different oscillation sectors, similar parallel stories unfolded: hints of neutrino flavor change in experiments studying  natural neutrinos were confirmed, and later refined, by experiments with artificial neutrinos.  The disappearance of atmospheric $\nu_\mu$ was unambiguously confirmed by several beam $\nu_\mu$ disappearance experiments, which have now achieved high precision on the driving ``atmospheric'' mixing parameters, i.e., the
mass-squared difference $|\Delta m^2_{32}|$ and the mixing parameter $\theta_{23}$. 
The observation of the disappearance of $\nu_e$ from the Sun, a decades-long mystery, was definitively confirmed as evidence for flavor change
using flavor-blind neutral-current interactions.  This ``solar'' oscillation was further confirmed, and the driving ``solar'' mixing parameters  ($\theta_{12}$ 
and the mass-squared difference $\Delta m^2_{21}$) were very well measured, using reactor antineutrinos and further solar data.  This complementarity illustrates the importance of exploring the diverse neutrino sources available (see Fig.~\ref{fig:sources}).

The current generation of detectors is now exploring oscillations in a three-flavor context, with both accelerator and reactor tour-de-force experiments having now measured, with good precision, the value of the third mixing angle, $\theta_{13}$, via positive searches for $\nu_\mu \rightarrow \nu_e$ appearance and $\bar{\nu}_e$ disappearance respectively.  

Furthermore, while most of the data fit the three-flavor paradigm very well,
some experiments have uncovered intriguing anomalies  that do not fit this simple picture.
These exceptions include apparent short-baseline  $\nu_\mu\rightarrow\nu_e$ and $\bar{\nu}_\mu\rightarrow\bar{\nu}_e$ transitions, and the anomalous disappearance of reactor and radioactive source electron-type antineutrinos and neutrinos.  Although these hints currently have only modest statistical significance, if confirmed they would be evidence for beyond-the-standard-model states or interactions.

The observation of neutrino oscillations implies that neutrinos have nonzero masses, a discovery of fundamental significance.  
We do not know the mechanism responsible for the generation of neutrino masses, but we can state 
with some certainty that new degrees of freedom are required. The number of options is enormous. The current data do not reveal, for example, whether the new physics scale is very low (say, 1~eV) or very high (say, $10^{15}$~GeV). The origin of neutrino masses is one of the biggest puzzles in particle physics today, and will only be revealed, perhaps only indirectly,  with more experimental information from different probes in the different frontiers of particle physics research. Furthermore, the pattern of lepton mixing is very different from that of quarks. We do not yet know what that means, but precision studies of lepton mixing via neutrino oscillations may reveal crucial information regarding the long-standing flavor puzzle.

\subsection{The Big Questions and physics opportunites}

We are now poised to answer some of the most fundamental and important questions of our time. \textbf{There is a clear experimental path forward}, which builds heavily on the recent successful history of this rapidly-evolving field of particle physics.

\noindent \textit{What is the pattern of neutrino masses? Is there CP violation in the lepton sector?  To what extent does the three-flavor paradigm describe nature?}

The current neutrino data allow for very large deviations from the three-flavor paradigm. New neutrino--matter interactions as strong as the standard-model weak interactions are not ruled out, and the existence of new ``neutrino'' states with virtually any mass is allowed, and sometimes expected from different mechanisms for generating neutrino masses. 

Even in the absence of more surprises, we still do not know how the neutrino masses are ordered: 
do we have two ``light'' and one ``heavy'' neutrino (the so-called normal mass hierarchy) or two ``heavy'' and one ``light'' neutrino (the inverted hierarchy)?
 The resolution of this issue is of the utmost importance, for both practical and fundamental reasons. As will become more clear below, resolving the neutrino mass hierarchy will allow one to optimize the information one can obtain from other neutrino experimental probes, including searches for leptonic CP invariance violation, searches for the absolute value of the neutrino masses, and searches for the violation of lepton number via neutrinoless double-beta decay. In addition, the mass hierarchy will also reveal invaluable information concerning the origin of neutrino masses. If the mass hierarchy were inverted, for example, we would learn that at least two of the three neutrino masses are quasi-degenerate, a condition that is not observed in the spectrum of charged leptons or quarks. 

Experimental neutrino oscillation data revealed that CP invariance can be violated in the lepton sector. The lepton sector accommodates up to three new sources of CP violation --- two Majorana phases and one Dirac phase. While the so-called Dirac phase can be explored in future oscillation experiments, the Majorana phases are physical only if neutrinos are Majorana fermions. Some of these may be constrained (depending on the physics of lepton-number violation) from the rate of neutrinoless double-beta decay, a determination of the mass hierarchy, and a direct measurement of the neutrino mass. Currently, two sources of CP violation are known: the CP-odd phase in the CKM matrix, and the QCD $\theta$ parameter. The former is known to be large, while the latter is known to be at most vanishingly small. Exploring CP violation in the lepton sector is guaranteed to significantly increase our understanding of this phenomenon. It is also likely that information regarding CP violation in the lepton sector will play a key role when it comes to understanding the mechanism for baryogenesis. 

The current generation of oscillation experiments, including Double Chooz, RENO, Daya Bay, T2K, and NOvA, will start to resolve the neutrino mass hierarchy and, especially with results combined, may provide a first glimpse at CP violation in the lepton sector. These will also provide improved measurements of almost all neutrino oscillation parameters. Next-generation experiments, along with very intense proton beams, will definitively resolve the neutrino mass hierarchy and substantially improve our ability to test CP invariance in the lepton sector. Next-generation reactor-neutrino experiments with intermediate (around 50~km) baselines and atmospheric-neutrino experiments may also independently shine light on the neutrino mass hierarchy. The former will also provide precision measurements of the ``solar'' parameters, $\Delta m^2_{21}$ and $\theta_{12}$. Different experiments with different energies, baselines and detector technologies will allow good constraints on physics beyond the three-flavor paradigm. For the farther future, a more definitive probe of the three-flavor paradigm and precision measurements of CP violation in the lepton sector (or lack thereof) will require long-baseline experiments with different neutrino beams. Leading candidates include 
neutrinos from pion decay at rest produced by high-intensity cyclotron proton sources,
and neutrinos from muon storage rings.
A muon-storage-ring facility should be able to measure the Dirac CP-odd phase with a precision on par with
the quark sector, and provide the most stringent constraints on the three-flavor paradigm, thanks to its capability to measure
several different oscillation channels with similar precision.

Comprehensive and detailed studies of neutrino-matter scattering not only serve as tests of the standard model and probes of nuclear structure but are also a definite requirement for precision neutrino oscillation experiments.  The convolution of an uncertain neutrino flux with imprecise scattering cross sections and roughly-estimated nuclear effects can result in large, even dominant, systematic errors in the measurement of neutrino oscillation parameters.  
More generally, we need to fully characterize
neutrino--matter interactions to enable deeper understanding of neutrino oscillations, supernova dynamics, and dark matter searches; this will require dedicated theoretical and experimental efforts.

\noindent \textit{Are neutrinos Majorana or Dirac particles?}  

Massive neutrinos are special. Among all known fermions, neutrinos are the only ones not charged under the two unbroken gauge symmetries: electromagnetism and color. This implies that, unlike all known particles, neutrinos may be Majorana fermions. Majorana neutrinos would imply, for example, that neutrino masses are a consequence of a new fundamental energy scale in physics, potentially completely unrelated to the electroweak scale. Dirac neutrinos, on the other hand, would imply that $U(1)_{B-L}$, or some subgroup, is a fundamental symmetry of nature, with deep consequences for our understanding of the laws of physics.  

If neutrinos are Majorana fermions, lepton number cannot be a conserved quantum number. Conversely, lepton number-violation indicates that massive neutrinos are Majorana fermions. Hence, the best (perhaps only) probes for the hypothesis that neutrinos are Majorana fermions are searches for lepton-number violation. By far, the most sensitive probe of lepton-number conservation is the pursuit of neutrinoless double-beta decay ($0\nu\beta\beta$), $(Z)\to (Z+2)e^-e^-$, where $(Z)$ stands for a nucleus with atomic number $Z$. Independent from the strict connection to the nature of the neutrino, the observation of $0\nu\beta\beta$ would dramatically impact our understanding of nature (similar to the potential observation of baryon number violation) and would provide clues concerning the origin of the baryon asymmetry.  

In many models for the origin of lepton-number violation, $0\nu\beta\beta$ is dominated by the exchange  of virtual massive Majorana neutrinos, in such a way that its amplitude is, assuming all neutrinos are light, proportional to $m_{ee}\equiv \sum_{i}U_{ei}^2m_i$, $i=1,2,3,\ldots$. Under these circumstances, the observation of $0\nu\beta\beta$ would not only reveal that neutrinos are Majorana fermions, but would also provide information concerning the absolute values of the neutrino masses. Conversely, given that we know the neutrino mass-squared differences and the magnitude of the relevant elements of the mixing matrix, one can predict the rate for $0\nu\beta\beta$ as a function of the unknown value of the lightest neutrino mass. In particular, if the neutrino mass hierarchy is inverted, there is a lower bound to $|m_{ee}|\gtrsim 20$~meV. 

The current generation of 100-kg-class $0\nu\beta\beta$ search experiments should reach effective masses in the 100~meV range; beyond that, there are opportunities for multi-ton-class experiments that will reach sub-10~meV effective mass sensitivity, pushing below the inverted hierarchy region. In order to fully exploit the relation between $0\nu\beta\beta$ and nonzero Majorana neutrino masses, it is imperative to understand in detail the associated nuclear matrix elements. These require detailed theoretical computations beyond those carried out to date. 

\noindent \textit{What is the absolute neutrino mass scale?}  

While the values of the neutrino mass-squared differences are known, their absolute values remain elusive. In order to properly understand particle physics in general, and neutrinos in particular, it is clear that knowledge of particle masses --- not just mass-squared differences --- is mandatory. The current neutrino data still allow for the possibility that the lightest neutrino mass is vanishingly small, or that all three known neutrino masses  are quasi-degenerate. These two possibilities are qualitatively different and point to potentially different origins for the nonzero neutrino masses.

Neutrino masses can only be directly determined via non-oscillation neutrino experiments. The most model-independent observable sensitive to sub-eV neutrino masses is the shape of the endpoint of beta decay spectra. Precision studies of tritium beta decay provide the most stringent bounds, and are expected to play a leading role in next-generation experiments. KATRIN, the most ambitious current-generation  tritium-beta-decay experiment, will directly probe neutrino masses a factor of 10 smaller than the best current bounds. Innovative new ideas may help to go beyond this level of sensitivity.

Other probes of the absolute value of the neutrino masses include $0\nu\beta\beta$, discussed above, and different maps of the large-scale structure of the Universe. Both are, in their own ways, much more model-dependent than precision studies of beta decay. Today, cosmological observables provide the most stringent bounds on the absolute values of the neutrino masses, constraining their sums to be below several tenths of an eV, and the prospects for the next several years are very exciting.

\noindent \textit{Are there already hints of new physics in existing data?}  

There are intriguing anomalies that cannot be accommodated within the three-flavor paradigm, and suggest new physics beyond it.  In particular, there is marginal (two to four $\sigma$) yet persistent evidence for oscillation phenomena at baselines not consistent with the well-established oscillation lengths associated to the ``solar'' and ``atmospheric'' mass-squared differences.  These anomalies, which are not directly ruled out by other experiments, include the excess of $\bar{\nu}_e$ events observed by the LSND experiment, the $\nu_e$ and $\bar{\nu}_e$ excesses observed by MiniBooNE (particularly at low-energies), the deficit of $\bar{\nu}_e$ events observed by reactor neutrino experiments and the deficit of $\nu_e$ events observed in the SAGE and GALLEX radioactive source experiments. Although there may be several possible ways to explain these anomalies by introducing new physics, the most credible ones, while not ruled out, do not provide a very good fit to {\sl all} available  neutrino data. 
Combined, the anomalies are often interpreted as evidence for one or more additional neutrino states, known as sterile neutrinos. 
The $3+N$ sterile neutrino model, in which there are three light mostly-active neutrinos and $N$ mostly-sterile neutrinos which mix with the active flavors, is often used to fit the existing data and gauge the reach of proposed next-generation experiments. 
For $N>1$, these models allow for CP-violating effects in short-baseline appearance experiments.  

Beyond particle physics, there are hints of additional neutrinos coming from cosmology.  Fits to astrophysical data sets (including the cosmic microwave background, large-scale structure, baryon acoustic oscillations and Big Bang nucleosynthesis) are sensitive to the effective number of light degrees of freedom ($N_{\rm eff}$).  In the standard model, $N_{\rm eff}$ is equivalent to the effective number of neutrino species, although in principle this could include other types of light, weakly-coupled states.  The recent Planck data are consistent with $N_{\rm eff}=3$ but still allow $N_{\rm eff}=4$. Potential connections between this hint and the short-baseline anomalies above are tantalizing but neither established nor excluded.

These anomalies may go away with more data; but if they are confirmed, the consequences would open up a whole new sector to explore experimentally and theoretically. The discovery of new neutrino states, for example, would revolutionize our understanding of particle physics. Definitive tests are clearly needed and concrete efforts are already underway. The MicroBooNE experiment, for example, aims at addressing the low-energy excesses observed at MiniBooNE. A variety of neutrino sources and flavor-changing observables  are being pursued as potential means to address the different anomalies.

\noindent \textit{What new knowledge will neutrinos from astrophysical sources bring?}  

Neutrinos come from natural sources as close as the Earth and Sun, to as far away as distant galaxies, and even as remnants from the Big Bang. They range in kinetic energy from less than one meV to greater than one PeV.  As weakly-interacting particles, they probe otherwise inaccessible properties of the astrophysical sources they come from.  Astrophysical neutrino sources furthermore shed light on the nature of neutrinos themselves, and on cosmology.  

At the very lowest energies, we can access information about the  $T_{\nu}=1.95$~K Big Bang relic neutrinos via cosmological observables; direct detection of these is extremely challenging but nevertheless can be pursued.

In the few to few-tens-of-MeV energy range, 
large underground liquid-scintillator, water-Cherenkov and liquid-argon detectors are the instruments of choice.
Solar neutrinos may have more to tell us about neutrino oscillations and other neutrino properties, and about solar physics.
Neutrinos from stellar core collapse  have the potential not only to shed light on the astrophysics of gravitational collapse, but provide a unique probe of neutrino properties.  It is now even possible to study the Earth via MeV geo-neutrinos from terrestrial radioactivity.

In the TeV-and-higher-energy region, fluxes of atmospheric neutrinos start to diminish,  and neutrinos of astrophysical origin should begin to dominate.  Unlike photons and charged particles, these cosmic neutrinos travel unimpeded from their sources and will bring information on the origin of ultra-high-energy cosmic rays.  Due to the very small expected fluxes,  instrumentation of enormous natural reservoirs of water as neutrino telescopes is required.
The recent detection of the first ultra-high-energy astrophysical neutrinos by IceCube has opened a crucial new window of investigation into the study of nature's highest-energy particle accelerators.

\subsection{The path forward}

Table~\ref{experimentstable} gives a summary of the many current and proposed experiments, in the U.S. and abroad, designed to address various physics questions. The number of possibilities is endless. We now describe a specific path forward, both in the U.S. and in an international context. 
Neutrino physics is a broad subfield of fundamental particle physics, and requires a multi-pronged approach in order to address all the outstanding questions and fully explore the new physics revealed by neutrino oscillation experiments. 
Investment in a range of large, medium and small-scale neutrino experiments (as well as in detector R\&D and theory) will ensure a healthy program.

\begin {itemize}
\item \textit{Comprehensive test of the three-flavor paradigm, via long-baseline, precision neutrino oscillation experiments:} The next-generation experiments will take full advantage of conventional neutrino beams from pion decay in flight. These will begin to over-constrain the parameter space, and will start to seriously explore CP-violating phenomena in the lepton sector.
The U.S., with the Long-Baseline Neutrino Experiment (LBNE) and a future multi-megawatt beam from Project
X, is uniquely positioned to lead an international campaign to measure
CP violation and aggressively test the three-flavor paradigm.
 
Complementary experiments with different energies, baselines and detector technologies (e.g., Hyper-K in Japan) are required in order to fully exploit conventional neutrino beams. 
The accompanying very-large detectors, if placed underground, also allow for the study of atmospheric neutrinos, nucleon decay, and precision measurements of neutrinos from a galactic supernova explosion.
PINGU, an upgrade of IceCube, provides a promising opportunity to measure the mass hierarchy using atmospheric neutrinos. 

Next-next generation experiments will require better (both more intense, and better understood) neutrino beams.  Promising possibilities include neutrinos from muon storage rings (e.g., NuMAX), and neutrinos from very intense cyclotron-based sources of pion decay at rest (e.g., DAE$\delta$ALUS). 
Muon-based neutrino beams in particular have strong synergies with
Project X and provide a necessary step in the R\&D for a high-energy muon collider. 
 While these large, ambitious projects are vigorously developed, the following medium and small-scale neutrino activities need to be pursued.
\begin{itemize}

\item \textit{Precision measurements and theories of neutrino cross sections and a detailed understanding of the neutrino flux from pion-decay-in-flight neutrino beams.} These activities can be pursued in the “near- detectors” associated with the large long-baseline projects or alongside R\&D projects related to next-next generation neutrino beams, as well as by small-scale dedicated experiments. A well-considered program of precision scattering experiments in both low- and high-energy regimes, combined with a renewed dedicated theoretical effort to develop a reliable, nuclear-physics-based description of neutrino interactions in nuclei is mandatory.   Scattering measurements may also be of intrinsic interest. 

\item \textit{Definite resolution of the current short-baseline anomalies.} These will (probably) require neutrino sources other than pion-decay-in-flight and the pursuit of different flavor-changing channels, including $\nu_{e,\mu}$ disappearance and $\nu_{\mu}\to\nu_e$ appearance, using a combination of reactor, radioactive source and accelerator experiments.  In addition to small-scale dedicated experiments, such experiments can be carried out as part of  R\&D projects related to next-next generation neutrino beams (e.g., nuSTORM, IsoDAR).
\item \textit{Vigorous pursuit of R\&D projects related to the development of next-next generation neutrino experiments.} As discussed above, these medium and small experiments will also address several key issues in neutrino physics.
\end{itemize} 
\item \textit{Searches for neutrinoless double-beta decay:} The current generation of experiments is pursuing different detector technologies with different double-beta decaying isotopes. The goals of these experiments are to (a) discover neutrinoless double-beta decay, which is guaranteed if the neutrinos are Majorana fermions and their masses are quasi-degenerate, (b) provide information regarding the most promising techniques for the next generation.

Next-generation experiments aim at discovering neutrinoless double-beta decay if neutrinos are Majorana fermions and if the neutrino mass hierarchy is inverted. In the case of a negative result, assuming oscillation experiments have revealed that the neutrino mass hierarchy is inverted, these experiments will provide strong evidence that the neutrinos are Dirac fermions. As with precision measurements of beta decay (see below), the information one can extract from the current and the next generation of neutrinoless double-beta decay experiments increases significantly if indirect evidence for neutrino masses is uncovered, e.g., with cosmological probes. 
\item \textit{Determination of the absolute values of the neutrino masses:} Precision measurements of beta decay remain the most promising model-independent probes. While the KATRIN experiment is taking data, vigorous R\&D efforts for next-generation probes (e.g., ECHo, Project 8, PTOLEMY) are required in order to identify whether it is possible to reach sensitivities to the effective ``electron-neutrino mass'' below $0.05$~eV. Nontrivial information is expected from different cosmological probes of the large-scale structure of the Universe. 

\end{itemize}

The relevance of neutrino science and technology extends well beyond the fundamental research community.  Neutrinos may be useful for monitoring reactors in the context of international nuclear nonproliferation (e.g., WATCHMAN).  The essential building blocks of neutrino science  ---  detectors  and accelerators  ---  have important spin-off applications for medicine and in industry. Finally,

the unusual, ghostlike properties of neutrinos are fascinating to the general public.  The success of our field depends on our ability to convey both the mystery and utility of neutrino science to the public, policy-makers and funding agencies.

The diversity of physics topics within the neutrino sector is enormous and the interplay between neutrino physics and other fields is rich.   Neutrinos have and will continue to provide important information on structure formation in the early universe, Earth, solar and supernova physics, nuclear properties, and rare decays of charged leptons and hadrons.  Conversely, information regarding neutrino properties and the origin of neutrino masses  is expected from the Energy and Cosmic frontiers, and from other areas of Intensity Frontier research (as well as nuclear physics). 

In the remainder of this document, we describe in more detail the many exciting possibilities for the future.  Section~\ref{sec:intro} is a pedagogical introduction to the basics of neutrino physics and experiments, and is intended primarily to be a guide for the non-expert to the neutrino physics that can be addressed using different kinds of neutrino sources and various experimental approaches.  The remaining sections provide more details of future opportunities, following our Neutrino Working Group substructure.  Section~\ref{sec:3nus} describes measurements addressing remaining unknowns and precision tests of the standard three-flavor paradigm. Section~\ref{sec:majorana} describes the $0\nu\beta \beta$ decay subfield, and Sec.~\ref{sec:mass} describes approaches for addressing the question of absolute neutrino mass.  Section~\ref{sec:scattering} describes neutrino scattering experiments. Section~\ref{sec:anomalies} describes existing anomalies and other beyond-the-standard-model tests, and the wide range of possible experiments to address them.  Section~\ref{nu6} describes physics and astrophysics that can be done using neutrinos from astrophysical sources.
Finally, Sec.~\ref{sec:society} describes direct and spin-off applications of neutrino physics, as well as relevant education and outreach.

\footnotesize
\begin{table}[h]
\footnotesize
\label{experimentstable}
\caption{Summary of the many current and proposed experiments, in the U.S. and abroad, designed to address various physics questions.  Rows refer to neutrino sources and columns refer to categories of physics topics these sources can address (roughly corresponding to the neutrino working groups).
The intent is not to give a ``laundry list'', but to give a sense of the activity and breadth of the field.  Some multipurpose experiments appear under more than one physics category.
Experiments based in the U.S. (or initiated and primarily led by U.S. collaborators) are shown in blue and underlined  (note that many others have substantial U.S. participation or leadership).  Proposed and future experiments are in bold; current experiments (running or with construction well underway) are in regular font. More details and references can be found in the subsections of the Neutrino Working Group report.}
\begin{center}
\begin{tabular}{|c|c|c|c|c|c|c|c|} \hline
Source & 3-flavor osc.& Maj./Dirac & Abs. Mass & Interactions & Anomalies/Exotic$^2$ & Astro/Cosmo \\  \hline \hline
Reactor  & KamLAND,& && {\color{blue}\textbf{\uline{RICOCHET}}}& \textbf{DANSS}, \textbf{STEREO}, & \\ 
  & Double Chooz,&&& &  {\color{blue}\textbf{\uline{PROSPECT}}}, &  \\
           & Daya Bay, \textbf{JUNO}, &&&&{\color{blue}\textbf{\uline{RICOCHET}}} & \\ 
            & RENO, \textbf{RENO-50}&&&&& \\ \hline   
Solar & Super-K, &&&&& Super-K,\\ 
              & Borexino, SNO+, &&&& & Borexino, SNO+,\\
            & \textbf{Hyper-K}, {\color{blue}\uline{\textbf{LENS}}} &&&&& \textbf{Hyper-K}, {\color{blue}\uline{\textbf{LENS}}}\\  \hline
Supernova$^1$  & Super-K, Borexino, &&&&  & Super-K, Borexino, \\  
            & KamLAND, LVD, &&&& & KamLAND, LVD\\
            & {\color{blue}\uline{IceCube/\textbf{PINGU}}},&&&&&{\color{blue}\uline{IceCube/\textbf{PINGU}}},\\
             & \textbf{Hyper-K}, {\color{blue}\uline{\textbf{LBNE}}},  &&&&& \textbf{Hyper-K}, {\color{blue}\uline{\textbf{LBNE}}},\\
            & SNO+, \textbf{LAGUNA},  &&&&& SNO+, \textbf{LAGUNA},\\
           & {\color{blue}\uline{\textbf{WATCHMAN}}} &&&&& {\color{blue}\uline{\textbf{WATCHMAN}}}\\ \hline
Atmospheric&Super-K, {\color{blue}\uline{MINOS+}} , &&&& &\\ 
& {\color{blue}\uline{IceCube/\textbf{PINGU}}}, &&&& &\\  
& {\color{blue}\uline{\textbf{LBNE}}}, \textbf{ICAL}, &&&&&\\ 
&\textbf{Hyper-K}, \textbf{LAGUNA}  &&&&&\\ \hline
Pion DAR & {\color{blue}\uline{\textbf{DAE$\delta$ALUS}}} &&&{\color{blue}\uline{\textbf{OscSNS}}}, {\color{blue}\uline{\textbf{CSI}}},& {\color{blue}\uline{\textbf{OscSNS}}}&\\ 
 &  &&& {\color{blue}\uline{\textbf{CENNS}}}, & &\\ 
 &  &&& {\color{blue}\uline{\textbf{CAPTAIN}}} & &\\ \hline  
Pion DIF  
          &  {\color{blue}\uline{MINOS+}}, T2K,&&&{\color{blue}\uline{MicroBooNE}},&{\color{blue}\uline{MicroBooNE}},&\\  
          &  {\color{blue}\uline{NOvA}}, \textbf{Hyper-K},&&&{\color{blue}\uline{MINER$\nu$A}}, & {\color{blue}\uline{\textbf{MiniBooNE+/II}}},&\\  
          & \textbf{LAGUNA-LBNO},&&& {\color{blue}\uline{NOvA}}, & \textbf{Icarus/NESSiE},&\\   
          &  {\color{blue}\uline{\textbf{RADAR}}}, {\color{blue}\uline{\textbf{CHIPS}}}, &&& {\color{blue}\uline{\textbf{SciNOvA}}}& {\color{blue}\uline{\textbf{LAr1}}, \uline{\textbf{LAr1-ND}}},&\\  
          &  {\color{blue}\uline{\textbf{LBNE}}}, \textbf{ESS$\nu$SB} &&&& {\color{blue}\uline{MINOS+}}&\\  \hline
$\mu$DIF & {\color{blue}\uline{\textbf{NuMAX}}} & & & {\color{blue}\uline{\textbf{nuSTORM}}} &{\color{blue}\uline{\textbf{nuSTORM}}} &  \\ \hline
Radioactive & & Many: see& KATRIN,&& \textbf{SOX}, \textbf{CeLAND},&\\ 
Isotopes &  & Nu2 report  &  {\color{blue} \textbf{\uline{Project 8}}},&& \textbf{Daya Bay Source}, &\\  
 &  & for table &  \textbf{ECHo}, && \textbf{BEST},  &\\  
 &  &  & {\color{blue}\uline{\textbf{PTOLEMY}}} & &  {\color{blue}\uline{\textbf{IsoDAR}}} &\\  \hline  
Cosmic & & & & & & {\color{blue}\uline{IceCube/\textbf{PINGU}}}, \\ 
 neutrinos & & & & & &  ANTARES/\textbf{ORCA}, \\ 
 & & & & & &  {\color{blue}\textbf{\uline{ARA},\uline{ARIANNA}}}, \\
 & & & & & &  {\color{blue}\uline{ANITA}, \uline{\textbf{EVA}}}, \\
& & & & & &  \textbf{KM3NET} \\ \hline
\end{tabular}
\end{center}
\footnotesize
\vspace{-0.11in}
$^1$Included are only kt-class underground detectors; many others would also record events.  
$^2$We note that nearly all experiments can address anomalies or exotic physics at some level; we include in this column only those with this as a primary physics goal.
\end{table}

\normalsize

%% file: intro.tex
\section{Introduction: physics of neutrinos}\label{sec:intro}

Neutrinos are the most elusive of the known fundamental particles. They are color-neutral and charge-neutral spin-one-half fermions.  To the best of our knowledge, they only interact with charged fermions and massive gauge bosons through the weak interactions. For this reason, neutrinos can only be observed and studied because there are very intense neutrino sources (natural and artificial) and only if one is willing to work with large detectors. 

The existence of neutrinos was postulated in the early 1930s, but they were only first observed in the 1950's~\cite{Cowan:1992xc}. The third neutrino flavor eigenstate, the tau-type neutrino $\nu_{\tau}$, was the last of the fundamental matter particles to be observed~\cite{Kodama:2000mp}, eluding direct observation six years longer than the top quark~\cite{Abe:1995hr,Abachi:1995iq}.  More relevant to this report, in the late 1990s the discovery of nonzero neutrino masses moved the study of neutrino properties to the forefront of experimental and theoretical particle physics.

Experiments with solar~\cite{Cleveland:1998nv,Hampel:1998xg,Ahmad:2002jz,Abdurashitov:2002nt,Fukuda:2001nj,Ahmed:2003kj}, atmospheric~\cite{Fukuda:1998mi,Ashie:2004mr}, reactor~\cite{Eguchi:2002dm,Araki:2004mb,Abe:2012tg,Ahn:2012nd,An:2012bu} and accelerator~\cite{Ahn:2002up,Michael:2006rx,Abe:2013xua} neutrinos  have established, beyond reasonable doubt, that a neutrino produced in a well-defined flavor state (say, a muon-type neutrino $\nu_{\mu}$) has a nonzero probability of being detected in a different flavor state  (say, an electron-type neutrino $\nu_e$). This flavor-changing probability depends on the neutrino energy and the distance traversed between the source and the detector. The simplest and only consistent explanation of almost all neutrino data collected over the last two decades is a phenomenon referred to as ``neutrino mass-induced flavor oscillation.'' These neutrino oscillations, which will be discussed in more detail in Sec.~\ref{sec:osc}, in turn imply that neutrinos have nonzero masses and neutrino mass eigenstates are different from neutrino weak eigenstates, i.e., leptons mix.  

In a nutshell, if the neutrino masses are distinct and leptons mix, a neutrino can be produced, via weak interactions, as a coherent superposition of mass-eigenstates, e.g., a neutrino $\nu_{\alpha}$ with a well-defined flavor, and has a nonzero probability to be measured as a neutrino $\nu_{\beta}$ of a different flavor ($\alpha,\beta=e,\mu,\tau$). The oscillation probability $P_{\alpha\beta}$ depends on the neutrino energy $E$, the propagation distance $L$, and on the neutrino mass-squared differences, $\Delta m^2_{ij}\equiv m_i^2-m_j^2$, $i,j=1,2,3,\ldots$, and the elements of the leptonic mixing matrix,\footnote{Often referred to as the Maki-Nakagawa-Sakata (MNS) Matrix, or the Pontecorvo-Maki-Nakagawa-Sakata (PMNS) Matrix.} $U$, which relates neutrinos with a well-defined flavor ($\nu_e,\nu_{\mu},\nu_{\tau}$) and neutrinos with a well-defined mass ($\nu_{1},\nu_2,\nu_3,\ldots$). For three neutrino flavors, the elements of $U$ are defined by
\begin{equation}
\left(\begin{array}{c}\nu_e \\ \nu_{\mu} \\ \nu_{\tau} \end{array} \right) =\left(\begin{array}{ccc} U_{e1} & U_{e2} & U_{e3} \\ U_{\mu1} & U_{\mu2} & U_{\mu3} \\ U_{\tau1} & U_{e\tau2} & U_{\tau3}\end{array}\right)
\left(\begin{array}{c}\nu_1 \\ \nu_2 \\ \nu_3 \end{array} \right).
\label{UMNS}
\end{equation}
Almost all neutrino data to date can be explained assuming that neutrinos interact as prescribed by the standard model, there are only three neutrino mass eigenstates, and $U$ is unitary.
Under these circumstances, it is customary to parameterize $U$ in Eq.~(\ref{UMNS}) with three mixing angles $\theta_{12},\theta_{13},\theta_{23}$ and three complex phases, $\delta,\xi,\zeta$, defined by
\begin{equation}
\frac{|U_{e2}|^2}{|U_{e1}|^2}\equiv \tan^2\theta_{12};
~~~~\frac{|U_{\mu3}|^2}{|U_{\tau3}|^2}\equiv \tan^2\theta_{23};~~~~
U_{e3}\equiv\sin\theta_{13}e^{-i\delta},
\end{equation}
with the exception of $\xi$ and $\zeta$, the so-called Majorana $CP$-odd phases. These are only physical if the neutrinos are Majorana fermions, and have  no effect in flavor-changing phenomena.

\begin{figure}[ht]
\centerline{\includegraphics[width=0.45\textwidth]{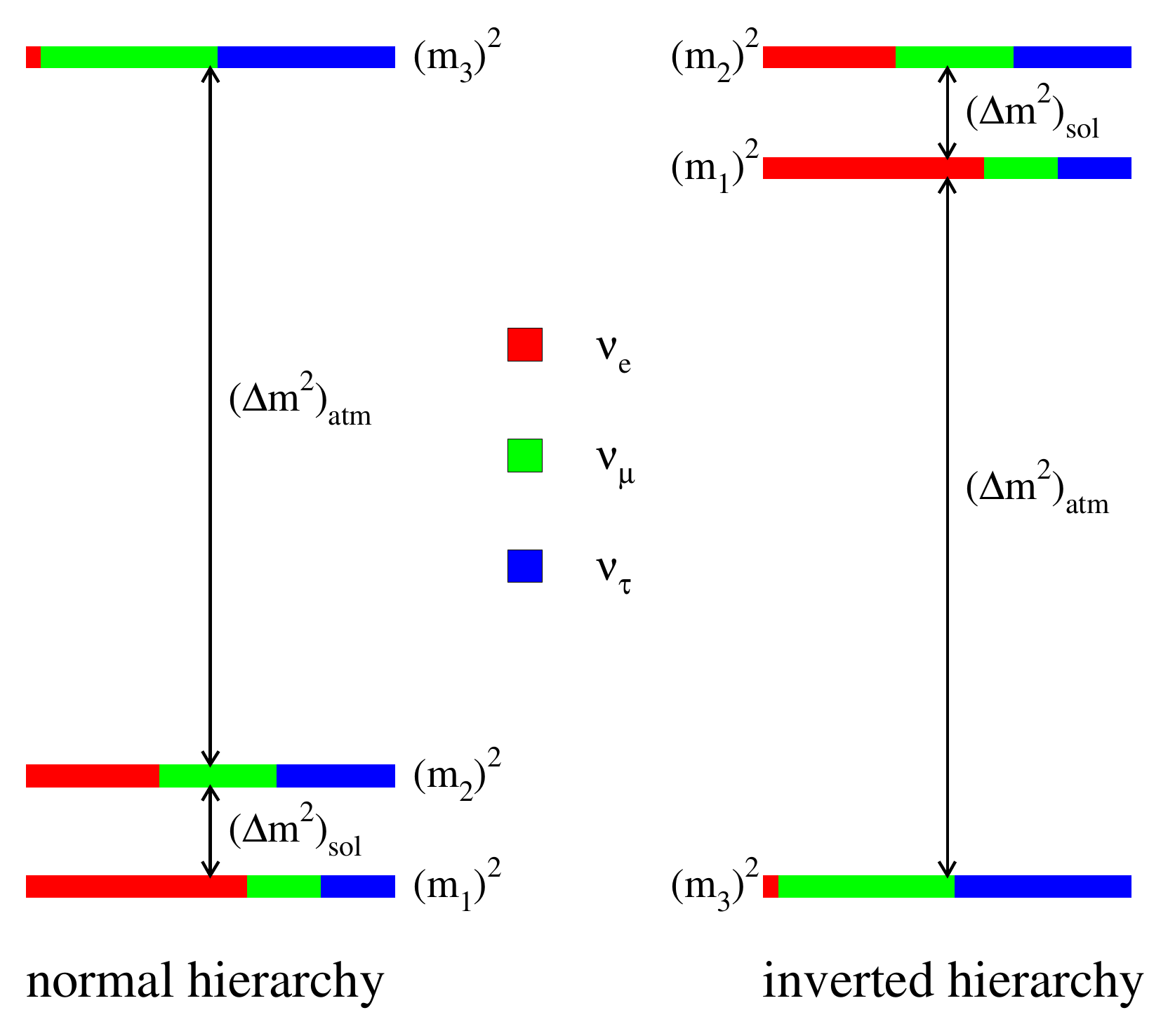}}
\caption{Cartoon of the two distinct neutrino mass hierarchies that fit nearly all of the current neutrino data, for fixed values of all mixing angles and mass-squared differences. The color coding (shading) indicates the fraction $|U_{\alpha i}|^2$ of each distinct flavor $\nu_{\alpha}$, $\alpha=e,\mu,\tau$ contained in each mass eigenstate $\nu_i$, $i=1,2,3$. For example, $|U_{e2}|^2$ is equal to the fraction of the $(m_2)^2$ ``bar'' that is painted red (shading labeled as ``$\nu_e$'').
\label{3nus_pic}}
\end{figure}

In order to relate the mixing elements to experimental observables, it is necessary to properly define the neutrino mass eigenstates, i.e., to ``order'' the neutrino masses. This is done in the following way: $m_2^2>m_1^2$ and $\Delta m^2_{21}<|\Delta m^2_{31}|$. In this case, there are three mass-related oscillation observables: $\Delta m^2_{21}$ (positive-definite), $|\Delta m^2_{31}|$, and the sign of $\Delta m^2_{31}$. A positive (negative) sign for $\Delta m^2_{31}$ implies $m_3^2>m_2^2$  ($m_3^2<m_1^2$) and characterizes a so-called normal (inverted) neutrino mass hierarchy. 
The two mass hierarchies are depicted in Fig.~\ref{3nus_pic}.

Our knowledge of neutrino oscillation parameters has evolved dramatically over the past two decades. As summarized in Sec.~\ref{sec:3nus}, all three mixing angles have been measured relatively well, along with (the magnitudes of) the mass-squared differences. On the other hand, we have virtually no information concerning $\delta$ (nor, for that matter, $\xi$ and $\zeta$) or the sign of $\Delta m^2_{32}$. We also don't know the value of the neutrino masses themselves --- only differences of the masses-squared. We can't rule out the possibility that the lightest neutrino is virtually massless ($m_{\rm lightest}\ll 10^{-3}$~eV) or that all neutrino masses are virtually the same (e.g., $m_1\sim m_2\sim m_3\sim 0.1$~eV). Probes outside the realm of neutrino oscillations are required to investigate the values of the neutrino masses. These are described in Sec.~\ref{sec:mass}.

One of the main goals of next-generation experiments is to test whether the scenario outlined above, the standard three-massive-neutrinos paradigm, is correct and complete. This can be achieved by next-generation experiments sensitive to neutrino oscillations via not simply determining all of the parameters above, but by ``over-constraining'' the parameter space in order to identify potential inconsistencies. This is far from a simple task, and the data collected thus far, albeit invaluable, allow for only the simplest consistency checks. Precision measurements, as will be discussed in Sec.~\ref{sec:3nus}, will be required.

In more detail, given all we know about the different neutrino oscillation lengths, it is useful to step back and appreciate what oscillation experiments have been able to measure. Solar data, and data from KamLAND, are, broadly speaking, sensitive to $|U_{e2}|$, $|U_{\mu 2}|^2+|U_{\tau 2}|^2$, and $|U_{e2}U_{e1}|$. Data from atmospheric neutrinos and long-baseline, accelerator-based experiments are sensitive to $|U_{\mu 3}|$ and, to a much lesser extent, $|U_{\mu 3}U_{\tau 3}|$ and $|U_{\mu 3}U_{e3}|$. Finally, km-scale reactor experiments are sensitive to $|U_{e3}|$. Out of the nine (known) complex entries of $U$, we have information, usually very limited, regarding the magnitude of around six of them. Clearly, we have a long way to go before concluding that the three-flavor paradigm is the whole story.

Life may, indeed, already be much more interesting. There are several, none too significant, hints in the world neutrino data that point to a neutrino sector that is more complex than the one outlined above. These will be discussed in Sec.~\ref{sec:anomalies}. Possible surprises include new, gauge singlet fermion states that manifest themselves only by mixing with the known neutrinos, and new weaker-than-weak interactions.

Another issue of fundamental importance is the investigation of the status of CP invariance in leptonic processes. Currently, all observed CP-violating phenomena are governed by the single physical CP-odd phase parameter in the quark mixing matrix. Searches for other sources of CP violation, including the so-called strong CP-phase $\theta_{QCD}$, have, so far, failed. The picture currently emerging from neutrino-oscillation data allows for a completely new, independent source of CP  violation. The CP-odd parameter $\delta$, if different from zero or $\pi$, implies that neutrino oscillation probabilities violate CP-invariance, i.e., the values of the probabilities for neutrinos to oscillate are different from those of antineutrinos! We describe this phenomenon in more detail in Secs.~\ref{sec:osc}, \ref{sec:3nus}. 

It should be noted that, if neutrinos are Majorana fermions, the CP-odd phases $\xi$ and $\zeta$ also mediate CP-violating phenomena~\cite{deGouvea:2002gf} (alas, we don't yet really know how to study these in practice). In summary, if neutrinos are Majorana fermions, the majority of CP-odd parameters  in particle physics --- even in the absence of other new physics --- belong to the lepton sector. These are completely unknown and can ``only'' be studied in neutrino experiments. Neutrino oscillations provide a  unique opportunity to revolutionize our understanding of CP violation, with potentially deep ramifications for both particle physics and cosmology.  
An important point is that all modifications to the standard model that lead to massive neutrinos change it qualitatively. For a more detailed discussion of this point see, e.g.,~\cite{DeGouvea:2005gd}.

\begin{figure}[ht]
\centerline{\includegraphics[width=0.65\textwidth]{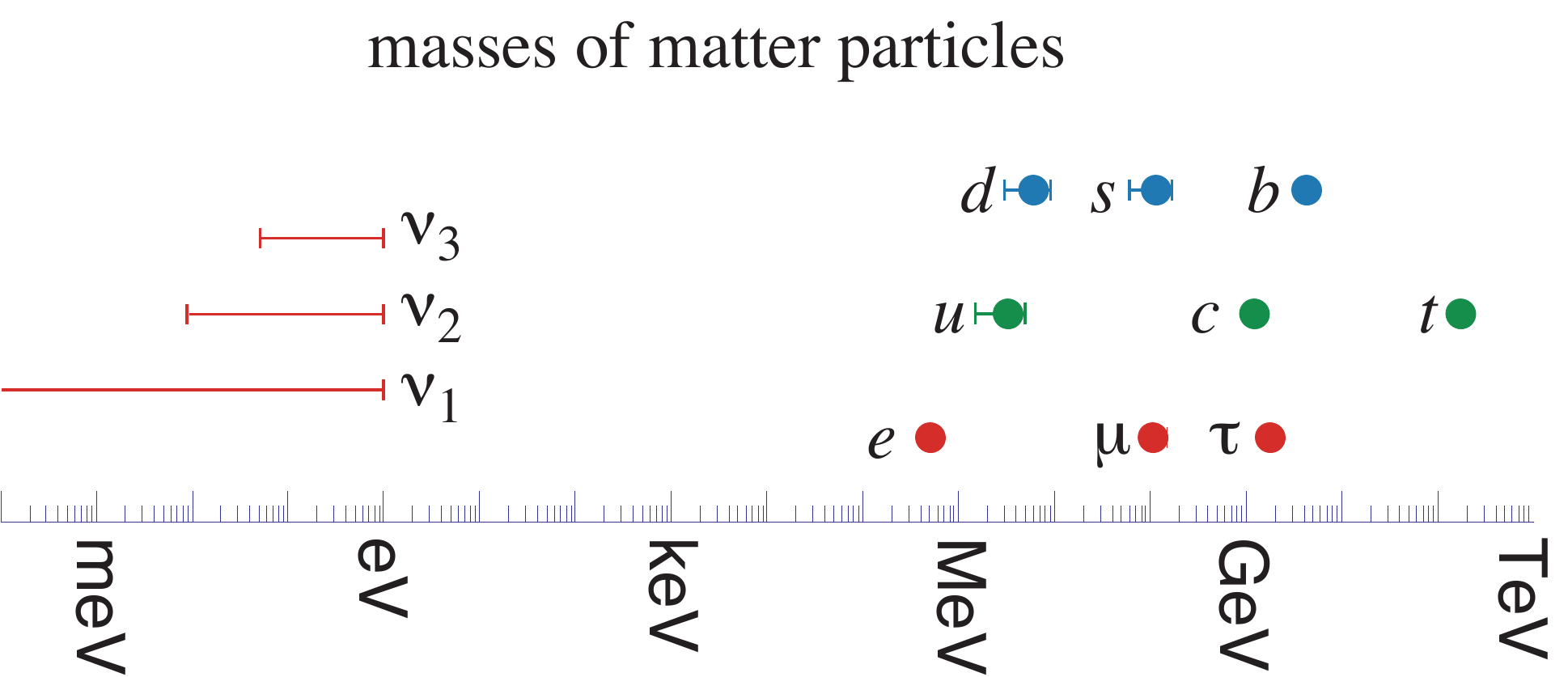}}
\caption{Standard model fermion masses. For the neutrino masses, the normal mass hierarchy was assumed, and a loose upper bound $m_i<1$~eV, for all $i=1,2,3$ was imposed.}
\label{fig:fermionmasses}
\end{figure}

Neutrino masses, while nonzero, are tiny when compared to all other known fundamental fermion masses in the standard model, as depicted in Fig.~\ref{fig:fermionmasses}. Two features readily stand out: (i) neutrino masses are at least six orders of magnitude smaller than the electron mass, and 
(ii) there is a ``gap'' between the largest allowed neutrino mass and the electron mass.
We don't know why neutrino masses are so small or why there is such a large gap between the neutrino and the charged fermion masses. We suspect, however, that this may be nature's way of telling us that neutrino masses are ``different.''

This suspicion is only magnified by the possibility that massive neutrinos, unlike all other fermions in the standard model, may be Majorana fermions. The reason is simple: neutrinos are the only electrically-neutral fundamental fermions and hence need not be distinct from their antiparticles. Determining the nature of the neutrino --- Majorana or Dirac --- would not only help to guide theoretical work related to uncovering the origin of neutrino masses, but could also reveal that the conservation of lepton number is not a fundamental law of nature. The most promising avenue for learning the fate of lepton number, as will be discussed in Sec.~\ref{sec:majorana},  is to look for neutrinoless double-beta decay, a lepton-number violating nuclear process. The observation of a nonzero rate for this hypothetical process would easily rival, as far as its implications for our understanding of nature are concerned, the first observations of parity violation and $CP$-invariance violation in the mid-twentieth century.

It is natural to ask what augmented, ``new'' standard model ($\nu$SM) leads to nonzero neutrino masses. The answer is that we are not sure. There are many different ways to modify the standard model in order to accommodate neutrino masses. While these can differ greatly from one another, all succeed --- by design --- in explaining small neutrino masses and all are allowed by the current particle physics experimental data. The most appropriate question, therefore, is not what are the candidate $\nu$SM's, but how can one identify the ``correct'' $\nu$SM? The answers potentially lie in next-generation neutrino experiments, which are described throughout this report.

Before discussing concrete examples, it is important to highlight the potential theoretical significance of nonzero neutrino masses. In the standard model, the masses of all fundamental particles are tied to the phenomenon of electroweak symmetry breaking and a single mass scale --- the vacuum expectation value of the Higgs field. Nonzero neutrino masses may prove to be the first direct evidence of a new mass scale, completely unrelated to electroweak symmetry breaking, or evidence that electroweak symmetry breaking is more complex than dictated by the standard model. 

Here we discuss one generic mechanism in more detail.  The effect of heavy new degrees of freedom in low-energy phenomena can often be captured by adding higher-dimensional operators to the standard model. As first pointed out in~\cite{Weinberg:1979sa}, given the standard model particle content and gauge symmetries, one is allowed to write only one type of dimension-five operator --- all others are dimension-six or higher:
\begin{equation}
 {1 \over \Lambda}\ (L H) (L H) + h.c.\quad  \Rightarrow \quad {v^{2} \over \Lambda}\nu\nu + h.c.,
\label{eq:seesaw}
\end{equation}
where $L$ and $H$ are the lepton and Higgs boson $SU(2)_L$ doublets, and the arrow indicates one of the components of the operator after electroweak symmetry is broken. $v$ is the vacuum expectation value of the neutral component of $H$, and $\Lambda$ is the effective new physics scale. If this operator is indeed generated by some new physics, neutrinos obtain Majorana masses $m_\nu \sim v^2/\Lambda$. For $\Lambda\sim 10^{15}$~GeV, $m_{\nu}\sim 10^{-1}$~eV, in agreement with the current neutrino data. This formalism explains the small neutrino masses via a seesaw mechanism: $m_{\nu}\ll v$ because $\Lambda\gg v$.

$\Lambda$ is an upper bound for the masses of the new particles that lead to Eq.~(\ref{eq:seesaw}). If the new physics is strongly coupled and Eq.~(\ref{eq:seesaw}) is generated at the tree-level, the new degrees of freedom are super-heavy: $M_{\rm new}\sim 10^{15}$~GeV. 
If that turns out to be the case, we will only be able to access the new physics indirectly through neutrino experiments and the study of relics in the Cosmic Frontier.
If, however,  the new physics is weakly coupled or Eq.~(\ref{eq:seesaw}) is generated at the loop level, virtually any value for $M_{\rm new}\gtrsim 1$~eV is allowed. There are many scenarios where the new physics responsible for nonzero neutrino masses can be probed at the Energy Frontier or elsewhere in the Intensity Frontier~\cite{Hewett:2012ns}.
In summary, if Eq.~(\ref{eq:seesaw}) is correct, we expect new physics to show up at a new mass scale $M_{\rm new}$ which lies somewhere between $10^{-9}$~GeV and $10^{15}$~GeV. Clearly, more experimental information is required!

Neutrino data also provide a new piece to the flavor puzzle: the pattern of neutrino mixing. The absolute values of the entries of the CKM quark mixing matrix are given by
\begin{equation}
|V_{\rm CKM}|\sim \left(\begin{array}{ccc} 1 & 0.2 & 0.004\\ 0.2 & 1 & 0.04 \\ 0.008 & 0.04 & 1\end{array} \right),
\end{equation}
while those of the entries of the PMNS matrix are given by
\begin{equation}
|U_{\rm PMNS}|\sim \left(\begin{array}{ccc} 0.8 & 0.5 & 0.2 \\ 0.4 & 0.6 & 0.7 \\ 0.4 & 0.6 & 0.7\end{array} \right).
\end{equation}
It is clear that the two matrices look very different. While the CKM matrix is almost proportional to the identity matrix plus hierarchically ordered off-diagonal elements, the PMNS matrix is far from diagonal and, with the possible exception of the $U_{e3}$ element, all elements are ${\cal O}(1)$. Significant research efforts are concentrated on understanding what, if any, is the relationship between the quark and lepton mixing matrices and what, if any, is the ``organizing principle'' responsible for the observed pattern of neutrino masses and lepton mixing. There are several different theoretical ideas in the market  (for summaries, overviews and more references see, e.g.,~\cite{Mohapatra:2005wg,Albright:2006cw}). Typical results include predictions for the currently-unknown neutrino mass and mixing parameters ($\theta_{23}$ octant, the mass hierarchy, CP-violating $\delta$) and the establishment of sum rules involving different parameters. Some of the challenges are discussed in Sec.~\ref{sec:3nus}.

Precision neutrino oscillation measurements are required to address the flavor questions above. That can only be achieved as the result of significant investments in intense, well-characterized neutrino sources and massive high-precision detectors. Some of these are summarized later in this section and spelled out in more detail throughout this report. Excellent understanding of neutrino interactions --- beyond the current state of the art ---  is also mandatory. This will require a comprehensive experimental program on neutrino scattering, as summarized in Sec.~\ref{sec:scattering}. These, of course, are not only supportive of neutrino oscillation experiments, but are also interesting in their own right. Neutrinos, since they interact only weakly, serve as a unique probes of nucleon and nuclear properties, and may reveal new physics phenomena at the electroweak scale, including some that are virtually invisible to the Tevatron and the LHC.

(Massive) neutrinos also serve as unique messengers in astrophysics and cosmology, as discused in Sec.~\ref{nu6}. Astrophysical neutrino searches may uncover indirect evidence for dark matter annihilation in the Earth, the Sun, or the center of the Galaxy. Neutrinos produced in supernova explosions contain information from deep within the innards of the exploding stars and their studies may also help reveal unique information regarding neutrino properties. Big Bang neutrinos play a definitive role in the thermal history of the universe. Precision cosmology measurements also may reveal neutrino properties, including the absolute values of the neutrino masses. Finally, the unique character of the neutrinos and the experiments used to study them provide unique opportunities outside the realm of particle physics research. More details along these lines are discussed in Sec.~\ref{sec:society}.

\subsection{Overview of neutrino oscillations}\label{sec:osc}

Physical effects of nonzero neutrino masses, to date, have been observed only in neutrino oscillation experiments. Those are expected to remain, for the foreseeable future, the most powerful tools available for exploring the new physics revealed by solar and atmospheric neutrino experiments at the end of the twentieth century.

The standard setup of a neutrino oscillation experiment is as follows. A detector is located a distance $L$ away from a source, which emits ultra-relativistic neutrinos or antineutrinos with, most often, a continuous spectrum of energies $E$, and flavor $\alpha=e,\mu$, or $\tau$. According to the standard model, the neutrinos interact with matter either via $W$-boson exchange charged-current (CC) interactions where a neutrino with a well-defined flavor $\nu_{\alpha}$ gets converted into a charged lepton of the same flavor ($\nu_e X\to eX'$, {\it etc.}) or via $Z$-boson exchange neutral-current (NC) interactions, which preserve the neutrino flavor ($\nu_{\mu}X\to\nu_{\mu}X'$). The occurrence of a neutral-current process is tagged by observing the system against which the neutrinos are recoiling. The detector hence is capable of measuring the flux of neutrinos or antineutrinos with flavor $\beta=e,\mu$, or $\tau$, or combinations thereof, often as a function of the neutrino energy. By comparing measurements in the detector with expectations from the source, one can infer $P_{\alpha\beta}(L,E)$ or $\bar{P}_{\alpha\beta}(L,E)$, the probability that a(n)  (anti)neutrino with energy $E$ produced in a flavor eigenstate $\nu_{\alpha}$ is measured in a flavor $\nu_{\beta}$ after it propagates a distance $L$.  In practice, it is often preferable to make multiple measurements of neutrinos at different distances from the source, which can be helpful for both the cancellation of systematic uncertainties and for teasing out effects beyond the standard three-flavor paradigm.

\begin{figure}[ht]
\begin{center}
\includegraphics[width=0.6\textwidth]{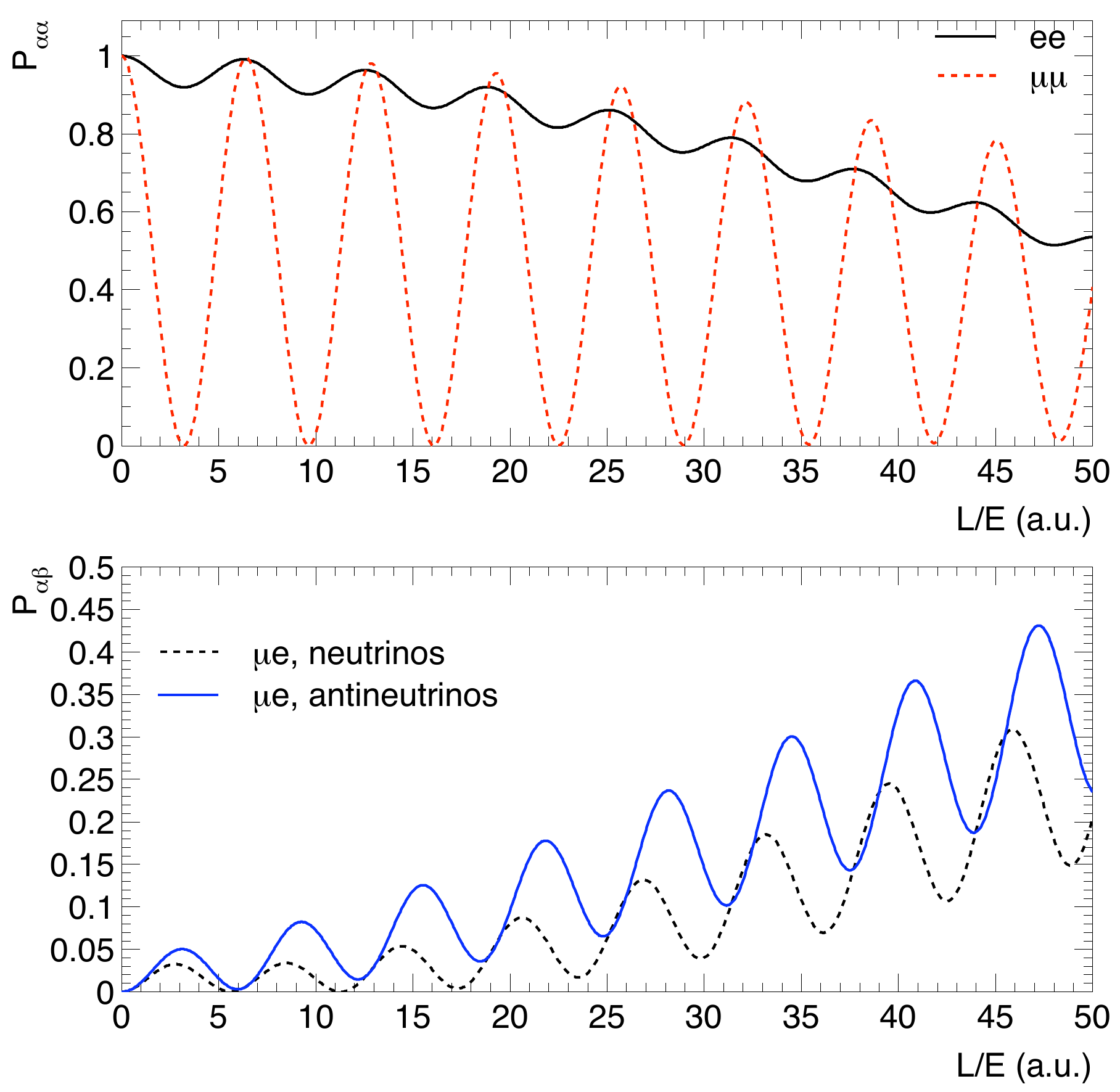}
\end{center}
%\vskip-0.8cm
\caption{Top: $P_{ee}$ and $P_{\mu\mu}$ in vacuum as a function of $L/E$ (in arbitrary units), for representative values of the neutrino oscillation parameters, including a nonzero value of $\delta$. Bottom: $P_{\mu e}$ and $\bar{P}_{\mu e}$ in vacuum as a function of $L/E$, for representative values of the neutrino oscillation parameters.}
\label{fig:oscillations}
\end{figure}

In the standard three-flavor paradigm, $P_{\alpha\beta}$ is a function of the mixing angles $\theta_{12,13,23}$, the Dirac $CP$-odd phase $\delta$, and the two independent neutrino mass-squared differences $\Delta m^2_{21,31}$.   Assuming the neutrinos propagate in vacuum, and making explicit use of the unitarity of $U$, one can express $P_{\alpha\beta}(L,E)=|A_{\alpha\beta}|^2$, where
\begin{eqnarray}
A_{\alpha\beta}=\delta_{\alpha\beta}+U_{\alpha2}U^*_{\beta2}\left(\exp\left(-i\frac{\Delta m^2_{21}L}{2E}\right)-1\right)+U_{\alpha3}U^*_{\beta3}\left(\exp\left(-i\frac{\Delta m^2_{31}L}{2E}\right)-1\right), \label{eq:Aab}
\\
\bar{A}_{\alpha\beta}=\delta_{\alpha\beta}+U^*_{\alpha2}U_{\beta2}\left(\exp\left(-i\frac{\Delta m^2_{21}L}{2E}\right)-1\right)+U^*_{\alpha3}U_{\beta3}\left(\exp\left(-i\frac{\Delta m^2_{31}L}{2E}\right)-1\right),
\label{eq:barAab}
\end{eqnarray}
up to an unphysical overall phase. $A$ ($\bar{A}$) is the amplitude for (anti)neutrino oscillations. It is easy to see that $P_{\alpha\beta}$ are oscillatory functions of $L/E$ with, in general, three distinct, two independent oscillation lengths proportional to  $\Delta m^2_{21}$, $\Delta m^2_{31}$ and $\Delta m^2_{32}\equiv\Delta m^2_{31}-\Delta m^2_{21}$, as depicted in Fig.~\ref{fig:oscillations}. Ideally, measurements of some $P_{\alpha\beta}$ as a function of $L/E$ would suffice to determine all neutrino oscillation parameters. These would also allow one to determine whether the standard paradigm is correct, i.e., whether Eqs.~(\ref{eq:Aab},\ref{eq:barAab}) properly describe neutrino flavor-changing phenomena.

For example, if one could measure both $P_{ee}$ and $P_{\mu\mu}$ as a function of $L/E$, one should be able to determine not only $\Delta m^2_{21}$ and $|\Delta m^2_{31}|$, but also $|U_{e2}|^2$,  $|U_{e3}|^2$, $|U_{\mu2}|^2$ and $|U_{\mu3}|^2$, and the sign of $\Delta m^2_{31}$. This in turn would translate into measurements of all mixing parameters, including the $CP$-odd phase $\delta$. One would also be able to determine, for example, whether there are other oscillation lengths, which would indicate there are new, yet-to-be-observed, neutrino states, or whether $P_{ee,\mu\mu}\neq 1$ in the limit $L\to 0$, which would indicate, for example, the existence of new, weaker-than-weak, charged-current type interactions.

In the real world, such measurements are, to say the least, very hard to perform, for several reasons. $\Delta m^2_{21}$ is much smaller than the magnitude of $\Delta m^2_{31,32}$, which in turn makes it challenging to observe two independent oscillation frequencies in the same experimental setup. For this reason, for all measurements of $P_{\mu\mu}$ performed to date 
the $L/E$ factors probed are too small to ``see'' the $\Delta m^2_{21}$-driven oscillations or distinguish $\Delta m^2_{31}$ from $\Delta m^2_{32}$. On the other hand, the magnitude of $|U_{e3}|$ is much smaller than that of the other entries of $U$. For this reason, measurements of ${P}_{ee}$ for solar neutrinos have only been precise enough to definitively observe $\Delta m^2_{21}$-driven oscillations and hence determine its magnitude, along with that of $U_{e2}$.

Another real-world issue is that, for any setup, it is not possible to measure any $P_{\alpha\beta}$ with perfect $L/E$ resolution. Furthermore, the available $L/E$ ranges are, in many cases, narrow. More realistically, one expects to measure, with decent statistics and small systematic errors, $P_{\alpha\beta}$ integrated over a few finite-sized $L/E$ bins. This discreteness of the data leads to ambiguities when it comes to measuring the different mixing parameters. For example, different pairs of $\theta_{13},\delta$ values lead to identical values for $P_{\alpha\beta}$ integrated over a fixed $L/E$. The same is true for pairs of $\theta_{13},\theta_{23}$, and so on. A so-called eight-fold degeneracy has been identified and studied in great detail in the neutrino literature (see, for example,~\cite{Cervera:2000kp,BurguetCastell:2001ez,Barger:2001yr}). The solution to this challenge is to perform several measurements of different $P_{\alpha\beta}$ at different values of $L$ and $E$ (and $L/E$). This is especially true if one is interested in not only measuring the three-flavor neutrino mixing parameters but also, much more importantly, over-constraining the standard paradigm and hence testing its validity. For example, one would like to precisely measure $\theta_{13}$ in different channels, for different values of $L$ and $E$, to find out if all of them agree.

Measurements of vacuum survival probabilities, $P_{\alpha\alpha}$ or $\bar{P}_{\alpha\alpha}$, do not violate $CP$ invariance: $P_{\alpha\alpha}=\bar{P}_{\alpha\alpha}$  is guaranteed by $CPT$-invariance. In order to directly observe $CP$-invariance violation, one needs to measure an appearance probability, say $P_{\mu e}$. $P_{\mu e}$ is different from $\bar{P}_{\mu e}$,\footnote{Note that T-invariance violation, $P_{e\mu}\neq P_{\mu e}$, is also present under the same conditions.} as depicted in Fig.~\ref{fig:oscillations} (bottom), if the following conditions are met, as one can readily confirm by studying Eqs.~(\ref{eq:Aab},~\ref{eq:barAab}): (i) all $U_{\alpha i}$ have nonzero magnitude, (ii) $U_{\alpha2}U^*_{\beta2}$ and $U_{\alpha3}U^*_{\beta3}$ are relatively complex, (iii) $L/E$ is large enough that both $\Delta m^2_{21,31}\times L/E$ are significantly different from zero. Given what is known about the oscillation parameters, condition (iii) can be met for any given neutrino source by choosing a large enough value for $L$. This, in turn, translates into the need for a very intense source and a very large, yet high-precision, detector, given that for all known neutrino sources the neutrino flux falls off like $1/L^2$ for any meaningful value of $L$. Whether conditions (i) and (ii) are met lies outside the control of the experimental setups. Given our current understanding, including the newly-acquired knowledge that $|U_{e3}|\neq0$,  condition (i) holds. That being the case, condition (ii) is equivalent to $\delta \neq 0,\pi$. In the standard paradigm, the existence of $CP$-invariance violation is entirely at the mercy of the value of $CP$-odd phase $\delta$, currently unconstrained.

High-energy (accelerator and atmospheric) neutrino data accumulated so far provide evidence for nonzero $P_{\mu\tau}$~\cite{Abe:2012jj,Agafonova:2013dtp} and $P_{\mu e}$~\cite{Abe:2013xua,Adamson:2013ue}.\footnote{Solar data translate into overwhelming evidence for $P_{e\mu}+P_{e\tau}\neq 0$. In the standard paradigm, this is indistinguishable from $1-P_{ee}\neq 1$ and hence cannot, even in principle, provide more information than a disappearance result.} Both results are only sensitive to one scale of mass-squared difference ($|\Delta m^2_{31}|\sim |\Delta m^2_{32}|$) and to $|U_{\mu3}U_{\tau3}|$ and $|U_{\mu3}U_{e3}|$, respectively. The goal of the current neutrino oscillation experiments NOvA and T2K is to observe and study $P_{\mu e}$ and $\bar{P}_{\mu e}$ governed by $\Delta m^2_{31}$, aiming at measuring $U_{e3}$ and, perhaps, determining the sign of $\Delta m^2_{31}$ through matter effects, as will be discussed promptly.

Eqs.~(\ref{eq:Aab}, \ref{eq:barAab}) are valid only when the neutrinos propagate in a vacuum. When neutrinos propagate through a medium, the oscillation physics is modified by so-called matter effects~\cite{Wolfenstein:1977ue,Mikheev:1986gs}, also known as Mikheyev-Smirnov-Wolfenstein (MSW) effects. These are due to the coherent forward scattering of neutrinos with the electrons present in the medium, and they create an additional contribution to the phase differences.  Notably, this additional contribution distinguishes between neutrinos and antineutrinos, since there are no positrons present in the Earth.\footnote{In fact, the electron background effectively violates $CPT$ symmetry. For neutrinos oscillating in matter, it is no longer true, for example, that $P_{\alpha\alpha}=\bar{P}_{\alpha\alpha}$.} Matter effects also depend on whether the electron neutrino is predominantly made out of the heaviest or lightest mass eigenstates, thus allowing one to address the ordering of the neutrino mass eigenstates. For one mass hierarchy, the oscillation of neutrinos for a certain range of $L/E$ values can be enhanced with respect to that of antineutrinos, while for the other mass hierarchy the effect is reversed. On the flip side, if the mass hierarchy is not known, matter effects lead to ambiguities in determining the oscillation parameters, as discussed briefly earlier. Matter effects have already allowed the determination of one ``mass hierarchy,'' that of $\nu_1$ and $\nu_2$. Thanks to matter effects in the Sun, we know that $\nu_1$, which is lighter than $\nu_2$, has the larger electron component: $|U_{e1}|^2>|U_{e2}|^2$. A similar phenomenon should be observable in the $\Delta m^2_{31}$ sector, given the recent discovery that $|U_{e3}|$ is not zero. Quantitatively, the importance of matter effects will depend on the density of the medium being traversed, which determines the so-called matter potential $A\equiv \sqrt{2}G_FN_e$, where $G_F$ is the Fermi constant and $N_e$ is the electron number-density of the medium, and on the value of $\Delta m^2_{21,31}/E$. Matter effects are irrelevant when  $A\ll\Delta m^2_{21,32}/E$. For $\Delta m^2_{31(21)}$, matter effects in the Earth's crust are significant for $E\gtrsim 1$~GeV (20~MeV).

\subsection{Neutrino experiments: sources and detectors}

Next-generation experiments have at their disposal a handful of neutrino sources, which we describe qualitatively here, concentrating on their prospects for neutrino oscillation searches.   The sources span many orders of magnitude in energy: see Fig.~\ref{fig:sources}.   Associated with each experiment is an appropriate detector. The requirements for the detectors depend on the neutrino source.

The Sun is a very intense source of $\nu_e$ with energies between $\sim$100~keV and 10~MeV. Precision measurements of the low-energy component of the solar neutrino flux (the so-called $pp$ neutrinos) may enable measurements of $\sin^2\theta_{12}$ ~\cite{Bahcall:2003ce} and probe solar physics. The detection of very low-energy solar neutrinos is very challenging, but R\&D related to building such detectors profits from significant synergy with efforts to look for dark matter and observe neutrinoless double-beta decay. Solar neutrinos in the few-MeV range are very sensitive to solar matter effects, and provide a unique opportunity to test the standard model. Indeed, data from the SNO experiment seem to hint at potential deviations from standard model expectations~\cite{Aharmim:2011vm}. During this decade, more (neutrino) light is expected to shine on this potentially very important matter, from the Borexino~\cite{Alimonti:2008gc} and the SNO+~\cite{Kraus:2010zzb} experiments.

Nuclear reactors are an intense, very pure source of $\bar{\nu}_e$ with energies between a few and several MeV. Due to the low neutrino energies, only $\bar{\nu}_e$ can be detected in the final state, which is done via inverse $\beta$-decay, $\bar{\nu}_e+p \rightarrow e^+ + n$.  The current generation of reactor experiments aims at percent-level measurements of the $\bar{\nu}_e$ spectrum, one or two kilometers away from the source, for which one is sensitive only to $\Delta m^2_{31,32}$-driven oscillations. The necessary precision has been achieved through the comparison of data obtained at near and far detectors.  The near detector measures the neutrino flux before oscillations have had time to act, while the far detector measures the effects of the oscillations~\cite{Abe:2012tg, Ahn:2012nd,An:2012bu,Anderson:2004pk}. Reactor neutrino experiments with much longer baselines (say, 50~km) have been considered: see, e.g.,~\cite{Minakata:2004jt,Petcov:2006gy}. These would be sensitive to both $\Delta m^2_{31,32}$ and $\Delta m^2_{21}$-driven oscillations, and, in principle, would allow much more precise measurements of $\Delta m^2_{21}$ and $|U_{e2}|$. A large reactor experiment with exquisite energy resolution may also be sensitive to the neutrino mass hierarchy (see, e.g.,~\cite{Ghoshal:2010wt}). A concrete proposal for a 60-km reactor neutrino experiment, JUNO, is currently under serious consideration in China~\cite{Kettell:2013eos}, as is a proposal, RENO-50, for South Korea~\cite{reno50}.

Meson decays are a very good source of $\nu_{\mu}$ and  $\bar{\nu}_{\mu}$  and also produce $\nu_{\tau}$ and their antiparticles. The heavy $\tau$-lepton mass, however, prevents any realistic means of producing anything that would qualify as a $\nu_{\tau}$-beam, so we will only discuss $\nu_{\mu}$ beams. Pions and, to a lesser extent, kaons are produced in large numbers through proton--nucleus interactions. These, in turn, can be sign-selected in a variety of ways to yield a mostly pure $\nu_{\mu}$ or $\bar{\nu}_{\mu}$ beam. The neutrino energy is directly related to the pion energy.

The lowest energy $\nu_{\mu}$ ``beams'' (really, isotropic sources) are achieved from pion decay at rest. A large sample of mostly $\pi^+$ at rest yields a very well-characterized flux of mono-energetic $\nu_{\mu}$ (from the $\pi^+$ decay), along with $\bar{\nu}_{\mu}$ and $\nu_e$ from the subsequent daughter muon decay. All neutrino energies are below the muon production threshold, so only $\nu_e$ and $\bar{\nu}_e$ can be detected via charged-current interactions. An interesting experimental strategy is to search for $\bar{\nu}_e$ via inverse $\beta$-decay, a very well understood physics process, and hence measure with good precision $\bar{P}_{\mu e}$~\cite{Conrad:2009mh}. Matter effects play an insignificant role for the decay-at-rest beams, rendering oscillation results less ambiguous. On the other hand, even very precise measurements of $\bar{P}_{\mu e}$ from pion decay at rest are insensitive to the neutrino mass hierarchy.

Boosted pion-decay beams are the gold standard of readily accessible neutrino oscillation experiments. A pion beam is readily produced by shooting protons on a target. These can be charge- and energy-selected, yielding a beam of either mostly $\nu_{\mu}$ or $\bar{\nu}_{\mu}$. Larger neutrino energies allow one to look for $\nu_e$, $\nu_{\mu}$ and, for energies above a few GeV, $\nu_{\tau}$ in the far detector. Large neutrino energies, in turn, require very long baselines\footnote{The oscillation phase scales as $L/E$. For a 1~GeV beam, one aims at $L$ values close to 1000~km.} and hence very intense neutrino sources and very large detectors. Intense neutrino sources, in turn, require very intense proton sources.
For this reason, these pion-decay-in-flight beams are often referred to as superbeams. Larger neutrino energies and longer baselines also imply nontrivial matter effects even for $\Delta m^2_{31}$-driven oscillations. A neutrino beam with energies around 1~GeV and baselines around 1000~km will allow the study of $P_{\mu\mu}$ and $P_{\mu e}$ (and, in principle, the equivalent oscillation probabilities for antineutrinos) as long as the far detector is sensitive to both $\nu_{\mu}$ and $\nu_e$ charged-current interactions.  One may choose to observe the neutrino flux a few degrees
off the central beam axis,  where the pion decay kinematics result in a narrowly-peaked neutrino spectrum. This is beneficial for optimizing sensitivity at the oscillation maximum and for reducing backgrounds outside the energy regime of interest.

The constant collision of cosmic rays with the atmosphere produces mesons (mostly pions and kaons) and, upon their decays, $\nu_{\mu}$, $\bar{\nu}_{\mu}$,  $\nu_e$, $\bar{\nu}_e$. These atmospheric neutrinos cover a very wide energy range (100~MeV to 100~GeV and beyond) and many different distances (15~km to 13000~km), some going through the core of the Earth and hence probing matter densities not available for Earth-skimming neutrino beams. This is, by far, the broadest (in terms of $L/E$ range) neutrino ``beam.''   However, uncertainties in the atmospheric neutrino flux are not small, and the incoming neutrino energy and direction must be reconstructed only with information from the neutrino detector.
In the past, atmospheric neutrinos have provided the first concrete evidence for  neutrino oscillations, and at present they are still a major contributor to the global fits to neutrino oscillation parameters. They will continue to be important in the future. They are also ubiquitous and unavoidable. IceCube DeepCore is already taking data and will accumulate close to a million events with energies above about 10 GeV over the next decade~\cite{FernandezMartinez:2010am}. Any other very large detector associated with the Intensity Frontier program, if sited underground, will also collect a large number of atmospheric neutrino events in various energy ranges, through different types of signatures. While atmospheric neutrino data suffer from larger systematic uncertainties, some of these can be greatly  reduced by studying angular and energy distributions of the very high-statistics data. The study of atmospheric neutrinos can  complement that of the high-precision measurements from fixed-baseline experiments. For example, non-standard  interactions of neutrinos, additional neutrino flavors and other new physics phenomena  affecting neutrinos could be present, and their effects are likely to be more important at higher energies or in the presence of matter (see, e.g.,~\cite{GonzalezGarcia:2011my}). Furthermore, a precise, very high statistics measurement of the atmospheric neutrino flux itself over a very  large range of energies will also contribute to a better understanding of cosmic-ray propagation through the atmosphere~\cite{Gaisser:2002jj,GonzalezGarcia:2006ay,Abbasi:2010ie}.

Muon decays are also excellent sources of neutrinos. The physics and the kinematics of muon decay are very well known and yield two well-characterized neutrino beams for the price of one: $\nu_{\mu}+\bar{\nu}_e$ in case of $\mu^-$ decays,  $\bar{\nu}_{\mu}+\nu_e$ in the case of $\mu^+$. A neutrino factory is a storage ring for muons with a well-defined energy.  Depending on the muon energy, one can measure, with great precision, $P_{\mu\mu}$ and $P_{e\mu}$, assuming the far detector can tell positive from negative muons, potentially along with $P_{\mu e}$ and $P_{ee}$, if the far detector is sensitive to electron charged-current events and can deal with the $\pi^0$ backgrounds, or $P_{\mu\tau}$ and $P_{e\tau}$, if the muon energy is large enough and if the far detector has the ability to identify $\tau$-leptons with enough efficiency. Neutrino factories are widely considered the ultimate sources for neutrino oscillation experiments~\cite{NF:2011aa}, and probably allow for the most comprehensive tests of the standard three-neutrino paradigm.

Finally, nuclei that undergo $\beta$-decay serve as a very well-characterized source of $\nu_e$ or $\bar{\nu}_e$. An intense, highly-boosted beam of $\beta$-decaying nuclei would allow one to study  $P_{e\mu}$. 
Such sources are known as ``$\beta$-beams''~\cite{Zucchelli:2002sa}.
Radioactive sources at rest can also be used for low-energy neutrino experiments (see Sec.~\ref{sec:anomalies}).

To do neutrino oscillation experiments, one must of course detect neutrinos.  Neutrino detectors span a huge range of technologies, some standard for particle physics and others highly specialized.  Detectors are typically quite large, up to multi-kt scale and higher, due to the smallness of neutrino-interaction cross sections.
Specific detector requirements  depend on neutrino energy and physics goals.  In general, good reconstruction capabilities, i.e., ability to reconstruct momenta and particle types of interaction products, are needed.  
For long-baseline beams and atmospheric neutrinos, for which energies are high ($\sim$GeV), a variety of tracking detector technologies can be used, each with advantages and disadvantages.    Commonly-employed detector technologies include segmented trackers (e.g., Soudan, MINOS, NOvA,  ICAL), some of which have magnetic fields to enable interaction-product sign selection, water-Cherenkov detectors (Super-K, Hyper-K), and liquid argon time projection chambers (Icarus, LBNE).  
At the very highest energies, astrophysical neutrino detectors employ enormous volumes of water or ice (IceCube, ANTARES).
For low-energy neutrinos (few to tens of MeV neutrinos from the Sun, reactors, supernovae, stopped-pion sources), homogeneous volumes of liquid scintillator are frequently employed (Borexino, KamLAND, SNO+,  Daya Bay, RENO, Double Chooz, JUNO, LENA).  
For the lowest-energy interaction products, dark-matter WIMP detector technology sensitive to nuclear recoils can be used (see Secs.~\ref{cenns},~\ref{cenns_app}).

Many R\&D activities related to neutrino detection are currently underway~\cite{instrumentation_writeup}.
For neutrino-beam experiments, for which neutrinos can be easily separated from cosmogenic  backgrounds because they tend to arrive in sharp bursts associated with beam pulses, surface detectors are possible.   However for physics involving natural neutrinos or steady-state sources, cosmogenic backgrounds become critical.  Siting underground, away from cosmic rays, then becomes essential~\cite{Bernstein:2009ms}.

Tables~\ref{experimentstable2} and \ref{experimentstable3} summarize the capabilities of current and future neutrino-oscillation experiments.

\begin{table}
\footnotesize
\caption{Types of current or proposed neutrino oscillation experiments along with their accessibility to different oscillation channels. $\surd\surd$ indicates the most important  oscillation channel(s) while $\surd$ indicates other accessible channels. `$\nu_{e,\mu}$ disapp' refers to the disappearance of $\nu_{e}$ or $\nu_{\mu}$ (neutrinos or antineutrinos) which are related to $P_{ee}$ and $P_{\mu\mu}$, respectively.   `$\nu_{\mu}\leftrightarrow\nu_e$' refers to the appearance of $\nu_e$ in a $\nu_{\mu}$ beam or vice versa, related to $P_{e\mu}$ or $P_{\mu e}$. `$\nu_{\tau}$ app' refers to the appearance of $\nu_{\tau}$ from an initial state $\nu_e$ or $\nu_{\mu}$, related to $P_{(e,\mu)\tau}$. `Pion DAR/DIF' refers to neutrinos from pion decay at rest or in flight. `$\mu$ DAR/DIF' and `$\beta$ Beam' refer to neutrinos from muon decay and nuclear decay in flight, respectively. In particular Pion DIF stands for a so-called conventional neutrino beam. For examples of experiments, see Table~\ref{experimentstable}.}
\label{experimentstable2}
\begin{center}
\begin{tabular}{|c|c|c|c|c|c|c|c|c|c|c|c|c|c|} \hline
Expt. Type & $\nu_{e}$~disappearance & $\nu_{\mu}$~disappearance & $\nu_{\mu}\leftrightarrow\nu_{e}$  & $\nu_{\tau}$~appearance$^1$  \\ \hline \hline 
Reactor & $\surd\surd$ & -- & -- & -- \\ \hline
Solar$^2$ & $\surd\surd$ & -- & $\surd$ & -- \\  \hline
Supernova$^3$ & $\surd\surd$ & $\surd$ & $\surd\surd$ & -- \\ \hline
Atmospheric & $\surd$ & $\surd\surd$ & $\surd$ & $\surd$ \\  \hline
Pion DAR & $\surd$ & -- & $\surd\surd$ & -- \\ \hline
Pion DIF & -- & $\surd\surd $ & $\surd\surd$ & $\surd$ \\  \hline
$\mu$ DIF & $\surd$ & $\surd\surd$ & $\surd\surd$ & $\surd$ \\ \hline
Isotope DAR & $\surd$ & -- & & --\\ \hline
$\beta$ beam & $\surd$ & -- & $\surd\surd$ & -- \\  \hline
\end{tabular}
\end{center}
$^1$In order to observe $\nu_{\tau}$ appearance, a dedicated detector or analysis is required, along with a high-enough neutrino energy. $^2$Solar neutrino experiments are sensitive, at most, to the $\nu_e$ and the $\nu_{e}+\nu_{\mu}+\nu_{\tau}$ components of the solar neutrino flux. $\nu_{\mu}\leftrightarrow\nu_{e}$ for this row refers to NC detection and includes oscillation to $\nu_\tau$. $^3$Signatures of neutrino oscillation occurring both in the collapsed star matter and in the Earth will be present in the spectra of observed fluxes of different flavors, and do not strictly fall in these categories; detectors are sensitive to $\nu_e$ and $\bar{\nu}_e$ fluxes, and to all other flavors by NC interactions.
\end{table}

\begin{table}
\footnotesize
\caption{Types of current or proposed neutrino oscillation experiments and their ability to address some of the outstanding issues in neutrino physics. `NSI' stands for non-standard neutrino interactions, while $\nu_s$ ($s$ for sterile neutrino) stands for the sensitivity to new neutrino mass eigenstates (see Sec.~\ref{sec:anomalies}). `$\star\star\star$' indicates a very significant contribution from the current or proposed version of these experimental efforts, `$\star\star$' indicates an interesting contribution from current or proposed experiments, or a significant contribution from a next-next generation type experiment, `$\star$' indicates a marginal contribution from the current or proposed experiments, or an interesting contribution from a next-next generation type experiment. See Table \ref{experimentstable} and text for more details.}
\label{experimentstable3}
\begin{center}
\begin{tabular}{|c|c|c|c|c|c|c|c|c|c|c|c|c|c|c|} \hline
Expt. Type & 
$\sin^2\theta_{13}$ & sign($\Delta m^2_{31}$) & $\delta$ & $\sin^2\theta_{23}$ & $\left|\Delta m^2_{31}\right|$ &  $\sin^2\theta_{12}$ & $\Delta m^2_{21}$ & NSI & $\nu_{s}$\\ \hline \hline 
Reactor &  $\star\star\star$ & $\star\star$& -- & -- & $\star$ & $\star\star$ & $\star\star$ & -- & $\star\star$ \\ \hline
Solar &  $\star$ & -- & -- & -- & -- & $\star\star\star$ & $\star$ & $\star\star$ & $\star\star$ \\ \hline  
Supernova &  $\star$ & $\star \star \star$ & -- & -- & -- & $\star$ & $\star$ & $\star\star$ & $\star\star$ \\ \hline  
Atmospheric & $\star\star$ & $\star\star$ & $\star\star$ & $\star\star$ & $\star\star$ & -- & -- & $\star\star\star$ & $\star\star$ \\ \hline  
Pion DAR &  $\star\star\star$ & -- & $\star\star\star$ & $\star$ & $\star\star$ & $\star$ & $\star$ & -- & $\star\star$ \\ \hline
Pion DIF &  $\star\star\star$ & $\star\star\star$ & $\star\star\star$ & $\star\star$ & $\star\star$ & $\star$ & $\star$ & $\star\star$ & $\star\star$ \\ \hline
$\mu$ DIF & $\star\star\star$ & $\star\star\star$ & $\star\star\star$ & $\star\star\star$ & $\star\star\star$ & $\star$ & $\star$ & $\star\star$ & $\star\star$ \\ \hline
Isotope DAR & -- & -- & & & & & & $\star\star$ & $\star\star$ \\ \hline
$\beta$ Beam & $\star\star\star$ & -- & $\star\star\star$ & $\star\star$ & $\star\star$ & $\star$ & $\star$ & -- & $\star\star$ \\ \hline
\end{tabular}
\end{center}
\end{table}

%% file: nu1.tex
%-----------------------------------------------------------------------
\section{The standard oscillation paradigm}
\label{sec:3nus}

The three-flavor oscillation framework is quite successful in
accounting for a large number of results obtained in very different
contexts: the transformation of $\nu_e$ into $\nu_{\mu,\tau}$ from the
Sun~\cite{Aharmim:2011vm}; the disappearance of $\nu_\mu$ and
$\bar\nu_\mu$ from neutrinos produced by cosmic-ray interactions in
the atmosphere~\cite{Wendell:2010md,Abe:2011ph}; the disappearance of $\nu_\mu$ and
$\bar\nu_\mu$ from neutrino beams
over distances from
200-740\,km~\cite{Ahn:2006zza,Adamson:2012rm,Abe:2012gx}; the
disappearance of $\bar\nu_e$ from nuclear reactors over a distance of
about 160\,km~\cite{Abe:2008aa}; the disappearance of $\bar\nu_e$
from nuclear reactors over a distance of about 2\,km~\cite{Abe:2012tg,Ahn:2012nd,An:2012bu}. Now also the appearance of
$\nu_e$~\cite{Abe:2013xua,Adamson:2011qu} and, at relatively low
significance, the appearance of $\nu_\tau$~\cite{Abe:2012jj,Agafonova:2013dtp} have
been observed.   All these experimental results can be succinctly
and accurately described by the oscillation of three active neutrinos
governed by the following parameters, including their $1\,\sigma$
ranges from a global fit~\cite{Fogli:2012ua}\footnote{See~\cite{Fogli:2012ua} for more details. When it comes to the ``large'' mass-squared difference, different experiments, in principle, are most sensitive to different linear combinations of $\Delta m^2_{31}$ and $\Delta m^2_{32}$. Throughout this document, however, we will refer to these different quantitates as $\Delta m^2_{32}$, unless otherwise noted, as the current data, and most of the data expected from near-future efforts, are not precise enough to be sensitive to the slight differences.}
\begin{eqnarray}
\label{eq:nu1:parameters}
\Delta m^2_{21}=7.54^{+0.26}_{-0.22}\times10^{-5}\,\mathrm{eV}^2\,,(3.2\%)&\quad&
\Delta m^2_{32}=2.43_{+0.1}^{-0.06}\times10^{-3}\,\mathrm{eV}^2\,,(3.3\%)\\
\sin^2\theta_{12}=3.07_{-0.16}^{+0.18}\times10^{-1}\,,(16\%)&\quad&\sin^2\theta_{23}=3.86_{-0.21}^{+0.24}\times10^{-1}\,,(21\%) \nonumber\\
\sin^2\theta_{13}=2.41\pm0.25\times10^{-1}\,,(10\%)&\quad&\delta/\pi=1.08_{-0.31}^{+0.28}\,\mathrm{rad}\,, \nonumber
\end{eqnarray}
where for all parameters whose value depends on the mass hierarchy, we
have chosen the values for the normal mass ordering. The choice of
parametrization is guided by the observation that for those parameters
the $\chi^2$ in the global fit is approximately Gaussian, except for $\delta$. The
percentages given in parentheses indicate the relative error on each
parameter. For the mass splitting we reach errors of a few percent;
however, for all of the mixing angles the errors are
in the 10-30\% range, while the CP-odd phase is unconstrained at the 2$\sigma$ level.   The mass hierarchy and octant of $\theta_{23}$ (i.e., whether
$\theta_{23}$ is smaller or larger than $\pi/4$) are not constrained at all.
Therefore, while three-flavor oscillation is
able to describe a wide variety of experiments, it would seem
premature to claim that we have entered the era of precision neutrino
physics or that we have established the three-flavor paradigm at a
high level of accuracy. This is also borne out by the fact that there
are interesting hints at short baselines for a fourth
neutrino~\cite{Abazajian:2012ys}. Also, more generally, so-called
non-standard interactions (NSI) are not well constrained by neutrino data;
for a recent review on the topic see Ref.~\cite{Ohlsson:2012kf}. The
issue of what may exist beyond three-flavor oscillations will be
discussed in detail in Sec.~\ref{sec:anomalies} of this report.

The next question is: how well do we want to determine the
various mixing parameters? The answer can be given on two distinct
levels. One is a purely technical one --- if I want know $X$ to a
precision of $x$, I need to know $Y$ with a precision of $y$.
For example, $Y$ could be given by $\theta_{13}$ and $X$ could be the
mass hierarchy.  At another level, the answer is driven by theory
expectations of how large possible phenomenological deviations from
the three-flavor framework could be. In order to address the technical
part of the question, one first has to define the target precision
from a physics point of view. Guidance from other subareas of particle
physics reveals that the target precision evolves over time.
For example, history shows that before the top quark discovery, theoretical estimates of the top quark mass from electroweak precision data
and other indirect observables
seem to have been, for the most part (and with very large uncertainties), only several GeV ahead of the experimental
reach --- at the time, there always was a valid physics
argument for why the top quark was ``just around the corner.'' A similar
evolution of theoretical expectations can be observed in, e.g., searches for
new phenomena in quark flavor physics. Thus, any argument based on
model-building-inspired target precisions is always of a preliminary
nature, as our understanding of models evolves over time. With this
caveat in mind, one argument for a target precision can be based on a
comparison to the quark sector. Based on theoretical guidance from
Grand Unification, one would expect that the answer to the flavor
question should find a concurrent answer for leptons and quarks. 
Therefore, tests of such models are most 
sensitive if the precision in the lepton and quark sector is
comparable. For instance, the CKM angle $\gamma$, which is a very close
analog of $\delta$ in the neutrino sector, is determined to
$(70.4^{+4.3}_{-4.4})^\circ$~\cite{Lenz:2012az} and thus, a precision
target for $\delta$ of roughly $5^\circ$ would follow.

Beyond those very general arguments, one has to look at specific
models, and each model presumably will yield a different answer. In the
context of neutrino physics this problem is exacerbated by the fact
that we currently have no experimental evidence for the scale of
physics responsible for neutrino masses and this, in turn, limits the
number of models which have clear, fully worked-out predictions. In
the following, we will show \emph{one} specific example, but the
example was chosen to also highlight certain general features. In
general, symmetries imply structure and structure implies well defined
relationships between the physical parameters of a theory. A
significant test of these relationships requires considerable
precision, especially if the goal is to distinguish between models or
to determine the underlying symmetries.
Neutrino sum rules~\cite{King:2005bj} arise, for example, in models where the neutrino mixing matrix
has a certain simple form or texture at a high energy scale and the
actual low-energy mixing parameters are modified by a non-diagonal
charged lepton mass matrix. The simplicity of the neutrino mixing
matrix is typically a result of a flavor symmetry, where the overall
Lagrangian possesses an overall flavor symmetry $G$, which can be
separated into two sub-groups $G_\nu$ and $G_l$ for the neutrinos and
charged leptons; it is the mismatch between $G_\nu$ and $G_l$ which
will yield the observed mixing pattern; see, 
  e.g.,~\cite{Altarelli:2010gt}. Typical candidates for $G$ are given
by discrete subgroups of SU(3) which have a three-dimensional
representation, e.g., $A_4$. In a model-building sense, these
symmetries can be implemented using so-called flavon fields which
undergo spontaneous symmetry breaking, and it is this symmetry breaking
which picks the specific realization of $G$; for a recent review
see~\cite{King:2013eh}. The idea of flavor symmetries is in stark
contrast to the idea that neutrino mixing parameters are anarchic,
i.e., random numbers with no underlying dynamics. For the most
recent version of this argument, see Ref.~\cite{deGouvea:2012ac}. To
find out whether the patterns observed in lepton mixing correspond to an underlying symmetry
is one of the prime tasks of neutrino physics.
Of course, distinguishing among the many candidate underlying symmetries is also a very high priority.

\begin{figure}[ht]
\begin{center}
\includegraphics[width=0.8\textwidth]{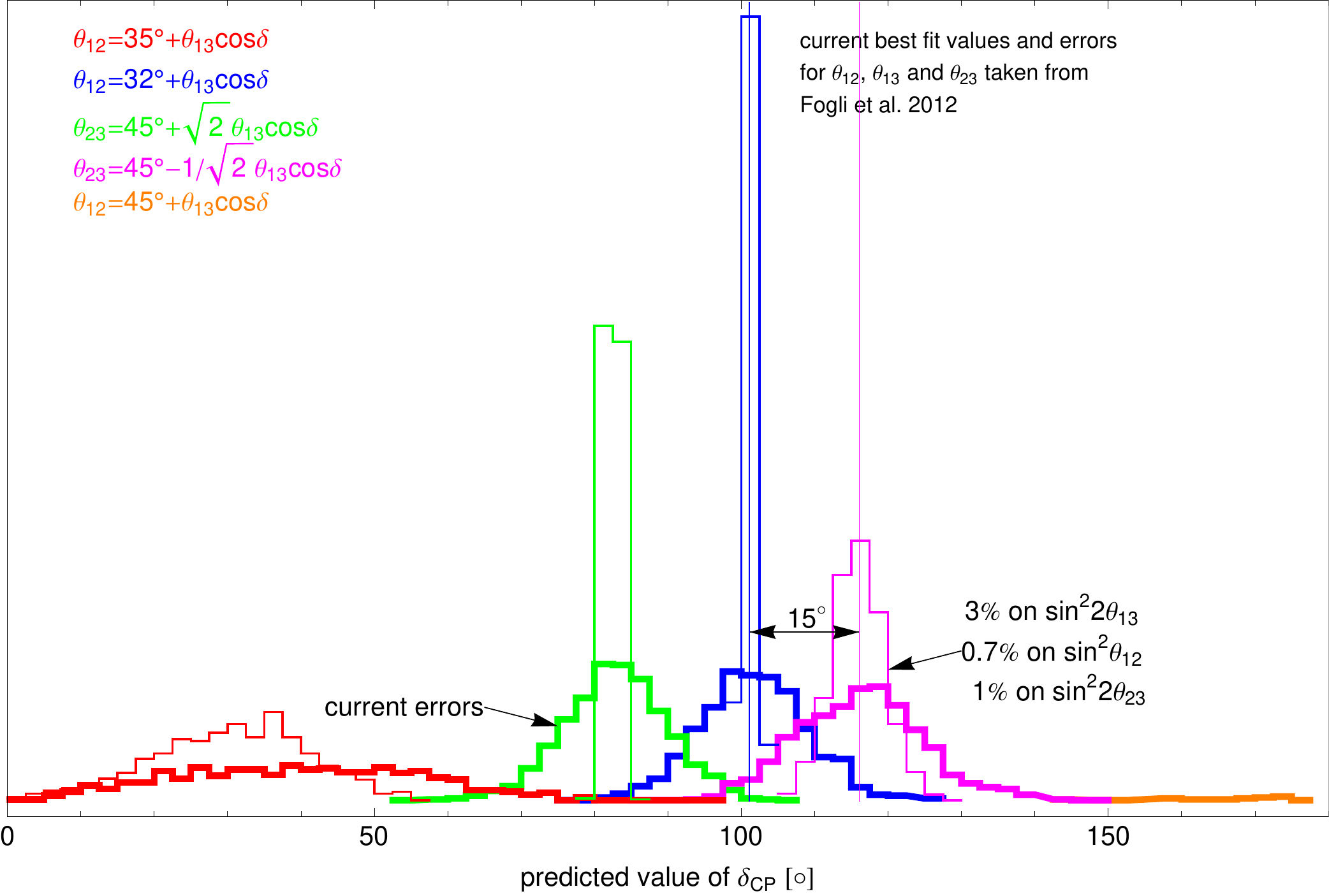}
\end{center}
\caption{\label{fig:nu1:sumrules} Shown are the distributions of
  predicted values for $\delta$ from various sum rules as denoted in
  the legend and explained in the text.}
\end{figure}

In practice, flavor symmetries will lead to relations between
measurable parameters, whereas anarchy will not. For example, if the
neutrino mixing matrix is of tri-bi-maximal form,
$|U_{e3}|=0$ is naively expected to vanish, which is clearly in contradiction to observations. In
this case, a non-diagonal charged lepton mass matrix can be used to
generate the right value of $|U_{e3}|$.  For one concrete model, the following sum rule
arises:
\begin{equation}
\label{nu1:eq:sumrule}
\theta_{12}-\theta_{13}\cos\delta=\arcsin \frac{1}{\sqrt{3}}\,,
\end{equation}
which can be tested if sufficiently precise measured values for the
three parameters $\theta_{12},\theta_{13},\delta$ are available.
Depending on the underlying symmetry of the neutrino mixing matrix,
different sum rules are found. In Fig.~\ref{fig:nu1:sumrules} several
examples are shown and for each case the values of $\theta_{13}$ and
$\theta_{12}$ or $\theta_{23}$ are drawn many times from a Gaussian
distribution where the mean values and ranges are taken from
Eq.~\ref{eq:nu1:parameters}. The resulting predictions of the value of
the CP phase $\delta$ are histogrammed and shown as colored lines.  The
width of the distribution for each sum rule arises from the finite
experimental errors on $\theta_{12}$ or $\theta_{23}$ and $\theta_{13}$.
Two observations arise from this simple comparison: first, the distance
between the means of the distributions is as small as $15^\circ$, and
second, the width of the distributions is significant compared to
their separation and a reduction of input errors is mandated.  The
thin lines show the results if the errors are reduced to the value
given in the plot, which would be achieved by Daya Bay for
$\sin^22\theta_{13}$, by JUNO for $\sin^2\theta_{12}$, and by
NOvA for $\sin^2\theta_{23}$. Assuming that the errors on
$\theta_{12}$, $\theta_{23}$ and $\theta_{13}$ are reduced to this
level, the limiting factor is the natural spread between models, which is
about $15^\circ$. A $3\,\sigma$ distinction between models
translates into a target precision for $\delta$ of $5^\circ$. A
measurement at this precision would allow one to obtain valuable
information on whether indeed there is an underlying symmetry behind
neutrino mixing. Moreover, it is likely to also provide
hints regarding which specific class of symmetries is realized.  This would
constitute a major breakthrough in our understanding of flavor.

For the parameter $\sin^22\theta_{13}$ the {\it status quo} is
determined by the results from the reactor experiments Double
Chooz~\cite{Abe:2012tg}, Daya Bay~\cite{An:2012bu} and
RENO~\cite{Ahn:2012nd} and their results agree well. It is expected
that Double Chooz will improve its systematic error by a significant
amount with the planned addition of a near detector by the end of
2013. Daya Bay started running in its full eight-detector
configuration only in the fall of 2012 and it is expected that a three-year run with all detectors will eventually reach a 3\% error on
$\sin^22\theta_{13}$, compared to currently about 12.5\% on this
parameter~\cite{TheDayaBay:2013kda}. Of all beam experiments, only a neutrino factory will be
able to match this precision~\cite{Coloma:2012wq}. A comparison of the
values of $\theta_{13}$ obtained in $\bar\nu_e$ disappearance at
reactors with the result of $\nu_e$ and $\bar\nu_e$ appearance in
beams will be a sensitive test of the three-flavor framework, which is
particularly sensitive to new physics, such as non-standard matter effects and sterile neutrinos (e.g.,~\cite{Qian:2013ora}).

For the atmospheric $\Delta m^2_{32}$, currently the most precise
measurement comes from MINOS~\cite{Adamson:2012rm} with an error of
$3.2\%$ and MINOS+~\cite{Plunkett:2013} will slightly improve on this
result. It is expected that both NOvA and T2K will contribute
measurements with errors of $\sim3\%$ and $\sim4\%$, respectively.
Daya Bay will provide a measurement of this parameter in $\bar\nu_e$
disappearance of about $4\%$. By increasing the size of the event
sample and going to an off-axis location, CHIPS~\cite{Adamson:2013xka} (see next section) has the
potential to reduce the current error by perhaps as much as a factor
2-3, which is of course subject to sufficient control of systematic
errors and needs further study. JUNO~\cite{Kettell:2013eos} ultimately
may have the potential to bring the error down to below one percent.
For $\theta_{23}$, two related but distinct questions arise.  First, what is the
precise value of $\sin^22\theta_{23}$ or how close it is to unity? 
Secondly, if $\sin^22\theta_{23}\neq1$, is $\theta_{23}$ smaller or
larger than $\pi/4$, i.e., what is the so-called octant of $\theta_{23}$?  An
experiment can be very good at determining the value of
$\sin^22\theta_{23}$ without obtaining any information on the octant
question. The resolution of the octant question can be either achieved
by comparing long-baseline data obtained at different baselines, like
NO$ \nu$A and T2K, or by comparing a precise $\nu_\mu\rightarrow\nu_e$
long-baseline measurement with a precise determination of
$\bar\nu_e\rightarrow\bar\nu_e$ oscillations from a reactor experiment
like Daya Bay. Within the U.S. program, the initial long-baseline pieces of
data can come from the NuMI beam, and NOvA is well positioned to provide information, as
would be potential extensions of the NuMI program in the form of
extended NOvA running~\cite{Plunkett:2013}, RADAR~\cite{Adamson:2013jsa} and CHIPS~\cite{Adamson:2013xka}.  
Eventually, the Long-Baseline Neutrino-Experiment (LBNE), with its
very long (1,300~km) baseline and wide beam spectrum, will provide good
sensitivity to the octant on its own. NOvA and T2K have the
potential to reduce the error on $\sin^22\theta_{23}$ to 1-2\% and
most likely further improvements in beam experiments will
require an improved understanding of systematics.

For the solar $\Delta m^2_{21}$, the current uncertainties are determined by
KamLAND, and a future improvement is necessary to measure the mass
hierarchy without using matter effects as proposed by JUNO~\cite{Kettell:2013eos} and RENO-50~\cite{reno50}.
Such experiments may able to reduce the error to below 1\%. The solar mixing
parameter $\sin^2\theta_{12}$ has been most accurately measured by SNO and Super-K, and some improvement may be possible with further solar neutrino measurements.
A promising method
relies on the observation of $\bar\nu_e$ disappearance at a distance
of about 60\,km as proposed in JUNO and RENO-50, with the potential to
bring the error to below 1\%~\cite{Kettell:2013eos}. The value of $\theta_{12}$ and its
associated error play an important role for sum rules, as explained
previously, and also for neutrinoless double $\beta$-decay.

\subsection{Towards the determination of the neutrino mass hierarchy}\label{mh}

The recently observed ``large'' value of $\theta_{13}$ has opened the possibility of determining, mostly using matter effects, the mass hierarchy through a variety of different experiments and observations. This includes accelerator-based neutrino oscillation experiments, atmospheric neutrino detectors, as well as reactor antineutrino experiments, and observations of astrophysical neutrinos from supernovae, as well as cosmology. A broad suite of experiments has been proposed to study the mass hierarchy using these possibilities 
and R\&D is underway to address the viability of these options.  It is possible that one or more of these experiments will be able to make an unambiguous determination of the mass hierarchy in the next decade. More likely, we will obtain a suite of results with indications that may point to the ordering of the neutrino mass eigenstates in a joint analysis. Now that we know the size of $\theta_{13}$, a measurement of the neutrino mass hierarchy is within reach and may well be one of the next big milestones in neutrino physics~\cite{Cahn:2013taa}. 

\subsubsection{Mass hierarchy from oscillations and other observables}\label{mhother}

The neutrino mass hierarchy manifests itself in different types of phenomena, most of which are potentially observable in neutrino oscillation experiments. We review them here, before discussing the reach of different types of experiments and opportunities for the near and intermediate future.

If all mixing angles are nonzero, the neutrino mass hierarchy manifests itself in all oscillation probabilities, including those associated with neutrinos propagating in vacuum. This can be quickly understood via a concrete example. The survival probability of, say, electron neutrinos in vacuum is given by
\begin{equation}
P_{ee}=1-\left[A^e_{21}\sin^2\left(\frac{\Delta m^2_{21}L}{4E}\right)+A^e_{31}\sin^2\left(\frac{\Delta m^2_{31}L}{4E}\right)+A^e_{32}\sin^2\left(\frac{\Delta m^2_{32}L}{4E}\right)\right],
\end{equation}
where $A^e_{ij}\equiv4|U_{ei}|^2|U_{ej}|^2$. A measurement of $P_{ee}$ capable of establishing that there are three (related) oscillation frequencies can determine the mass hierarchy as long as the three $A^e_{ij}$ are nonzero and distinct (and known). This comes from the fact that under these circumstances one can tell whether $|\Delta m^2_{31}|>|\Delta m^2_{32}|$ or vice-versa. For the normal mass hierarchy $|\Delta m^2_{31}|>|\Delta m^2_{32}|$ as one can readily see from Fig.~\ref{3nus_pic}, with the situation reversed for the inverted mass hierarchy. For a more detailed discussion see, e.g.,~\cite{deGouvea:2005hk}. The fact that $|\Delta m^2_{31}|\gg\Delta m^2_{21}$ and $\sin^2\theta_{13}\ll 1$ renders such a measurement, in practice, very hard as, for almost all experimental setups, observations are very well-described by an effective two-flavor oscillation scheme, completely blind to the mass hierarchy. A large reactor neutrino experiment with exquisite energy resolution and an intermediate baseline (around 50~km) should be able to see the interplay of all oscillation terms with  $\Delta m_{31}^2$ and  $\Delta m_{32}^2$ and would be sensitive to the mass hierarchy.

Matter effects allow one to probe the mass hierarchy in a different way, as already discussed in Sec.~\ref{sec:osc}. Electron-type  neutrinos interact with electrons differently from muon-type and tau-type neutrinos. As neutrinos propagate inside a medium filled with electrons, the neutrino dispersion relation, and hence the oscillation probabilities, are modified in a way that can distinguish electron-type neutrinos from muon-type or tau-type neutrinos. This translates into a sensitivity to whether the mass eigenstates containing ``more'' electron-type neutrinos --- $\nu_1$ and $\nu_2$ --- are lighter (normal hierarchy) or heavier (inverted hierarchy) than the eigenstates containing ``less'' electron-type neutrinos --- $\nu_3$. Such a measurement is possible even for very small $\Delta m^2_{12}$, as long as $\theta_{13}$ is not vanishingly small and one is probing oscillations of or into electron-type neutrinos. In practice, sensitivity to matter effects requires small values of $|\Delta m^2_{32}|/E$ and, since one requires $L$ such that $|\Delta m^2_{32}|L/E$ is large enough, long distances. For neutrino energies around 1~GeV, $L$ values of order at least several hundred kilometers are required.

Core-collapse supernovae (SN) from massive stars are an abundant source of neutrinos of all flavors: see Sec.~\ref{nu6_supernova}, and matter effects are abundant and qualitatively different from the ones encountered anywhere else (except, perhaps, for the very early universe).
There are multiple possible signatures sensitive to mass hierarchy in the supernova neutrino flux, e.g.,~\cite{Takahashi:2002cm, Lunardini:2003eh, Dasgupta:2008my, Chakraborty:2011ir,Dighe:2003be,Choubey:2010up}.
During neutrino emission from the SN core the MSW effects are encountered twice at high and low density,  and the resulting flavor conversion depends on the neutrino mass hierarchy in addition to the star's density, neutrino energy,  and the oscillation parameters. In addition, shock waves in the SN envelope and Earth matter effects can impact the observed neutrino spectra.  Shock waves change the adiabatic to non-adiabatic conversion and multiple MSW effects take place. They occur either in the $\nu_e$ or $\bar{\nu}_{e}$ channel and depend on the mass hierarchy. Turbulence can have similar effects as shock waves. In addition, neutrino conversion can take place near the neutrinosphere due to $\nu$-$\nu$ interactions. The conversion probability is energy dependent and may introduce a spectral split.
Model-dependent effects in the emitted SN spectrum will have to considered  in the use of SN data for a mass hierarchy determination.

Finally, observables outside of neutrino oscillations sensitive to the neutrino masses themselves, as opposed to only mass-squared differences, are also in principle sensitive to the neutrino mass hierarchy. Some of these are discussed in Secs.~\ref{sec:majorana}, \ref{sec:mass}, \ref{nu6}. For example, if the sum of all neutrino masses were constrained to be less than around 0.1~eV, the inverted mass hierarchy hypothesis would be ruled out. Such a sensitivity (or better) is expected from several next-generation probes of the the large-scale structure of the universe, as will be discussed in more detail in Sec.~\ref{nu6}.

\subsubsection{Experimental approaches}

\begin{figure}[ht]
\begin{center}
\includegraphics[width=0.28\textwidth]{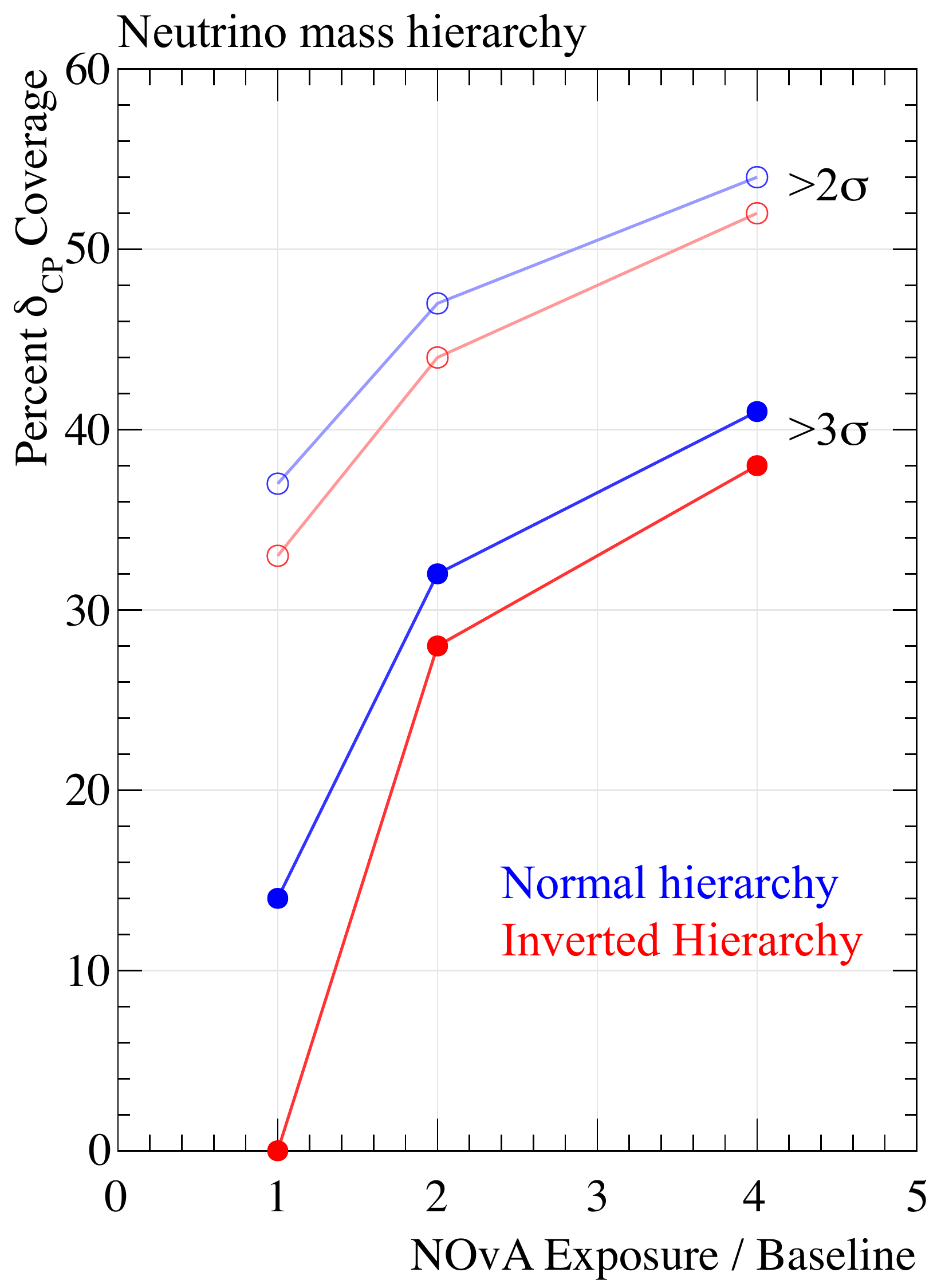}
\includegraphics[width=0.42\textwidth]{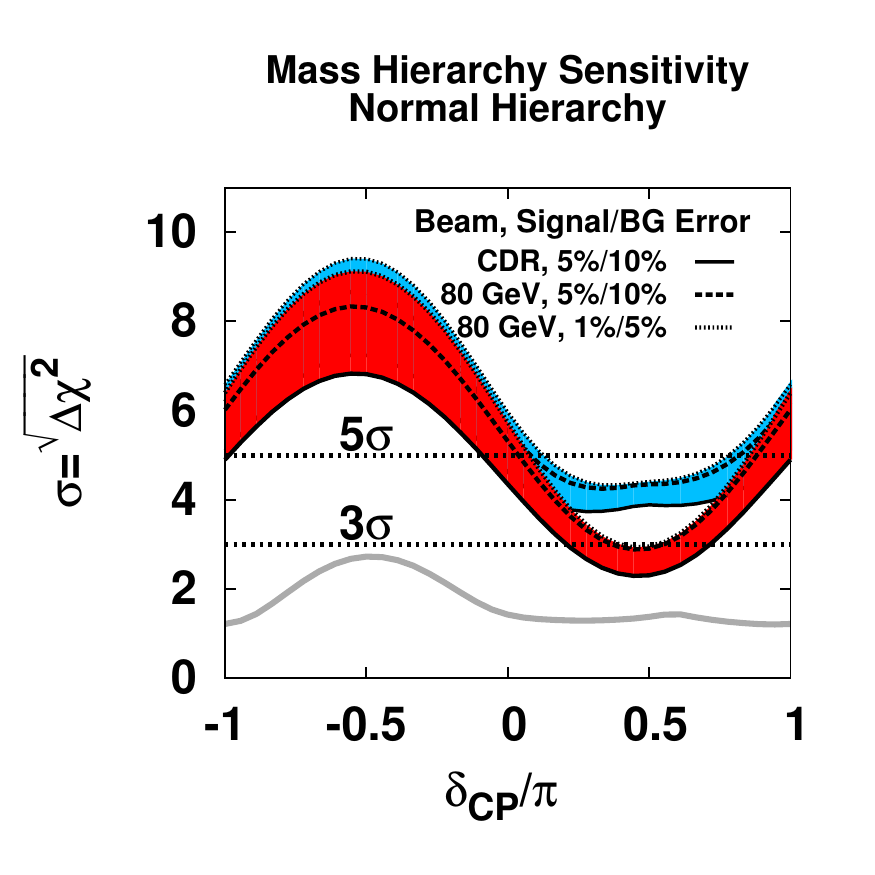}
\end{center}
\caption{\label{fig:nu1:MHacc} 
Left: Percent of $\delta_{CP}$  values for which NOvA can resolve the neutrino mass hierarchy at 2 and 3 $\sigma$ CL, as a function of factor multiplying baseline exposure~\cite{Messier:2013sfa}. NOvA is in construction and has started data taking with a partial detector configuration. Right: Significance with which mass hierarchy can be determined as a function of $\delta_{CP}$ (assuming normal hierarchy), for different combinations of experiments. Grey curve (bottom): T2K+NOvA; red curve (middle): LBNE10; blue curve (topmost): LBNE10+T2K+NOvA.  The beam exposure assumed is 5+5 years ($\nu+\bar{\nu}$) in a 708-kW beam for LBNE10; for NOvA the assumption is 3+3 ($\nu+\bar{\nu}$) and for T2K the assumption is $5\times10^{21}$~protons on target.
  T2K is operational and taking data. NOvA is in the commissioning phase and will finish construction in 2014. LBNE10 is the 10-kt detector currently in preliminary design and R\&D and preparing for Critical Decision 2. Figure from~\cite{Adams:2013qkq}. 
}
\end{figure}

{\bf Accelerator experiments:}
Ongoing and future accelerator experiments are a key element in a program to determine the neutrino mass hierarchy. Very intense beams of muon neutrinos from pion sources can be used to search for electron neutrino appearance. For intermediate and long baselines the appearance probability will depend on the ordering of the neutrino mass states. The upcoming NOvA experiment together with T2K will have a chance of determining the neutrino mass hierarchy  with accelerator neutrinos for a range of oscillation parameters. In the long term, the Long-Baseline Neutrino Oscillation experiment (LBNE) or experiments at neutrino factories will allow the definitive measurement of the neutrino mass hierarchy. See Fig.~\ref{fig:nu1:MHacc}. The proposed Long-Baseline Neutrino Oscillation (LBNO) experiment in Europe would also have very good sensitivity.  The CHIPS and RADAR proposals seek to exploit the NuMI beam from FNAL with new detectors at baselines similar to MINOS and NOvA. The experimental advantages of LBNE
include an optimum baseline from the neutrino source to the detector,  a large and sophisticated far detector, a high-power, broadband, sign-selected muon neutrino beam, and a highly-capable near neutrino detector. If placed underground, the LBNE far detector may even allow the possibility of atmospheric neutrino studies and oscillation measurements through a channel with different systematics than the accelerator-based experiments.  Optimization of the LBNE baseline to determine the mass hierarchy with no ambiguities depends only on the known oscillation parameters. To achieve mass hierarchy sensitivity over all phase space requires a baseline $>$1000~km.
It has been proposed~\cite{Qian:2013nhp} that a second detector at an off-axis location for LBNE will further enhance mass hierarchy determination capability.

\begin{figure}[ht]
\begin{center}
\includegraphics[width=1.0\textwidth]{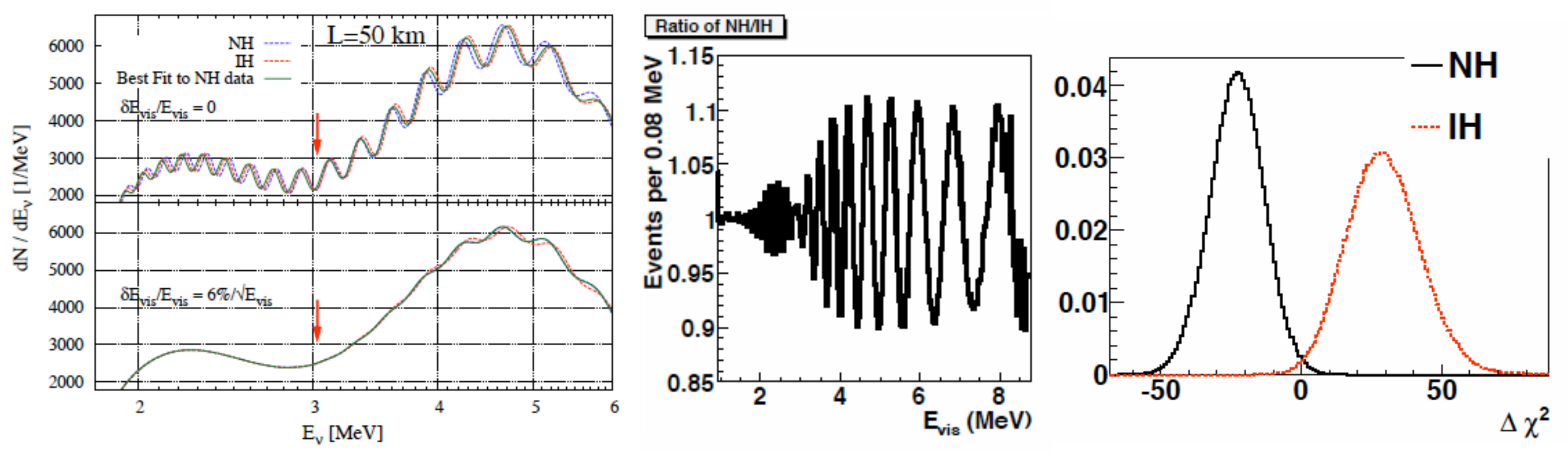}
\end{center}
\caption{\label{fig:nu1:MHreactors}
Left: Energy distribution of reactor antineutrinos with baseline length of 50~km. The solid line shows  the best fit of IH assumption to the NH data. The red arrow points out the energy at which the difference due to the mass hierarchy vanishes. The lower panel shows the effect of 6\% energy resolution. Figure from~\cite{Ge:2012wj}. Middle: Ratio of reactor antineutrino spectra for NH and IH case for the ideal energy spectrum without fluctuation and fixed $\Delta m^2_{32}$ . Statistical fluctuations,  the unknown true value of  $\Delta m^2_{32}$, as well as experimental effects such as energy scale uncertainty, will degrade the observable effect.  Right: The $\Delta \chi^2$ spectrum  from Monte Carlo simulation. The probability of the mass hierarchy being NH is calculated as $P_{NH}/(P_{NH}+P_{IH})$ and found to be 98.9\% for 100-kt-year exposure. Figures from~\cite{Qian:2012xh}.}
\end{figure}

{\bf Reactor experiments:}  The success of recent reactor experiments in the measurement of $\theta_{13}$ at baselines of $\sim$1~km has resulted in proposals for the precision study of neutrino oscillation at medium baselines of 50-60~km. 
A measurement of the neutrino mass hierarchy with reactor antineutrinos requires excellent detector energy resolution of $\sim$3\% and  absolute energy scale calibration at $<$1\% Degeneracies caused by the current experimental uncertainty of 
$|\Delta m^2_{32}|$ have to be considered.  Several recent studies have considered the sensitivity of reactor experiments at medium baseline~\cite{Ghoshal:2010wt,Kettell:2013eos,Capozzi:2013psa,Ciuffoli:2013pla,Ciuffoli:2013rza,Ghosh:2012px,Qian:2012xh,Qian:2012zn}.
See Fig.~\ref{fig:nu1:MHreactors}. 
Two experiments are currently proposed to make this measurement: JUNO in China and RENO-50 in South Korea. The current design of RENO-50 includes a 18-kt liquid scintillator detector $\sim$47~km from a $\sim$17-GWth power plant. JUNO proposes a 20~kt liquid scintillator detector $\sim$700~m underground and $\sim$60~km from two nuclear power plants with $\sim$40~GWth power. 

\begin{figure}[ht]
\begin{center}
\includegraphics[width=0.33\textwidth]{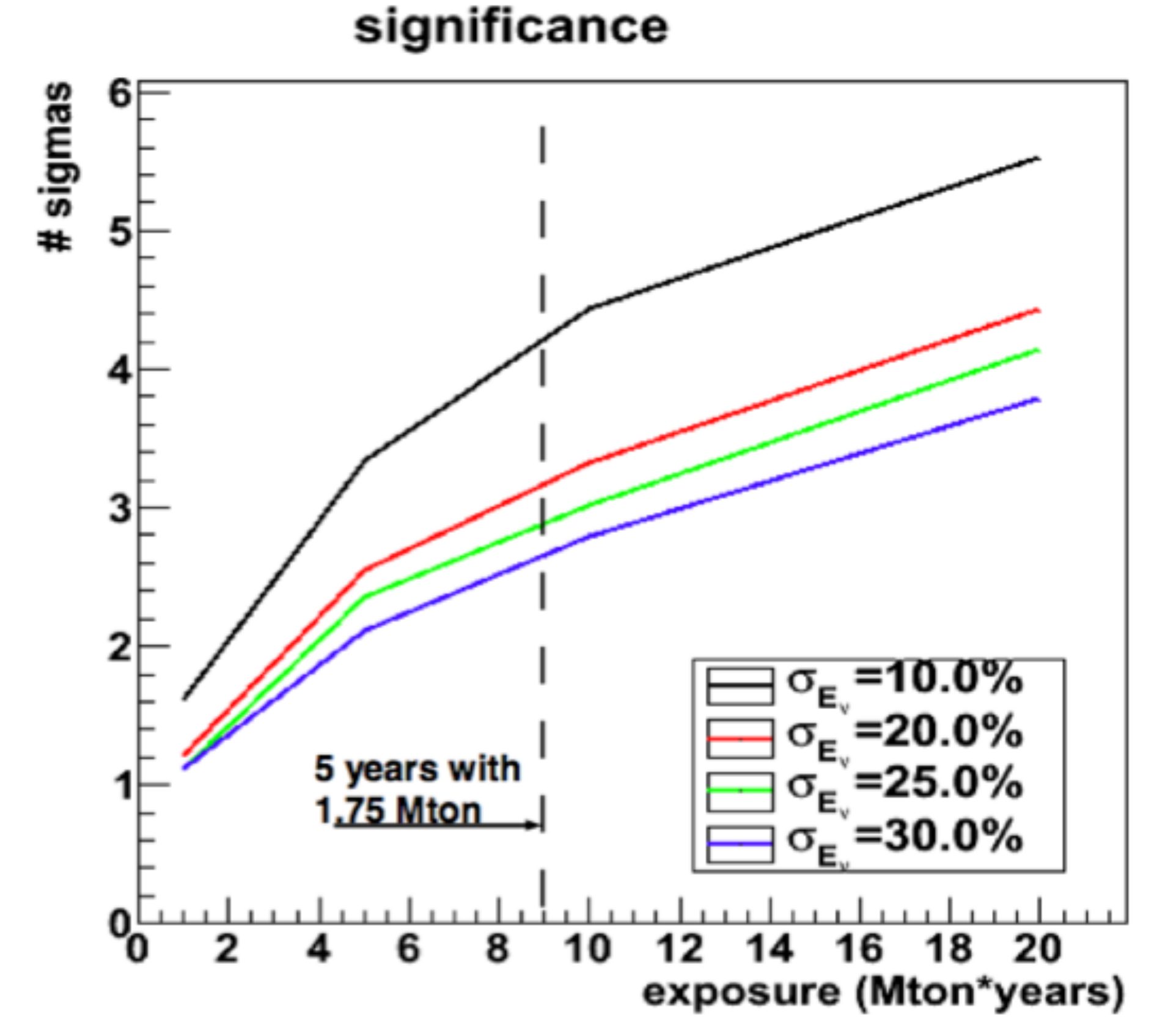}
\includegraphics[width=0.42\textwidth]{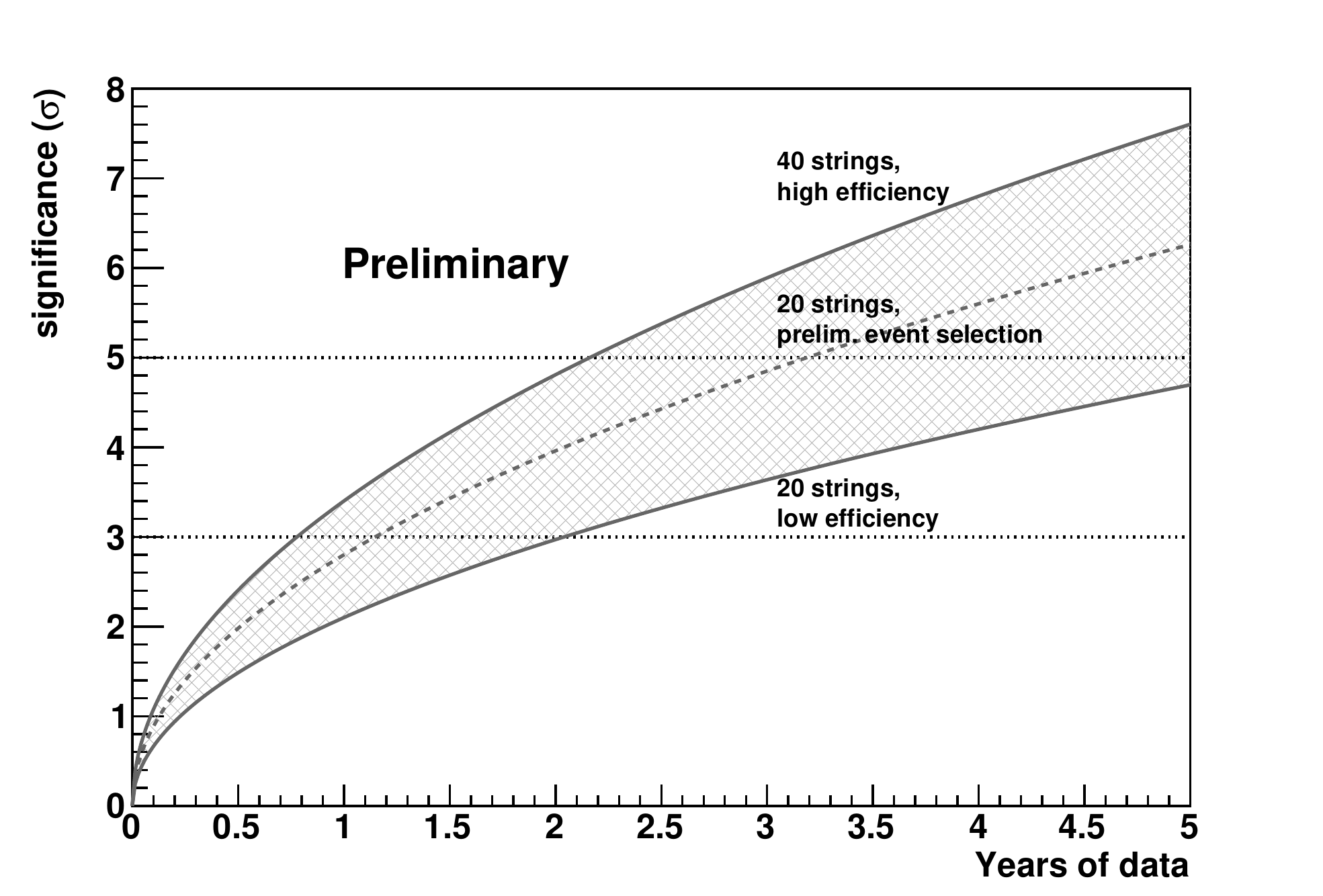}
\end{center}
\caption{\label{fig:nu1:MHatm} 
Preliminary sensitivities of the ORCA~\cite{ORCA-Aspen} (left) and PINGU~\cite{IceCube:2013aaa} (right) proposals to mass hierarchy as a function of exposure.  Independent evaluations of the proposals sensitivities indicate varying levels of significance depending on the choice of the true mass hierarchy and experimental input assumptions (e.g.,~\cite{Cahn:2013taa}).
}
\end{figure}

{\bf Atmospheric neutrino experiments:} 
Atmospheric neutrinos remain an important probe of neutrino oscillations and the large statistics that can be collected by large Cherenkov detectors at the Mton-scale such as Hyper-K, MEMPHYS, PINGU, and ORCA will offer an an unprecedented opportunity to study them in detail. Atmospheric neutrinos exist in both neutrino and antineutrino varieties in both muon and electron flavors. Up to 10$^6$  events are expected to be collected in a 10-year period in half-megaton detectors such as Hyper-K.  There are two experimental approaches to the study of the mass hierarchy with atmospheric neutrinos. One approach is based on charge discrimination and distinguishes between  neutrinos and antineutrinos. Large magnetized calorimeters such as ICAL at INO~\cite{Thakore:2013xqa} with good energy and angular resolution and thresholds of 1-2 GeV are an example of this type of detector. The second approach uses water Cherenkov detectors and makes use of the different cross-sections and different $\nu$ and $\overline{\nu}$ fluxes.   Examples of future water Cherenkov detectors include Hyper-K~\cite{Abe:2011ts,Kearns:2013lea}, a larger version of the successful water-based Super-K detector, MEMPHYS~\cite{Agostino:2012fd}, ORCA~\cite{ORCA-Aspen}, an extension of ANTARES in the Mediterranean Sea, and PINGU, an upgrade of the IceCube DeepCore detector at the South Pole~\cite{Winter:2013ema,IceCube:2013aaa}.  Atmospheric neutrino measurements are also possible in large liquid argon TPCs such as that being planned for LBNE~\cite{Adams:2013qkq}. Key to the measurement of the mass hierarchy with these experiments will be a large statistical sample collected in a large fiducial volume,  good energy and angular resolution for the study of the L/E oscillation effects and discrimination of backgrounds. See Fig.~\ref{fig:nu1:MHatm} and references~\cite{IceCube:2013aaa,Ghosh:2013mga,Ribordy:2013xea,Franco:2013in,Barger:2012fx}  for a number of independent studies on the mass hierarchy sensitivity of atmospheric neutrino experiments. 

{\bf Supernova neutrinos:}  A suite of neutrino observatories is currently operational worldwide with a variety of target materials including water or ice (Super-K, IceCube), liquid scintillator (KamLAND, Borexino, Daya Bay, LVD), and lead (HALO)~\cite{Scholberg:2012id}. They offer several detection channels through the scattering of $\bar{\nu}_{e}$ with protons, the $\nu_{e}$ scattering with nuclei and $\nu_x$ interactions with electrons and nucleons. Together they have the ability to measure the SN flux at different thresholds and different flavor sensitivities, although most current detectors are primarily  sensitive to $\bar{\nu}_e$.
Future detectors will have broader flavor sensitivity; in particular liquid argon will be valuable for observation of the $\nu_e$ component of the flux, which may have unique sensitivity to the hierarchy~\cite{Adams:2013qkq}.   

\subsubsection{Summary of experimental status and opportunities for mass hierarchy measurement}

The measurement of large $\theta_{13}$ has opened a broad range of possibilities for the determination of the neutrino mass hierarchy. Several experiments with complementary approaches have been proposed that will allow us to determine the neutrino mass hierarchy in oscillation experiments using neutrinos from accelerators, reactors, or the atmosphere.   NOvA is the only funded oscillation experiment underway to start an experimental investigation of the neutrino mass hierarchy in a range of the allowed parameter space.  T2K is taking data; it has relatively low sensitivity
due to its short baseline, although a combination of T2K and NOvA will improve mass hierarchy sensitivity. For some of the recent proposals  under consideration significant R\&D and design work is still required.  A dedicated experiment to measure the neutrino mass hierarchy with atmospheric or reactor neutrinos may be feasible by 2018. After 2022, with data from the LBNE experiment it should be possible to determine the neutrino mass hierarchy for the entire range of CP values.  In the meantime, $0\nu\beta\beta$ and direct neutrino mass experiments combined with data from cosmology may also tell us about the hierarchy if $\sum m_{\nu}$ is measured to be less than 0.1~eV~\cite{Abazajian:2013oma}.  A supernova event detected in one or several of the existing large neutrino observatories would enable a rich physics program and may allow the determination of the ordering of the neutrino mass states, although astrophysics and uncertainties in the supernova models may make this challenging.  Table~\ref{tab:nu1:MHexp} summarizes the status of the ongoing and proposed experiments.

\begin{table}[htbp]
   \centering
  \begin{tabular}{l  l  l  l}
      \hline
     Category & Experiment  & Status & Oscillation parameters\\
    \hline
      Accelerator & MINOS+~\cite{Plunkett:2013} & Data-taking & MH/CP/octant \ \\ 
      Accelerator & T2K~\cite{Abe:2013xua} & Data-taking & MH/CP/octant \ \\ 
    Accelerator  & NOvA~\cite{Ayres:2007tu} & Commissioning  &  MH/CP/octant  \\
      Accelerator  & RADAR~\cite{Adamson:2013jsa} & Design/ R\&D  & MH/CP/octant \\
      Accelerator  & CHIPS~\cite{Adamson:2013xka} & Design/ R\&D & MH/CP/octant\\
     Accelerator  & LBNE~\cite{Adams:2013qkq} & Design/ R\&D & MH/CP/octant \\
      Accelerator & Hyper-K~\cite{Kearns:2013lea} & Design/ R\&D & MH/CP/octant  \\ 
      Accelerator  & LBNO~\cite{Stahl:2012exa} & Design/ R\&D & MH/CP/octant \\
      Accelerator  & ESS$\nu$SB~\cite{Baussan:2013ema} & Design/ R\&D & MH/CP/octant \\
  Accelerator  & DAE$\delta$ALUS~\cite{Aberle:2013ssa} & Design/ R\&D & CP \\
\hline
     Reactor & JUNO~\cite{Kettell:2013eos} & Design/R\&D  &  MH  \\
      Reactor & RENO-50~\cite{reno50} & Design/R\&D  &  MH  \\
 \hline
      Atmospheric & Super-K~\cite{Wendell:2010md} & Data-taking &  MH/CP/octant \\
      Atmospheric & Hyper-K~\cite{Kearns:2013lea} & Design/R\&D &  MH/CP/octant \\
      Atmospheric & LBNE~\cite{Adams:2013qkq} & Design/R\&D &  MH/CP/octant \\
  Atmospheric & ICAL~\cite{Thakore:2013xqa} & Design/R\&D & MH/octant   \\
      Atmospheric & PINGU~\cite{IceCube:2013aaa} & Design/R\&D &  MH \\
      Atmospheric & ORCA~\cite{ORCA-Aspen} & Design/R\&D &  MH \\
      Atmospheric & LAGUNA~\cite{Rubbia:2010zz}  & Design/R\&D &  MH/CP/octant \\
\hline
      Supernova & Existing and future~\cite{Scholberg:2012id}  & N/A &   MH \\
      \hline
   \end{tabular}
   \caption{Ongoing and proposed oscillation experiments for the measurement of neutrino oscillation parameters.  The last column indicates sensitivity to unknown oscillation parameters.  (Note that many of these experiments can improve precision on known parameters as well.)}
   \label{tab:nu1:MHexp}
\end{table}

\subsection{Towards the determination of CP violation in neutrinos}

The standard approach to measuring CP violation in neutrinos is to use 
long-baseline beams of both neutrinos and antineutrinos. As for the mass hierarchy
determination, nature provides beams of atmospheric neutrinos and antineutrinos
free of charge, over a wide range of energies and baselines--- the catch is 
that one has no control over their distribution and so one must measure their 
properties precisely, and/or gather immense statistics in order to extract
information on CP violation from these sources. Alternate approaches include 
using well-controlled, well-understood accelerator-based beams of $\sim$GeV neutrinos 
or else lower-energy neutrinos from pion decay-at-rest sources. Here, we will 
discuss the CP reach of all three possibilities: accelerator-based long-baseline
neutrinos, atmospheric neutrinos, and pion decay-at-rest sources.

\subsubsection{CP violation with accelerator-based long-baseline neutrinos}

The study of $\nu_{\mu}\to\nu_e$ and $\bar{\nu}_{\mu}\to\bar{\nu}_e$ transitions using accelerator-based beams is
sensitive to CP-violating phenomena arising from the CP-odd phase $\delta$ in the
neutrino mixing matrix. The evidence for CP violation (assuming $\delta\neq 0,\pi$) manifests itself 
both as an asymmetry in the oscillation of neutrinos and antineutrinos and as a distortion in the electron-type (anti)neutrino energy spectrum. For experiments with neutrino energies above muon-production threshold and that need to tag the muon-type neutrino flavor at production or detection, baselines longer than 100~km are required. For long-enough baseline (see Sec.~\ref{mh}), the matter effects also induce an asymmetry in the oscillation of neutrinos and antineutrinos. The matter asymmetry, however, is largest for higher neutrino energies and hence maximal at the first oscillation maximum, whereas the CP asymmetry induced by $\delta$ is more significant at the secondary oscillation nodes and is constant as a function of baseline. An
experiment with a wide-band beam of neutrinos and antineutrinos that can
cover at least two oscillation nodes over a long enough baseline ($>1000$~km)
can unambiguously determine both the mass hierarchy and the CP phase
simultaneously. This is the philosophy behind the Long-Baseline Neutrino
Experiment (LBNE)~\cite{Adams:2013qkq}. Additionally, the study of $\nu_\mu \rightarrow \nu_e$ 
oscillations can help determine the $\theta_{23}$ octant since the 
oscillation probability is also proportional to $\sin^2\theta_{23}$.

\begin{figure}[ht]
\begin{center}
\includegraphics[width=0.4\textwidth]{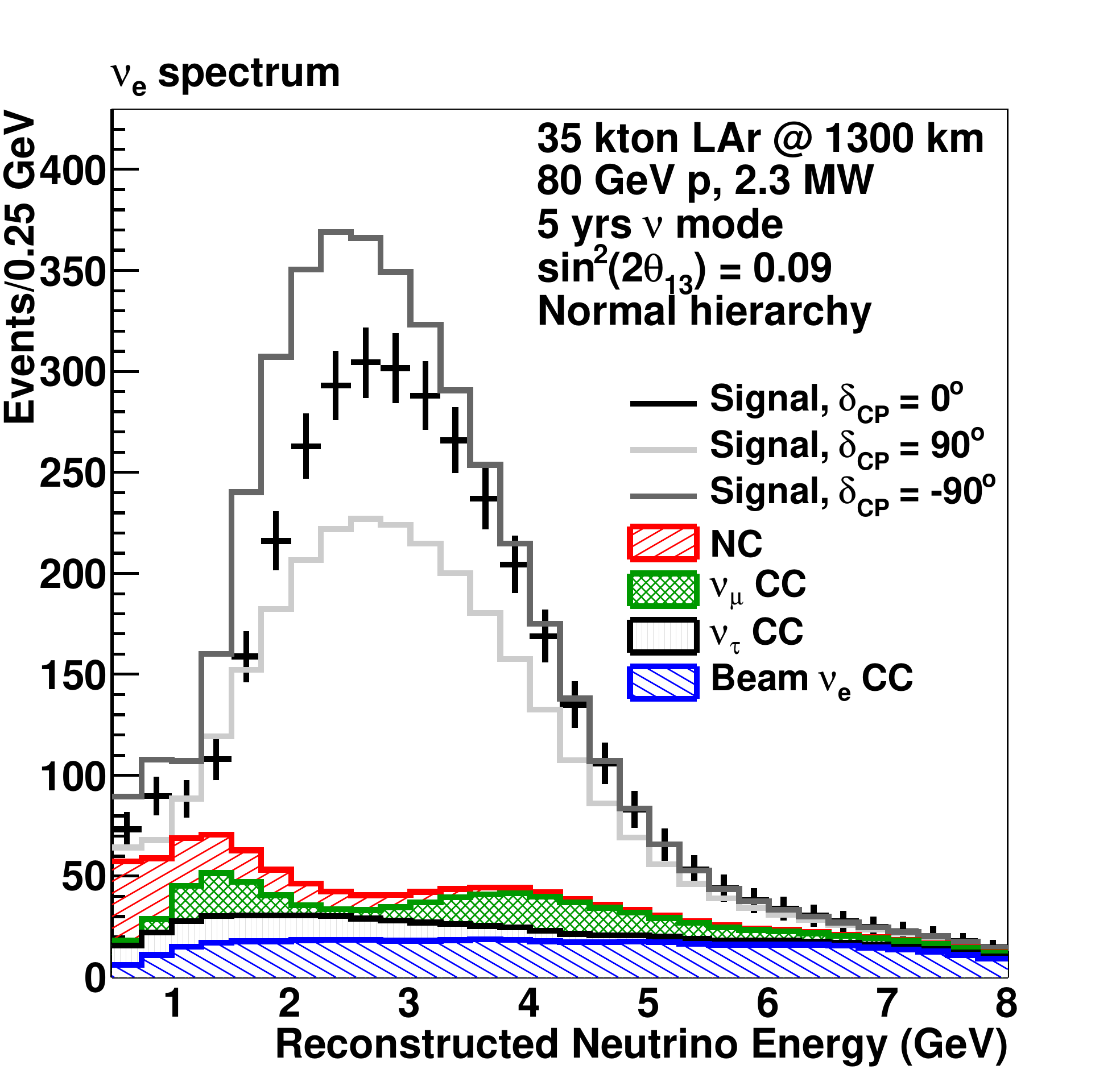}
\includegraphics[width=0.4\textwidth]{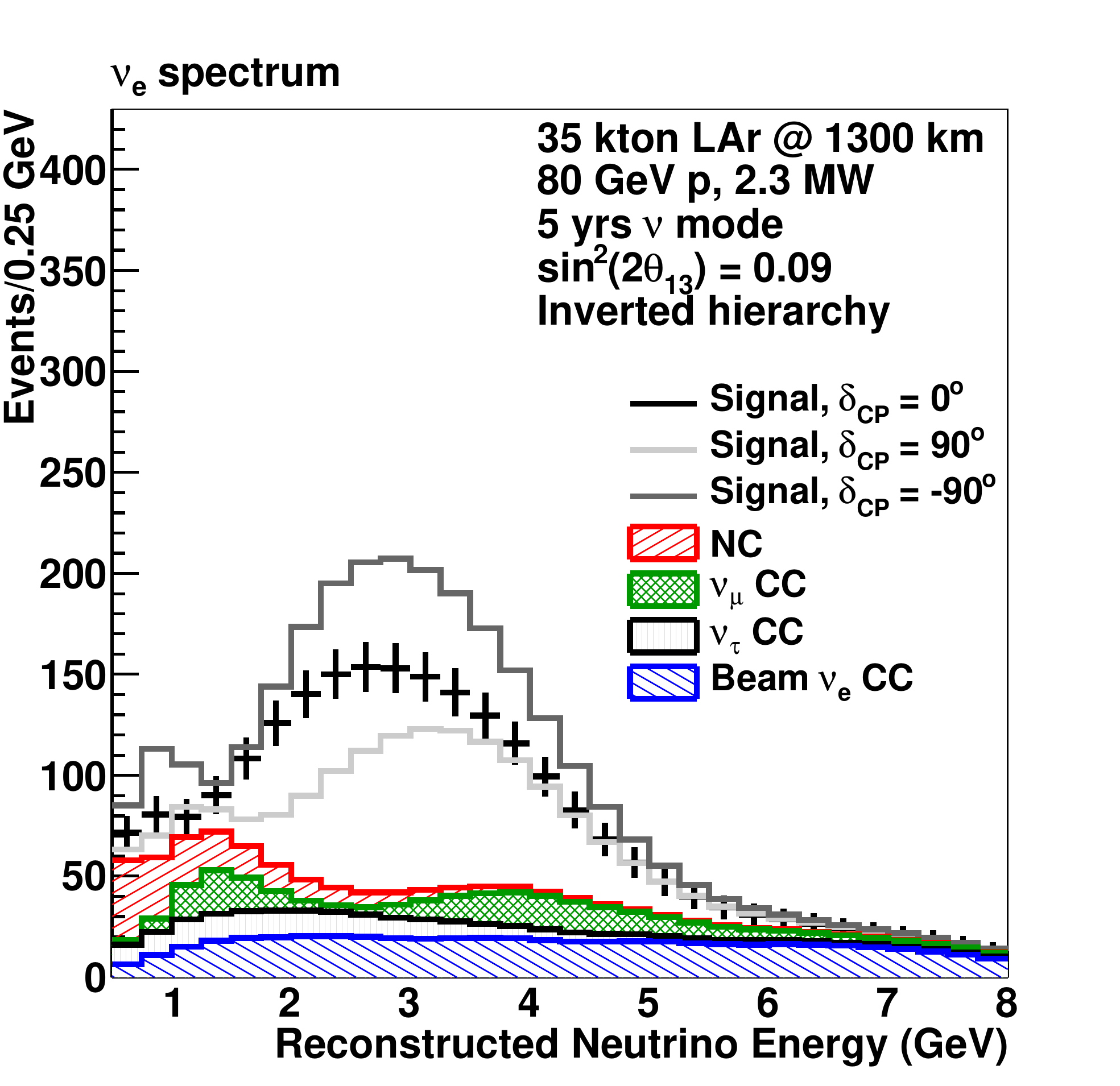}
\includegraphics[width=0.4\textwidth]{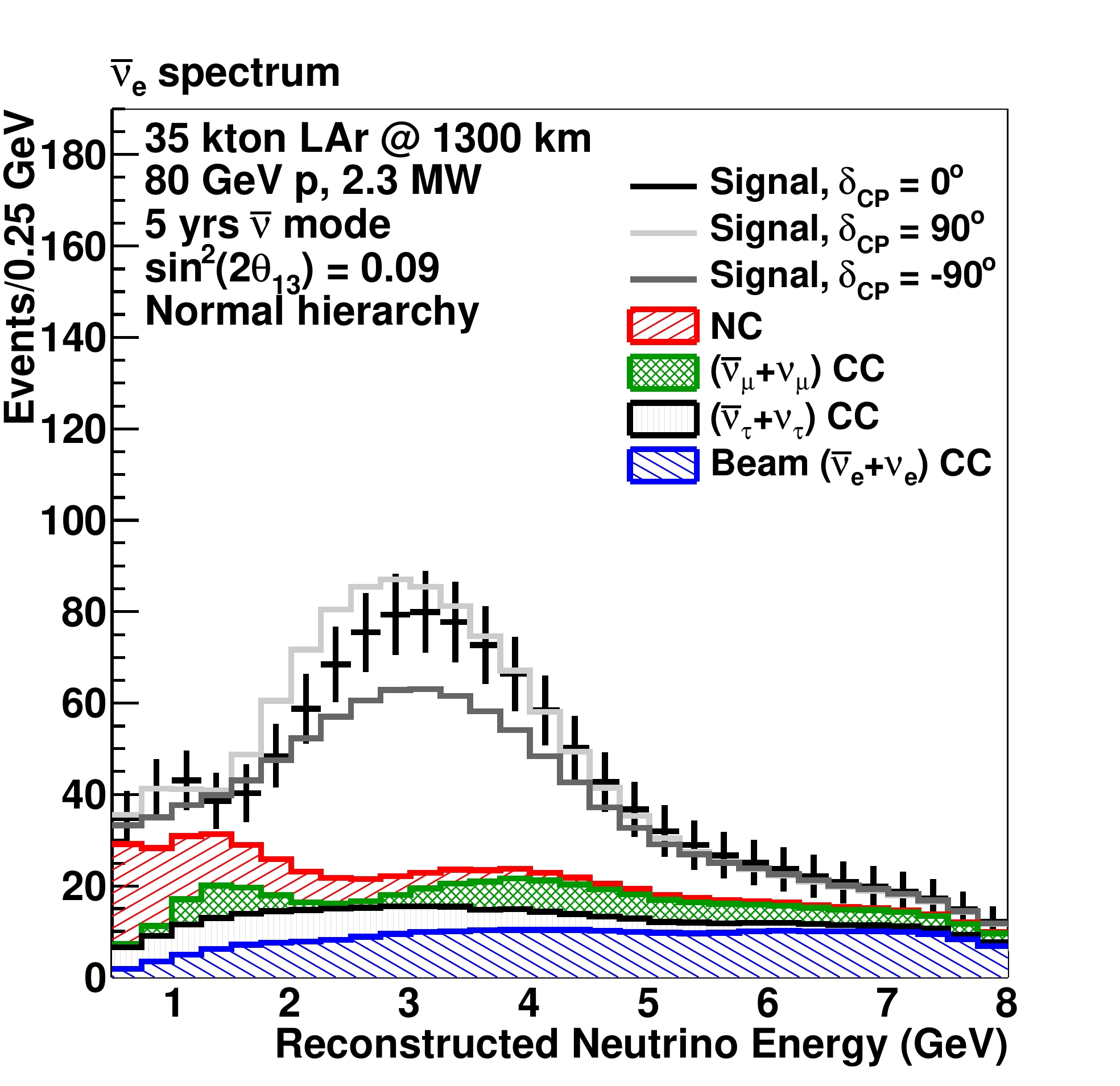}
\includegraphics[width=0.4\textwidth]{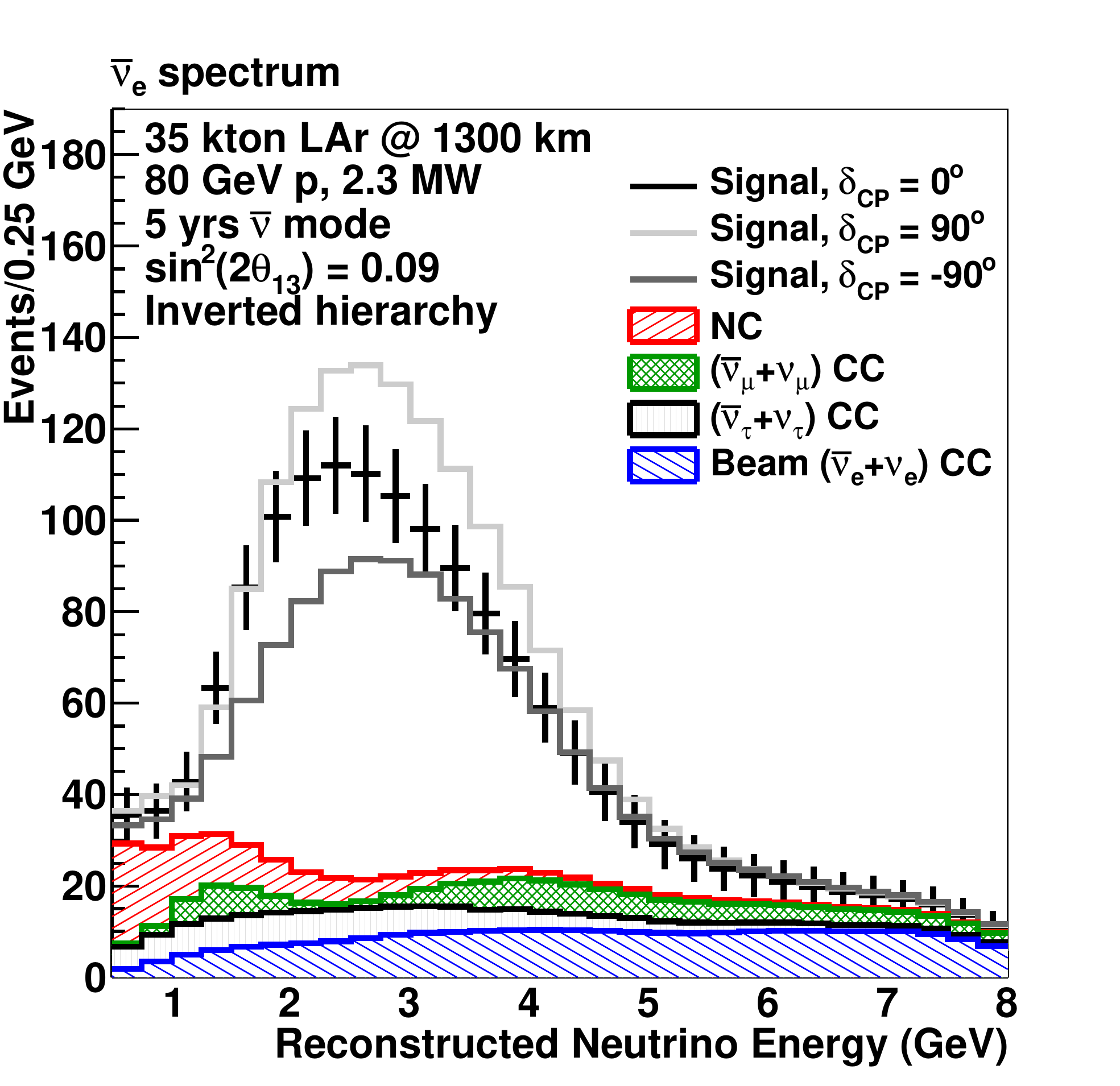}
\end{center}
\caption{\label{fig:nue_signal} {The expected appearance of $\nu_e$ (top)
and $\bar{\nu}_e$ (bottom) signals for the possible mass orderings 
(left: normal hierarchy, right: inverted hierarchy) and varying values of
CP $\delta$ for the example of LBNE/Project X. Figures from~\cite{Adams:2013qkq}.}}
\end{figure}

Figure~\ref{fig:nue_signal} shows examples of observed spectra for a 
1300-km baseline and a beam of a few GeV (the LBNE/Project X configuration 
with a LAr TPC far detector) for $\nu_e$ and $\bar{\nu}_e$ appearance.   
Different values of $\delta_{CP}$ correspond to different spectral shapes 
for neutrinos versus antineutrinos; also, the $\nu_e$ signal is larger in 
neutrinos for the normal mass hierarchy and in antineutrinos for the inverted 
hierarchy. Good event reconstruction and rejection of background are critical 
for this measurement. In the case of LBNE, a LAr TPC was chosen as the far
detector technology, given its excellent 3D position resolution and superior
particle identification in large volumes. In addition to detailed
event topologies and measurements of particle kinematics, such detectors
can also unambiguously distinguish electrons from photons over a wide range
of energies, an important asset in the precision measurement of CP-violating
effects in $\nu_\mu \rightarrow \nu_e$ oscillations.
\begin{figure}[ht]
\begin{center}
\includegraphics[width=0.35\textwidth]{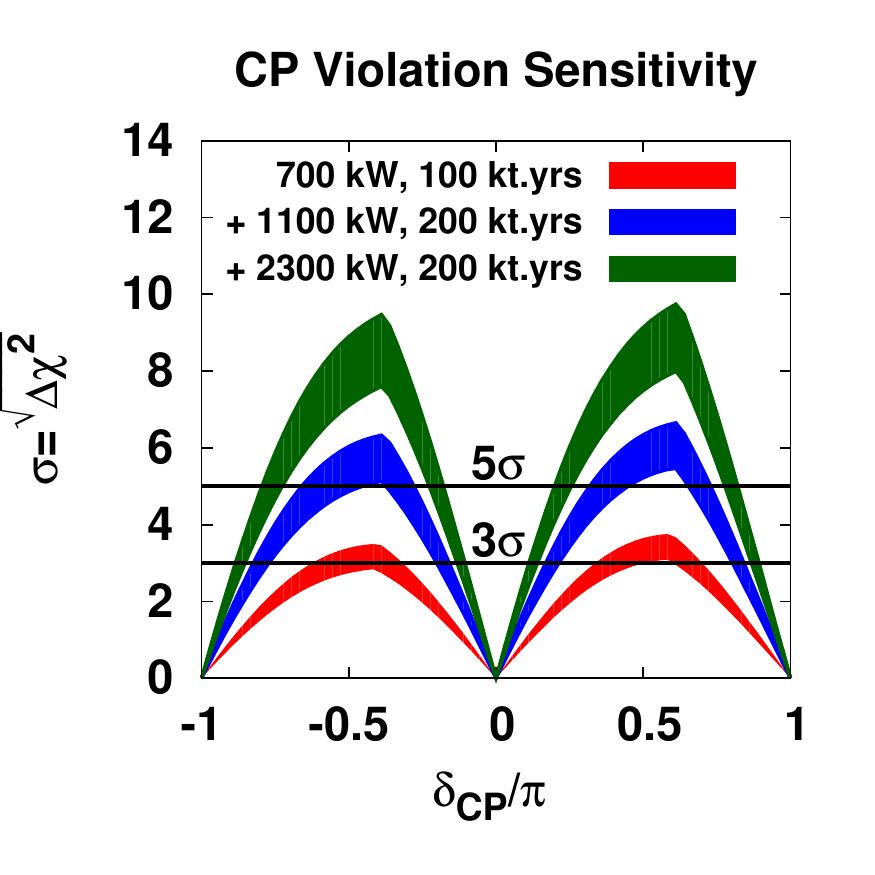}
\includegraphics[width=0.35\textwidth]{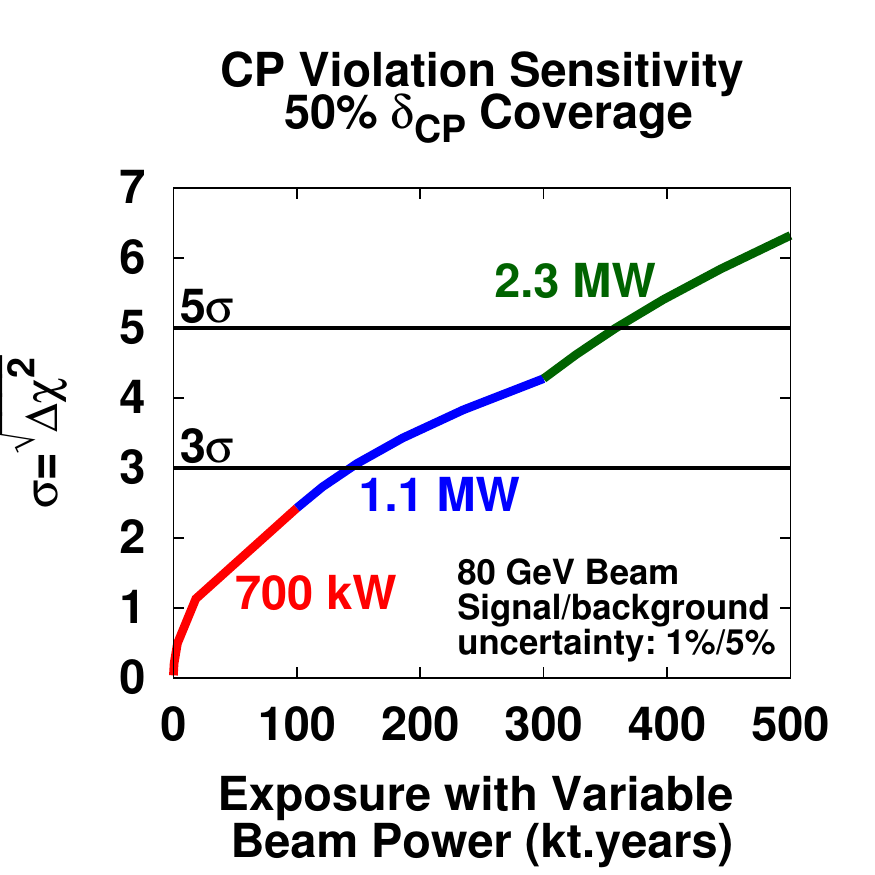}
\includegraphics[width=0.35\textwidth]{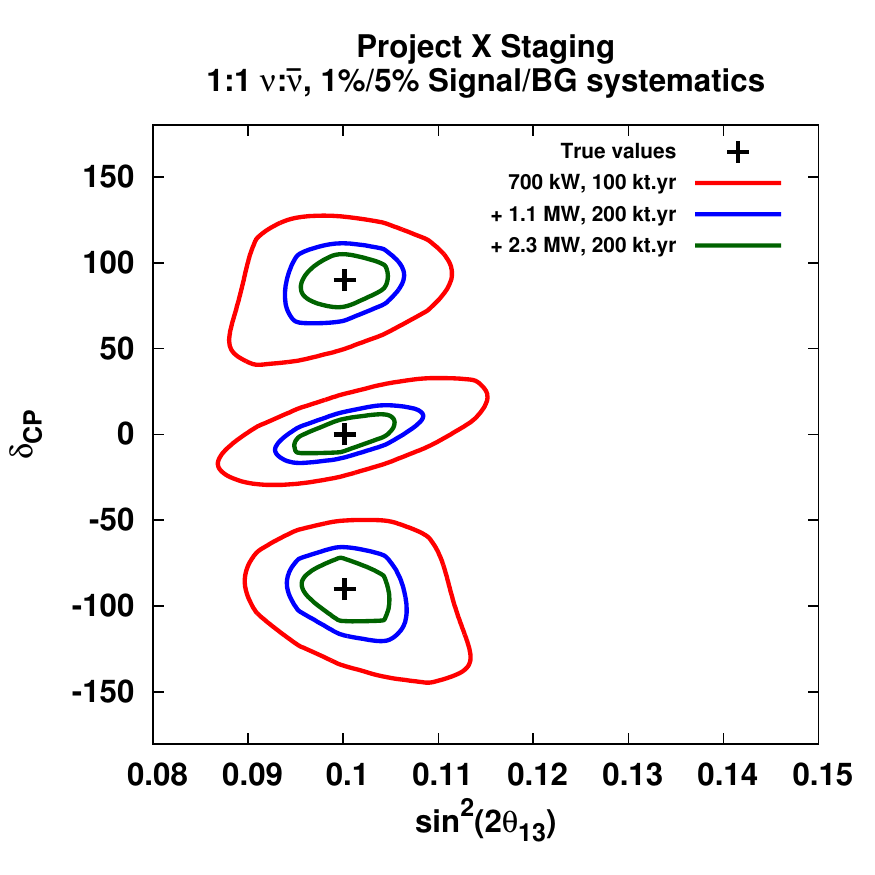}
\includegraphics[width=0.35\textwidth]{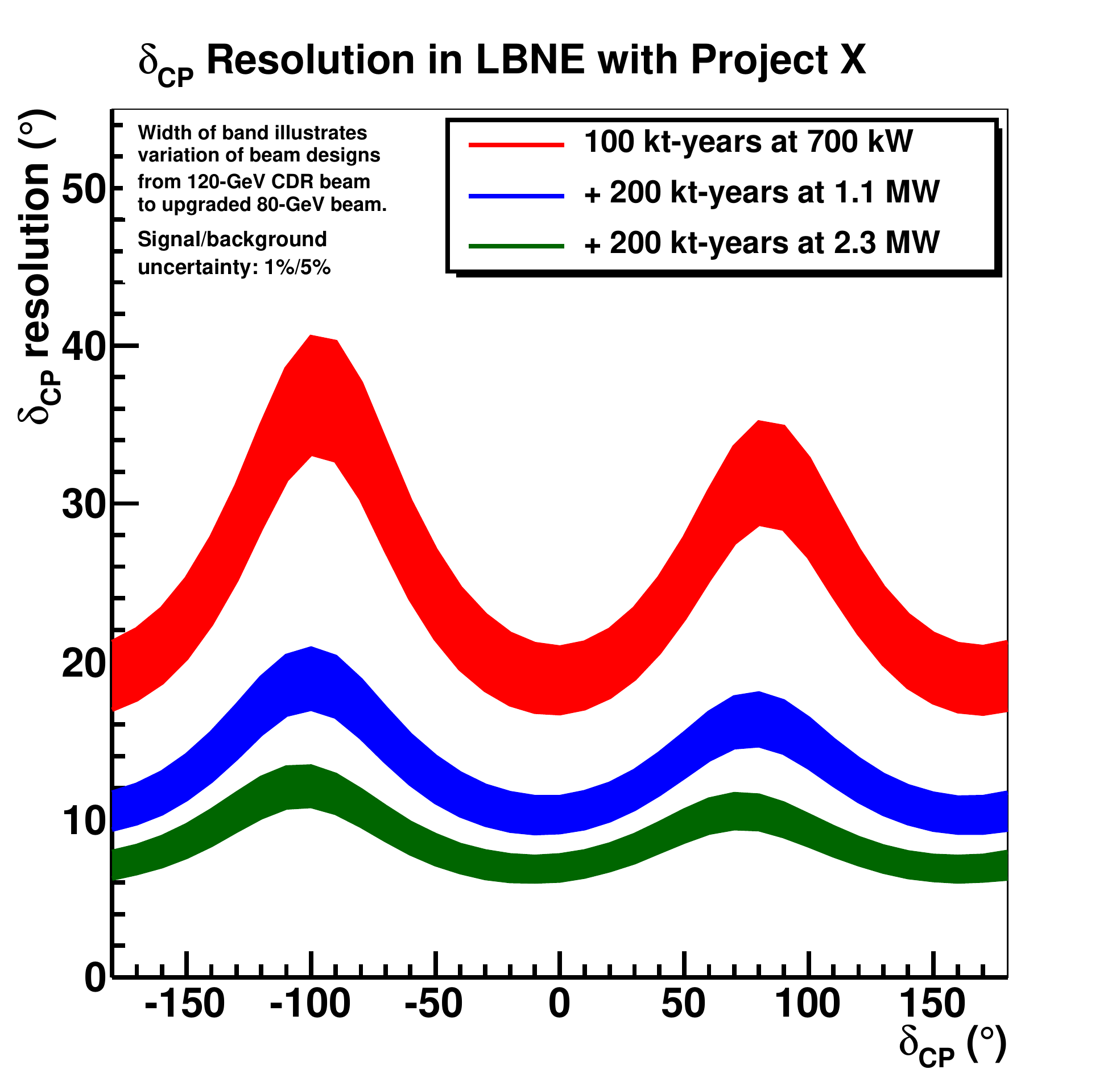}
\end{center}
\caption{\label{fig:cpsens} {CP-violation sensitivity as a function of 
$\delta_{CP}$ (top left) and exposure for 50\% coverage of the 
full $\delta_{CP}$ range (top right). Also shown are the projected precision 
on the measurement of $\delta_{CP}$ for various true points in the 
$\delta_{CP}$-$\sin^22\theta_{13}$ plane (bottom left) and as a function 
of $\delta_{CP}$ (bottom right). All plots show the increasing precision
possible in a staged long-baseline neutrino program in LBNE starting from 
nominal 700-kW running (red), through 1.1 MW using Project X Stage 1 (blue), 
to 2.3 MW with Project X Stage 2 (green).} Figures from~\cite{Adams:2013qkq,Kronfeld:2013uoa}.}
\end{figure}
Figure~\ref{fig:cpsens} illustrates the significance with which measurements
of CP violation and the unknown CP phase can be made with a staged 
long-baseline neutrino program in LBNE~\cite{Adams:2013qkq}. 
Ultimately, a $5\sigma$ 
determination of CP violation and a $\leq 10^\circ$ measurement of the CP 
violating phase are possible with such an experimental program.

LBNE plays a central role in the future U.S. program, and while being
the most advanced of all the proposals to measure CP violation in the
neutrino sector, there is a large number of alternative proposals in the U.S. 
and abroad. In this document, we will not be able to provide an in-depth 
comparison of the scientific merit of each of these proposals, which vary in maturity. Nonetheless, 
we can give an impression of how their performance for specific measurements 
might look. The most challenging measurement within the framework of
oscillation of three active neutrinos for long-baseline experiment is
the search for leptonic CP violation and a precise measurement of the
associated CP phase, $\delta_{CP}$. Therefore, apart from the value of
a determination of $\delta_{CP}$, as outlined in Sec.~\ref{sec:3nus},
the ability to measure the CP phase with precision is a reasonable
proxy for the overall potential to have a major scientific impact.

\begin{figure}[!h]
\begin{center}
\includegraphics[width=0.6\textwidth]{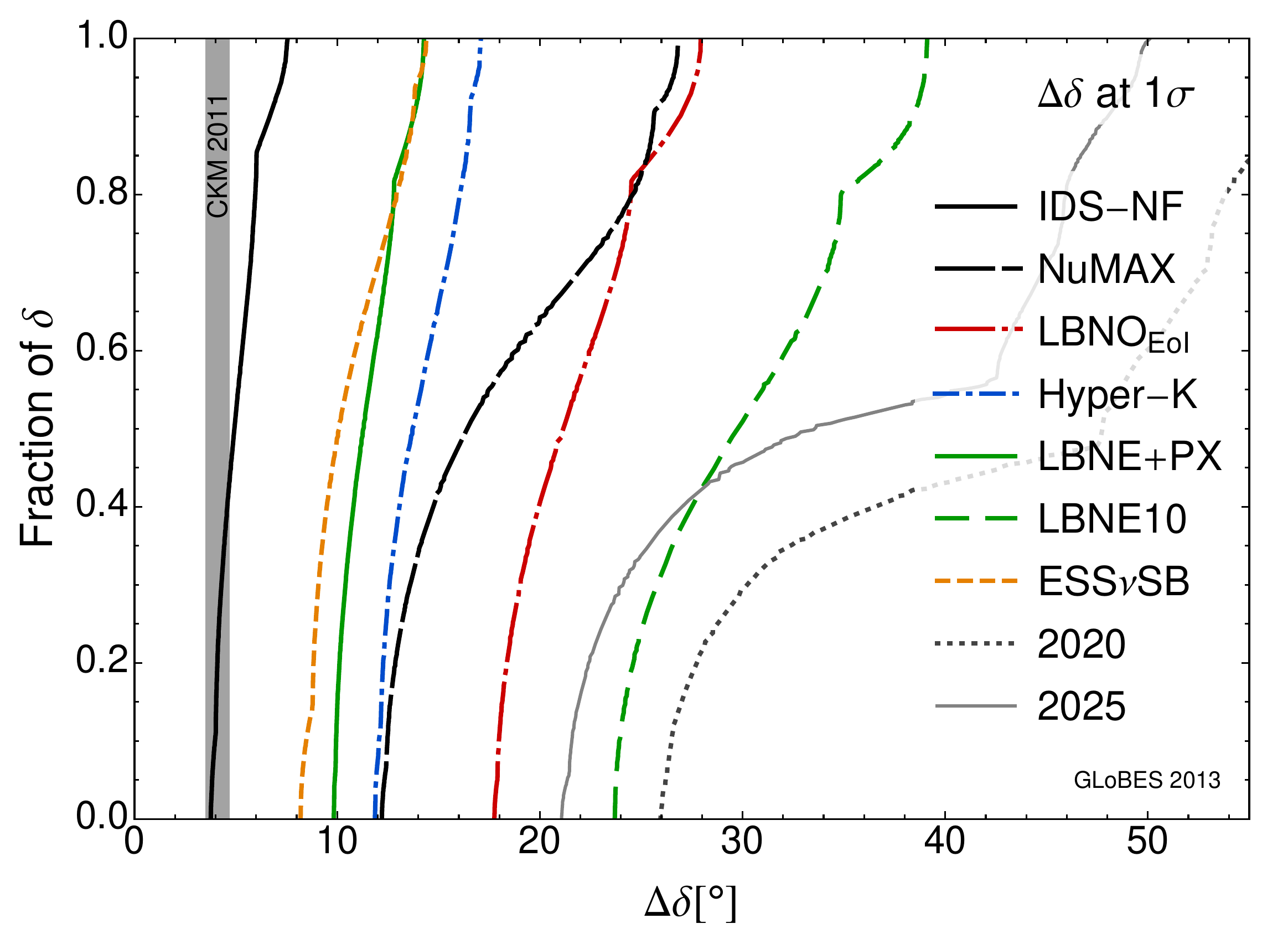}
\end{center}
\caption{Expected precision for a measurement of $\delta$ at present and future long-baseline oscillation experiments. Results are shown as a function of the fraction of possible values of $\delta$ for which a given precision (defined as half of the confidence interval at $1\sigma$, for 1 d.o.f.) is expected. All oscillation parameters are set to their present best-fit values, and marginalization is performed within their allowed intervals at $1\sigma$, with the exception of $\theta_{13}$ for which marginalization is done within the allowed interval expected at the end of the Daya Bay run. Matter density is set to the value given by the PREM profile, and a 2\% uncertainty is considered. The hierarchy is assumed to be normal, and no sign degeneracies are accounted for. Systematic uncertainties are implemented as in~\cite{Coloma:2012ji}. All facilities include an ideal near detector, and systematics are set to their ``default'' values from Table~2 in~\cite{Coloma:2012ji}.
 The different lines correspond to the following configurations.
 \textbf{2020} shows the expected combination of NOvA and T2K by the year 2020, simulated following~\cite{Patterson:2012zs} and~\cite{Huber:2009cw}, respectively. NOvA is assumed to run for three years per polarity while T2K is run for five years only with neutrinos. The line labeled as \textbf{2025} is an extrapolation of \textbf{2020}, where NOvA is run for a longer period and five years of $\bar\nu$ running at T2K are added following~\cite{Huber:2009cw}.
 \textbf{ESS$\nu$SB} corresponds to the performance of a 500-kt water Cherenkov detector placed at 360 km from the source; see~\cite{Baussan:2012cw}. The beam would be obtained from 2-GeV protons accelerated at the ESS proton linac. Migration matrices from Refs.~\cite{Agostino:2012fd,Agostino:2013ilw} have been used for the detector response.
 \textbf{LBNE10} corresponds to the first phase of the LBNE project. The CDR~\cite{CDR} beam flux has been used. The detector performance has been simulated as in~\cite{CDR} as well, using migration matrices for NC backgrounds from~\cite{Akiri:2011dv}. The exposure corresponds to 70 MW$\cdot$kt$\cdot$years.
 \textbf{LBNE+PX} corresponds to an upgrade of the previous setup, but exposure is set in this case to 750 MW$\cdot$kt$\cdot$years.
 \textbf{Hyper-K} stands for a 750-kW beam aiming from Tokai to the Hyper-Kamiokande detector (560-kt fiducial mass) in Japan. The baseline and off-axis angle are the same as for T2K. The detector performance has been simulated as in~\cite{Coloma:2012ji}.
 \textbf{LBNO$_\textrm{EoI}$} stands for the LBNO Expression of Interest~\cite{Stahl:2012exa} to place a 20-kt LAr detector at a baseline of 2,300 km from CERN. The results shown here correspond to the same statistics used in Fig.~75 therein. Neutrino fluxes corresponding to 50 GeV protons (from~\cite{Longhin:2012ae}) have been used, rescaling the number of protons on target to match the beam power in~\cite{Stahl:2012exa}. A similar detector performance as for \textit{LBNE10} is assumed, and five years of data taking per polarity are assumed in this case.
 \textbf{NuMAX} corresponds to a low-luminosity neutrino factory obtained from the decay of 5 GeV muons, simulated as in~\cite{Christensen:2013va}. The beam luminosity is set to $2\times10^{20}$ useful muon decays per year, and the flux is aimed to a 10-kt magnetized LAr detector placed at 1300 km from the source. \textbf{IDS-NF} corresponds to the IDS-NF setup. It considers a 100-kt MIND detector placed at 2000 km from the source, and $2\times10^{21}$ useful muon decays per year. Migration matrices, kindly provided by R.~Bayes (see also~\cite{Bayes:2012ex}), are used to simulate the detector response.
\label{fig:cp-precision}}
\end{figure}

The results of this comparison are shown in
Fig.~\ref{fig:cp-precision} using the methods and common systematics
implementation including near detectors as in
Ref.~\cite{Coloma:2012ji}.  The lines labeled 2020 and 2025 show what
can be achieved by those dates using a combination of the existing
experiments T2K and NOvA and Daya Bay, where the implementation of
all three follows Ref.~\cite{Huber:2009cw} and the NOvA
description has been updated for this report~\cite{Patterson:2013}.
This is the precision that can be reached without any new
experiments. Furthermore, we will compare two phases of LBNE: LBNE-10
with a 10-kt detector and a 700-kW beam and LBNE-PX with a 34-kt
detector and the 2.3-MW beam from Project X; both phases do include a
near detector and the other details can be found in~\cite{Adams:2013qkq}. 
After sufficient exposure, LBNE operating in the intense beams from Project X  could approach a precision for the CP-odd phase in the lepton sector comparable to that achieved for the CP-odd phase in the quark sector.
 In order
to accomplish this, however, systematic uncertainties on the signal and the
background need to be controlled at the percent level --- 
almost an order of magnitude improvement. No studies of the feasibility of
this increase in systematics control have been performed to date.

Beyond LBNE, we compare several different superbeam experiments.
LBNO plans to use liquid
argon TPC, based on dual-phase readout in contrast to LBNE, and a
baseline of 2\,300\,km. The initial detector size is assumed  be 20-kt
(labeled LBNO$_\mathrm{EOI}$) as described in detail in
Ref.~\cite{Stahl:2012exa}; 
the beam power assumed is around 700\,kW derived
from the CERN SPS. The Hyper-K setup~\cite{Abe:2011ts,Kearns:2013lea} in Japan will use
a 560-kt (fiducial) water-Cherenkov detector and a $\sim$ 1~MW beam.
A more recent European proposal (ESS$\nu$SB)~\cite{Baussan:2013ema} is to upgrade the superconducting 5-MW and 14-Hz-pulse-rate proton linac of the European Spallation Source linac, which is under construction in Lund in Sweden, and use it in conjunction with a 600-kt water Cherenkov detector (MEMPHYS~\cite{Agostino:2012fd,deBellefon:2006vq}) at a 500-km-baseline site in Sweden.
Finally, we also show the results obtained from a neutrino factory
(NF) --- in a neutrino factory an intense beam of muons is put in a
storage ring with long straight sections and a neutrino beam
consisting of equal numbers of $\nu_\mu$ and $\bar\nu_e$ results. The
current standard design of a neutrino factory will produce $10^{21}$
useful muon decays (summed over both stored $\mu^-$ and $\mu^+$) per
$10^7\,\mathrm{s}$ at a muon energy of 10\,GeV aimed a 100-kt
magnetized iron detector (MINOS-like) at a distance of
$\sim$2,000~km~\cite{Choubey:2011zzq}. This facility requires a 4\,MW
proton beam at around 8\,GeV, muon phase-space cooling and subsequent
muon acceleration. This considerable technical challenge should be
contrasted with the resulting advantages: a neutrino beam with known
flux, better than 1\%, beam spectrum and flavor composition with an
easy to identify final state in the far detector. 
The NF offers a unique
level of systematics control paired with very high-intensity beams;
 therefore they are considered the ultimate tool for precision neutrino
physics, see, e.g.,~\cite{Edgecock:2013lga}. The NF facility would provide the
most stringent tests of the standard three-flavor paradigm. 

Several new proposals have been submitted in the form of white papers,
notably a series of ideas on how to use the existing Main Injector
neutrino beam line (NuMI) by adding new detectors. RADAR~\cite{Adamson:2013jsa}
proposes to add a 6-kt liquid-argon TPC following the proposed LBNE TPC design in the NOvA far
detector hall at a baseline of 810\,km, to act as an R\&D stepping-stone that also advances the physics reach of the overall U.S. program. CHIPS~\cite{Adamson:2013xka} proposes
to build off-beam-axis water Cherenkov detectors in shallow, flooded mine pits, which
could provide potentially large fiducial masses in the range of
100 kt, first in the NuMI beam and potentially later in the LBNE beam.  According to the CHIPS proponents, in terms of physics reach, this
would be equivalent to about 20 kt of liquid argon TPC. 

A staged approach to a neutrino factory is
proposed~\cite{Christensen:2013va}, where an initial stage called the
low-luminosity low-energy neutrino factory is built on the basis of
existing accelerator technology and Project~X Phase~2. In this
facility, which does not require muon cooling and which starts with a
target power of 1\,MW, $10^{20}$ useful muon decays per polarity and
year can be obtained. The muon energy is chosen to be 5\,GeV as to
match the baseline of 1,300\,km. In combination, this allows one to
target the LBNE detector, maybe with the addition of a
magnetic field.  This approach would allow for a step-wise development
from nuSTORM (see Sec.~\ref{sec:anomalies}), via the low-luminosity
low-energy neutrino factory to a full neutrino factory, and if
desired, to a multi-TeV muon collider. This phased muon-based program
is well aligned with the development of Project~X~\cite{Kronfeld:2013uoa,Holmes:2013hfa}.

 In summary, a measurement of the leptonic CP phase at levels of
precision comparable to those of the CP phase in the quark sector will ultimately be possible in long-baseline oscillation experiments, given that $\theta_{13}$ has been
measured to be nonzero.  To do so will require a product of very high
proton beam intensity and very large detector mass --- nominally
beams in excess of 1~MW, paired with detectors in
the 100-kt range or larger, and running times of order one decade ---
regardless of the specifics of the chosen technology or proposal.
Experiments with long baselines 
and wide-band neutrino
beams that cover the first two oscillation maxima are best
positioned to exploit the rich spectral information contained in the
oscillation patterns and therefore have optimal sensitivity to 
CP violation for the minimal required exposure.
Wide-band very long-baseline experiments such as LBNE and LBNO 
can reach better than $10^\circ$ precision on $\delta$ with
exposures under 1000 kt$\cdot$MW$\cdot$years --- provided that systematic uncertainties
can be controlled to the level of a few percent or better. 
A neutrino factory with similar exposure --- a next-next generation project --- 
should be able to measure $\delta$ at the $5^\circ$ level, 
and provide the most stringent constraints on the three-flavor paradigm, thanks to its capability to measure
several different oscillation channels with similar precision. 

\subsubsection{CP violation with atmospheric neutrinos}

As noted previously, neutrinos and antineutrinos  from the atmosphere come with a range of baselines 
and energies, and in principle similar CP-violating observables are accessible as for beams, so long as the detectors have sufficient statistics and resolution.    Water Cherenkov detectors have relatively low resolution in energy and direction, and have difficulty distinguishing neutrinos from antineutrinos, although some information is to be had via selection of special samples~\cite{Abe:2011ph} and using statistical differences in kinematic distributions from $\nu$ and $\bar{\nu}$. In spite of worse resolution, water Cherenkov detectors have potentially vast statistics and reasonable sensitivity~\cite{Abe:2011ts,Kearns:2013lea}.
Large long-string ice and water-based detectors, while sensitive to mass hierarchy if systematics can be reduced, lack resolution for CP studies.
LArTPC detectors, in contrast, should have significantly improved resolution on both neutrino energy and direction, and even in the absence of a magnetic field can achieve better $\nu$ vs $\bar{\nu}$ tagging than water Cherenkov detectors~\cite{Adams:2013qkq}.  
Atmospheric neutrino information can be combined with beam information in the same or different detectors to improve overall sensitivity.

\subsubsection{CP violation with pion decay-at-rest sources}

A different approach for measuring CP violation is
DAE$\delta$ALUS~\cite{Conrad:2009mh,Aberle:2013ssa,Alonso:2010fs,Agarwalla:2010nn}.  The idea is to use muon antineutrinos produced by cyclotron-produced stopped-pion decay ($\pi^+\to\mu^+\nu_{\mu}$) at rest (DAR) neutrino sources, and to vary the baseline by having sources at different distances from a detector site.
For DAR sources, the neutrino energy is a few tens of MeV.  For baselines ranging from 1 to 20 km, both $L$ and $E$ are smaller than for the conventional long-baseline beam approach, and the ratio of $L/E$ is similar. Matter effects are negligible at short baseline.  This means that the CP-violating signal is clean;  however there is a degeneracy in oscillation probability for the two mass hierarchies.   This degeneracy can be broken by an independent measurement of the hierarchy.

\begin{figure}[ht]
\begin{center}
\includegraphics[width=0.6\textwidth]{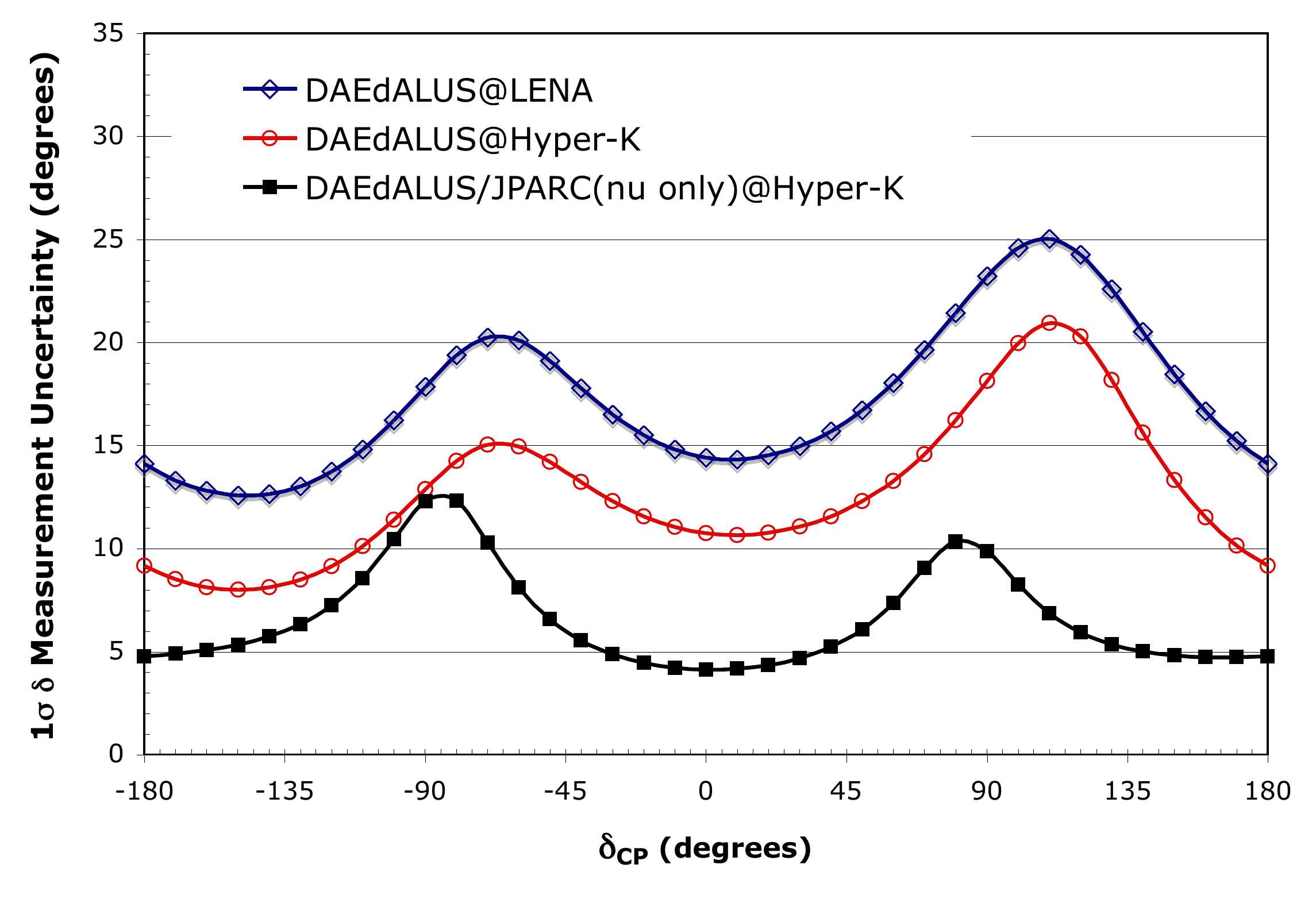}
\end{center}
\caption{\label{fig:daedalus_cp} 
{Sensitivity of a CP search for DAE$\delta$ALUS combined with LENA or Hyper-K~\cite{Aberle:2013ssa}, and combined with an independent J-PARC beam to Hyper-K.}}
\end{figure}

The electron-type antineutrino appearance signal from the oscillation of muon-type antineutrinos from pion DAR is detected via inverse beta-decay ($\bar{\nu}_e p \to e^+n$). Consequently very large detectors with free protons are required. The original case was developed for a 300-kt Gd-doped water detector concept at Homestake~\cite{Alonso:2010fy}.  Possibilities currently being explored for the detector include LENA~\cite{Wurm:2011zn} or  Super-K/Hyper-K~\cite{Abe:2011ts,Kearns:2013lea}.
Figure~\ref{fig:daedalus_cp} shows the projected CP sensitivity of DAE$\delta$ALUS.

The DAE$\delta$ALUS collaboration proposes a phased approach~\cite{Aberle:2013ssa}, with early phases involving IsoDAR (see Sec.~\ref{sec:nu5_accproj}) with sterile neutrino sensitivity.
The phased program offers also connections to applied cyclotron research (see Section~\ref{sec:accelerator}).

%% file: nu2.tex
%-----------------------------------------------------------------------
\section{The nature of the neutrino -- Majorana versus Dirac}
\label{sec:majorana}

Understanding
the neutrino mass generation mechanism, the absolute neutrino mass
scale, and the neutrino mass spectrum are essential topics to be addressed
by future neutrino experiments.   Whether neutrinos are Dirac fermions
(i.e., exist as separate massive neutrino and  
antineutrino states) or Majorana fermions (neutrino and antineutrino
states are equivalent) is a key experimental
question, the answer to which will guide the theoretical description of neutrinos.

All observations involving leptons are consistent with their appearance and
disappearance in flavor-matched particle anti-particle pairs. This property is expressed in the
form of lepton
number, $L$, being conserved by all fundamental forces.
We know of no fundamental symmetry relating to this empirical conservation law.
Neutrinoless double-beta decay, a weak nuclear decay process in which
a nucleus decays to a different nucleus 
emitting two beta-rays and no neutrinos,
 violates lepton number conservation by two units and thus,
if observed, requires a revision of our current understanding of particle physics.
In terms of field theories, such as the
Standard Model, neutrinos are assumed
to be massless and there is no chirally right-handed neutrino field.
The guiding principles
for extending the Standard Model are the conservation of electroweak isospin and renormalizability,
which do not preclude each neutrino mass eigenstate $\nu_i$ to be identical to its
antiparticle $\overline{\nu}_i$, or a Majorana particle.
However, $L$ is no longer conserved if $\nu = \overline{\nu}$.
Theoretical models, such as the seesaw mechanism that can explain the smallness of neutrino
mass, favor this scenario. Therefore, the discovery of Majorana neutrinos would
have profound theoretical implications in the formulation of a new Standard Model while
yielding insights into the origin of mass itself. If neutrinos are Majorana particles, they
may fit into the leptogenesis scenario for creating the baryon asymmetry, and hence ordinary
matter, of the Universe.

As of yet, there is no firm experimental evidence to confirm
or refute this theoretical prejudice. Experimental evidence of neutrinoless double-beta decay
($0\nu\beta\beta$) decay would establish the Majorana nature of neutrinos.
It is clear that $0\nu\beta\beta$ experiments sensitive at least
to the mass scale indicated by the atmospheric neutrino oscillation results are needed.
\begin{figure}[ht]
\begin{center}
\includegraphics[width=0.65\textwidth]{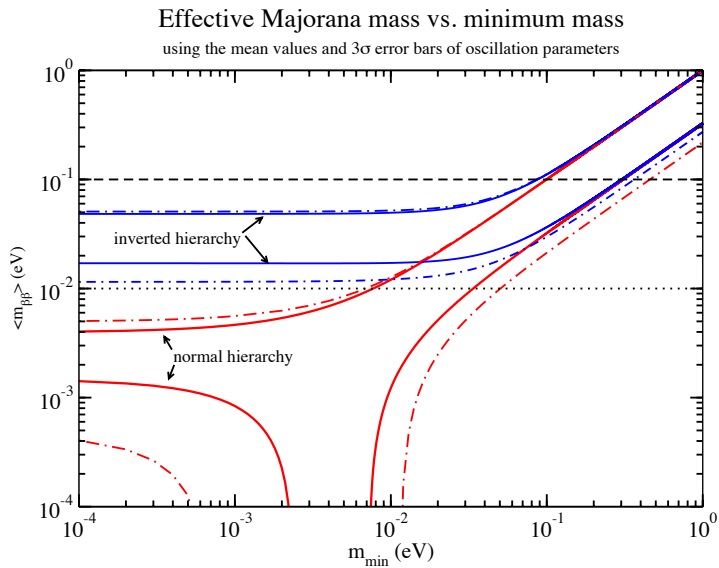}
\end{center}
%\vskip-0.8cm
\caption{Allowed values of $\langle m_{\beta \beta} \rangle$ as a function of the lightest neutrino mass for the inverted and normal hierarchies. The  regions defined by the solid curves correspond
to the best-fit neutrino mixing parameters from~\cite{Forero:2012}  and account for the degeneracy
due to the unknown Majorana phases. The regions defined by the dashed-dotted curves correspond to the maximal allowed regions
including mixing parameter uncertainties as evaluated in~\cite{Forero:2012}.
The dashed line shows expected sensitivity of next-generation $\sim$100~kg class experiments and the dotted line shows potential reach of multi-ton scale future experiments.}
\label{fig:SensitivityExposure}
\end{figure}

For $0\nu\beta\beta$ decay the summed energy of the emitted electrons takes a single value.
Observation of a sharp peak at the $\beta\beta$ endpoint would thus quantify the $0\nu\beta\beta$
decay rate,
demonstrate that neutrinos are Majorana particles, indicate that lepton number is not
conserved, and, paired with nuclear structure calculations, provide a measure of an
effective Majorana mass, $\langle m_{\beta\beta} \rangle$. 
There is consensus within the neutrino physics community that such a decay peak
would have to be observed for at least two different decaying isotopes at two different
energies to make a credible claim for $0\nu\beta\beta$ decay.

In more detail, the observed half-life can be related to an effective Majorana mass
according to 
$(T_{1/2,0\nu\beta\beta})^{-1} 
= G_{0\nu}|M_{0\nu}|^2\langle m_{ \beta \beta} \rangle^2$,
where $\langle m_{ \beta \beta} \rangle^2 \equiv | \sum_i U_{ei}^2 m_i|^2$.
$G_{0\nu}$ is a phase space factor, $m_i$ is the mass
of neutrino mass eigenstate $\nu_i$, and $M_{0\nu}$ is the transition nuclear matrix
element.  The matrix element has significant nuclear theoretical uncertainties,
dependent on the nuclide under consideration.  

In the standard three-massive-neutrinos paradigm,  
\begin{equation}
\langle m_{\beta\beta}\rangle= |\cos^2\theta_{12}\cos^2\theta_{13}e^{-2i\xi}m_1+\sin^2\theta_{12}\cos^2\theta_{13}e^{-2i\zeta}m_2+\sin^2\theta_{13}e^{-2i\delta}m_3|.
\end{equation}
If none of the neutrino masses vanish, $\langle m_{\beta\beta} \rangle$ is a function of not only the oscillation parameters $\theta_{12,13}$ and the neutrino masses $m_{1,2,3}$ but also the two Majorana phases $\xi,\zeta$~\cite{Rodejohann:2012xd}.
Neutrino oscillation experiments indicate that at least one
neutrino has a mass of $\sim 45$~meV or more. As a result and as shown in Fig.~\ref{fig:SensitivityExposure}, in the
inverted hierarchy mass spectrum with $m_3 = 0$~meV, $\langle m_{\beta\beta} \rangle$ is
between 10 and 55 meV
depending on the values of the Majorana phases. This region is sometimes referred to as the
atmospheric mass scale region. Exploring this region requires a sensitivity to half-lives exceeding
$10^{27}$ years. This is a challenging goal requiring several ton-years of exposure
and very low backgrounds. The accomplishment of this goal requires a detector at the ton
scale of enriched material and a background level below 1 count/(ton y) in the spectral
region of interest (ROI).  Very good energy resolution is also required.

There is one controversial result from a subset of collaborators
of the Heidelberg-Moscow experiment, who claim a measurement
of the process in $^{76}$Ge, with 70 kg-years of data~\cite{KlapdorKleingrothaus:2001ke}.
These authors interpret the observation as giving an $\langle m_{\beta\beta}\rangle$ of
440~meV.   Recent limits using the isotope $^{136}$Xe from EXO-200 and KamLAND-Zen (see below) are in tension
with this $\langle m_{\beta\beta}\rangle$ regime.

There is a large number of current neutrinoless double-beta decay search efforts, employing very different techniques; a recent review is~\cite{Avignone:2007fu}.  Here we will highlight some for which there is a component
of effort from physicists based in the U.S..  These represent different kinds of detectors and experimental approaches~\cite{SNO+:2013,SuperNEMO:2013,NEXT:2013,Winslow:2013qia,MOON:2013,CUORE:2013,Majorana:2013,nid:2013,EXO:2013,cubed:2013}.

The {\sc Majorana}~\cite{Majorana:2013, Aguayo:2011sr,Schubert:2011nm,Wilkerson:2012ga} experiment employs the germanium isotope $^{76}$Ge.  The current phase of the experiment is the ``{\sc Demonstrator}''.  This will employ 30~kg of Ge enriched to 86\% $^{76}$Ge and 10~kg of Ge P-type point contact detectors, and is being constructed underground at the Sanford Underground Research Facility (SURF). It will have first data in 2013 with data from enriched detectors in 2014.    The {\sc Majorana} collaboration is planning a ton-scale effort in collaboration with its European counterpart GERDA~\cite{Agostini:2013mba}.

The ``bolometric'' CUORE experiment~\cite{CUORE:2013, Alessandria:2011rc}, located at Gran Sasso National Laboratory in Italy, employs $^{130}$Te in the form of natural TeO$_2$ crystals.  This is a cryogenic setup, operated at temperatures around 10 mK, that determines the energy deposit via temperature rise measured with thermistors. 
The prototype of this experiment, Cuoricino, ran from 2003-2008 with 11.3~kg of $^{130}$Te mass.  The first stage of CUORE, CUORE-0, is currently operating with a $^{130}$Te mass of 11~kg, and the full CUORE detector plans commencing operations in 2014 with 206~kg. CUORE aims at the sensitivity to the $0\nu\beta\beta$ lifetime of $2\times 10^{26}$ after five years of operation.

The EXO experiment~\cite{EXO:2013} makes use of $^{136}$Xe, which double-beta
decays as $^{136}\rm{Xe} \rightarrow ^{136}{}\rm{Ba}^{++} + e^- + e^-$.
The first version of EXO, EXO-200, is currently taking data at the Waste Isolation Pilot
Plant in New Mexico with 200~kg of xenon enriched to 80\% in the isotope 136.
A liquid-phase time projection chamber is used to detect both scintillation light
from the interaction and ionization energy deposited by the electrons
in the xenon.  EXO-200 reported the
first observation of the two-neutrino double-beta decay~\cite{Ackerman:2011gz} in $^{136}$Xe (later improved~\cite{Albert:2013gpz}) as well as a limit on the
neutrinoless double beta decay~\cite{Auger:2012ar} in $^{136}$Xe.  
The
EXO collaboration is planning a 5-ton detector called nEXO
that builds on the success of the EXO-200 detector. The expected nEXO
sensitivity to the $0\nu\beta\beta$ half-life is $2.5 \times 10^{27}$
years after 10 years of operation.  
The EXO collaboration's novel idea for an upgrade
is the use of barium tagging: the principle is to reduce backgrounds 
by identifying the resulting nucleus by laser spectroscopy~\cite{Danilov:2000pp}.

Another ambitious idea for a double-beta decay experiment is SNO+~\cite{Kraus:2010zzb,SNO+:2013}.
SNO+ is an experiment at SNOLAB in Canada which plans to refill
the acrylic vessel of SNO with liquid scintillator.  This experiment would
in addition provide a rich physics program of solar, supernova, and geo-neutrino physics (see Sec.~\ref{nu6}). SNO+ plans to load the scintillator with 0.3\%~Te, which after five years of data should give them a 90\% CL sensitivity of approximately $7.2\times10^{25}$ years (neutrino mass sensitivity of $\sim$140 meV). 

KamLAND-Zen~\cite{KamLANDZen:2012aa} (the Kamioka Liquid Anti-Neutrino Detector,
ZEro Neutrino double-beta decay experiment) is an extension of the
KamLAND~\cite{Gando:2013nba} liquid scintillator experiment. 
In 2011, the collaboration added an additional low-background
mini-balloon into the inner sphere that contains 13 tons of liquid
scintillator loaded with 330 kg of dissolved Xe gas enriched to 91\%
in $^{136}${Xe}. The initial results include an improved
limit on neutrinoless double-beta decay for $^{136}${Xe} and a
measurement of two-neutrino double-beta decay that agrees with the
recent EXO-200 result~\cite{Gando:2012zm}. 
The collaboration has an
additional 400 kg of enriched Xe in hand and is considering options to
upgrade the detector with a larger-size internal balloon.

NEXT~\cite{NEXT:2013,Alvarez:2011my,Yahlali:2010zz} (Neutrino Experiment with Xenon TPC) intends to use $>$100 kg of Xe enriched to $\sim$90\% in $^{136}${Xe}. The detector will be a moderate-density gas TPC
that will detect primary and secondary scintillation light. By operating at low pressures ($\sim$15 bar), the design should not only provide good energy resolution, but also permit tracking that allows fairly detailed track reconstruction to confirm that candidate events involve two electrons moving in opposite directions. 
Construction started in 2012 with commissioning scheduled to start in 2014. It will operate at the Laboratorio Subterr\'{a}neo de Canfranc in Spain.

The LUX-ZEPLIN (LZ) experiment is a proposed two-phase (liquid/gas) Xe detector, containing 7 tons of natural Xe instrumented as a time projection chamber, with readout of direct scintillation and readout of charge via proportional scintillation. While LZ is primarily designed to perform a world-leading direct dark matter search, it is also sensitive to the neutrinoless double beta decay of $\rm ^{136}Xe$. LZ will replace the currently operating LUX experiment~\cite{Akerib:2012ys}
at SURF, and is planned to be commissioned in 2017. 
After three years of LZ operation, the $0\nu\beta\beta$ half-life sensitivity is projected to be 2.2 $\times$ $10^{26}$ years.

The SuperNEMO~\cite{SuperNEMO:2013,Arnold:2010tu} proposal builds on the great success of the NEMO-3 (Neutrino Ettore Majorana Observatory) experiment, which measured two-neutrino double-beta decay rates and set some of the most stringent constraints for $0\nu\beta\beta$ transitions for seven isotopes~\cite{Bongrand:2011ei}. 
The design uses calorimetry to measure energies and timing, and tracking to provide topological and kinematical information about the individual electrons. SuperNEMO will improve on NEMO-3 by using a larger mass of isotope, lowering backgrounds, and improving the energy resolution. 
The complete experiment will be ready by the end of the decade in a recently-approved extension of the Modane laboratory in the Fr\'{e}jus Tunnel in France. Its design  sensitivity for the $0\nu\beta\beta$ half-life of $^{82}${Se}  is $10^{26}$\,yr, in a 500\,kg$\cdot$yr exposure.

The current and next-generation experiments are of 10-100~kg masses;
these have sensitivities down to about 100~meV.  Further ton-scale
experiments are planned for the generation beyond that: these should
have sensitivities reaching the 10 meV or smaller scale.  Reaching
this regime will be very interesting in its complementarity with
oscillation experiments: if the mass hierarchy is independently determined to be inverted,
and there is no $0\nu\beta\beta$ decay signal at the 10 meV scale,
then neutrinos must be Dirac (assuming nature has not been so diabolical as to contrive a fine-tuned suppression from e.g., nuclear matrix elements).
If a signal is observed at the few meV scale, then not only will we
know that neutrinos are Majorana, but we will also know that the
hierarchy must be normal, even in the absence of an independent
determination.

\begin{table}[h]
\hspace{-1cm}
\begin{tabular}{|c|c|c|c|c|c|}
\hline
Experiment &   Isotope
& Mass &  Technique
& Status  & Location  \\ \hline
AMoRE~\cite{Kim11,Lee:2011zzc}
&$^{100}${Mo}  & 50 kg &CaMoO$_4$ scint. bolometer crystals
& Devel.  & Yangyang \\
CANDLES~\cite{Kishimoto:2009zz}  &  $^{48}${Ca} & 0.35 kg
& CaF$_2$ scint. crystals
& Prototype &  Kamioka \\
CARVEL~\cite{Zdesenko:2005by}  &$^{48}${Ca} & 1 ton
& CaF$_2$ scint. crystals
& Devel.  & Solotvina        \\
COBRA~\cite{Zuber:2001vm}  &  $^{116}${Cd} & 183 kg
& $^{enr}${Cd} CZT semicond. det.
& Prototype             &  Gran Sasso   \\
CUORE-0~\cite{Alessandria:2011rc}  &  $^{130}${Te} & 11 kg
& TeO$_2$ bolometers  & Constr. (2013)  &  Gran Sasso            \\
CUORE~\cite{Alessandria:2011rc}  &  $^{130}${Te} & 206 kg
& TeO$_2$ bolometers  & Constr. (2014)  &  Gran Sasso            \\
DCBA~\cite{Ishihara:1999vp}  & $^{150}${Ne} & 20 kg
&$^{enr}${Nd} foils and tracking
& Devel.  & Kamioka  \\
EXO-200~\cite{Ackerman:2011gz,Albert:2013gpz,Auger:2012ar}  &  $^{136}${Xe} & 200 kg
&Liq. $^{enr}${Xe} TPC/scint.
& Op. (2011)  & WIPP             \\
nEXO~\cite{Danilov:2000pp}  &  $^{136}${Xe} & 5 t
&Liq.  $^{enr}${Xe} TPC/scint.
& Proposal              &  SNOLAB             \\
GERDA~\cite{Agostini:2013mba,Schonert:2005zn}  &  $^{76}${Ge} & $\sim$35 kg
&$^{enr}${Ge} semicond. det.  
& Op. (2011)  &  Gran Sasso             \\
GSO~\cite{Danevich:2000tk}  & $^{160}${Gd} & 2 t
&Gd$_2$SiO$_5$:Ce crys. scint. in liq. scint.
& Devel. &  \\
KamLAND-Zen~\cite{KamLANDZen:2012aa,Gando:2012zm}
& $^{136}${Xe}  & 400 kg & $^{enr}${Xe} dissolved in liq. scint.          
& Op. (2011)    & Kamioka \\
LZ~\cite{Akerib:2012ys} & $^{136}$Xe & 600 kg 
& Two-phase $^{nat}$Xe TPC/scint & Proposal & SURF \\ 
LUCIFER~\cite{Giu10,Arnaboldi:2010jx}
& $^{82}${Se} & 18 kg
& ZnSe scint. bolometer crystals
& Devel.  & Gran Sasso \\
MAJORANA~\cite{Aguayo:2011sr,Schubert:2011nm,Wilkerson:2012ga}
&  $^{76}${Ge}  & 30 kg &$^{enr}${Ge} semicond. det. & Constr. (2013)
&  SURF             \\
MOON~\cite{Ejiri:2007zza}
&  $^{100}${Mo}  & 1 t &$^{enr}${Mo} foils/scint.
& Devel.  &               \\
SuperNEMO-Dem~\cite{Arnold:2010tu}
&  $^{82}${Se}  & 7 kg & $^{enr}${Se} foils/tracking
& Constr. (2014)       &  Fr\'{e}jus             \\
SuperNEMO~\cite{Arnold:2010tu}  &  $^{82}${Se} & 100 kg
& $^{enr}${Se} foils/tracking
& Proposal (2019) &  Fr\'{e}jus             \\
NEXT~\cite{Alvarez:2011my,Yahlali:2010zz}
&  $^{136}${Xe} & 100 kg
& gas TPC                                    
& Devel. (2014)     &   Canfranc            \\
SNO+~\cite{Kraus:2010zzb,Chen:2005yi,Chen:2008un}
&  $^{130}${Te} & 800 kg
& Te-loaded liq. scint.                       
& Constr. (2013)    &  SNOLAB             \\
\hline
\end{tabular}\label{tab:FutureExperiments} 
\caption{A summary list of neutrinoless double-beta decay proposals and experiments.}
\end{table}%

A key point is that several experiments using different isotopes are in order, at each step of sensitivity.  First, different isotopes involve different matrix elements with their uncertainties.  In addition, unknown small-probability $\gamma$ transitions may occur at or near the endpoint of a particular isotope, but it is very unlikely that they occur for {\it every} double-beta-decay emitter.    Finally, and maybe most importantly, different isotopes generally correspond to radically different techniques, and since $0\nu\beta\beta$ searches require exceedingly low backgrounds, it is virtually impossible to decide {\it a priori} which technique will truly produce a background-free measurement.
The long-term future for $0\nu\beta\beta$ experiments will depend on what is observed: if no experiments, or only some experiments, see a signal at the 100-kg scale, then ton-scale experiments are in order.  If a signal is confirmed, the next generation of detectors 
will need to better investigate the $0\nu\beta\beta$ mechanism by separately measuring the energies of each electron as well as their angular correlations.

%% file: nu3.tex
%-----------------------------------------------------------------------
\section{Absolute neutrino mass}
\label{sec:mass}

\subsection{Kinematic neutrino mass measurements}

The neutrino's absolute mass cannot be determined by oscillation experiments, which give information only on mass differences.
The neutrino's rest mass has a small but potentially measurable
effect on its kinematics, in particular on the phase space available
in low-energy nuclear beta decay.  The effect is indifferent to the
distinction between Majorana and Dirac masses, and independent of nuclear matrix element calculations.

Two nuclides are of major importance to current experiments: tritium
($^3$H or T) and $^{187}$Re.  The particle physics is the same in both
cases, but the experiments differ greatly.  Consider the superallowed decay
$^3\mathrm{H}\rightarrow \mathrm{^3He} + e^- + \bar{\nu}_e$.  The
electron energy spectrum has the form:

\begin{equation}
dN/dE \propto F(Z,E) p_e (E+m_e) (E_0 - E)\sqrt{(E_0-E)^2 - m_\nu^2}
\end{equation}

where $E$, $p_e$ are the electron energy and momentum, $E_0$ is the Q-value,
and $F(Z,E)$ is the Fermi function.   If the neutrino is massless, the
spectrum near the endpoint is approximately parabolic around $E_0$.   A finite neutrino mass makes
the parabola ``steeper'', then cuts it off $m_\nu$ before
the zero-mass endpoint. The value of  $m_\nu$ can be extracted from the shape without knowing $E_0$ precisely, and without resolving the cutoff.

The flavor state $\nu_e$ is an admixture of at least three mass states
$\nu_1$, $\nu_2$, and $\nu_3$ (more than three if there are sterile neutrinos).  Beta decay yields a superposition of three
spectra, with three different endpoint shapes and cutoffs,
whose relative weights depend on the magnitude of elements of the mixing matrix.   Unless the three endpoint steps are fully resolved, the spectrum is
well approximated by the single-neutrino spectrum with an effective
mass $m_\beta^2 = \Sigma_{i}|U_{ei}|^2m_i^2$.  Past tritium experiments have determined $m_\beta < 2.0$ eV~\cite{Kraus:2004zw,Lobashev:2003kt,Aseev:2011dq}.

To measure this spectrum distortion, any experiment must have
the following properties.  First, it must have high energy resolution --- in particular, a resolution function
lacking high-energy tails --- to isolate the near-endpoint electrons
from the more numerous low-energy electrons.  Second, it must have extremely
well-known spectrometer resolution.  The observed neutrino mass
parameter depends very strongly on the detector resolution.  Finally, it must have the ability to observe
a very large number of decays, with high-acceptance
spectrometers and/or ultra-intense sources, in order to collect adequate
statistics in the extreme tail of a rapidly-falling spectrum.

\subsection{Upcoming experiments}\label{nu2:expt}
 
\noindent
\textbf{KATRIN:} The KATRIN experiment
~\cite{Angrik:2005ep, Drexlin:2013lha,Robertson:2013aaaa},  
now under construction, will attempt to extract
$m_{\beta}$  from decays of gaseous T$_2$.  KATRIN achieves 
high energy resolution using a MAC-E (Magnetic Adiabatic
Collimation-Electrostatic) filter.  In this technique, the T$_2$
source is held at high magnetic field. Beta-decay
electrons within a broad acceptance cone are magnetically guided towards a
low-field region; the guiding is adiabatic and forces the electrons' momenta nearly parallel to $B$ field lines.  In the parallel region, an electrostatic field serves as a sharp energy filter.  Only the highest-energy electrons can pass
the filter and reach the detector, so MAC-E filters can tolerate huge
low-energy decay rates without encountering detector rate problems.
In order to achieve high statistics, KATRIN
needs a very strong source, supplying $10^{11}$ e$^-/s$ to the spectrometer acceptance.   This cannot be done by increasing the source thickness, which is
limited by self-scattering, so the cross-sectional area of the
source and spectrometer must be very large --- 53 cm$^2$ and 65 m$^2$
respectively.   
KATRIN anticipates achieving a neutrino mass exclusion limit down to $0.2$ eV at 90\% confidence, or $0.35$ eV for a 5$\sigma$
discovery.
Data-taking for KATRIN is expected to begin in late 2015.

\noindent
\textbf{Project 8:} 
Project 8 is a new technology for pursuing the tritium endpoint~\cite{Monreal:2009za,Doe:2013jua};  it anticipates providing a
roadmap towards a large tritium experiment with new neutrino mass
sensitivity, via a method with systematic errors largely independent of the MAC-E filter method.   In Project 8, a low-pressure gaseous tritium source is
stored in a magnetic bottle.  Magnetically-trapped
decay electrons undergo cyclotron motion for $\sim 10^6$ orbits.  This motion emits microwave radiation at frequency $\omega = qB/\gamma m$, where $\gamma$ is the Lorentz factor.  A measurement of the frequency can be translated into an electron energy.  A prototype, now operating at the University of
Washington, is attempting to detect and characterize single conversion
electrons from a $^{83m}$Kr conversion electron calibration source.  The prototype is intended to help answer a number of technical questions, regarding the merits of various magnetic-trap configurations for the electrons, waveguide vs. cavity configurations for the microwaves, and data analysis techniques.   
A first experiment would aim for few-eV neutrino mass sensitivity while precisely measuring other parameters of the decay spectrum.  A larger followup experiment would extend the sensitivity down to the limits of the technique.   

\noindent
\textbf{Microcalorimeter methods:}
While most of the neutrino-mass community is focused on tritium, there are several other nuclides of potential experimental interest.  Tritium (18.6~keV endpoint) is the only low-energy beta-decay nuclide whose decay rate (and low Z) permit the creation of thin, high-rate sources.  If one can detect decays in a cryogenic microcalorimeter, the requirement of a thin source is removed, and one can explore lower-energy decays.   For a neutrino mass $m_\nu$ and a beta-decay energy $E_0$, the fraction of decays in the signal region scales as $(m_\nu/E_0)^3$.   The best-known candidate is $^{187}$Re, whose beta-decay endpoint is unusually low at 2.469~keV.  However, the long lifetime of $^{187}$Re forces any such experiment to instrument a very large total target mass, and the low-temperature properties of Re are unfavorable.   
Another candidate, $^{163}$Ho, is somewhat more promising.  In the electron-capture decay $^{163}$Ho $\rightarrow$ $^{163}$Dy, the inner bremsstrahlung spectrum is sensitive to the neutrino mass.  Speculation~\cite{PhysRevC.81.045501} that atomic effects might enhance the endpoint phase space has been largely resolved.   At the moment, however, 
%even ambitious 
microcalorimeter proposals require long data-taking periods to accumulate statistics with sub-eV sensitivity, and the systematic errors are underexplored.

\noindent
\textbf{PTOLEMY:}
The PTOLEMY experiment~\cite{Betts:2013uya} at Princeton is attempting to combine many different technologies in a single tritium-endpoint spectrometer.   While its primary goal is the detection of relic neutrinos, as discussed in Sec.~\ref{bbrelic},
its measurements would certainly be relevant to a direct search for neutrino masses.   
PTOLEMY installed a small technology-validation prototype at the Princeton Plasma Physics Laboratory in February 2013.   
Several of PTOLEMY's methods are untested and may present serious practical challenges.  The use of their solid-state source will require a careful roadmap towards answering systematic-error questions.  

\noindent
\textbf{Cosmological probes:}
Information on the absolute neutrino masses can also come from the Cosmic Frontier~\cite{Abazajian:2013oma}. 
Global fits to the recent wealth of cosmological data --- large-scale
structure, high-redshift supernovae, cosmic microwave background, and Lyman
$\alpha$ forest measurements --- yield limits on the sum of the three
neutrino masses of less than about
0.3-0.6~eV. 
These also constrain possible heavier neutrino states.
Specific results depend on assumptions.  
Future
cosmological measurements will further constrain the absolute mass
scale.
References~\cite{Abazajian:2013oma,Lesgourgues:2006nd,Abazajian:2011dt,Wong:2011ip} are recent reviews.  The Planck experiment has very recently published new global cosmology fits, including strong neutrino mass constraints~\cite{Ade:2013zuv}.

\subsection{The future of absolute mass measurements, and implications}

There is substantial complementarity between kinematic measurements, $0\nu\beta\beta$ measurements, and cosmological constraints.  
Kinematic measurements are sensitive to $m_\beta$, a simple mixing-weighted sum with a nonzero lower bound.   $0\nu\beta\beta$ is either (a) insensitive to $m_{\beta\beta}$, if neutrinos are Dirac particles, or (b) if neutrinos are Majorana, sensitive to $m_{\beta\beta}$, a quantity which incorporates masses, mixing angles, and complex phases, and may in certain cases be zero.  Cosmological probes are sensitive to the simple sum of masses, independent of mixing angles and symmetries, 
but this sensitivity 
correlates with changes to the 
cosmological assumptions, including (but not limited to) new fundamental physics.    

\begin{figure}[ht]
\begin{center}
\includegraphics[width=0.60\textwidth]{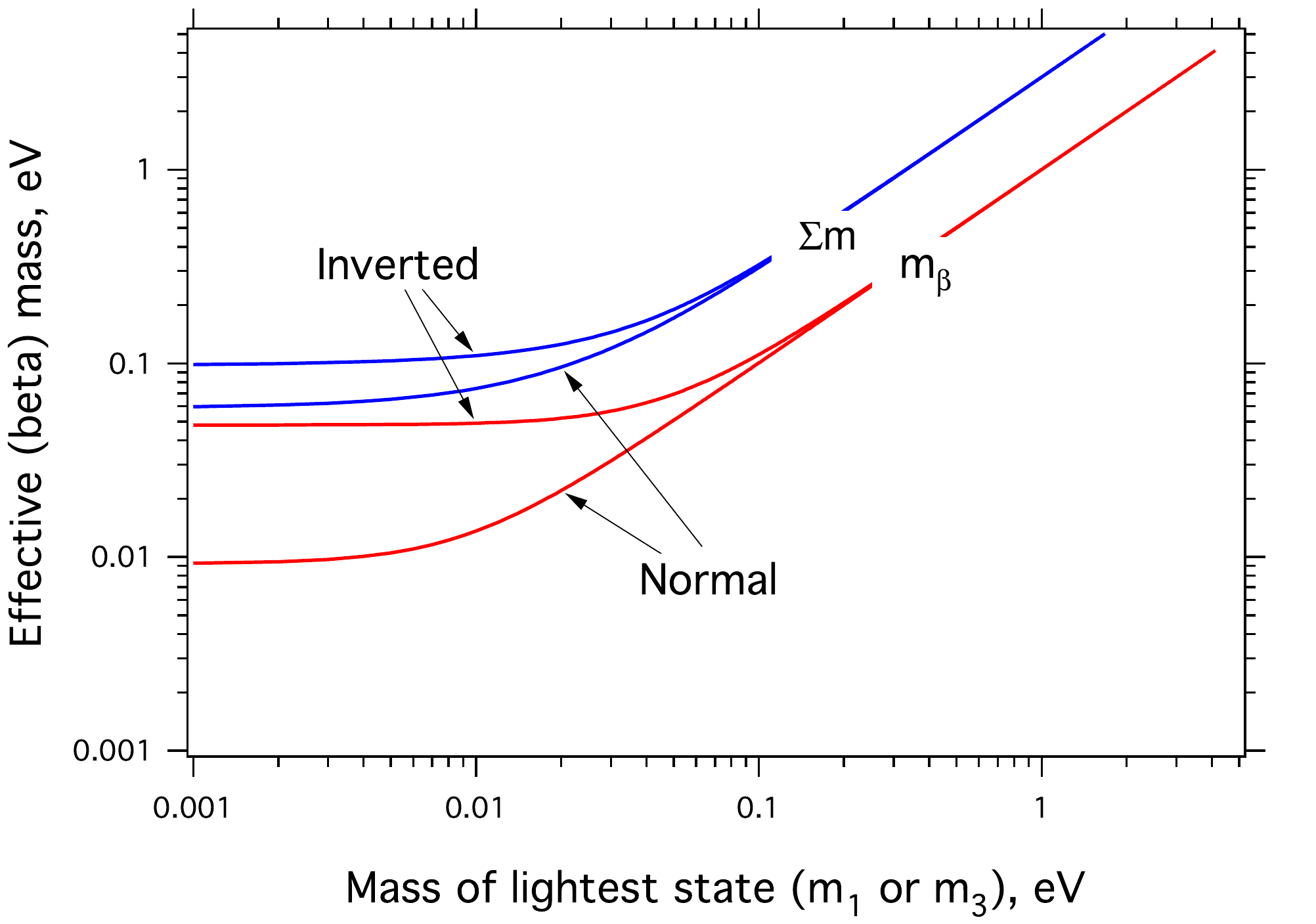}
\caption{Dependence of the effective mass $m_\beta$ on the mass of the lightest eigenstate $m_1$ or $m_3$ for the normal and inverted hierarchies.  Also shown are the sums of the eigenmasses.
%for each hierarchy.  
The oscillation parameters are $\Delta m_{21}^2=7.54\times10^{-5}$ eV$^2$, $|\Delta m_{32}^2|=2.42\times10^{-3}$ eV$^2$, $\theta_{12}=34.1$ degrees, and $\theta_{13}=9.1$ degrees~\cite{Fogli:2012ua}. }
\label{fig:mbeta}
\end{center}
\end{figure}

One worthwhile question is, under what circumstances do direct measurements resolve the neutrino mass hierarchy?  See Fig.~\ref{fig:mbeta}.  Direct measurements based on $\beta$-decay are capable of unambiguous determination of the hierarchy because they can identify the three masses weighted by their electron-flavor content.  However, 
such a measurement is well beyond present capabilities for any choice of mass or hierarchy.    A measurement at the achievable sensitivity represented by KATRIN, 200~meV, would show that neutrinos have a nearly-degenerate hierarchy, perhaps even more interesting from the theoretical standpoint than the level ordering.   In the foreseeable future, new ideas such as Project 8 may be able to reach the 50~meV level.  Non-observation of the mass at this level would show that the hierarchy is normal.

The field of direct neutrino mass determination, with KATRIN leading the push to $\sim$0.2 eV sensitivity, is balancing both statistical and systematic errors.   Experiments aiming for lower masses, including Project 8 and PTOLEMY, take it for granted that large statistical power is needed.  However, attention must be paid to systematics.  One systematic error in particular, the molecular excited-state distribution of the daughter ion (in T$_2$ $\rightarrow$ (T $^3$He)$^{+*}$ + e$^-$ + $\bar{\nu}_e$) produces an irreducible smearing of all T$_2$ decay spectra; this smearing is currently unmeasured, and only known from quantum theory with an uncertainty difficult to quantify.   The effect is present in common in KATRIN, Project 8, and any future T$_2$-based experiment.  The field would benefit from an experimental verification or a theory cross-check on these excited-state spectra.  Technologies allowing high-purity atomic %T$_0$ 
tritium sources would remove this uncertainty.   Most other systematic errors in T$_2$ experiments are technology-specific, which is important for robust comparisons between experiments.

%% file: nu4.tex
%-----------------------------------------------------------------------
\def \ereco {E_{\rm rec}}
\def\parenbar#1{{\null\!                        % left-hand spacing
   \mathop#1\limits^{\hbox{\tiny (--)}}       % (--) in 5pt type
% for LaTeX I changer \fiverm to \tiny
   \!\null}}                                    % right-hand spacing

\section{Neutrino scattering}
\label{sec:scattering}

Predictions for the rates and topologies of neutrino interactions with matter 
are a crucial component in many current investigations within nuclear and 
astroparticle physics. Ultimately, we need to measure neutrino-matter interactions 
precisely to enable adequate understanding of high-priority physics including 
neutrino oscillations, supernova dynamics, and dark matter searches. Precise 
knowledge of such neutrino interactions is an absolute necessity for future
measurements of the masses and mixings mediating neutrino oscillations. To enable
further progress, we eventually need to understand, fairly 
completely, the underlying physics of the neutrino weak interaction within a 
nuclear environment. This completeness is required so that we can reliably apply 
the relevant model calculations across wide energy ranges and varying nuclei.

Neutrino cross-section uncertainties are already becoming a limiting factor in 
the determination of neutrino oscillation parameters in many experiments. 
Furthermore, experiments using heavier nuclear targets to increase their signal 
yields have to contend with the presence of significant nuclear effects impacting 
both the interaction cross sections and observed final states. Such nuclear 
effects also impact the reconstruction of the incoming neutrino energy, a key 
quantity in the determination of neutrino oscillation parameters. Understanding 
these neutrino-nucleus scattering processes directly affects how well one can 
separate signal from background. Uncertainties in both the neutrino interaction 
cross sections and associated nuclear effects must be understood to maximize 
the sensitivity of an experiment to neutrino oscillations. Of course, depending 
on the detector, the scientific question being asked, and the oscillation 
parameters, different cross-section uncertainties can take on different levels 
of importance. For example, careful control of neutrino/antineutrino cross section 
differences will be particularly important in establishing CP violation in the 
neutrino sector~\cite{jorge}. In fact, since $|U_{e3}|$ is relatively large, such systematic uncertainties become even more important because 
the expected $\nu$/$\bar{\nu}$ asymmetry becomes increasingly smaller for 
larger $|U_{e3}|$.

In addition, we need better understanding of neutrino-nucleus interactions 
for understanding the dynamics of supernovae. The physics of core-collapse supernova 
is not yet well-understood, and neutrinos are valuable probes into their inner workings.
Furthermore, we will need to understand neutrino-nucleus interactions in the few-tens-of-MeV regime in order to interpret a supernova neutrino burst observation.

These and related physics topics are most easily categorized according to the 
energy of the incident neutrino. The 0.2-10~GeV energy range (called 
``intermediate-energy'' here) is of most relevance to current and planned meson 
decay-in-flight (DIF) neutrino beams such as those being used currently for 
long-baseline experiments.
In 
addition, a beam from stored muons (e.g., 
the proposed nuSTORM facility~\cite{Adey:2013pio}) would also elucidate this regime.
The 10-100~MeV range (``low-energy'') is relevant for supernova neutrino studies.
A summary of 
current and future experiments relevant for these topics is given in
Table~\ref{table:xsection}.

\newcommand{\pdif}{$\pi$ DIF}
\newcommand{\pdar}{$\pi$ DAR}
\begin{table}[bt]
\centerline{
\begin{threeparttable}
\caption{\label{table:xsection} Current and proposed experiments for
$\nu$ cross section measurements or related studies. The upper (lower) part of table summarizes the 
intermediate- (low-) energy regime.}
\begin{tabular}{l|llcp{1.5cm}lcc} \hline
Experiment           & Physics\tnote{1}         &    $\nu$ Source          &  Energy (GeV)  & Target & Detector\tnote{2}       & Host       & Status         \\ \hline
%^Experiment &  &  $\nu$ Source &  (GeV)   & & type\tnote{2}  & Host     & Status  \\\hline
MiniBooNE~\cite{mb-xsec}  & MedE 	         & \pdif\        & 0.4-2    & CH$_2$ & Ch/calo     & Fermilab & Current \\
T2K~\cite{t2k-xsec}        & MedE	         & \pdif\        & 0.3-2    & CH     & Scitrk/        & J-PARC    &Current \\
           &               &               &          &        & TPC/calo         &          &        \\
MINERvA~\cite{minerva-xsec}    & MedE            & \pdif\        & 1-20     & many\tnote{3}   
                                                                 & Scitrk/calo    & Fermilab & Current \\
MINOS~\cite{minos-xsec}  & MedE            & \pdif\        & 1-20  & CH     & Scitrk         & Fermilab & Current \\   
ArgoNeuT~\cite{argoneut-xsec}  & MedE            & \pdif\        & 1-10  & Ar     & TPC         & Fermilab & Current \\   
NOvA NDOS~\cite{Betancourt:2013mba}  & MedE            & \pdif\        & 1        & CH$_2$     & Scitrk         & Fermilab &Current \\
NOvA near~\cite{Ayres:2007tu}  & MedE            & \pdif\        & 1.5-2.5  & CH$_2$     & Scitrk         & Fermilab & In constr. \\
MicroBooNE~\cite{ub-xsec} & MedE            & \pdif\        & 0.2-2    & Ar     & TPC            & Fermilab & In constr. \\
LArIAT~\cite{Adamson:2013/02/28tla} & MedE &  N/A$^4$ & 0.2-2 & Ar & TPC & Fermilab & In constr. \\
MINERvA~\cite{minerva-D2-onepager}    & MedE, PDFs      & \pdif\      & 1-10     & H,D    & Scitrk/calo    & Fermilab & Proposed \\
nuSTORM~\cite{Adey:2013pio} & MedE, $\nu_e$ xs & \pdif\       & 0.5-3.5  & TBD    & TBD            & Fermilab & Proposed \\ 
SciNOvA~\cite{cooper:2013}    & MedE            & \pdif\        & 1.5-2.5  & CH     & Scitrk         & Fermilab & Proposed \\
MiniBooNE+~\cite{Dharmapalan:2013yla}   & MedE            & \pdif\        & 0.3-0.5  & CH$_2$ & Ch/calo     & Fermilab & Proposed \\
CAPTAIN~\cite{Berns:2013usa}    & MedE            & \pdif\        & 1-10     & Ar     & TPC            & Fermilab & Proposed \\
LBNE near~\cite{Adams:2013qkq}  & MedE            & \pdif\        & 0.5-5      & TBD    & TBD            & Fermilab & Proposed \\
\hline
CAPTAIN~\cite{Berns:2013usa}    & LowE            & \pdar\        & 0.01-0.05  & Ar     & TPC         & ORNL      & Proposed \\
OscSNS~\cite{Elnimr:2013wfa}     & LowE            & \pdar\        & 0.01-0.05  & CH$_2$  & Ch/calo  & ORNL      & Proposed \\
IsoDAR~\cite{Aberle:2013ssa}     & LowE            & $^8$Li DAR 
                                             & 0.002-0.05 & TBD  & TBD            & TBD       & Proposed \\
CENNS~\cite{Brice:2013}      & $\nu A$ coh.    & \pdar\        & 0.01-0.05  & Ar   & Calo           & Fermilab  & Proposed \\
CSI~\cite{Akimov:2013fma}     & $\nu A$ coh.    & \pdar\        & 0.01-0.05  & TBD  & TBD            & ORNL      & Proposed \\
\hline
\end{tabular}
\begin{tablenotes}
\footnotesize
\item $^1$ Physics topics:
``MedE'' = quasi-elastic scattering, $\pi$ production, etc; 
``LowE'' = $\nu$-nucleus inelastic scattering and processes relevant for supernovae;
``$\nu N$ coh.'' = $\nu N$ coherent scattering
\item $^2$ Detector types: ``Ch'' = Cherenkov, ``Scitrk'' = scintillation tracker; ``Calo'' = calorimeter; ``TPC'' = time projection chamber
\item $^3$ many = He, CH, H$_2$O, Pb, Fe
\item $^4$ Charged-particle test beam (e,$\pi$,K,p).
\end{tablenotes}
\end{threeparttable}
}
\end{table}

\normalsize
\subsection{Intermediate-energy regime}

\begin{figure}[ht]
\begin{center}
       \includegraphics[width=0.48\textwidth]{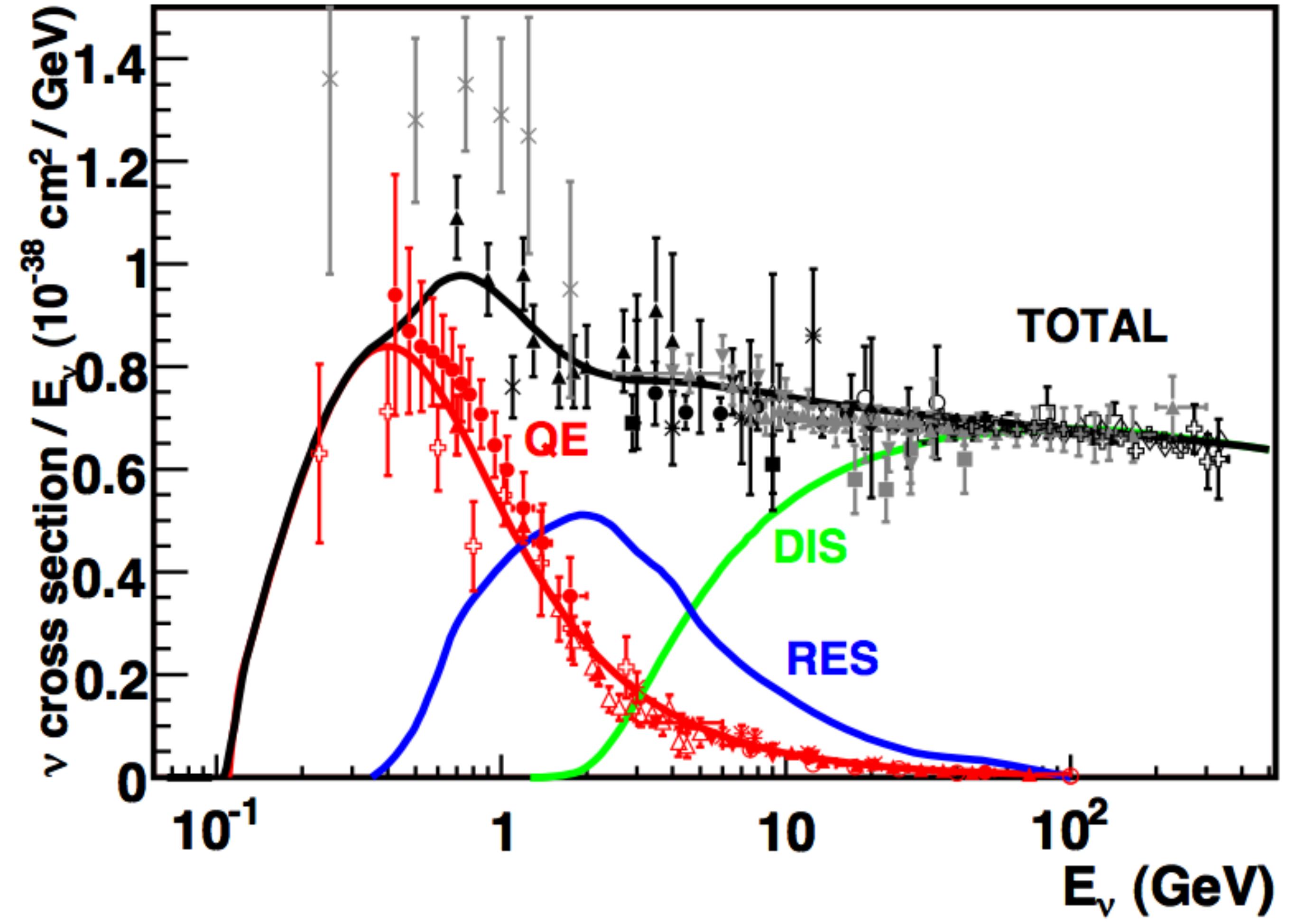}
       \includegraphics[width=0.48\textwidth]{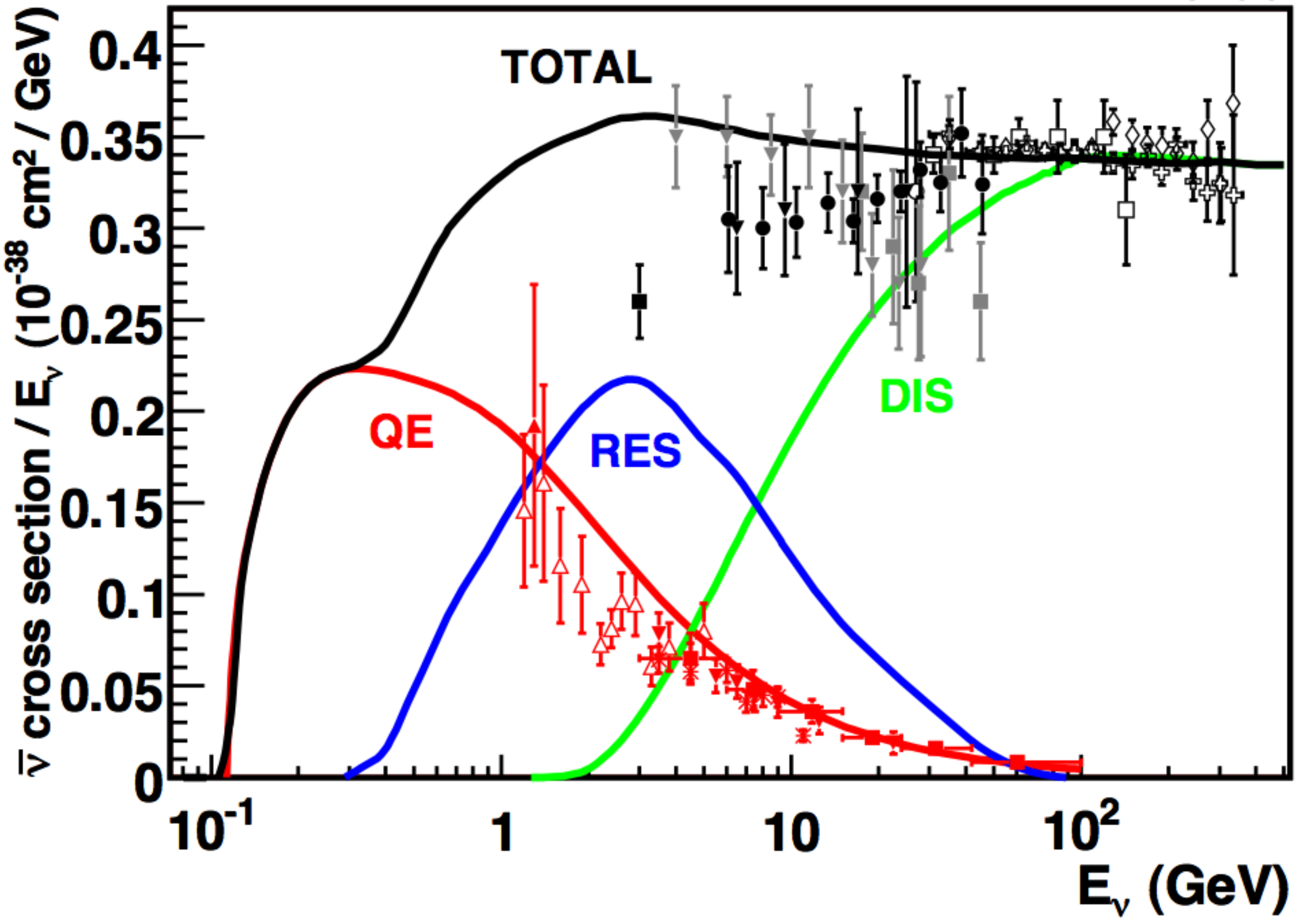}
\end{center}
\caption{Existing muon neutrino (left) and antineutrino (right) CC 
         cross section measurements~\cite{Formaggio:2013kya}
         and predictions~\cite{Casper:2002sd} as a function of neutrino energy.
         The contributing processes in this energy region include 
         quasi-elastic (QE) scattering, resonance production (RES), 
         and deep inelastic scattering (DIS). The error bars in the 
         intermediate energy range reflect the uncertainties in these
         cross sections (typically $10-40\%$, depending on the 
         channel).}
\label{fig:neutrino-xsecs}
\end{figure}

In the 0.2-10~GeV neutrino energy  regime, neutrino interactions are a 
complex combination  of quasi-elastic (QE) scattering, resonance production, and 
deep inelastic scattering processes, each of which has its own model and associated 
uncertainties. Solar and reactor oscillation experiments operating at very 
low neutrino energies and scattering experiments at very high energies have enjoyed 
very precise knowledge of their respective neutrino interaction cross sections 
(at the few-percent level) for the detection channels of interest.  However, the 
same is not true for the relevant intermediate energy regime. In this region, 
the cross sections even off free nucleons are not very well measured (at the 
$10 - 40\%$ level) and the data are in frequent conflict with theoretical predictions.  
Furthermore, the nuclear effects ranging from multi-nucleon-target initial states 
to complex final-state interactions are still quite poorly known.
Figure~\ref{fig:neutrino-xsecs} shows 
existing measurements of CC neutrino cross sections in the 
relevant energy range. Such measurements form the foundation of our knowledge 
of neutrino interactions and provide the basis for simulations in present use. 

There has been renewed interest and progress in neutrino interaction physics in the 
last ten years because of recent efforts to understand and predict signal and background 
rates in neutrino oscillation searches in few-GeV beams.  
One of several intriguing results from these new data 
comes from recent measurements of QE scattering. QE scattering
is a simple reaction historically thought to have a well-known cross section;
this is one reason why it is chosen as the signal channel in many neutrino 
oscillation experiments. Interestingly, the neutrino QE cross section recently 
measured on carbon at low energy by the MiniBooNE experiment is about $40\%$ higher 
than the most widely used predictions~\cite{Smith:1972xh} and is even larger than 
the free nucleon scattering cross section in some energy 
regions~\cite{AguilarArevalo:2010zc}. Similar effects are seen for 
antineutrinos~\cite{AguilarArevalo:2013hm}. These results are surprising because 
nuclear effects have always been expected to reduce the cross section, not enhance 
it. A recent QE cross section measurement from NOMAD at higher energies does not exhibit 
such an enhancement~\cite{Lyubushkin:2008pe}. A possible reconciliation between 
the two classes of measurements has suggested that previously-neglected 
nuclear effects could in fact significantly increase the QE cross section 
on nuclei at low energy~\cite{Martini:2009uj}. A similar enhancement has 
been observed in electron-nucleus scattering~\cite{Carlson:2001mp}. 
If true, this radically changes our thinking on nuclear effects and their 
impact on low-energy neutrino interactions. This revelation has been the 
subject of intense theoretical scrutiny and experimental investigation 
over the past few years (see, e.g.,~\cite{Fiorentini:2013ezn,Sobczyk:2012ms,Martini:2010ex, Fields:2013zhk}).

In the so-called resonance/transition region, the channels of interest are mainly 
hadronic resonances with the most important being the $\Delta(1232)$. 
Typical final states are those with a single pion. During the last five years,
several new pion production measurements have been performed. In all of them, 
the targets were nuclei (most often carbon). % and interpretation of the data 
As one example, the MiniBooNE experiment recently measured a comprehensive suite 
of CC $1\pi^+$, CC $1\pi^0$, and NC $1\pi^0$ production cross 
sections~\cite{AguilarArevalo:2010bm}. 
A variety of flux-integrated differential cross sections, often double differential, 
were reported for various final-state particle kinematics. The cross-section results 
differ from widely-used predictions at the 20\% level or more.

There are several efforts currently producing results that will add significantly to 
the available data and to the underlying physics understanding.  The MINERvA experiment in 
the 1-10~GeV NuMI beam at Fermilab has very recently published results on QE scattering
measured with a precise tracking detector from both neutrino and antineutrinos on 
carbon~\cite{Fiorentini:2013ezn,Fields:2013zhk}.  The near detectors of the 
T2K~\cite{Abe:2013jth} experiment are also measuring neutrino-nucleus 
interactions as part of their oscillation measurement program. T2K has recently reported 
total cross sections for neutrino CC inclusive scattering~\cite{Abe:2013jth}. 
Additional results on exclusive channels from MINERvA and the T2K and NOvA near 
detectors will be forthcoming in the near future. 
The MINERvA experiment will also perform the first studies of nuclear effects in
neutrino interactions using a suite of nuclear targets including He, C, O (water), 
Fe, and Pb in addition to a large quantity of scintillator CH. Analysis of neutrino
scattering processes from these varying nuclei are already underway. Another possible
step in the MINERvA  program is the addition of a deuterium 
target~\cite{minerva-D2-onepager} which is currently under review.  This is 
an intriguing, albeit challenging, possibility as it will allow nuclear effects 
in these processes to be separated from the bare-nucleon behavior.

All current accelerator-based neutrino experiments use a meson-decay beam either
on-axis, or off-axis to narrow the energy spread of the beam.  The uncertainty 
in the neutrino flux normalization and spectral shape will ultimately limit our 
understanding of the underlying physics of neutrino interactions and the ability to
conduct precision neutrino oscillation measurements.  Because of these 
uncertainties, an improved understanding of our neutrino beams is paramount. For 
these beams, some improvement in the knowledge of the neutrino flux is possible through 
meson-production experiments that determine the underlying meson momentum and angular 
distributions.  These can then be combined with detailed simulations of the neutrino 
beamline optics.  This procedure has been performed for the 
MiniBooNE~\cite{AguilarArevalo:2008yp}, 
K2K~\cite{Catanesi:2005rc}, and
T2K~\cite{Abe:2012av} experiments yielding predicted fluxes with $\sim$ 10\% errors.  
New experiments will require similar efforts with associated hadroproduction 
experiments~\cite{Sorel:2008zz} 
to push to a goal of 5\% errors.

Additional experiments in beams of different energies provide a valuable cross-check 
on the underlying energy dependence of physics models as well as the background 
calculations of the experiments.  For example, the NOvA experiment, which will soon 
run in the NuMI off-axis neutrino beam, offers a unique opportunity to add to the 
world's neutrino interaction data by measuring cross sections with its near detector 
as well as with a possible upgrade to a relatively-inexpensive fine-grained detector 
such as the proposed SciNOvA experiment~\cite{Paley:2010zza,cooper:2013}.

A potentially transformative next step 
would be the use of circulating muon beams.  The muons may be either uncooled and 
unaccelerated as in the case of nuSTORM~\cite{Adey:2013pio} or both cooled and accelerated 
as in the case of a Neutrino Factory. These facilities will yield a flux 
of neutrinos known to better than 1\%.
Another significant advantage of these muon-decay-based 
neutrino sources would be the availability, for the first time, of an intense and 
well-known source of electron-(anti)neutrinos.  Such beams would allow the measurement 
of $\nu_e$-nucleus cross sections, which are not measured and
are of great importance to future $\nu_\mu\rightarrow \nu_e$ oscillation experiments
since lepton universality may be broken due to nuclear effects in nuclei.

In addition to beam improvements, up-and-coming detector technologies such as LAr TPCs 
will both provide increased tracking precision for better final-state exclusivity as 
well as measurements specifically on argon (e.g.,~\cite{Palamara:2013maa}). Understanding interactions on argon is 
crucial for oscillation measurements in LBNE given that the far detector
of choice is a LAr TPC. New neutrino scattering measurements on argon are
already being reported by ArgoNeuT, which ran in the NuMI beam in 
2009--2010~\cite{Anderson:2011ce}. The near-future MicroBooNE experiment, which will
begin taking data
starting in 2014,
will further boost this effort in the next few years.
The LArIAT test-beam experiment at Fermilab  will study final-state particles in a LAr TPC  beginning in 2014~\cite{Adamson:2013/02/28tla}.
In addition, other efforts 
with $\sim 10$~ton LAr TPCs~\cite{Berns:2013usa} in an existing neutrino beam 
such as NuMI can also provide more information on reconstruction and final-state 
topology to further this effort. 

In order to adequately map out the complete nuclear dependence of the physics,  
there is need for multiple nuclear targets
combined 
with a precision tracker.  For this, an attractive follow-on to MINERvA would be a 
straw-tube/transition-radiation detector that employs multiple nuclear targets 
(including argon) simultaneously in the same beam such as that proposed for one of 
the LBNE near-detector options~\cite{Adams:2013qkq}. 

\subsection{Low-energy regime}

The 10-100~MeV neutrino energy range addresses a varied set of topics at the forefront 
of particle physics such as supernovae, dark matter, and nuclear structure. Low-energy
neutrino scattering experiments are possibilities at currently-existing high-intensity 
proton sources such as the ORNL SNS or the Fermilab Booster neutrino beam line. They 
should also be considered at future facilities such as Project X at Fermilab.

\noindent
\textbf{Supernova neutrino physics:} 
The multiple physics signatures and expected neutrino fluxes from a core-collapse signature are described in Secs.~\ref{mhother}, \ref{nu6_supernova}.  To get the most from the next supernova neutrino observation, it will be critical to understand the interactions of neutrinos with matter in the tens-of-MeV energy range~\cite{Berns:2013usa,nusns}.
A stopped-pion source provides a monochromatic source of 30 MeV $\nu_\mu$'s
from pion decay at rest, followed on a 2.2 $\mu$s timescale by
$\bar{\nu}_\mu$ and $\nu_e$ with a few tens of MeV from $\mu$ decay.
The $\nu$ spectrum matches the expected supernova spectrum reasonably
well. % (see Fig.~\ref{fig:SN_nu_spectra}).  
A $\sim1$~GeV, high-intensity,
short-pulse-width, proton beam is desirable for
creating such a $\nu$ source.  Prior examples used for neutrino
physics include LANSCE and ISIS.  A rich program of physics is possible 
with such a stopped-pion $\nu$ source, including measurement 
of neutrino-nucleus cross sections in the few tens of MeV range in a variety 
of targets relevant for supernova neutrino physics.  This territory is almost 
completely unexplored: so far only $^{12}$C has been measured at the 10\% level.  
A pion DAR neutrino source such as that currently available at the ORNL SNS would be an excellent source of neutrinos for this
physics on a variety of nuclei relevant for supernova~\cite{Bolozdynya:2012xv}.  In
addition, this source would allow specific studies to better understand the
potential of a large LAr detector such as that proposed for LBNE.  In particular,
low-energy neutrino-argon cross sections, required for supernova detection in
a large LAr detector could be measured with a near-future prototype 
LAr detector (CAPTAIN)~\cite{Berns:2013usa}.
In the farther future, 
the high-intensity FNAL Project X 1-3~GeV linac would also provide a potential site 
for these experiments.

\noindent
\textbf{Coherent elastic neutrino-nucleus  scattering (CENNS):}\label{cenns}
CENNS is a process in which the 
target nucleus recoils coherently via a collective neutral current exchange 
amplitude with a neutrino or antineutrino, is a long-sought prediction of the 
standard model. 
Although the process is well predicted by the standard model and has a comparatively 
large cross section ($10^{-39}~\mathrm{cm}^2$) in the relevant energy region 
($0-50$~MeV), CENNS has never been observed before as the low-energy nuclear 
recoil signature is difficult to observe. Numerous groups world-wide are now working 
to detect this elusive process~\cite{CNSworkshop}. Only a few sources, in particular nuclear reactors and 
spallation neutron sources~\cite{Akimov:2013fma, Bolozdynya:2012xv}
(as well as potential existing sources, such as the  FNAL 8~GeV proton source at a far off-axis location~\cite{Brice:2013})
produce the required 1-50 MeV energies of the neutrinos 
in sufficient quantities for a definitive first measurement. 
A modest sample of a few hundred events collected with a
keV-scale-sensitive dark-matter-style detector could improve upon
existing non-standard neutrino interaction parameter sensitivities by
an order of magnitude or more. A deviation from the $\sim$5\%
predicted cross section could be an indication of new physics~\cite{Scholberg:2005qs,Barranco:2005yy}.
The cross section is relevant for understanding the evolution of
core-collapse supernovae, characterizing future burst
supernova neutrino events collected with terrestrial detectors, and
a measurement of the process will ultimately set the background
limit to direct WIMP searches with detectors at approximately the ten-ton scale~\cite{Monroe:2007xp,Gutlein:2010tq}.
Proposals have arisen to probe nuclear structure~\cite{PhysRevC.86.024612} owing to the sensitivity of the coherent scatter process to the number of neutrons in the nucleus, and to search for sterile neutrinos~\cite{Formaggio:2011jt, Anderson:2012pn} by exploiting the flavor-blind nature of the process.   There are also potentially practical applications, as described in Sec.~\ref{cenns_app}.

\subsection{Required theoretical/phenomenological work}

A strong effort in theory/phenomenology/modeling is requisite to profit from improved
measurements in neutrino experiments.  While there is a healthy community working on 
the subject of neutrino-nucleus interactions in Europe, there is a dearth of 
phenomenologists in the  U.S. able to address the pressing theoretical questions needed 
to fully understand this subject and apply it to the interpretation of experimental data.  
There is a critical need within the U.S. physics community to devote 
time and resources to a theoretical/phenomenological understanding of neutrino-nucleus 
scattering. This naturally directly calls for a united effort of both the particle
and nuclear physics communities to better support these efforts~\cite{deGouvea:2013ita}.
There are numerous ideas that have been put forth by both experimentalists and theorists for
how best to proceed~\cite{Mariani:2013,Mosel:2013}.  They include suggestions for improvements 
to neutrino event generators with more sophisticated underlying calculations for neutrino
interactions on nucleons within nuclei, as well as considerations of the formation length of pions and nucleons and final-state interactions of the hadronic shower.

%% file: nu5.tex
\section{Beyond the standard paradigm --- anomalies and new physics}
\label{sec:anomalies}
Neutrinos moved beyond the standard model years ago with the discovery of neutrino oscillations, which 
implied the existence of neutrino mass.  Much of the oscillation data can be described by a three-neutrino 
paradigm.  However, there are intriguing anomalies that cannot be accommodated within this paradigm, and 
suggest new physics beyond it.  In particular, the marginal yet 
persistent evidence of oscillation phenomena around $\Delta m^2 \sim 1$~eV$^2$, which is 
not consistent with the well-established solar and atmospheric $\Delta m^2$ scales, is 
often interpreted as evidence for one or more additional neutrino states, known as 
sterile neutrinos.  Beyond the sterile neutrino, new physics may appear through a broad 
array of mechanisms collectively known as non-standard interactions (NSI).  Typically, 
searches for these effects occur in experiments designed to study standard 
phenomena.  One type of NSI that has been the subject of dedicated searches in the past
and may play a role in the future program is the neutrino magnetic moment. 
There are other ways that neutrino experiments can probe exotic physics.
For example, the possibility that neutrino oscillations may violate, to some degree, the very fundamental principles of Lorentz and CPT invariance has been considered; see e.g.,~\cite{Diaz:2009qk,Diaz:2013wia}.
In the 
following subsections we will discuss the prospects for neutrino experiments sensitive to
anomalies and new physics over the next several years.

\subsection{Sterile neutrinos}

Data from a variety of short-baseline experiments, as well as astrophysical 
observations and cosmology, hint at the existence of additional neutrino mass 
states beyond the three active species in the standard model (see, e.g.~\cite{Abazajian:2012ys}). 
The implications of these putative sterile neutrino states would be profound, 
and would change the paradigm of the standard model of particle physics. As a result, 
great interest has developed in testing the hypothesis of sterile neutrinos and 
providing a definitive resolution to the question: do light sterile neutrinos 
exist? 

Recently, a number of tantalizing results (anomalies) have emerged from 
short-baseline neutrino oscillation experiments that cannot be explained by the current 
three-neutrino paradigm.  These anomalies, which are not directly ruled out by other 
experiments, include the excess of $\bar{\nu}_e$ events ($3. 8\sigma$) observed 
by the LSND experiment~\cite{Aguilar:2001ty},  the $\nu_e$ ($3.4\sigma$) and 
$\bar{\nu}_e$  ($2.8\sigma$) excesses observed by MiniBooNE~\cite{Aguilar-Arevalo:2013pmq}, 
particularly at low-energy in $\nu_e$ mode~\cite{AguilarArevalo:2008rc},
the deficit of $\bar{\nu}_e$ events ($0.937\pm0.027$) observed 
by reactor neutrino experiments~\cite{Mention:2011rk}, and the deficit 
of $\nu_e$ events ($0.86\pm0.05$) observed in the SAGE and 
GALLEX radioactive source experiments~\cite{Giunti:2010zu}. 

Although there may be several possible ways to explain these anomalies, a simple explanation
is the $3+N$ sterile neutrino model, in which there are three light, mostly active neutrinos and 
$N$, mostly sterile neutrinos which mix with the active flavors. For $N>1$, these models 
allow for CP-violating effects in short-baseline appearance experiments.  The world's 
oscillation data can be fit to these 3+$N$ models resulting in allowed regions that close at 95\% CL or 
better, as shown in Fig.~\ref{fig:sterile1} and \ref{fig:sterile2} for the 3+1 model. Still, significant 
tension exists between the appearance and disappearance data~\cite{Kopp:2013vaa}, particularly  
due to the absence of $\nu_{\mu}$ disappearance in the $\Delta m^2 \sim 1$ eV$^2$ 
region~\cite{Mahn:2011ea,Adamson:2011ku}, a key prediction of the 3+$N$ models. 

Beyond particle physics, there are hints of additional neutrinos coming from cosmology.  Fits 
to astrophysical data sets (including the cosmic microwave background (CMB), large scale 
structure, baryon acoustic oscillations and Big Bang nucleosynthesis) are sensitive to the 
effective number of light degrees of freedom ($N_{\rm eff}$)  (which in the standard model is equivalent 
to saying the effective number of neutrino families, although in principle this could include
other types of light, weakly-coupled states).   Prior to the release of the Planck data in 2013, 
there was an astonishing trend that such fits, conducted by different groups and involving 
differing mixes of data sets and assumptions, tended to favor $N_{\rm eff}$ closer to 4 than 
3~\cite{Abazajian:2012ys}.  With the release of Planck data~\cite{Ade:2013zuv} new, more precise 
fits to $N_{\rm eff}$ are now more consistent with 3.  The Planck collaboration fit values range from $3.30\pm0.52$ 
(95\% CL) to $3.62\pm0.49$ (95\% CL) depending on which other data sets are included in the fit.  
The pre-Planck fits used the full-sky WMAP~\cite{Komatsu:2010fb} data set for the first three peaks of the the CMB angular power spectrum, but typically relied 
on narrow-sky, high angular resolution observations by the South Pole Telescope~\cite{Keisler:2011aw}, 
or the Atacama Cosmology Telescope~\cite{Dunkley:2010ge} for the next four peaks.  The Planck mission 
combined a full-sky survey with high angular resolution, and was, for the first time, able to measure 
the first seven peaks in the spectrum with one apparatus.  The Planck Collaboration believes 
that a miscalibration in the stitched together spectra was responsible for the anomalously high 
value of $N_{\rm eff}$  found in the earlier fits~\cite{Ade:2013zuv}, but the issue is not yet resolved. 
There is tension between the value of the Hubble constant extracted from Planck data, and that measured
by the Hubble Space Telescope. The resolution of this issue may impact the extracted value of $N_{\rm eff}$. Nonetheless, 
while the new fits to $N_{\rm eff}$ are now more consistent with three light degrees of freedom, they are still high and allow $N_{\rm eff}=4$ at less than, at most, the 2$\sigma$ level.  
Finally, it is important to keep in mind that cosmological constraints on the existence of light sterile neutrinos depend on the masses of the mostly sterile states, and on whether they are in thermal equilibrium with the rest of the Universe. Even at face value, the Planck data are still consistent with one or more massless sterile neutrino states that were not fully thermalized~\cite{Ade:2013zuv}.

For a comprehensive review of light sterile neutrinos including the theory, the cosmological evidence, 
and the particle physics data, see Ref.~\cite{Abazajian:2012ys}.

In order to determine if these short-baseline anomalies are due to neutrino oscillations in 
a $3+N$ sterile neutrino model, future short-baseline experiments are needed. Table~\ref{table:sterile} lists many proposals for such experiments. These experiments 
should have robust signatures for electron and/or muon neutrino interactions and they should be capable of 
measuring the $L/E$ dependence of the appearance or disappearance effect.  Several ways of 
measuring $L/E$ dependence have been proposed including: 1) placing a large detector close to a 
source of low-energy neutrinos from a reactor, cyclotron or intense radioactive source and measuring the $L/E$ 
dependence of the $\boss{\nu}{e}$ disappearance with a single detector, 2) positioning detectors at 
two or more baselines from the neutrino source, and 3) measuring the $L/E$ dependence of high 
energy atmospheric neutrinos, where strong matter effects are expected, in particular close 
to the matter resonance expected for the sterile $\Delta m^2$ in the Earth's core.  In addition, experiments 
sensitive to neutral current interactions, in which active flavor disappearance would be a direct test of 
the sterile hypothesis, are needed.  

Finally, it is important to note that satisfactorily resolving these short-baseline anomalies, even if unrelated to sterile neutrinos, is 
very important for carrying out the three-flavor neutrino oscillation program described earlier.  The 
2 to $3\ \sigma$ effects reported at the sub-percent 
to the several-percent level are similar in scale and effect to the  CP-violation and mass 
hierarchy signals being pursued in long-baseline experiments. 

Independent from the short-baseline anomalies, new, mostly sterile, neutrino mass eigenstates with different masses can be searched in a variety of different ways, ranging from weak decays of hadrons and nuclei, charged-lepton flavor violating processes, to searches at lepton and hadron colliders. For details, see, e.g.,~\cite{Smirnov:2006bu,deGouvea:2007uz,Atre:2009rg,Gninenko:2013tk}. In particular, a next-generation $e^-e^+$ collider would provide very stringent bounds on sterile neutrinos with masses around tens of GeV and other new neutrino phenomena (see, e.g.,~\cite{Carena:2003aj}).

\begin{table}[bt]
\centerline{
\begin{threeparttable}
\caption{\label{table:sterile}Proposed sterile neutrino searches.}
\begin{tabular}{l|ccccc} \hline
Experiment                            & $\nu$ Source          & $\nu$ Type        & Channel                & Host           & Cost Category\tnote{1} \\ \hline
CeLAND~\cite{Gando:2013zla}                & $^{144}$Ce-$^{144}$Pr & $\bar{\nu}_e$     & disapp.                & Kamioka, Japan & small\tnote{2} \\
Daya Bay Source~\cite{DayaBaySource}  & $^{144}$Ce-$^{144}$Pr & $\bar{\nu}_e$     & disapp.                & China          & small \\ 
SOX~\cite{SOX}                        & $^{51}$Cr             & $\nu_e$           & disapp.                & LNGS, Italy    & small\tnote{2} \\
                                      & $^{144}$Ce-$^{144}$Pr & $\bar{\nu}_e$     & disapp.                &                & \\ 
BEST~\cite{Abazajian:2012ys}  & $^{51}$Cr & $\nu_e$     & disapp.                & Russia          & small \\ \hline
PROSPECT~\cite{Hans1:2013}          & Reactor               & $\bar{\nu}_e$     & disapp.                & US\tnote{3}    & small \\
STEREO                                & Reactor               & $\bar{\nu}_e$     & disapp.                & ILL, France    & N/A\tnote{4} \\
DANSS~\cite{DANSS}                    & Reactor               & $\bar{\nu}_e$     & disapp.                & Russia         & N/A\tnote{4} \\ \hline
OscSNS~\cite{Elnimr:2013wfa}               & $\pi$-DAR             & $\bar{\nu}_{\mu}$ & $\bar{\nu}_{e}$ app.   & ORNL, U.S.       & medium \\
LAr1~\cite{Fleming:2013uaa}                      & $\pi$-DIF             & $\boss{\nu}{\mu}$ & $\boss{\nu}{e}$ app.   & Fermilab       & medium \\  
LAr1-ND~\cite{Fleming:2013uaa}                      & $\pi$-DIF             & $\boss{\nu}{\mu}$ & $\boss{\nu}{e}$ app.   & Fermilab       & medium \\  
MiniBooNE+~\cite{Dharmapalan:2013yla}          & $\pi$-DIF             & $\boss{\nu}{\mu}$ & $\boss{\nu}{e}$ app.   & Fermilab       & small  \\
MiniBooNE II~\cite{MiniBooNEII}       & $\pi$-DIF             & $\boss{\nu}{\mu}$ & $\boss{\nu}{e}$ app.   & Fermilab       & medium \\ 
ICARUS/NESSiE~\cite{Antonello:2012hf} & $\pi$-DIF             & $\boss{\nu}{\mu}$ & $\boss{\nu}{e}$ app.   & CERN           & N/A\tnote{4} \\
IsoDAR~\cite{Aberle:2013ssa}                  & $^8$Li-DAR            & $\bar{\nu}_e$     & disapp.                & Kamioka, Japan & medium \\ 
nuSTORM~\cite{Adey:2013pio}        & $\mu$ Storage Ring    & $\boss{\nu}{e}$   & $\boss{\nu}{\mu}$ app. & Fermilab/CERN  & large \\ \hline
\end{tabular}
\begin{tablenotes}
\item $^1$ Rough recost categories: small: $<$\$5M, medium: \$5M-\$50M, large: \$50M-\$300M.
\item $^2$ U.S. scope only.
\item $^3$ Multiple sites are under consideration~\cite{Hans2:2013}.
\item $^4$ No U.S. participation proposed.  
\end{tablenotes}
\end{threeparttable}}
\end{table}

\subsubsection{Projects and proposals with radioactive neutrino sources}\label{radsources}
Proposals to use radioactive neutrino sources to search for sterile neutrino oscillations actually predate the 
``gallium anomaly''~\cite{Grieb:2006mp}.  Perhaps the most intriguing opportunity with the source experiments is the 
possibility of precision oscillometry --- the imaging, within one detector, the oscillation over multiple wavelengths in $L/E$. 
Therefore this approach would likely be the best way to deconvolve the multiple frequencies expected if there are two or 
more sterile neutrino states.  Typically these proposals are built around existing detectors with well-measured 
backgrounds, where the new effort involves creating a source and delivering it to the detector.  There are two types 
of sources actively under consideration: 1) $^{51}$Cr, an electron capture isotope which produces $\nu_e$ of 
750~keV, and 2) $^{144}$Ce-$^{144}$Pr, where the long-lived $^{144}$Ce ($\tau_{1/2}=285$~days) $\beta$-decays 
producing a low energy $\bar{\nu}_e$ of no interest, while the daughter isotope, $^{144}$Pr, rapidly $\beta$-decays 
producing a $\bar{\nu}_e$ with a 3~MeV endpoint.   Since $^{51}$Cr neutrinos are monoenergetic, with no need to reconstruct 
the neutrino energy, they can be detected by CC, NC or elastic scattering interactions.  $^{144}$Pr neutrinos, on 
the other hand, are emitted with a $\beta$ spectrum and must be detected 
via a charged-current process such as inverse $\beta$-decay.  

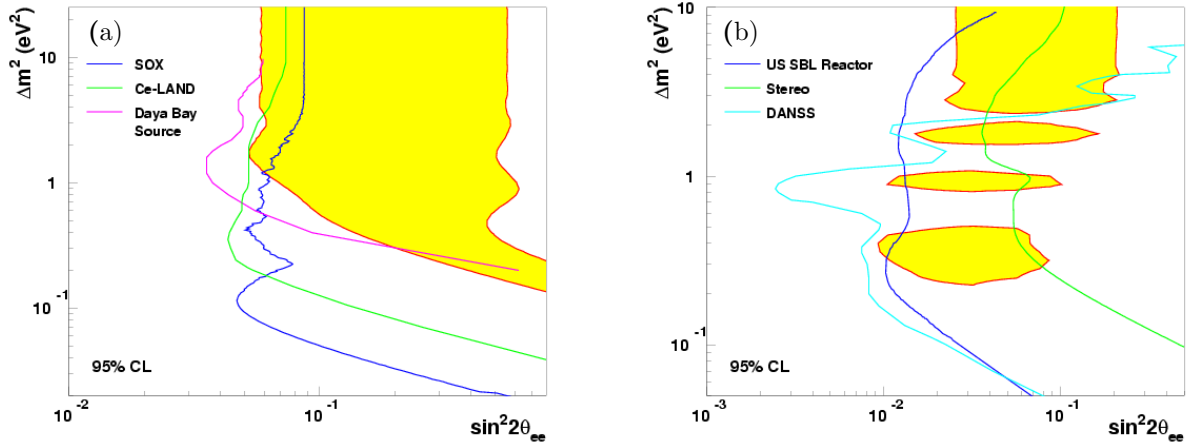
\begin{figure}[tbp]
\begin{center}
     \includegraphics[width=0.45\textwidth]{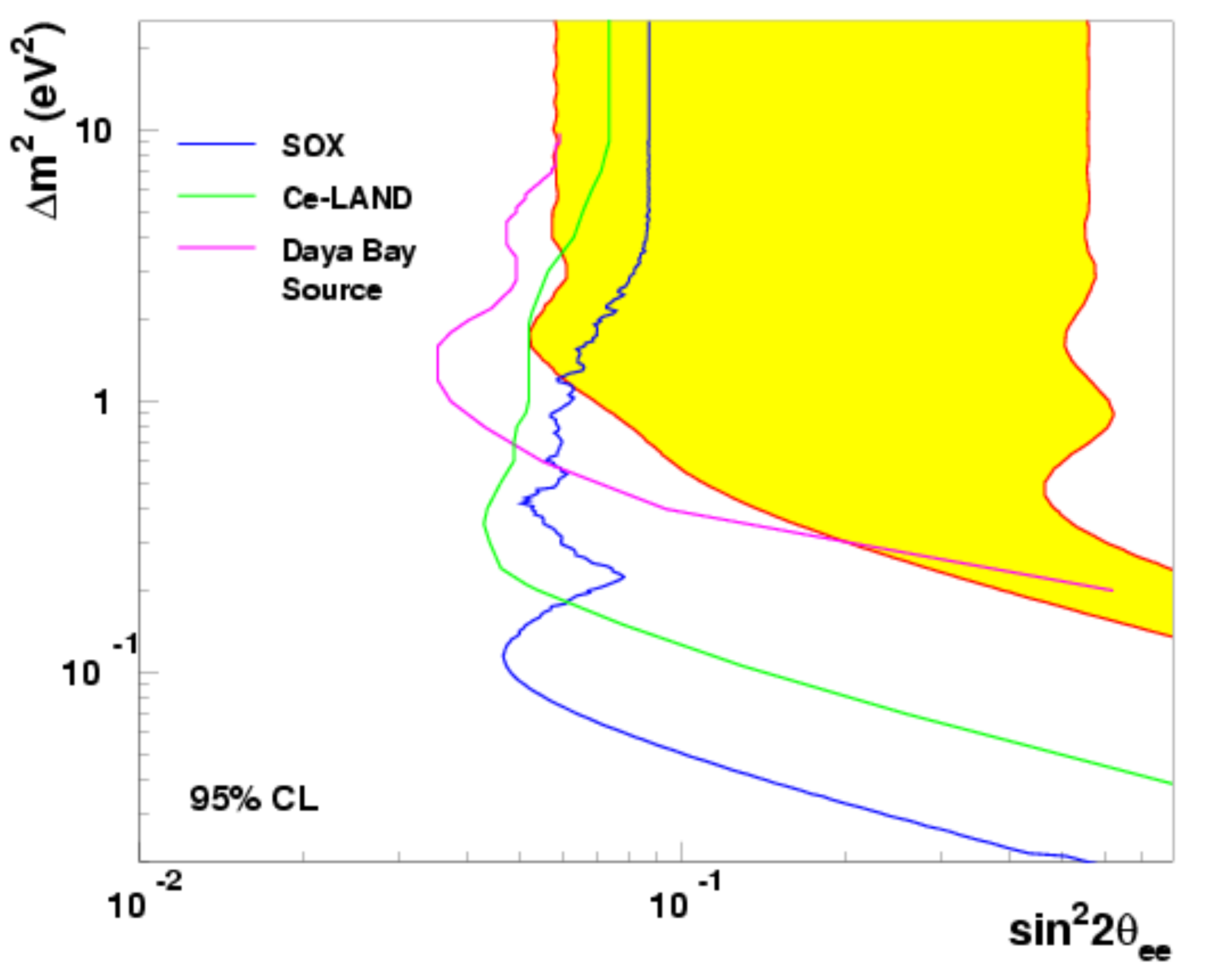}
     \hspace{0.05\textwidth}
     \includegraphics[width=0.45\textwidth]{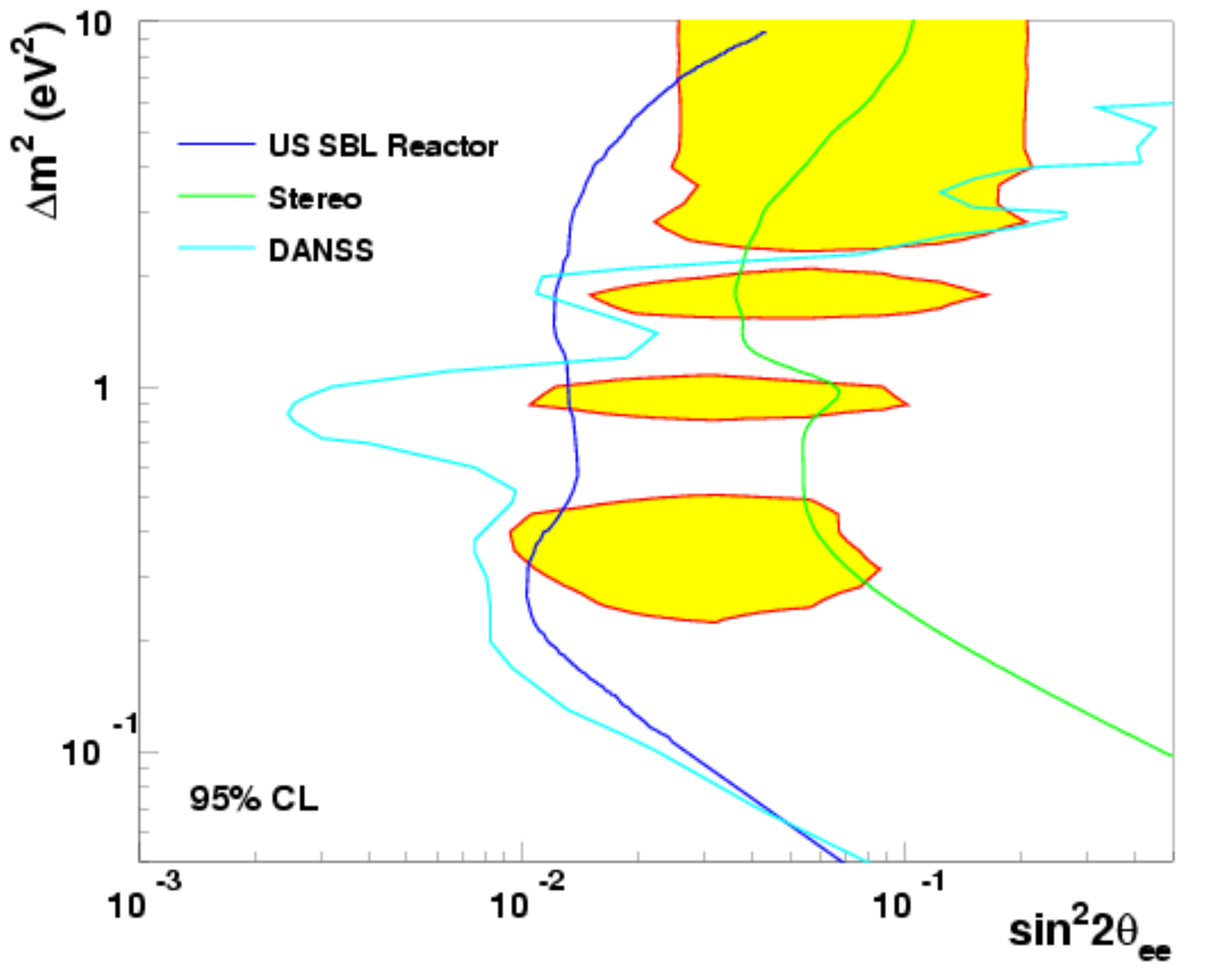}
\end{center}
\begin{picture}(0,0)
  \put (40,177) {\textbf (a)}
  \put (280,177) {\textbf (b)}
\end{picture}
%\vskip-0.8cm
\caption{\label{fig:sterile1}  Collaboration-reported sensitivity curves for proposed source (a) and
reactor (b) experiments plotted against the global fits~\cite{Kopp:2013vaa} for the gallium 
anomaly and reactor anomaly respectively.}
\end{figure}

Proposals actively under consideration include SOX~\cite{SOX} based on the Borexino detector, 
CeLAND~\cite{Gando:2013zla} based on the KamLAND detector, and a Daya Bay Source 
experiment~\cite{DayaBaySource}.  SOX is considering both $^{51}$Cr and $^{144}$Ce-$^{144}$Pr phases.  In the $
^{51}$Cr phase, a source of up to 10~MCi is placed about 8~m from the center of the detector.  This phase takes 
advantage of Borexino's demonstrated ability to see the $\nu_e-e$ elastic scattering of 861~keV, $^7$Be solar
neutrino~\cite{Arpesella:2008mt}.  Later phases may involve a $^{144}$Ce-$^{144}$Pr source which could be located 
either inside or outside the detector, the former requiring major modifications to the Borexino detector.  The CeLAND and the Daya 
Bay Source proposals are both based on $^{144}$Ce-$^{144}$Pr.  In the Daya Bay Source proposal, a 500~kCi source 
is placed in between the four 20-ton antineutrino detectors at the Daya Bay far site.  With CeLAND, a 75~kCi 
source could be placed either outside the detector, 9.5~m from the center, or inside the detector (only after 
the KamLAND-Zen $0\nu\beta\beta$ run is complete).  The sensitivity for these proposals is shown in 
Fig.~\ref{fig:sterile1}a.  BEST, in Russia~\cite{Abazajian:2012ys}, proposes a 3-MCi $^{51}$Cr source with a 50-t Ga metal detector.

There is also the possibility of a sterile neutrino measurement based on the 
combination  of a $^{51}$Cr source with cryogenic solid state bolometers, to detect all active neutrino flavors 
through neutral current CENNS~\cite{Formaggio:2011jt} (see Sec.~\ref{cenns}).  This proposal, known as 
RICOCHET, would be a direct test of the sterile hypothesis since the neutral current is equally 
sensitive to all active flavors, but blind to sterile neutrinos.  

\subsubsection{Projects and proposals that directly address the reactor anomaly}\label{reactorsterile}
The apparent deficit of neutrinos in short-baseline reactor neutrino experiments, known as the reactor anomaly, is
result of two distinct lines of analysis: the theoretical calculations of the reactor antineutrino 
flux~\cite{Schreckenbach:1985ep,Hahn:1989zr,Mueller:2011nm,Huber:2011wv} (see also~\cite{Hayes:2013wra}), which are based on measurements 
of the $\beta$-spectra from the relevant fission isotopes~\cite{Schreckenbach:1985ep,Hahn:1989zr}, and 
the reactor antineutrino measurements~\cite{Kwon:1981ua,Declais:1994ma,Declais:1994su,
Zacek:1986cu,Afonin:1988gx,Kuvshinnikov:1990ry,Vidyakin:1987ue,Vidyakin:1994ut,Greenwood:1996pb}.  The 
anomaly~\cite{Mention:2011rk} emerges in the comparison of these two analyses, and as such, both improved flux 
calculations (and the underlying $\beta$-spectra measurements) and new reactor antineutrino measurements are needed.  

The most direct proof of a sterile neutrino solution to the reactor anomaly would be to observe a spectral 
distortion in the antineutrino rate that varies as a function of distance from the reactor core.  There are 
several projects and proposals from all over the world to search for this effect, including 
STEREO~\cite{Abazajian:2012ys} at ILL in France and DANSS~\cite{DANSS} at the Kalinin 
Power Plant in Russia, to name two.  
In the U.S., the PROSPECT collaboration~\cite{Hans1:2013} is preparing for a precision measurement of the reactor antineutrino spectrum at very short baselines at the NIST, HFIR, or ATR research reactors~\cite{Hans2:2013}. Two segmented detectors with several tons of target mass will be used to measure the antineutrino spectrum at distances ranging from 4-20~m and to search for short-baseline oscillations. 
A compact research reactor core is highly desirable to reduce the smearing and uncertainty in $L$.  A segmented detector design with sufficient spatial resolution, improved background rejection and better neutron tagging is under development~\cite{Hans3:2013}.

For antineutrinos, the existing reactor $\theta_{13}$ experiments, such as 
Daya Bay~\cite{TheDayaBay:2013kda}, with their high-statistics near detectors, at baselines far enough to 
average out any spectral distortions from sterile oscillations, will provide the world's best data on reactor 
fluxes, ensuring that the uncertainty on the reactor anomaly is dominated by the flux calculation.  New 
measurements of the $\beta$-spectra of the fission isotopes~\cite{Asner:2013nda} would be helpful in further 
reducing the uncertainty on the flux calculation, but theoretical uncertainties from effects such as weak 
magnetism~\cite{Huber:2011wv} will ultimately limit this approach.    

\subsubsection{Projects and proposals with accelerator-induced neutrinos}\label{sec:nu5_accproj}
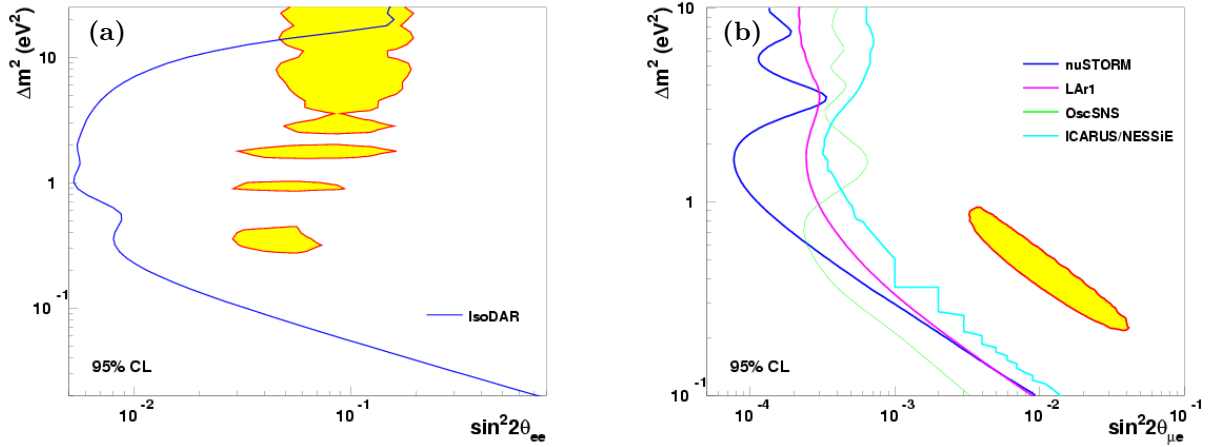
\begin{figure}[tbp]
\begin{center}
     \includegraphics[width=0.45\textwidth]{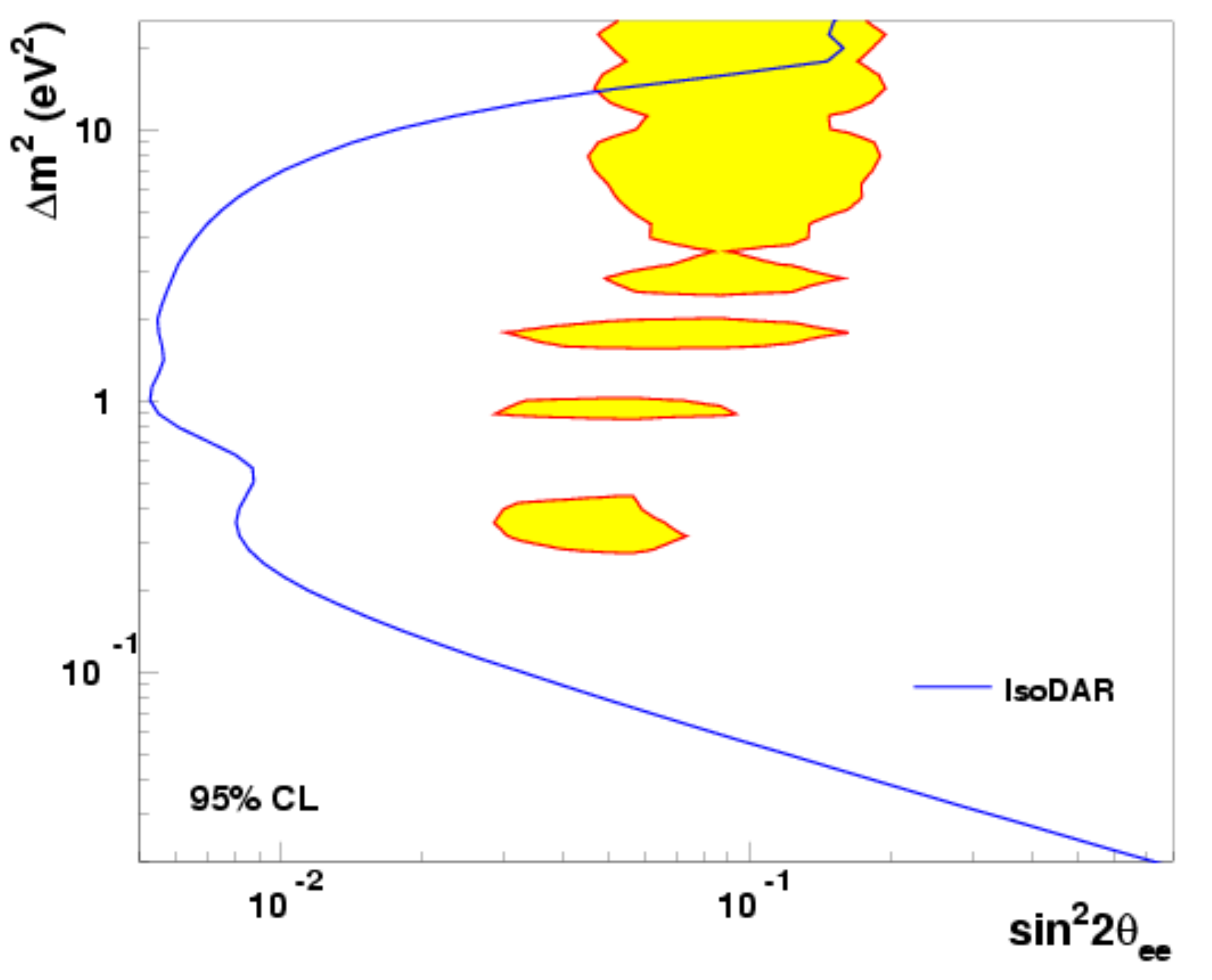}
     \hspace{0.05\textwidth}
     \includegraphics[width=0.45\textwidth]{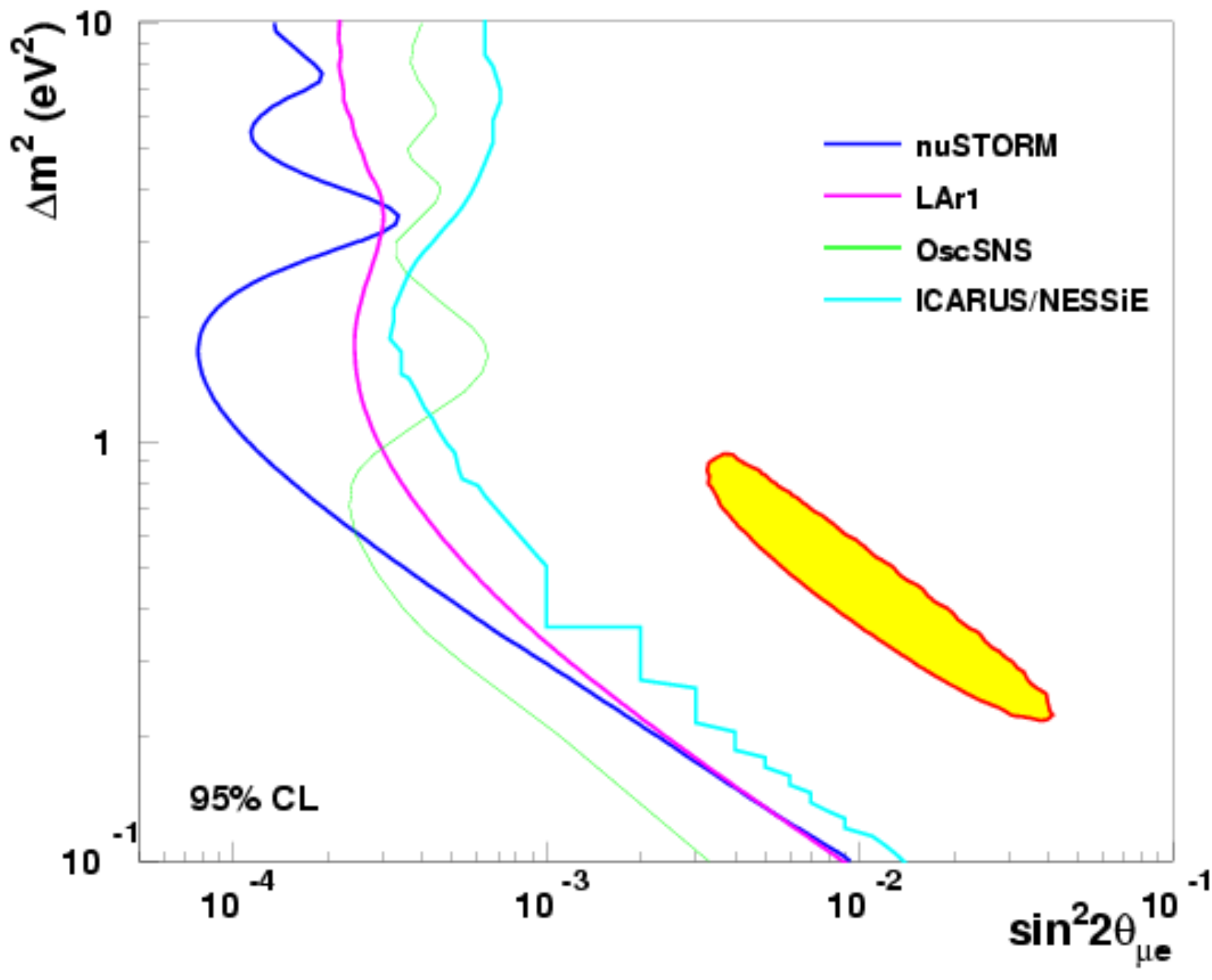}
\end{center}
\begin{picture}(0,0)
  \put (40,177) {\textbf{(a)}}
  \put (280,177) {\textbf{(b)}}
\end{picture}
%\vskip-0.8cm
\caption{\label{fig:sterile2} Collaboration-reported sensitivity curves for proposed accelerator-based
experiments sensitive to (a) $\nu_e$ and $\bar{\nu}_e$ disappearance (IsoDAR)~\cite{Aberle:2013ssa} and (b) appearance which includes 
$\nu_{\mu}\to \nu_e$ and $\nu_e \to \nu_{\mu}$ in both neutrinos and antineutrinos, plotted against 
the global fits~\cite{Kopp:2013vaa}.}
\end{figure}

There are a number of proposals involving Fermilab's Booster Neutrino Beam (BNB) which are 
relevant to the sterile neutrino question.  The MicroBooNE experiment~\cite{Ignarra:2011yq}, which is currently under construction 
just upstream of MiniBooNE,  will use the fine grain 
tracking of its 170-ton LAr TPC to study, in detail, the interaction region of events 
corresponding to the MiniBooNE low-energy excess, and may help to determine if these 
$\nu_{\mu}\to\nu_e$  oscillation candidates are really $\nu_e$ charged current quasielastic events 
as assumed by MiniBooNE.   Similarly, the proposed MiniBooNE+~\cite{Dharmapalan:2013yla} 
would look for neutron captures following $\nu_e$ candidate events.  In the MiniBooNE energy 
range, the production of free neutrons in a neutrino interaction is five times more likely for charged-current than for neutral-current events.  MiniBooNE+ would attempt to detect these neutrons by adding scintillator to 
the MiniBooNE detector, making it sensitive to the 2.2~MeV gammas produced when a neutron
captured on hydrogen.  This neutron tagging capability would be used to study whether the 
MiniBooNE low-energy excess events are truly $\nu_e$ events as the oscillation hypothesis 
requires.  The MiniBooNE II proposal~\cite{MiniBooNEII}, to either build a new 
near detector or move the existing MiniBooNE detector to a near location, is also intended 
as a test of MiniBooNE excess.  The presence of a near detector may help to confirm or refute the 
baseline dependence of the excess. 
The LAr1 proposal~\cite{Fleming:2013uaa} is a multi-baseline proposal 
for the BNB which is based on LAr.  It would add a 40-ton-fiducial ``LAr1-ND''
detector at 100~m and eventually a 1-kt-fiducial, ``LAr1", detector at 700~m to the existing MicroBooNE detector, 
which is at a baseline of 470~m.  The projected sensitivity of this three-detector combination 
is shown in Fig.~\ref{fig:sterile2}b.  
In Fermilab's NuMI beam line the MINOS+
experiment~\cite{Plunkett:2013} will search for $\nu_\mu$ disappearance caused by oscillations to $\nu_s$.
There is also a proposal at CERN for a two detector LAr TPC known as ICARUS/NESSiE~\cite{Antonello:2012hf}. 
In this proposal, the ICARUS T600 LAr TPC would be moved from Gran Sasso and set 1600~m downstream
from a new neutrino beam extracted from the CERN-SPS.  A second, smaller LAr TPC would be built at 300~m.  Additionally a muon 
spectrometer would be installed behind each TPC.  The projected sensitivity of ICARUS/NESSiE is shown in 
Fig.~\ref{fig:sterile2}b.

The Spallation Neutron Source (SNS) at Oak Ridge National Laboratory is also an intense and 
well-understood source of neutrinos from $\pi^+$ and $\mu^+$ decays-at-rest in much the same way that LAMPF 
produced neutrinos for LSND~\cite{Athanassopoulos:1996ds}.  As such, it is an excellent place to make a 
direct test of LSND.   The OscSNS~\cite{Elnimr:2013wfa} 
proposal would build an 800-ton detector approximately 60~m from the SNS beam dump.  OscSNS could improve 
upon LSND in at least three specific ways: 1) the lower duty factor of the SNS significantly reduces cosmic 
backgrounds, 2) the detector would be placed upstream of the beam lowering the possibility of non-neutrino, 
beam-correlated backgrounds, and 3) gadolinium-doped scintillator may be used to capture neutrons, providing a more robust tag 
of inverse $\beta$-decay.  In addition to $\bar{\nu}_e$ appearance, OscSNS would search 
for $\nu_{\mu}$ disappearance (via the $\nu_{\mu} + {}^{12}{\rm C} \rightarrow \nu_{\mu}+ {}^{12}{\rm C}^*$ channel using 30-MeV $\nu_\mu$)  and $\nu_e$ disappearance.  The projected sensitivity of the OscSNS $\bar{\nu}_e$ appearance 
search is shown in Fig.~\ref{fig:sterile2}b.

IsoDAR~\cite{Aberle:2013ssa} is a proposal to use a low-energy, high-power cyclotron to produce $^8$Li, which $\beta$-decays 
producing a $\bar{\nu}_e$ with an endpoint of 13~MeV; it is potentially a precursor to DAE$\delta$ALUS~\cite{Aberle:2013ssa}.  This cyclotron would be placed near the KamLAND detector which 
would detect the $\bar{\nu}_e$ via inverse $\beta$-decay.  This arrangement would be sensitive to the disappearance of $\bar{\nu}_e$,
and, given the low energy of the neutrinos and 13-m diameter detector, it should be capable of precision oscillometry.  The 
projected sensitivity of IsoDAR is shown in Fig.~\ref{fig:sterile2}a.

The nuSTORM~\cite{Adey:2013pio} proposal is to build a racetrack-shaped muon storage ring, to provide clean 
and well-characterized  beams of $\nu_e$ and $\bar{\nu}_{\mu}$ (or $\bar{\nu}_e$ and $\nu_{\mu}$ if
$\mu^-$ are stored).  These beams would enable extremely precise searches for sterile neutrino oscillations in four
neutrino types, in both appearance and disappearance channels.  The most powerful and unprecedented capability 
of nuSTORM would be to search for $\boss{\nu}{\mu}$ appearance.  The nuSTORM beams are essentially free of 
intrinsically-produced wrong sign/wrong flavor neutrinos which are unavoidable in pion decay-in-flight beams.  On the other hand 
muon storage rings simultaneously produce $\nu_e$ and $\bar{\nu}_\mu$, so it essential to have magnetic detectors to 
distinguish  between $\bar{\nu}_{\mu}$ from oscillation and $\nu_{\mu}$ from the beam.  
The proposed nuSTORM project has near and far magnetized iron detectors, but future upgrades could include magnetized 
LAr TPCs.  NuSTORM is a facility which, in addition to sterile neutrino searches, would make neutrino cross-section 
measurements critical to the long-baseline program (see Sec.~\ref{sec:scattering}) and conduct neutrino factory R\&D, yet it is based on existing accelerator 
technology.  Proposals for nuSTORM are currently being considered by both Fermilab~\cite{Kyberd:2012iz} and 
CERN~\cite{Adey:2013afh}.  The projected sensitivity of the nuSTORM $\boss{\nu}{e}\to\boss{\nu}{\mu}$ search is shown 
in Fig.~\ref{fig:sterile2}b.

\subsubsection{Sensitivity from atmospheric neutrinos}
The disappearance of atmospheric $\nu_{\mu}$ in the 0.5 to 10~TeV energy range can be enhanced by matter effects in the 
Earth's core for the case of a sterile neutrino with $\Delta m^2 \sim 1$~eV$^2$~\cite{Nunokawa:2003ep,Choubey:2007ji}.  
Such neutrinos are observed by the IceCube experiment~\cite{Barger:2011rc,Esmaili:2012nz} at the South Pole, which can measure or set 
limits on the muon to sterile mixing amplitude by studying the zenith angle (effectively $L$) and energy dependence of any 
disappearance effect.

\subsection{Non-standard interactions of neutrinos} 
 
Neutrino experiments in general, and neutrino oscillation experiments in particular, are also 
very sensitive to new, heavy degrees of freedom that mediate new 
neutral-current interactions or to modifications of the standard-model weak interactions. These so-called non-standard interactions (NSI) between neutrinos and 
charged fermions modify not only neutrino production and detection, but also neutrino 
propagation through matter effects. 

Different types of new physics lead to NSI (see, e.g.,~\cite{Gavela:2008ra,Biggio:2009nt}, and references therein). These can be parametrized in terms of the effective operators 
\begin{equation}
G_F\epsilon^f_{\alpha\beta}\nu_{\alpha}\gamma_{\mu}\nu_{\beta}\bar{f}\gamma^{\mu}f,
\end{equation} 
where $\nu_{\alpha,\beta}=\nu_{e,\mu,\tau}$, $f$ are charged fermions ($e, u, d, \mu, s$, 
\ldots), $G_F$ is the Fermi constant, and $\epsilon$ are dimensionless 
couplings.\footnote{$\epsilon\sim 1$ ($\ll 1$) implies that the new physics effects are on 
the order of (much weaker than) those of the weak interactions.} When $f$ is a 
first-generation fermion, the NSI contribute to neutrino detection and production at order 
$\epsilon^2$ (ignoring potential interference effects between the standard model and the NSI). 
On the other hand, the NSI also contribute to the forward-scattering amplitude for neutrinos 
propagating in matter, modifying the neutrino dispersion relation and hence its oscillation 
length and mixing parameters. These modified matter effects are of order $\epsilon^1$ and 
potentially more important than the NSI effects at production or detection. Furthermore, for 
$\alpha\neq\beta$, the NSI-related matter effects lead to 
$P_{\alpha\beta}\neq\delta_{\alpha\beta}$ in the very short baseline limit ($L\to 0$);
these are not present in the standard model case. More information -- including relations to 
charged-lepton processes -- current bounds, and prospects 
using different neutrino sources
are discussed in detail in, for 
example,~\cite{Gavela:2008ra,Biggio:2009nt}, and references therein. 

\subsection{Neutrino magnetic moment}
In the minimally-extended standard model, the neutrino magnetic moment (NMM) is expected to be very 
small ($\mu_{\nu}\sim 10^{-19}-10^{-20}\ \mu_B$)~\cite{Fujikawa:1980yx,Shrock:1982sc}. This makes the 
NMM an attractive place to look for new physics.  
The current best terrestrial limit of 
$\mu_{\nu}<2.9\times10^{-11}\ \mu_B$ at 90\% CL comes from the GEMMA experiment
~\cite{Beda:2013mta}.  
Many models for new physics 
allow for a NMM just below the current limit.  The NMM can be related to the Dirac 
neutrino mass scale by naturalness arguments such that the mass scale is proportional to the 
product of $\mu_{\nu}$ and the energy scale of new physics, which implies that 
$|\mu_{\nu}|\leq 10^{-14}\ \mu_B$ for Dirac neutrinos~\cite{Bell:2005kz}.  NMM for Majorana 
neutrinos (which can have transition magnetic moments) suffer from no such constraint.  Therefore a discovery of NMM of as much as a few 
orders of magnitude below the current limit would suggest that neutrinos are Majorana particles.  
Laboratory searches for NMM are based on neutrino-electron elastic scattering~\cite{Vogel:1989iv}.
Future reactor and radioactive source experiments for sterile searches, such as those discussed in Secs.~\ref{radsources} and ~\ref{reactorsterile}, can in many cases push the NMM bounds further.
Astrophysical processes also provide very stringent bounds to neutrino electromagnetic properties~\cite{Raffelt:1999gv}.  
Majorana neutrino transition magnetic moments in the oscillation of supernova neutrinos reveal that moments as small as $10^{-24}\mu_B$ may leave a potentially observable imprint on the energy spectra of neutrinos and antineutrinos from supernovae~\cite{deGouvea:2012hg,deGouvea:2013zp}.

%% file: nu6.tex
\section{Neutrinos in cosmology and astrophysics}\label{nu6}

Neutrinos come from astrophysical sources as close as the Earth and Sun, to as far away as distant galaxies, and even as remnants from the Big Bang. They range in kinetic energy from less than one meV to greater than one PeV, and can be used to study properties of the astrophysical sources they come from, the nature of neutrinos themselves, and cosmology.

\subsection{Ultra-low-energy neutrinos}\label{bbrelic}

The Concordance Cosmological Model predicts the existence of a relic neutrino background, currently somewhat colder than the cosmic microwave background (CMB), $T_{\nu}=1.95$~K. While relic neutrinos have never been directly observed, their presence is corroborated by several cosmological observables that are sensitive to the amount of radiation in the universe at different epochs. For example, precision measurements of the CMB, and measurements of the relic abundances of light elements, independently require relativistic degrees of freedom other than photons that are compatible with the three known neutrino species of the standard model of particle physics~\cite{Abazajian:2013oma,Izotov:2010ca,Hinshaw:2012aka}.  Interestingly, a number of recent measurements --- although well consistent with the standard model --- seem to slightly favor a larger amount of radiation, compatible with four light neutrinos.  This suggests a connection with the fact that a number of anomalies at neutrino experiments also favor the existence of a fourth ``sterile" light neutrino (see Sec.~\ref{sec:anomalies}). While any conclusion is premature, the question of a possible excess of cosmic radiation will be clarified by future, more precise, measurements of this quantity.  

The cosmological relic neutrinos constitute a component of the dark matter, and their properties determine the way they contribute, with the rest of the dark matter, to the formation of large-scale structures such as galactic halos. In particular, their mass has a strong impact on structure formation. This is because, being so light, neutrinos are relativistic at the time of decoupling and their presence dampens the formation of structure at small distance scales. The heavier the neutrinos, the more they influence structure formation, and the less structure is expected at small scales. Data are  consistent with 100\% cold dark matter and therefore  give an upper bound on the total mass of the three neutrino species: $\sum m_i < 0.7$~eV, approximately (see e.g.,~\cite{Hinshaw:2012aka}).   This bound should be combined with the lower limit from oscillation experiments: $\sum m_i>0.05$~eV (Sec. \ref{sec:3nus}), which sets the level of precision that next-generation cosmological probes must have  to observe effects of the relic neutrino masses. At this time, prospects are encouraging for answering this question.

The ``holy grail'' of neutrino astrophysics/cosmology is the direct detection of the relic neutrino background. This is extremely cold ($1.95~{\rm K}=1.7\times 10^{-5}$~eV) and today, at least two of the neutrino species are nonrelativistic. Several ideas have been pursued, and a clear path towards successfully measuring relic neutrinos has yet to emerge.  Recently, the idea, first discussed in~\cite{Weinberg:1962zz}, of detecting relic neutrinos through threshold-less inverse $\beta$ decay --- e.g., $\nu_e+{}^3{\rm H}\rightarrow {}^3{\rm He}+e^-$ --- has received some attention (e.g.,~\cite{Cocco:2007za}). 
Specific experimental setups have been proposed recently (e.g., PTOLEMY~\cite{Betts:2013uya}; also see Sec.~\ref{nu2:expt}).

\subsection{Low-energy neutrinos}

Sources of low energy (MeV to tens-of-MeV range) astrophysical neutrinos include the Earth, the Sun, and core-collapse supernovae.  Since neutrinos only interact weakly, they are unique messengers from these sources allowing us to probe deep into the astrophysical body. The following three distinct detector types proposed in the near future would be broadly sensitive to low-energy neutrino physics: liquid-scintillator detectors, water-Cherenkov detectors, and liquid argon time projection chambers. Each detector type has particular advantages.  Especially in the case of supernova neutrinos, a combination of all types would allow for a better exploration of all the potential science.

\subsubsection{Physics and astrophysics with low-energy neutrinos}

\noindent
\textbf{Solar neutrinos:} Despite the tremendous success of previous solar-neutrino experiments there are still many unanswered questions, e.g., such as: What is the total luminosity in neutrinos~\cite{Bahcall:2003ce}?  What is the metallicity of the Sun's core~\cite{Serenelli:2011py}? The answers to these questions could change our understanding of the formation of the Solar System and the evolution of the Sun. Precise measurements of solar $pep$ or $pp$ neutrinos are required to answer the first question, and precise measurements of CNO neutrinos could answer the second question. Solar neutrinos are also ideal probes for studying neutrino oscillation properties. The importance of previous solar neutrino experiments for understanding neutrino properties has been described in Sec.~\ref{sec:3nus}. New experiments, particularly at the energy of the $pep$ neutrinos, would be very sensitive to nonstandard physics. An observation of a day-versus-night difference in the solar neutrino rate would conclusively demonstrate the so-called MSW effect~\cite{Wolfenstein:1977ue,Mikheev:1986gs}.

\noindent
\textbf{Geo-neutrinos:}
Closer to home, the Earth is also a potent source of low-energy neutrinos produced in the decay of U, Th, K. Precise measurements of the flux of these neutrinos would allow for the determination of the amount of heat-producing elements in the Earth (see, e.g.,~\cite{Fiorentini:2007te}), which is currently only estimated through indirect means. Knowing the amount of heat-producing elements is important for our understanding of convection within the Earth, which is ultimately responsible for earthquakes and volcanoes. The most recent measurements from KamLAND~\cite{Gando:1900zz} and Borexino~\cite{Bellini:2013nah} are reaching the precision where they can start to constrain Earth models. However, these detectors are not sensitive to the neutrino direction and are therefore sensitive to local variations. Ultimately, we are interested in knowing the amount of heat-producing elements in the Earth's mantle, and hence a detector located on the ocean floor away from neutrinos produced in continental crust would be ideal.

\noindent
\textbf{Supernova neutrinos:}\label{nu6_supernova}
Supernovae are thought to play a key role in the history of the Universe and in shaping our world. For example, modern simulations of galaxy formation cannot reproduce the structure of the galactic disk without taking the supernova feedback into account. Shock waves from ancient supernovae triggered further rounds of star formation and dispersed heavy elements, enabling the formation of stars like our Sun. Approximately 99\% of the energy released in the explosion of a core-collapse supernova is emitted in the form of neutrinos.  The mechanism for supernova explosion is still not understood. Supernova neutrinos record the information about the physical processes in the center of the explosion during the first several seconds, as the collapse happens. Extracting the neutrino luminosities, energy spectra, and cooling timescale would also allow us to study the equation of state of the nuclear/quark matter in the extreme conditions at the core of the collapse.  

Supernovae also provide an incredibly rich source for the understanding of neutrino interactions and oscillations. As neutrinos stream out of the collapse core, their number densities are so large that their flavor states become coupled due to the mutual coherent scattering. This ``self-MSW" phenomenon results in non-linear, many-body flavor evolution and has been under active exploration for the last five years, as supercomputers caught up with the physics demands of the problem (see, e.g.,~\cite{Duan:2005cp,Fogli:2007bk,Raffelt:2007cb,Raffelt:2007xt,EstebanPretel:2008ni,Duan:2009cd,Dasgupta:2009mg,Duan:2010bg,Duan:2010bf}). While the full picture is yet to be established, it is already clear that the spectra of neutrinos reaching Earth will have spectacular nonthermal features. Neutrino flavor evolution is also affected by the moving front shock and by stochastic density fluctuations behind it, which may also imprint unique signatures on the signal. All of these features will give new large detectors a chance to observe neutrino oscillations in qualitatively new regimes, inaccessible on Earth, and will very likely yield information on the neutrino mass hierarchy (see Sec.~\ref{mhother}). 
Last but not least, the future data will allow us to place significant constraints on many extensions of particle physics beyond the standard model. This includes scenarios with weakly interacting particles, such as axions, Majorons, Kaluza-Klein gravitons, and others (see, e.g.~\cite{Raffelt:1999tx,Hannestad:2001jv}). These new particles could be produced in the extreme conditions in the core of the star and could modify how it evolves and cools.

Compared to the 1987A event, when only two dozen neutrinos were observed, future detectors may register tens --- or even hundreds --- of thousands of neutrino interactions from a core-collapse supernova in or nearby the Milky Way.
The burst will consist of neutrinos of all flavors with energies 
in the few tens of MeV range~\cite{Scholberg:2012id}.  
Because of their weak interactions, the neutrinos are
able to escape on a timescale of a few tens of seconds after core collapse (the promptness 
 enabling a supernova early warning for astronomers~\cite{Antonioli:2004zb}). 
From the point of view of maximizing physics harvest from a burst observation, flavor sensitivity --- not only interaction rate but the ability to tag different interaction channels --- is critical.

While a single supernova in our Galaxy could be expected to produce a large signal in a next-generation neutrino detector, such events are relatively rare (1-3 per century). However, it could also be possible to measure the flux of neutrinos from all the supernovae in cosmic history. The flux of these ``diffuse supernova neutrino background'' (DSNB) depends on the historical rate of core collapse, average neutrino production, cosmological redshift effects and neutrino oscillation effects~\cite{Beacom:2010kk,Lunardini:2010ab}. 

\subsubsection{Low-energy neutrino detectors}

In this subsection we describe the leading large-detector technologies for detection of low-energy neutrinos.

\noindent
\textbf{Liquid scintillator detectors:}
Depending on the depth, radiogenic purity, and location, large liquid scintillator detectors could be sensitive to geo-neutrinos, $pep$, $pp$, CNO, ${\rm ^{8}{B}}$ solar neutrinos, and supernova neutrinos. The majority of the liquid scintillator experiments consist of large scintillator volumes surrounded by light detectors. The Borexino~\cite{Alimonti:2008gc} and KamLAND~\cite{Mitsui:2011zz}  experiments continue to operate. The SNO+ experiment~\cite{Kraus:2010zzb} is currently under construction at SNOLAB, in Sudbury, Canada, and the JUNO experiment~\cite{Kettell:2013eos} is currently approved in China. The Hanohano experiment~\cite{Maricic:2011zz} to be located on the ocean floor, and the LENA experiment~\cite{Wurm:2011zn} to be located in Europe have been proposed.

The Borexino Collaboration recently announced the first positive measurement of $pep$ neutrinos~\cite{Collaboration:2011nga}, along with a nontrivial upper bound on neutrinos from the CNO cycle, which are yet to be observed. Because of its greater depth, the SNO+ experiment could make a precise measurement of the $pep$ neutrinos~\cite{Kraus:2010zzb}. Unlike the other experiments, the LENS experiment~\cite{Raghavan:2008zz} currently being planned consists of a segmented detector doped with In, which would allow precise measurement of the low-energy solar neutrino energy spectrum.
Geo-neutrinos were first observed in liquid scintillator detectors~\cite{Araki:2005qa, Bellini:2010hy} and all planned scintillator experiments would be sensitive to geo-neutrinos, although the location of the JUNO experiment next to nuclear power plants would make such a measurement very difficult. The Hanohano experiment located on the ocean floor would be the ideal geo-neutrino experiment.

All of the scintillator detectors would be sensitive to supernova neutrinos, primarily $\bar{\nu}_e$ through inverse $\beta$ decay (IBD), $\bar{\nu}_e+p \rightarrow e^+ + n$, but also $\nu_x$ neutrinos through proton scattering provided their thresholds are low enough~\cite{Dasgupta:2011wg}. The Hanohano and LENA detectors would also allow a measurement of the DSNB.

\noindent
\textbf{Water Cherenkov detectors:} Depending on the depth and radiogenic purity, large water-Cherenkov detectors could be sensitive to ${\rm ^{8}{B}}$ solar neutrinos and supernova neutrinos. The Super-K~\cite{Abe:2010hy} ($\sim 50,000$ tons, still operating) and SNO~\cite{Aharmim:2011vm} experiments ($\sim 1000$ tons, completed operation) dominate the $<3$\% measurement
of ${\rm ^{8}{B}}$ neutrinos from global fits to solar neutrino data.  Super-K has recently presented a 2.7$\sigma$ day-night asymmetry~\cite{renshaw}; an improved measurement
will require increased statistics. The proposed Hyper-K detector~\cite{Abe:2011ts,Kearns:2013lea} ($\sim 990,000$ tons) would allow for a measurement of the day versus night asymmetry with a significance better than $4\sigma$.

The tremendous size of the Hyper-K detector would result in $\sim 250,000$ interactions from a core-collapse supernova at the Galactic center, and $\sim 25$ interactions from a core-collapse supernova at Andromeda. The large number of events in a Galactic supernova would allow for very sensitive study of the time evolution of the neutrino signal. Although the IceCube detector could not detect individual events from a core-collapse supernova,  the large volume of ice visible to the photomultiplier tubes would result in a detectable change in the photomultiplier hit rates, allowing for a study of the time evolution of a supernova~\cite{Abe:2011ts,Kearns:2013lea,Mori:2013isa}.
The addition of Gd to the Super-K~\cite{Mori:2013isa} or Hyper-K detectors would allow for the study of the DSNB within the range of most predictions for the total flux.

\noindent
\textbf{Liquid argon time projection chambers:}
A liquid argon time projection chamber located underground could provide invaluable information about a Galactic core-collapse supernova.  Unlike other detectors,
which are primarily sensitive to $\bar{\nu}_e$,  the principal signal would be due only to $\nu_e$ interactions, for which unique physics and astrophysics signatures are expected~\cite{Choubey:2010up,Akiri:2011dv}.   For a supernova at 10~kpc, approximately 1000 events would be expected per 10~kt of liquid argon~\cite{Bueno:2003ei}.  It will be critical to site LBNE underground in order to take
advantage of the exciting and unique physics a core-collapse supernova will bring~\cite{Adams:2013qkq}.

\begin{table}[htdp]
\caption{Summary of low-energy neutrino astrophysics detectors. ** indicates significant potential, and * indicates some potential but may depend on configuration.  Here total mass is given; fiducial mass may be smaller.}
\begin{center}
\begin{tabular}{llllllll}
\hline\hline
Detector Type&		Experiment& Location &	Size (kt)&	Status		&Solar	&Geo	&Supernova\\
\hline
Liquid scintillator&	Borexino~\cite{Alimonti:2008gc} & Italy &	0.3&			Operating		&**		&**		&*\\
Liquid scintillator&	KamLAND~\cite{Eguchi:2002dm} & Japan &	1.0&			Operating		&**		&**		&*\\
Liquid scintillator&	SNO+~\cite{Kraus:2010zzb}&	Canada &	1.0&			Construction	&**		&**		&*\\
Liquid scintillator&	RENO-50~\cite{reno50} & South Korea &		10&			Design/R\&D	&*		&*		&**\\
Liquid scintillator&	JUNO~\cite{Kettell:2013eos}  & China &		20&			Design/R\&D	&*		&*		&**\\
Liquid scintillator&	Hanohano~\cite{Maricic:2011zz} & TBD (USA) &	20&			Design/R\&D	&*		&**		&**\\
Liquid scintillator&	LENA~\cite{Wurm:2011zn} & TBD (Europe) &		50&			Design/R\&D	&*		&**		&**\\
Liquid scintillator&	LENS~\cite{Raghavan:2008zz} & USA&		0.12&		Design/R\&D	&**		&		&*\\
\hline
Water Cherenkov&		Super-K~\cite{Fukuda:2002uc} & Japan&		50&			Operating		&**		&		&**\\
Water Cherenkov&		IceCube~\cite{halzenwp} & South Pole&		$>$2000&		Operating		&		&		&**\\
Water Cherenkov&		MEMPHYS~\cite{deBellefon:2006vq} & Sweden	&	685&		Design/R\&D	&**		&		&**\\

Water Cherenkov&		Hyper-K~\cite{Kearns:2013lea} & Japan	&	990&		Design/R\&D	&**		&		&**\\
\hline
Liquid argon&		LBNE~\cite{Adams:2013qkq} & USA	&	35&			Design/R\&D	&*		&		&**\\
\hline\hline
\end{tabular}
\end{center}
\label{default}
\end{table}%

\subsection{Neutrinos of GeV to PeV energies}

One of the most tantalizing questions in astrophysics, namely the origin and the evolution of the cosmic accelerators that produce the observed spectrum of cosmic rays, which extends to astonishingly high energies, may be best addressed through neutrinos. Because neutrinos only interact via the weak force, they travel from their source undeflected by magnetic fields and unimpeded by interactions with the CMB, unlike photons and charged particles.   Due to the low fluxes expected, the construction of high-energy neutrino telescopes requires the instrumentation of large natural reservoirs, a concept demonstrated by AMANDA, Baikal and ANTARES.  With the completion of the IceCube Neutrino Telescope~\cite{Achterberg:2006md} in the South Polar icecap in 2010, the era of kilometer-scale neutrino telescopes dawned, and plans for a complementary telescope in the Mediterranean are under development.  
Already, IceCube has reported evidence for an astrophysical flux at a significance in excess of 4$\sigma$~\cite{ipa2013}, including 
cascade events in excess of 1~PeV~\cite{Aartsen:2013bka}. Further, IceCube has placed severe constraints on favored mechanisms for 
gamma-ray bursts~\cite{Abbasi:2012zw}. Physics and astrophysics from future IceCube measurements are detailed in~\cite{halzenwp}.

As with previous generations of neutrino telescopes, these instruments are expected to provide insight into the nature of the messengers themselves.  The backgrounds for astrophysical fluxes include atmospheric neutrinos, which are collected by IceCube at a rate of about 100,000 per year in the 0.1 to 100 TeV range.  Atmospheric neutrinos probe neutrino physics and interactions at previously-unexplored energies.  At TeV energies, IceCube sensitivity  to sterile neutrinos in the eV mass range potentially exceeds that of any other experiment and is only limited by systematic errors (see Sec.~\ref{sec:anomalies}).  With the addition of IceCube's low-energy infill array, DeepCore~\cite{Tyce2011}, which extended its sensitivity down to 10 GeV, conventional neutrino oscillations have been observed with more than 5$\sigma$ significance, and such instruments could provide competitive precision measurements of neutrino oscillation parameters.  The atmospheric neutrino flux may someday also provide a glimpse into our Earth via neutrino radiography~\cite{Reynoso:2004dt}.
These instruments may also shed light on one of the most puzzling questions facing particle physics and cosmology: the nature of the dark matter.  Dark matter annihilations in the Sun and the Galactic center could be indirectly detected in neutrino telescopes, covering parameter space inaccessible at the LHC, and masses inaccessible to direct-detection experiments.  Neutrino telescopes are also sensitive to other exotica, such as magnetic monopoles.

\subsection{Neutrinos at energies over 1 PeV}

At ultra high energies~\cite{Greisen, Zatsepin}, neutrinos could be detected in dense, radio frequency (RF) transparent media via the Askaryan effect~\cite{Saltzberg:2000bk}.  The abundant cold ice covering the geographic South Pole, with its exceptional RF clarity, has been host to pioneering efforts to develop this approach, including RICE~\cite{rice_flux} and ANITA~\cite{anita1}. Currently, two discovery-scale instruments are in the prototyping phase: the Askaryan Radio Array (ARA)~\cite{Allison:2011wk}, which is envisioned to instrument a 100-square-km area near the South Pole with 200-m deep antenna clusters, and ARIANNA~\cite{Kleinfelder:2013zya}, which would be installed on the surface of the Ross Ice Shelf.  Efforts are underway to characterize the ice in Greenland, to determine its suitability as a future 
neutrino telescope site.

The fact that cosmic rays have been observed at energies in excess of $10^{20}$ eV makes the search for neutrinos at these energies particularly interesting.  These energies are above the threshold for pion photoproduction on the CMB, which would seem to guarantee a flux of ultra-high-energy (UHE) neutrinos.  However, the neutrino flux expectations are sensitive to the composition of the UHE cosmic rays, making the spectrum of UHE cosmic rays a sensitive probe of the heavy ion content.   In addition, for a sufficient sample of UHE neutrinos, it would be possible to measure the neutrino cross section at high energies.

%% file: nu7.tex
\section{Neutrinos and society}\label{sec:society}

 In this section we discuss the direct and spin-off applications, and the rich opportunities for outreach and education, offered by fundamental and applied antineutrino science. 

\subsection{Applied antineutrino physics}

Direct application of neutrinos to other domains falls into two categories. In geology, they may enable study of Earth's composition on largest scales (see Sec.~\ref{nu6}), and in nonproliferation, they offer the prospect of improved monitoring or discovery of operating nuclear reactors.  Since the signal in both cases arises from antineutrinos only, it is appropriate to refer to these studies as ``Applied Antineutrino Physics''.

Concerning nonproliferation, the main likely user of antineutrino-based reactor monitoring is the  International Atomic Energy Agency (IAEA). IAEA is responsible for monitoring the international fuel cycle, to detect attempts to divert fissile materials and production technologies to nuclear weapons programs. The international monitoring regime administered by the IAEA is referred to as the Safeguards regime~\cite{IAEASG}. Antineutrino detectors may play a role in this regime, which focuses on timely detection of illicit removal of fissile material from known and declared reactors and other fuel cycle facilities. They may also be useful in future expanded regimes,  such as the proposed Fissile Material Cutoff Treaty~\cite{FMCT}, which will seek to verify the non-existence of an undeclared fissile material production capability in a country or geographical region.  In a recent report, the IAEA encouraged continued research into antineutrino-detection-based applications for safeguards and other cooperative monitoring of nuclear reactors~\cite{IAEANote}.  In addition, the U.S. National Nuclear Security Administration has included a demonstration of remote reactor monitoring (1 km and  beyond) as an element of its 2011 Strategic Plan~\cite{NNSAStratPlan}. 

Nonproliferation applications are enabled by three features of reactor antineutrinos. First, reactors emit a copious flux of $\sim$ \hspace{-0.05 em}0\textendash10 MeV electron antineutrinos resulting from beta decay of neutron-rich fission fragments. Second, the antineutrino inverse-$\beta$-decay (IBD) cross section is high enough to allow detectors of tractable (cubic meter) sizes to be deployed at tens-of-meter standoff from a reactor.
Third, the detected antineutrino flux and energy spectrum both correlate with the core-wide content of fission fragments, and therefore bring information on the inventories of the main fissile isotopes used in weapons.

Concerning applications for existing or future reactor safeguards, cubic-meter-scale antineutrino detectors now make it possible to monitor the operational status, power levels, and fissile content of nuclear power reactors in near-real-time with stand-off distances of roughly 25-100 meters from the reactor core. This capability has  been demonstrated at civil power reactors in Russia and the United States, using antineutrino detectors designed specifically for reactor monitoring and safeguards~\cite{Klimov:1994ae,Bernstein:2008ap}. This  near-field monitoring capability may be of use within the International Atomic Energy Agency's (IAEA) Safeguards Regime, and other cooperative monitoring regimes.   

With respect to future missions related to remote discovery or exclusion of reactors, current kt-scale antineutrino detectors, exemplified by the KamLAND and Borexino liquid-scintillator detectors,  can allow monitoring, discovery or exclusion of small (few megawatt thermal, MWt) reactors at standoff distances up to 10 kilometers. In principle, reactor discovery and exclusion is also possible at longer ranges. 
More information on this topic may be found at~\cite{AAP}.

\subsubsection{ IBD detectors for near-field safeguards, and for short-baseline experiments}

As discussed in section \ref{sec:anomalies}, and in numerous Snowmass white papers~\cite{SNOWMASSWP}, short-baseline neutrino oscillation experiments are being planned by US and overseas groups. These experiments seek to deploy 1\textendash10-t scale antineutrino detectors from 5\textendash15 meters from a nuclear reactor core. The purpose of the experiments is to search for a possible sterile neutrino signal, and to measure the reactor antineutrino energy spectrum as precisely as possible.  The physics goals greatly constrain the experimental configuration. The need for close proximity to the reactor requires that the detector overburden  is necessarily minimal, at most $\sim$45 meters water equivalent (mwe). The physical dimension of the core must be as small as possible, to avoid smearing the oscillation-related spectral distortions with multiple baselines arising from different locations in the core. To be competitive with  experiments using strong single-element radioactive sources, this requires that a relatively low power ($\sim$20-50 MWt) research reactor be used for the experiment, greatly constraining the number of possible sites. 
The above requirements impose stringent constraints on detector design as well.
The technology goals for reactor short-baseline experiments and for nonproliferation applications are  similar in many respects. In both cases, R\&D is required to improve background rejection at shallow depths, while maintaining high efficiency and good energy resolution. 
To improve specificity for the two-step IBD signature, segmented designs~\cite{DANSS} are being contemplated for both cooperative monitoring and short-baseline detectors, as well as the use of Li-doped plastic or liquid-scintillator technologies~\cite{Hans3:2013}.  
A key difference between the fundamental and applied technology needs is that the detectors for nonproliferation must also be simple to operate, and may have additional cost constraints compared to the single-use detectors needed for the short-baseline physics experiments. 

\subsubsection{CENNS detection for nonproliferation and fundamental science}\label{cenns_app}

Numerous physics motivations for the measurement of coherent elastic neutrino-nucleus scattering (CENNS) are described in Sec.~\ref{cenns} (and see~\cite{CNSworkshop,Akimov:2013fma,CNSreactor}).
For monitoring applications, the process holds considerable interest, since the 100-1000 fold increase in cross section  compared with the next-most-competitive antineutrino interaction may enable a ten-fold or more reduction in detector volume, even with shielding accounted for. This could simplify and expand the prospects for deployment of these detectors in a range of cooperative monitoring contexts.   Furthermore, CENNS has important connections to the searches for WIMPs, due to similarity in the nuclear-recoil event signature.  Advances in CENNS technology will potentially improve the prospects for WIMP detection, and CENNS backgrounds from natural neutrino sources will eventually limit dark matter searches.

For CENNS detection, both phonon and ionization channel approaches are being pursued. Detector thresholds must be made sufficiently low, while maintaining effective background suppression, so as to allow good collection statistics above background in tractably-sized detectors. 
In the last few years, several groups worldwide have made significant progress in reducing thresholds in noble-liquid~\cite{Sangiorgio:2013ib,Santos:2011ju}, and germanium detectors~\cite{Akerib:2010pv}, with the intent of improving both CENNS and dark matter detectors. 

\subsubsection{Large IBD detectors and remote reactor monitoring}

One-hundred-kt to Mt-scale  liquid scintillator and water detectors have been proposed  as far detectors for long-baseline accelerator-based neutrino oscillation and  CP-violation experiments~\cite{Abe:2011ts,Kearns:2013lea,Autiero:2007zj}. If they can be made sensitive to few-MeV antineutrinos, such giant detectors offer an even more diverse physics program, including sensitivity to extra-galactic supernovae,  measurement of the diffuse supernova background (see Sec.~\ref{nu6}),  proton decay, and in the case of liquid scintillator detectors, sensitivity to reactor neutrino oscillations at several tens-of-kilometer baseline. 
The same types of detector could enable discovery, exclusion, or monitoring of nuclear reactors at standoff distances from one to as many as several hundred kilometers. With sufficient suppression of backgrounds, remote detectors (25-500 km standoff) on the 50-kt to one-Mt scale would provide a 25\% statistically accurate measurement of the power of a 10-MWt reactor in several months to a year~\cite{Bernstein:2009ab}.
 Water Cherenkov detectors are one promising approach to achieving detector masses on the scale required to meet  the above physics and nonproliferation goals.
While the water Cherenkov approach is currently disfavored in the United States' LBNE planning process, it nonetheless retains considerable interest for the global community, in particular in Japan~\cite{Abe:2011ts,Kearns:2013lea}. 
To allow sensitivity to low-energy antineutrinos through the IBD process, the water would be doped with gadolinium, so that final-state neutrons can be detected by the $\sim$4 MeV of measurable Cherenkov energy deposited in the gamma-ray cascade that follows capture of neutrons on gadolinium~\cite{Dazeley:2008xk,Watanabe:2008ru}.
A kt-scale demonstration of this detector type is now being proposed by the WATCHMAN collaboration in the United States~\cite{WATCHMAN}.
Scaling of pure liquid-scintillator designs such as KamLAND or Borexino is another approach to megaton-class detectors.  This approach is exemplified by the LENA collaboration in Europe~\cite{Wurm:2011zn,Autiero:2007zj}.

\subsubsection{Applications of neutrino-related technologies}

A high degree of synergy is evident in technology developments related to neutrino physics experiments.  
Close collaboration between laboratory, university and industry has  been fruitful, solving immediate needs of the neutrino community, and providing spinoff applications in quite different fields with broad societal impact. Examples are provided here.

\noindent
\textbf{Detectors:} Neutrino/antineutrino detection has motivated significant work on detection technology, the benefits of which extend well beyond the physics community. Examples include plastic and liquid scintillator doped with neutron-capture agents,  high-flashpoint scintillators with reduced toxic hazards compared to previous generators of scintillator, and low-cost flat-panel photomultiplier tubes. 
Doped organic plastic and liquid scintillator detectors are now being pursued in the United States~\cite{ Zaitseva201288},  as a means to improve sensitivity to the reactor antineutrino signal. 
In a similar way, companies such as Bicron Technologies and Eljen Technologies have devoted resources to reducing the biohazards and improving the optical clarity of their scintillation cocktails, in order to facilitate neutrino detection. These improvements clearly benefit other customers, such as the medical and pharmaceutical communities, which use scintillator detectors for radio-assay in nuclear medicine applications.  The overall product lines of these companies have benefited considerably from  research that has focused on making better neutrino detectors. 
Another area of research with  important spinoff potential is the development of low cost, high efficiency photomultiplier tubes.  Cutting-edge research that focused on low-cost PMTs is exemplified by the Large Area Pico-second Photo-Detectors project~\cite{LAPPD,Djurcic:2013}. Beyond enabling lower-cost neutrino detectors at every scale, such detectors would lower costs and improve performance of medical imaging devices such as Positron Emission Tomography systems, for which the photo-detector element is often a dominant cost and critical component.  Emerging nuclear security applications that demand PMT-based imaging, such as three-dimensional reconstruction of the locations and inventories of fissile material in a reprocessing or enrichment plant,  also greatly benefit from lower-cost PMTs. 

\noindent
\textbf{Accelerators:}\label{sec:accelerator} 
The National Academy of Engineering (NAE) declared one of the Grand Challenges for the new century to be, ``the Engineering of Tools for Scientific Discovery"~\cite{NAE} 
then muses, ``Perhaps engineers will be able to devise smaller, cheaper, but more powerful atom smashers, enabling physicists to explore realms beyond the 
reach of current technology"~\cite{NAE}. The current generation of high-power accelerators have rapidly advanced the boundary of ``current technology" and 
are accomplishing many breakthroughs in these new realms.  In the particular case of the rapidly-evolving field of neutrino studies, sources produced from 
the FNAL Main Injector, CERN's SPS, and J-PARC are enabling very rapid progress.  Future experiments with the SNS, and new capabilities at ESS-Lund, Project X, 
and the high-power DAE$\delta$ALUS cyclotrons will go a long way towards realizing the visions of the NAE.  
But, as important 
are the understandings in fundamental science, even greater are the societal impacts of the technologies 
being developed for these new accelerators.  
The indirect spinoffs are numerous:  advances in engineering with superconducting materials and magnets, high-volume cryogenics, 
sophisticated control systems and power converters,  and many more.  
A very direct connection with neutrinos is provided by the DAE$\delta$ALUS project, which is  
based on a cascade of compact cyclotrons capable of sending multi-megawatt beams onto neutrino-producing targets for CP-violation studies and searches for sterile neutrinos. 
Development of this technology, based on accelerating H$_2^+$ ions, pushes the performance of cyclotrons to new levels, and 
is being pursued by a broad collaboration of U.S. and foreign laboratories, universities, and industry~\cite{Conrad:2009mh, Alonso:2010fs}.  
As new, cost-effective sources of high-power beams, these cyclotrons will have a significant impact on ADS (Accelerator-Driven Systems) technology for critical nuclear energy-related applications such as driving thorium reactors and burning nuclear waste~\cite{ADS}.  On a nearer timescale, industry is quite interested in the application of this technology for isotope production.  One of the test prototypes being developed with the assistance of Best Cyclotron Systems Inc. is a 28-MeV cyclotron designed for H$_2^+$ injection studies.  This cyclotron is also suitable for acceleration of He$^{++}$, and is directly applicable to the production of $^{211}$At, a powerful therapeutic agent whose ``...use for [targeted $\alpha$ particle therapy] is constrained by its limited availability"~\cite{At211}.

\subsection{Education and outreach}

\noindent
\textbf{Educating physicists about nonproliferation:}
In order to reach out to the public effectively, physicists themselves should be made aware of the potential utility of neutrinos for nuclear security. 
As revealed by the growing field of applied antineutrino physics, awareness of these connections has grown over the last ten years in the physics community. However, relatively few physicists ---  including many actively engaged in applied research ---  have much, if any, formal education in the structure of the global nonproliferation regime, or in the history of the atomic era that led to the current state of affairs in nuclear security. This is especially unfortunate, since at least in the U.S., as this history is closely intertwined with the development of the large-scale accelerator and underground experiments that employ many of these same physicists.   In the last five years or so, a few physics departments have worked to develop courses that introduce physicists to both the relevant technology and policy of nonproliferation and nuclear security. Nuclear engineering departments have a closer connection to the nonproliferation regime, and 
 have developed explicit course elements targeting the connection between nuclear security and nuclear science.  These developments and connections should be nurtured.

\noindent
\textbf{Educating the general public about neutrino science:}
An aware and enthusiastic general public is the best way to ensure support and funding for basic research.  
Each one of us should accept our responsibility for conveying the message whenever possible that investments in our field are of benefit to the nation.  
Neutrino physics offers a wealth of fascinating and counter-intuitive concepts (e.g., oscillations, high fraction of the Sun's energy emitted as neutrinos, and extremely low cross sections enabling neutrinos to easily penetrate the Earth).
In addition, our field sports some highly photogenic experiments (e.g., IceCube, Borexino, Super-K).  A suggestion could be made that a reservoir of material be collected, updated and made available for persons to use in outreach talks and activities:  lecture outlines, lists of talking points, graphics, etc. 
The interesting practical applications of neutrinos described earlier 
provide highly relevant and compelling topics to be communicated to the public.

The importance of Education and Outreach is recognized in the establishment of a whole (Snowmass) ``Frontier" dedicated to this topic.  Our community should embrace this effort, looking for ways of coordinating and contributing to their activities for furtherance of our mutually-compatible goals.

%% file: glossary.tex
\section*{Glossary}
\label{glossary}

Below is a glossary of acronyms and experiment names:

\begin{itemize}
\item $0\nu\beta\beta$ -- neutrinoless double beta decay
 \item ADS: Accelerator-Driven Systems -- technology for driving nuclear reactors with beams
  \item AMANDA: Antarctic Muon and Neutrino Detector Array -- first-generation neutrino telescope experiment in Antarctica
  \item ANITA: Antarctic Transient Antenna -- neutrino radio antenna balloon experiment in Antarctica
  \item ANTARES: Astronomy with a Neutrino Telescope and Abyss environmental RESearch -- neutrino telescope experiment in the Mediterranean sea
 \item ARA: Askaryan Radio Array -- radiofrequency neutrino antenna experiment at the South Pole
\item ArgoNeuT: Argon Neutrino Test -- mini LAr TPC exposed in NuMI beam at Fermilab
  \item ARIANNA: Antarctic Ross Ice-shelf ANtenna Neutrino Array -- radiofrequency neutrino antenna experiment in Antarctica
\item ATR: Advanced Test Reactor -- research reactor at the Idaho National Laboratory
\item Baikal: neutrino telescope in Lake Baikal in Siberia
\item Baksan: underground laboratory in the Caucasus mountains in Russia
 \item BEST: Baksan Experiment on Sterile Transitions --proposed radioactive source experiment in Russia with a gallium detector
  \item BNB: Booster Neutrino Beam  -- neutrino beamline at Fermilab using the Booster
  \item Borexino: scintillator solar neutrino experiment in Gran Sasso National Laboratory
  \item CAPTAIN: Cryogenic Apparatus for Precision Tests of Argon INteractions -- LAr R\&D detector 
% CAPTAIN one could go to SNS
  \item CC: Charged Current
  \item Ce-LAND: $^{144}$Ce source to be placed in KamLAND to study the reactor neutrino anomaly
  \item CENNS: Coherent Elastic Neutrino-Nucleus Scattering -- refers to the NC process as well as a proposed experiment to be sited at the BNB
  \item CHIPS: CHerenkov detectors In mine PitS -- proposed experiment to use the Fermilab beams and a massive Cherenkov detectors in flooded mine pits
  \item CHOOZ: first-generation reactor neutrino experiment in France
  \item CKM: Cabibbo-Kobayashi-Maskawa matrix
  \item CL: Confidence Level
  \item CMB: Cosmic Microwave Background
  \item CP: Charge Parity
  \item CSI: Coherent Scattering Investigations at the SNS -- proposed CENNS search experiment for the SNS
  \item CUORE: Cryogenic Underground Observatory for Rare Events -- neutrinoless double beta decay search experiment at Gran Sasso National Laboratory
  \item DAE$\delta$ALUS: Decay At rest Experiment for $\delta_{CP}$ studies At the Laboratory for Underground Science -- proposed cyclotron-based neutrino oscillation experiment 
  \item DANSS: Detector of the reactor AntiNeutrino based on Solid-state plastic Scintillator -- reactor neutrino experiment in Russia
  \item DAR: Decay At Rest
  \item Daya Bay: reactor neutrino experiment in Daya Bay, China
  \item DIF: Decay In Flight
  \item DIS: Deep Inelastic Scattering
  \item Deep Core: PMT infill for low-energy extension to the IceCube experiment
  \item Double Chooz: reactor neutrino experiment in Chooz, France
  \item DSNB: Diffuse Supernova Neutrino Background
  \item ECHo: Electron Capture $^{163}$Ho experiment -- proposed neutrino mass microcalorimeter experiment
  \item ESS: European Spallation Source -- future facility in Lund, Sweden
  \item ESS$\nu$SB: European Spallation Source Neutrino Super Beam -- proposal to use the European Spallation Source (ESS) proton linac to generate a neutrino superbeam
  \item EXO: Enriched Xenon Observatory -- neutrinoless double beta decay experiment at WIPP (Waste Isolation Pilot Plant) in Carlsbad, New Mexico
  \item EVA: ExaVolt Antenna -- proposed balloon-based neutrino antenna experiment in Antarctica
  \item FNAL: Fermi National Accelerator Laboratory
  \item GALLEX: Gallium-based radiochemical neutrino detector -- radiochemical solar neutrino experiment at Gran Sasso National Laboratory
  \item GEMMA:  Germanium Experiment for measurement of the Magnetic Moment of Antineutrino -- neutrino magnetic moment experiment at the Kalinin nuclear power plant in Russia
  \item GERDA: Ge experiment searching for neutrinoless double beta decay 
 \item GLACIER: Giant Liquid Argon Charge Imaging Experiment-- proposed large liquid argon detector in Europe
  \item GNO: Gallium Neutrino Observatory -- radiochemical solar neutrino experiment at Gran Sasso National Laboratory (successor to GALLEX)
  \item HALO: Helium and Lead Observatory -- lead-based supernova neutrino detector at SNOLAB 
\item HFIR: High Flux Isotope Reactor -- reactor facility at Oak Ridge National Laboratory
  \item Hyper-K: Hyper-Kamiokande -- proposed large water Cherenkov detector in Japan
  \item IAEA: International Atomic Energy Agency
   \item IBD: Inverse Beta Decay (usually refers to $\bar{\nu}_e+p \rightarrow e^+ + n$)
  \item ICARUS: Imaging Cosmic And Rare Underground Signals -- LAr TPC-based neutrino oscillation experiment at Gran Sasso National Laboratory
  \item IceCube: neutrino telescope located at the Amundsen-Scott South Pole station in  Antarctica
  \item ICAL: iron calorimeter atmospheric neutrino experiment at INO
  \item IDS: International Design Study (for the Neutrino Factory)
 \item IH: Inverted Hierarchy
 \item INO: India-based Neutrino Observatory -- future underground laboratory in India
  \item ISIS: research center at Rutherford Appleton Laboratory near Oxford
  \item IsoDAR: Isotope Decay At Rest experiment -- proposed cyclotron-based sterile neutrino experiment 
  \item J-PARC: Japan Proton Accelerator Research Complex in Tokai, Japan
  \item JUNO: Jiangmen Underground Neutrino Observatory -- proposed reactor-based large scintillator experiment in China
  \item K2K: KEK to Kamioka -- first-generation long-baseline oscillation experiment using beam from KEK to Super-K
  \item KamLAND: Kamioka Liquid scintillator ANtineutrino Detector -- reactor neutrino experiment in Japan
  \item KamLAND-Zen: Zero neutrino double beta decay search -- neutrinolesss double beta decay experiment in Japan (Xe-doped balloon deployed in KamLAND).
  \item KATRIN: KArlsruhe TRitium Neutrino experiment -- neutrino mass spectrometer in Germany
  \item KEK: accelerator laboratory in Tsukuba, Japan
  \item KM3NET: multi-km$^3$ Neutrino Telescope -- future deep-sea neutrino telescope in the Mediterranean sea
 \item kt: kilotonne (metric unit; $10^3$~kilograms)
  \item LAGUNA: Large Apparatus studying Grand Unification and Neutrino Astrophysics -- collaborative project to assess the possibilities for a deep underground neutrino observatory in Europe; includes GLACIER (liquid argon), MEMPHYS (water Cherenkov) and LENA (liquid scintillator) concepts
  \item LANSCE: Los Alamos Neutron Science Center 
 \item LAr: liquid argon
  \item LAr1: proposal to add additional liquid argon TPCs to the Fermilab Booster neutrino beamline
  \item LAr1-ND: proposal to add a liquid argon TPC near detector in the Fermilab Booster neutrino beamline
\item LArIAT: Liquid Argon In A Testbeam -- liquid argon TPC test beam experiment at Fermilab
  \item LAr TPC: Liquid Argon Time Projection Chamber
  \item LBNE: Long-Baseline Neutrino Experiment -- proposed accelerator-based neutrino oscillation experiment in the U.S.
  \item LBNO: Long-Baseline Neutrino Oscillation experiment -- proposed accelerator-based neutrino oscillation experiment in Europe
  \item LENA: Low Energy Neutrino Astronomy -- proposed next-generation liquid scintillator detector  
  \item LENS: Low Energy Neutrino Spectroscopy -- low-energy indium-based solar neutrino experiment
  \item LSND: Liquid Scintillator Neutrino Detector -- sterile neutrino experiment at Los Alamos National Laboratory
  \item LUX: Large Underground Xenon -- xenon TPC dark matter detector
 \item LZ: LUX-Zeplin -- next-generation xenon TPC dark matter detector
  \item LVD: Large Volume Detector -- neutrino observatory in Gran Sasso National Laboratory studying low-energy neutrinos from gravitational stellar collapse
  \item MAJORANA: Ge experiment searching for neutrinoless double beta decay 
\item MEMPHYS: MEgaton Mass PHYSics –- proposed  large water Cherenkov  detector for CERN SPL or for ESS (Sweden).
  \item MicroBooNE: liquid argon TPC experiment in the Booster neutrino beamline at Fermilab
  \item MINER$\nu$A: Main Injector Experiment for $\nu$-A -- neutrino scattering experiment in the NuMI beamline at Fermilab
  \item MiniBooNE: short-baseline neutrino oscillation experiment using a mineral oil-based Cherenkov detector in the Booster neutrino beamline at Fermilab
\item MIND: Magnetised Iron Neutrino Detector -- proposed neutrino factory detector
  \item MINOS: Main Injector Neutrino Oscillation Search  -- neutrino oscillation experiment in the NuMI beamline at Fermilab
 \item MNS: Maki-Nakagawa-Sakata -- neutrino mixing matrix (see PMNS)
\item MSW: Mikheyev-Smirnov-Wolfenstein effect -- matter effect modifying oscillation probability as neutrinos pass through matter
\item mwe: meters water-equivalent -- a measure of overburden
  \item NC: Neutral Current
  \item NESSiE: Neutrino Experiment with SpectrometerS in Europe -- proposed experiment to search for sterile neutrinos using the CERN SPS beam and the ICARUS detector
  \item NEXT: Neutrino Experiment with Xenon TPC -- neutrinoless double beta decay experiment at the Canfranc Underground Laboratory
  \item NF: common abbreviation for the Neutrino Factory 
 \item NH: Normal Hierarchy
\item NIST: National Institute of Standards and Technology
  \item NMM: Neutrino Magnetic Moment
  \item NOMAD: Neutrino Oscillation MAgnetic Detector -- neutrino oscillation experiment at CERN
  \item NOvA: NuMI Off-Axis electron-neutrino Appearance experiment -- neutrino oscillation experiment in the NuMI beamline at Fermilab
  \item NSI: Non-Standard Interactions (of neutrinos)
  \item NuMAX: Neutrinos from Muon Accelerators at Project X -- proposed neutrino oscillation experiment using a muon-storage ring as a source of neutrinos
  \item NuMI: Neutrinos at the Main Injector -- neutrino beamline at Fermilab using the Main Injector, extending to Soudan and Ash River
  \item nuSTORM: neutrino from STORed Muons -- proposed short-baseline neutrino experiment to study sterile neutrinos using a muon storage ring as a source of neutrinos
  \item OPERA: Oscillation Project with Emulsion-tRacking Apparatus -- emulsion- and tracker-based neutrino oscillation experiment at Gran Sasso National Laboratory
  \item ORCA: Oscillation Research with Cosmics in the Abyss -- proposed experiment to measure the neutrino mass hierarchy using the KM3NeT neutrino telescope
\item ORNL: Oak Ridge National Laboratory in Tennessee
  \item OscSNS: oscillations at the Spallation Neutrino Source -- proposed sterile neutrino search using the SNS facility
  \item PINGU: Precision Icecube Next Generation Upgrade -- proposed low-energy extension to IceCube
  \item PMNS: Pontecorvo-Maki-Nakagawa-Sakata matrix 
 \item PMT: photomultiplier tube
 \item PREM: Preliminary Reference Earth Model -- model for Earth density distribution
  \item Project 8: proposed tritium-based neutrino mass experiment
  \item Project X: proposed proton accelerator complex at Fermilab
\item PROSPECT: Precision Reactor Neutrino Oscillation and Spectrum Experiment -- U.S.-based reactor short-baseline oscillation search experiment
  \item PTOLEMY: Princeton Tritium Observatory for Light Early-universe Massive neutrino Yield -- proposed relic Big Bang neutrino background experiment
  \item QCD: Quantum chromodynamics
  \item QE: Quasi-Elastic 
  \item RADAR: R\&D Argon Detector at Ash River -- proposal to add a LAr TPC to the NOvA far detector building in Ash River, Minnesota
  \item RENO: Reactor Experiment for Neutrino Oscillations -- reactor neutrino experiment in South Korea 
  \item RENO-50: proposed reactor-based experiment to measure the neutrino mass hierarchy with a large scintillator detector
  \item RICE: Radio Ice Cherenkov Experiment -- neutrino telescope experiment in Antarctica
  \item RICOCHET: proposed bolometric sterile neutrino search using CENNS
\item ROI: Region of Interest
  \item SAGE: Soviet American Gallium Experiment -- solar neutrino experiment in the Baksan Neutrino Observatory in Russia
  \item SciNOvA: proposed neutrino scattering experiment adding a fine-grained scintillator detector at the NOvA near site
  \item SOX: chromium and/or cesium source at Borexino to study the reactor neutrino anomaly
  \item SNO: Sudbury Neutrino Observatory -- heavy water solar neutrino experiment at SNOLAB in Canada
  \item SNO+: scintillator experiment at SNOLAB in Canada (using SNO acrylic vessel)
  \item SNOLAB: underground science laboratory in the Vale Creighton Mine located near Sudbury Ontario Canada
  \item SNS: Spallation Neutron Source -- facility at Oak Ridge National Laboratory
 \item Soudan: underground laboratory in northern Minnesota, housing MINOS and low-background experiments
  \item STEREO:  Search for Sterile Neutrinos at ILL reactor -- reactor short-baseline oscillation search in France
  \item SPS: Super Proton Synchotron at CERN
  \item Super-K: Super-Kamiokande experiment -- water Cherenkov detector in the Kamiokande mine in Japan studying proton decay as well as solar, atmospheric, and accelerator-based (T2K) neutrinos 
  \item Super-NEMO: super Neutrino Ettore Majorana Observatory -- neutrinoless double beta decay experiment in Europe
  \item SURF: Sanford Underground Research Laboratory  -- underground research laboratory in Lead, South Dakota
  \item T2K: Tokai to Kamiokande experiment -- neutrino oscillation experiment using the JPARC beam in Japan
  \item TPC: Time Projection Chamber
  \item UHE: Ultra High Energy
  \item WATCHMAN: WATer CHerenkov Monitoring of Anti-Neutrinos -- collaboration of U.S.-based universities and laboratories conducting a site search for a kton-scale advanced water detector demonstration
\item WIMP: Weakly Interacting Massive Particle -- dark matter candidate particle
\end{itemize}